\documentclass[10pt]{article}

\usepackage{amsmath}

\usepackage{array}

\usepackage{appendix}

\usepackage{tocloft}                   

\usepackage{graphicx}

\usepackage{amsfonts}

\usepackage{amssymb}

\usepackage{mathrsfs}

\usepackage{yfonts}

\usepackage{euscript}

\usepackage{centernot}                 

\usepackage{ifsym}                     

\usepackage{upgreek}

\usepackage{mathtools}

\usepackage{color}

\usepackage{slantsc}
\usepackage{calligra}

\usepackage{bbold}          

\usepackage[T1]{fontenc}

\usepackage{epsf}

\usepackage{latexsym}

\usepackage{tipa}

\usepackage{makeidx}

\makeindex

\def\b{\begin{equation}}

\def\e{\begin{equation}}

\def\be{\begin{equation}}              

\def\ee{\end{equation}}

\def\beq{\begin{equation}}

\def\eeq{\end{equation}}

\def\bea{\begin{eqnarray}}

\def\eea{\end{eqnarray}}

\def\m{\mbox{ }}

\def\mma {\m , \m \m }

\def\!{\hspace{-1.6667em}}

\def\Proof{{\n{\u{Proof}}} \m}

\def\Rightarrows{\m \m \Leftrightarrow \m \m}

\def\Leftrightarrows{\m \m \Leftrightarrow \m \m}

\def\c{\cite}

\def\l{\label}

\def\r{\ref}

\def\n{\noindent}

\def\f{\footnote}

\def\u{\underline}

\def\w{\widetilde}

\def\s{\stackrel}

\def\mB{\mbox{B}}  

\def\mC{\mbox{C}}                        

\def\mE{\mbox{E}}                        

\def\mF{\mbox{F}}

\def\mG{\mbox{G}}

\def\mH{\mbox{H}} 

\def\mI{\mbox{I}}                        

\def\mM{\mbox{M}}                        

\def\mP{\mbox{P}}

\def\mR{\mbox{R}}                        

\def\mS{\mbox{S}}                        

\def\mT{\mbox{T}} 

\def\mU{\mbox{U}}                        

\def\mW{\mbox{W}}

\def\mo{\mbox{o}}

\def\mp{\mbox{p}}

\def\ms{\mbox{s}}

\def\uR{\u{R}}

\def\uq{\u{\mbox{q}}}

\def\urho{{\u{\rho}}}

\def\fP{\mbox{\sffamily P}}

\def\fT{\mbox{\sffamily T}}

\def\fX{\mbox{\sffamily X}}

\def\sa{\mbox{\scriptsize a}}

\def\si{\mbox{\scriptsize i}}

\def\sll{\mbox{\scriptsize l}}  

\def\sm{\mbox{\scriptsize m}}

\def\sn{\mbox{\scriptsize n}}

\def\sss{\mbox{\scriptsize s}}  

\def\sx{\mbox{\scriptsize x}}

\def\sA{\mbox{\scriptsize A}}

\def\sC{\mbox{\scriptsize C}}

\def\sD{\mbox{\scriptsize D}}

\def\sF{\mbox{\scriptsize F}}

\def\sG{\mbox{\scriptsize G}}

\def\sJ{\mbox{\scriptsize J}}

\def\sM{\mbox{\scriptsize M}}

\def\sQ{\mbox{\scriptsize Q}} 

\def\sR{\mbox{\scriptsize R}}

\def\sS{\mbox{\scriptsize S}}

\def\sT{\mbox{\scriptsize T}}

\def\bigtau{\mbox{\Large$\tau$}}

\def\es{\m = \m}

\def\:={\m := \m}

\def\=:{\m =: \m}

\def\leqs{\m \leq \m}

\def\geqs{\m \geq \m}

\def\gs{\m > \m}

\def\ls{\m < \m}

\def\approxs{\m \approx \m}

\def\cr{\mbox{\scriptsize{\bf $\m \times \m$}}}

\def\sumi2{\sum\mbox{}_{\mbox{}_{\mbox{\scriptsize $i$=1}}}^2}

\def\sumi3{\sum\mbox{}_{\mbox{}_{\mbox{\scriptsize $i$=1}}}^3}

\def\sumABcycles3{\sum\mbox{}_{\mbox{}_{\mbox{\scriptsize cycles $A,B$=1}}}^{3}}

\def\sumCDcycles3{\sum\mbox{}_{\mbox{}_{\mbox{\scriptsize cycles $C,D$=1}}}^{3}}

\def\sumj3{\sum\mbox{}_{\mbox{}_{\mbox{\scriptsize $j$=1}}}^3}

\def\sumk3{\sum\mbox{}_{\mbox{}_{\mbox{\scriptsize $k$=1}}}^3}

\def\prodiA1{\prod\mbox{}_{\mbox{}_{\mbox{\scriptsize $i$=1}}}^{A - 1}}

\def\caps{\, \cap \, \,}

\def\coprods{\, \coprod \,}

\def\d{\textrm{d}}                                                  

\def\pa{\partial}                                                   

\def\sin{\mbox{sin}}

\def\cos{\mbox{cos}}

\def\tan{\mbox{tan}}

\def\cosec{\mbox{cosec}}

\def\cot{\mbox{cot}}

\def\arcsin{\mbox{arcsin}}

\def\arctan{\mbox{arctan}}

\def\FrI{\mbox{$\mathfrak{I}$}}                                

\def\FrT{\mathfrak{T}}                                         
                                                       
\def\sFrP{\mbox{\scriptsize $\mathfrak{P}$}}
												  
\def\sFrR{\mbox{\scriptsize $\mathfrak{R}$}}                   
\def\FrS{\mbox{\Large $\mathfrak{s}$}}                         
\def\sFrS{\mbox{\large$\mathfrak{s}$}}                         
   
\def\tFrS{\mbox{\footnotesize$\mathfrak{s}$}} 

\def\lFrg{\mbox{\Large$\mathfrak{g}$}}                         
\def\FrT{\mbox{\boldmath$\mathfrak{T}$}}                       
\def\FrG{\mathfrak{G}}                                         
\def\Hilb{\mbox{{\boldmath$\mathfrak{H}$}ilb}}                 

\def\bFrL{\mbox{\boldmath$\mathfrak{L}$}}                            

\def\Phase{\mbox{{\boldmath$\mathfrak{P}$}hase}}                     

\def\bFrR{\mbox{\boldmath$\mathfrak{R}$}}                            
\def\Rig-Phase{\bFrR\mbox{ig-}\Phase}                                
                                                                													   
\def\FrP{\mbox{\Large $\mathfrak{p}$}}                                 
\def\sFrP{\mbox{\large $\mathfrak{p}$}}                                

\def\FrR{\mbox{\boldmath$\mathfrak{R}$}}                             

\def\sFrR{\mbox{\scriptsize\boldmath$\mathfrak{R}$}}                 
\def\acute{\angle}

\def\obtuse{\backslash\u{\m}}

\def\Area{\bigalpha rea}

\def\Tau{\mbox{\Large$\tau$}}

\def\Aniso{\bigalpha niso}

\def\Ellip{\bigepsilon llip}

\def\LinEllip{\Lambda in\bigepsilon llip}

\def\area{\alpha rea}

\def\aniso{\alpha niso}

\def\ellip{\epsilon llip}

\def\linellip{\lambda in \ellip}

\def\Imat{\u{\u{I}}}                                                 

\def\1mat{\u{\u{1}}}                                                 

\def\Bmat{\u{\u{B}}}                                                 

\def\Mmat{\u{\u{M}}}                                                 

\def\invol{\u{\u{J}}}                                                

\def\Heron{\u{\u{H}}}                                                

\def\Ellmat{\u{\u{E}}}                                               

\def\Animat{\u{\u{A}}}                                               

\def\Leib{\bFrL\mbox{eib}}                                           

\def\Positive-Modespace{\mbox{{\boldmath$\mathfrak{M}$}odespace$^+$}}

\def\POSITIVE-MODESPACE{\mbox{{\boldmath$\mathfrak{M}$}ODESPACE$^+$}}

\def\scN{\mbox{\scriptsize ${\cal N}$}}

\def\Top{\FrT\mo\mp}

\def\RepBun{\mbox{{\boldmath$\mathfrak{R}$}ep-{\boldmath$\mathfrak{B}$}un}} 

\def\Kin-Hilb{\mbox{{\boldmath$\mathfrak{K}$}in-\Hilb}}                     

\def\Mid-Hilb{\mbox{{\boldmath$\mathfrak{M}$}id-\Hilb}}                     

\def\Dyn-Hilb{\mbox{{\boldmath$\mathfrak{D}$}yn-\Hilb}}                     

\def\5Star{\mbox{\Large$\star$}}              

\def\Frr{\mbox{$\mathfrak{r}$}}

\def\bigiota{\mbox{\Large $\iota$}}                 

\def\bigsigma{\mbox{\Large $\sigma$}}      

\def\bigmu{\mbox{\Large$\mu$}}

\def\bigalpha{\mbox{\Large$\alpha$}}

\def\bigepsilon{\mbox{\Large$\epsilon$}}

\begin{document}

\begin{titlepage}

\begin{center}

\large{\bf The Smallest Shape Spaces. III.} \normalsize

\vspace{0.1in}

\large{\bf Triangles in 2- and 3-$d$} \normalsize

\vspace{0.1in}

{\large \bf Edward Anderson$^*$}

\vspace{.2in}

\end{center}

\begin{abstract}

\n This is an innovative treatise on triangles, resting upon 
1) 3-body problem techniques including mass-weighted relative Jacobi coordinates.
2) Part I's detailed layer-by-layer topological and geometrical study of Kendall-type shape spaces -- configuration spaces of all possible shapes --  
which, for triangles, are (pieces of) spheres.
3) Hopf mathematics. 
Triangles are moreover prototypical through being the smallest models which carry relative-angle as well as length-ratio information.  
Both 1) and 3) produce insightful new versions of Heron's formula, 3)'s moreover simultaneously providing new foundations for 2). 
Medians, and regular triangles bounding between tall and flat triangles, also play prominent roles.   
Right triangles form three kissing cap-circles on the shape sphere, 
from which a shape-theoretic answer to the well-known conundrum of what is the probability that a triangle is obtuse very readily follows: 3/4.  
The differential-geometric aspects of this answer moreover generalize to numerous variant problems: 
conditioning on tallness and/or isoscelesness,
generalizing from right angles to arbitrary maximal angles via the maximal angle flow on the shape sphere...

\m

\n Hopf mathematics moreover gives a general bundle section interpretation to Kendall's iconic spherical blackboard of vertex-unlablelled mirror-image-identified triangles, 
and of its two variants where one of these two conditions are dropped. 
We attribute a monopole to each of these spaces and to the full shape sphere, one due to Dirac, one to Iwai and the other two are new to this paper.  
We finally make insightful comparison of triangles in 2-$d$ with 
a) Part II's 4 points on the line.
b) Triangles in 3-$d$, which are particularly significant as the smallest model exhibiting stratification. 
Stratified manifold--sheaf pairs \c{Pflaum, Kreck} -- sheaves \cite{Sheaves1, Sheaves} adding useful local and global structure to general bundles -- 
lie at the heart of Shape Theory's future development. 

\end{abstract}

\n PACS: 04.20.Cv, Physics keywords: background independence, inhomogeneity, configuration space, Killing vectors, monopoles, 3-body problem. 

\m

\n Mathematics keywords: triangles, spaces of triangles, Shape Geometry, Shape Statistics, Applied Topology, Applied Geometry, 
applications of fibre and general bundles, Hopf map, simple but new applications of Graph Theory to Shape Theory, simple examples of stratified manifolds.

\vspace{0.1in}
  
\n $^*$ Dr.E.Anderson.Maths.Physics@protonmail.com

\m

\end{titlepage}

\section{Introduction}\l{Intro-III}

We continue this treatise on the smallest shape spaces by considering the triangle shapes made by 3 points in 2- and 3-$d$.  
Some of the ways in which these are critical models are through providing the smallest models exhibiting the following features. 

\m

\n 1) They contain relative angle information (whereas Parts I and II's shapes consisted solely of side-ratio information).

\m

\n 2) They manifest nontrivial Configurational Relationalism (a concept outlined in Part I; see \cite{APoT3, ABook} for more details).   

\m

\n 3) They exhibit nontrivial bundle mathematics \cite{Husemoller}.

\m

\n 4) The 3-$d$ version moreover exhibits nontrivial stratification \cite{Whitney46, Thom55, Whitney65, Thom69} of configuration space.

\m
  
\n 5) Thereby, the 3-$d$ version furthermore requires a general bundle \cite{Husemoller} or sheaf \cite{Sheaves1, Sheaves} treatment, 
given that fibre bundles do not suffice to study stratified manifolds. 
In fact, as the current paper first reveals, the 2-$d$ version also requires this as regards quotienting out mirror image identification and point-or-particle distinguishability.  

\m

\n 6) The 3-body problem is much more typical of $N$-body problems \cite{Marchal, LR97} than the 2-body problem is.  
  
\m  
  
\n 7) Triangular shapes are also the smallest model which makes contact with widely studied flat geometry problems \cite{Court, PS70, Portnoy, Guy, Silvester, IMO}. 
For instance, Shape Theory provides new versions of Heron's formula \c{2-Herons}, 
and a nice way of addressing Lewis Carroll's pillow problem \c{Pillow} of what is the probability that a triangle is obtuse, $\mbox{Prob(Obtuse)}$ 
\c{Small, MIT, A-Pillow, Max-Angle-Flow}.    
  
\m
  
\n In the above ways, for instance, (3, 2) and (3, 3) models already support a complex theory of Background Independence 
\c{A64, A67, BB82, I93, Giu06, FORD, APoT2, APoT3, BI, ABeables, ABeables2, AMech, AConfig, ABeables3, ABook, Affine-Shape-1, Affine-Shape-2, Project-1, Project-2} 
and thus, as argued in Part I and \c{APoT2, AKendall, APoT3, ABook}, an in many ways nontrivial realization of the Problem of Time 
\c{DeWitt67, Battelle, Kuchar81, Kuchar91, Kuchar92, I93, Kuchar99, RovelliBook, KieferBook, APoT1, APoT2, APoT3, ABook}.  
These are foundational topics in General Relativistic and Quantum Gravitational Physics, 
for which the current treatise' detailed mastery of configuration space topology and geometry is a crucial and substantial first step.  
(3, 2) and (3, 3) moreover have distinct elements of the eventual ($N$, $d$) theory of inhomogeneities from those in Part II.  
In the case of (3, 2)'s features, these and (4, 1) features are first combined in Paper IV's treatment of (4, 2).  

\m

\n For 3 points in 2-$d$, the shape space is a 2-sphere $\mathbb{S}^2$ \c{Kendall} or some quotient thereof.
In particular, this is the hemisphere $\mathbb{RP}^2$ \c{Kendall} for the mirror image identified shapes, 
a lune covering 1/3 of the sphere for the indistinguishable shapes \c{AF, FileR}, 
or an isosceles spherical triangle covering 1/6 of the shape sphere the mirror image identified {\sl and} indistinguishable particle case \c{AF, FileR}. 
This last space is the {\sl Leibniz space} \c{I} for 3 points in 2-$d$, $\Leib_{\sFrS}(3, 2)$, alias 
Kendall's spherical blackboard \c{Kendall84, Kendall89, Small, Kendall} from Shape Statistics (see additionally \c{Bhatta, PE16}).  
The shape sphere for 3 points in 2-$d$ is tessellated by each of the aforementioned isosceles and scalene spherical triangles \c{AF}; 
these are two of the simpler tessellations of the sphere \c{Magnus}.       

\m 

\n In the current treatise, we move from outlay of techniques and (3, 1) example in Part I to (3, 2) example. 
This is in many ways an alternative to Part II's increased complexity in going from (3, 1) to (4, 1) models.
Triangles \c{Schaum, Silvester, Coxeter}, the space of triangles \c{Kendall89, Small, Kendall, HsiangStraume, ArchRat, Montgomery2, +Tri, FileR, MIT, ABook} 
and corresponding constellational viewpoints: the 3-body problem \c{Gronwall, Dragt, Iwai87, LR97, EoT, ArchRat, Montgomery2, Marchal, TriCl, 08I, 08II, +Tri, FileR} 
in $d \geq 2$ are comparatively often studied relative to 4 points-or-particles on a line \c{AF, Scale, FileR, II}.  

\m

\n We first consider topological shapes and topological graph shape spaces in Sec \r{Top-(3,2)}. 
These are simpler than for (4, 1) models.

\m

\n We next give a Lagrangian-level metric treatment of triangular shapes and shape spaces in Secs \r{Metric-(3,2)}, \r{Right} and \r{Met-Shape-Spaces} respectively, 
and extend these to the Jacobian level in Secs \r{Jac-Shape-in-Space} and \r{Jac-Str}.    

\m

\n Our Lagrangian-level treatment includes where collinearity, isoscelesness and equilaterality are located in the shape spaces. 
Medians, associated radii, and area and perimeter variables are also considered. 
We give furthermore a simple differential-geometric characterization of rightness on the shape sphere.
These turn out to trace three cap-circles kissing each other in pairs, the cap interiors of which are the obtuse triangles.
This is an example of the shapes-in-space to shape space correspondence tackling shapes-in-space flat geometry problems by mapping them to 
Differential Geometry problems in shape space.
This far more general and widely applicable scheme is moreover rendered particularly simple in the current case 
by both spheres and their caps having particularly simple Differential Geometry that is widely studied years before some people progress to study Differential Geometry.
A more complicated example we consider is the maximal angle flow over the shape sphere \c{Max-Angle-Flow}.

\m

\n The Jacobian-level treatment (Secs 13-14), on the other hand, introduces the concepts of tall and flat triangles, as separated by regular triangles: 
those whose base-side and median partial moments of inertia coincide. 
We also show here that regularity affords further characterization of equilateral triangles.
We additionally give the Jacobian versions of congruence and similarity conditions, now involving a side and its median (or the ratio thereof, respecctively) 
along with the `Swiss-army-knife' angle between these. 
We furthermore solve for the triangle's other sides and angles there-between in terms of such `Jacobi data'.
We consider the forms taken by mergers and Jacobi-uniformity for triangles, locate these notions in each type of shape space, 
and finally count out the number of qualitatively distinct types of triangle afforded by inclusion of these Jacobi notions (and significant subsets).  
Secs 3 to 7's levels of structure on shape space are furthermore supported by summary and inter-relation webs in Appendix B.   
Finding the cap-circles of rightness gives access to a very simple solution to $\mbox{Prob(Obtuse)}$; 
tche value Shape Theory allots to this is moreover well-known from a priori distinct methods: 3/4. 
Numerous variants of this problem are also formulated and solved: obtuse given isosceles, obtuse or acute conditioned on (or vice versa) the triangle being tall or flat, 
use of other critical angles in place of rightness, such as $2 \, \pi/3$, below which triangles have a nontrivially-located Fermat point.

\m

\n We also take advantage (Sec \r{Hopf}) of some `Hopf' structures and techniques well-known in Dynamics and Theoretical Physics 
which however has largely yet to enter the Shape Statistics literature. 
From this point of view, the Jacobian notion of regularity is placed on the same footing as the Lagrangian and directly geometrical notions of 
collinearity and isoscelesness. 
All are mutually perpendicular planes, whose perpendiculars form an axis system in the halfway shape, equilateral shape and B directions. 
Hopf structures being bundle-theoretic gives a deep differential-geometric underpinning to considering the notions in each of these triples of notions as coprimary. 
In this way, Shape Theory provides a unified model of both geometrical and dynamical triangles.  
This provides differential-geometrical means by which a shape-theoretic unified study 
goes further than studies rooted on just one of traditional geometrical practise or $3$-body problem considerations.  
There is moreover an analogous `generalized Hopf structure' for arbitrary $N$-a-gons (Part IV), by which this unified approach has many fertile extensions 
(it also persists under at least some further choices of $\lFrg$ \c{Affine-Shape-2}).  
Its partly involving Jacobian structure further motivates the current treatise' inclusion of Jacobian considerations from the geometrical point of view 
as well as from the $N$-body problem point of view in which these were a priori considered to be more primary.

\m

\n Sec \r{Hopf} moreover provides novel interpretation of the Hopf quantities as ellipticity, anisoscelesness and area, 
showing what functions of the triangle's sides and angles these correspond to.
Furthermore, Hopf congruence and similarity conditions are now given, and the triangle's sides and angles there-between are solved for in terms of `Hopf data'.

\m

\n We next consider (Sec 16) Killing vectors for each (3, 2) triangle space in Sec \r{Killing-(3,2)}, including similarity Killing vectors for the scaled version. 
The axes produced by the Hopf map are useful in phrasing and interpreting this.  
\n Subsequently, in Sec 17, we interpret the space of spaces of representatives over the full $\w{\FrS}$ \cite{+Tri, FileR}, mirror-image-identified$ \FrS$ \cite{+Tri, FileR}, 
indistinguishable $\FrI\FrS$, and Leibniz $\Leib_{\sFrS}$ \cite{Kendall84, Kendall89} shape spaces. 
For the shape sphere $\FrS(3, 2)$ this is the section of a principal fibre bundle -- the Hopf fibre bundle -- 
but in the other cases a general bundle is needed to define a section.  
This is because quotienting out by $S_3$ and/or $\mathbb{Z}_2$ creates a discrete version of the stratification effect,  
turning everywhere-equivalent fibres into in places inequivalent classes of bundle objects.  
We then identify Kendall's spherical blackboard picture locating triangles on $\Leib_{\sFrS}$ as a {\it section construct} which is a fortiori 
not at the level of fibre bundles but at the more broader level of general bundles or (with more tools available) of Sheaf Theory.
We provide various sections which are globally valid over $\Leib_{\sFrS}$ and others of interest which are however singular on one corner. 
On the other hand, fibre bundle sections will do for the $\FrS(3, 2)$ sphere, though of course this problem admits no global section. 
In this case, we provide a section which only goes bad at the equilateral triangle poles. 
So there is a trade-off: $\Leib_{\sFrS}(3, 2)$ is simpler in admitting a global section, but is less simple in that it is a general bundle rather than a fibre bundle.  
In Sec 18, we add -- to the Dirac monopole's realization in the $\mathbb{R}^3$ relational space, and Iwai's monopole in the corresponding half-space -- 
two new monopoles corresponding to $\FrI\FrS$, and $\Leib_{\sFrS}$.
See the Conclusion and \c{A-Monopoles} for yet further monopoles realized by further models of triangleland.  

\m

\n In the Conclusion (Sec 19), after summarizing Part III's results, we end with firstly a (3, 2) versus (4, 1) comparison, i.e.\ Part III versus Part II.    
This reveals that complex features of (4, 1) and (3, 2) are largely distinct; Part IV's (4, 2) then combines these to create considerable further complexity than either (Paper IV).  
Secondly, we compare (3, 2) and (3, 3), noting that in particular, stratification plays a substantial further role for (3, 3).   
In each case in which stratification appears in Part III, general bundles are required, 
and it looks to be further useful and powerful to model this with sheaf mathematics.  
In this way, (3, 2) and especially (3, 3) make for a useful testing ground for GR's Background Independence, especially at the global level \c{ABook, Project-2, Project-3}. 

\vspace{10in}

\section{Topological level of structure}\l{Top-(3,2)}

\subsection{Topological shapes}

{\bf Proposition 1} There are three topological types of 3-point configuration: Fig \r{(3, 2)-Top-Shapes}.
These coincide with the 1-$d$ case's 3 topological classes, albeit with the G class now more broadly interpreted.  
%
{            \begin{figure}[!ht]
\centering
\includegraphics[width=0.40\textwidth]{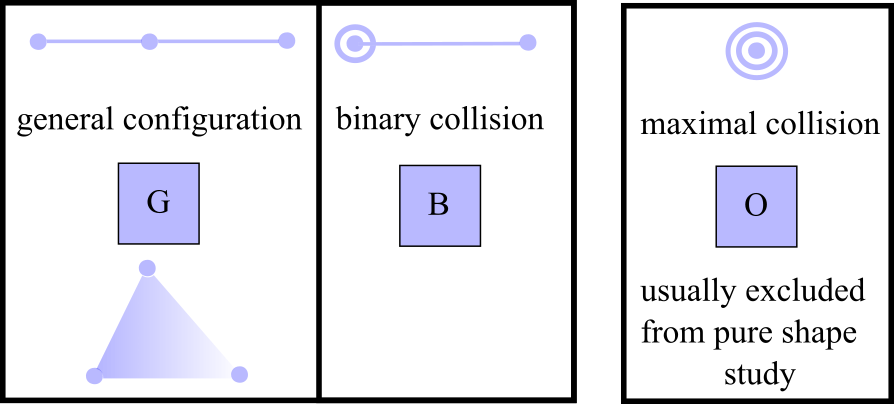}
\caption[Text der im Bilderverzeichnis auftaucht]{        \footnotesize{Topological classes of configurations for 3 particles in 1-$d$. 
Namely, the coincidence-or-collision-less generic configuration G, now covering both triangular and collinear metrically distinguished subcases, 
the binary coincidence-or-collisions B, and 
the maximal coincidence-or-collision O usually excluded from pure shape study.} }
\l{(3, 2)-Top-Shapes} \end{figure}          }

\m

\n{\bf Remark 1} As for (3, 1), these are all partitions, unless labels or mirror image distinctions further discern between types of B.  

\m

\n{\bf Remark 2} Regardless of whether the points-or-particles are labelled or mirror image configurations are held to be distinct, 
\be
\mbox{\#(G)} = 1 \m : 
\ee
the generic rubber triangle (including collinear cases but excluding binary or maximal collisions) can be deformed from any labelling to any other.   
Thus topologically there is only one face or 2-cell.  
For a labelled triangle, regardless of whether mirror images are identified, 
\be 
\mbox{\#(B)} = \mbox{(ways of leaving one particle out)} 
             = 3                                            \m .  
\ee
The above represent two salient differences with the 1-$d$ case.

\subsection{Topological shape spaces}
%
{            \begin{figure}[!ht]
\centering
\includegraphics[width=0.51\textwidth]{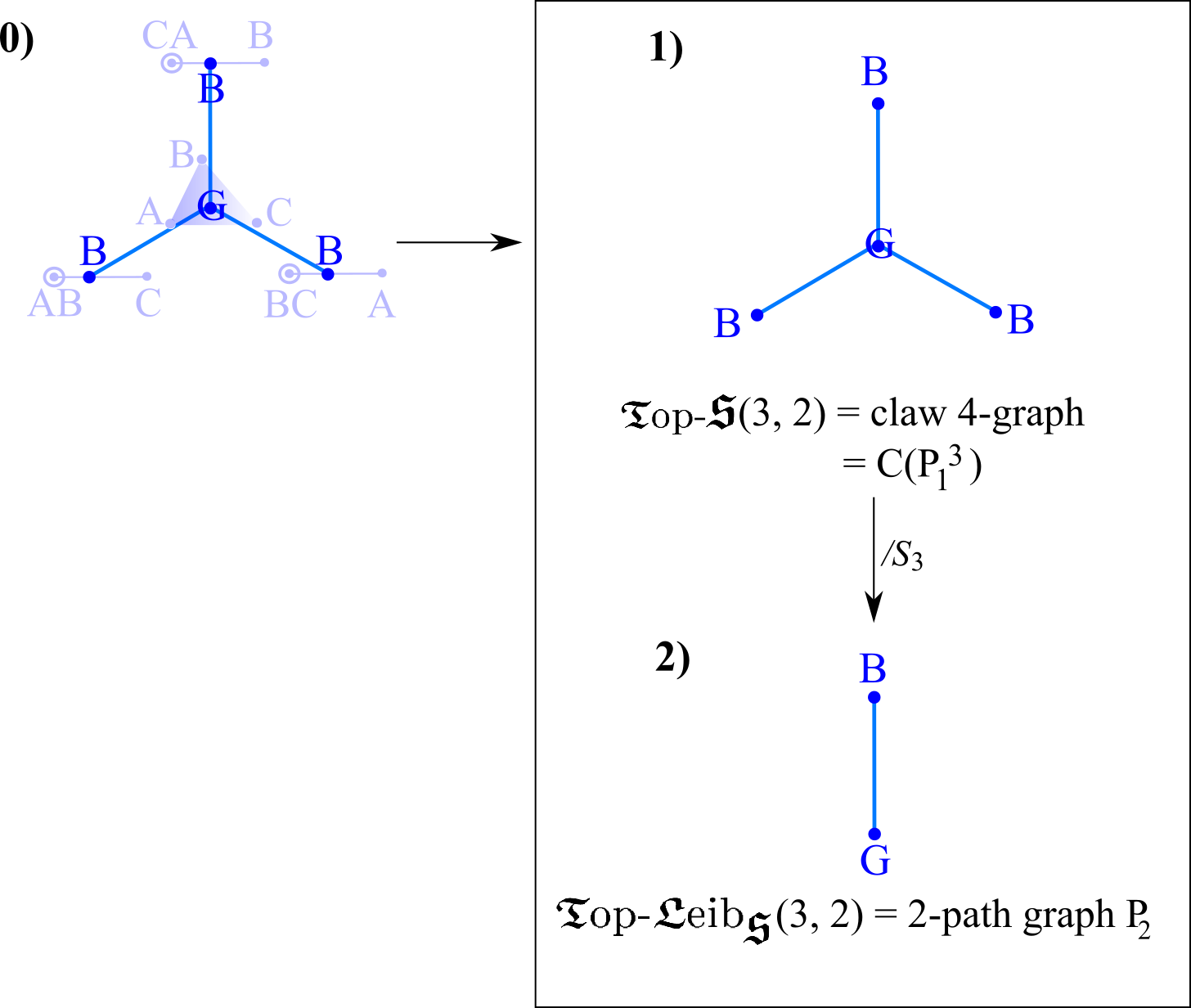}
\caption[Text der im Bilderverzeichnis auftaucht]{        \footnotesize{0) Continuity method for determining the topology of
$\Top\mbox{-}\sFrS(3, 2)$ to be the claw graph alias 3-star (Appendix I.A); the 3 `talons' are labelled equally with B's (up to cluster labelling choice).
2-$d$ gives no choice of regarding mirror images to be distinct at the topological level: the `left' and `right' topological shapes can be deformed into each other.
Thus $\Top\mbox{-}\sFrS(3, 2)$ is the same; $\Top\mbox{-}\FrI\sFrS(3, 2)$ likewise coincide and are the 2-path graph depicted.} }
\l{S(3, 2)-Top} \end{figure}          }
 
\n{\bf Proposition 1} For distinguishably labelled points-or-particles, the topological shape space is
\be
\Top\mbox{-}\w{\FrS}(3, 2) = \mbox{claw} 
                           = \Top\mbox{-}\FrS(3, 2)     \m :
\ee
the 4-vertex {\it claw graph} with equally labelled `talons' as per Fig \r{S(3, 2)-Top}.1).     

\m

\Proof Continuity considerations show that these shapes fit together in the manner of Fig \r{S(3, 2)-Top}.0), which closes up to form Fig \r{S(3, 2)-Top}.1).  $\Box$

\m  

\n{\bf Proposition 2} For indistinguishable points-or-particles,   
\be
\Top\mbox{-}\FrI\FrS(3, 2)  \es  \frac{\Top\mbox{-}\FrS(3, 2)}{S_3} 
                            \es  \mP_{2} 
					        \es  \Top\mbox{-}\Leib_{\sFrS}(3, 2) \m : 
\l{Leib(3, 2)}
\ee
the 2-vertex path graph with distinct end-point labels as per Fig \r{S(3, 2)-Top}.2).

\m

\n{\bf Remark 1} Acting on the topological shape space claw graph, identifying mirror images has no separate effect.   
Thus 
\be
S_3 \times \mathbb{Z}_2 \m \mbox{ acts as } \m  D_3 \mbox{ (dihaedral group of order 6) } , 
\ee
permitting us to rewrite the labelling and mirror image based definition (\r{Leib(3, 2)}) of the topological configuration in space as 
\be
\Top\mbox{-}\Leib_{\sFrS}(3, 2)  \es  \frac{\Top\mbox{-}\FrS(3, 2)}{D_3}              \m . 
\ee
This is very natural in configuration space since the unlabelled claw graph has automorphism group $D_3$.

\subsection{The topological coincidence-or-collision structure}
%
{            \begin{figure}[!ht]
\centering
\includegraphics[width=0.42\textwidth]{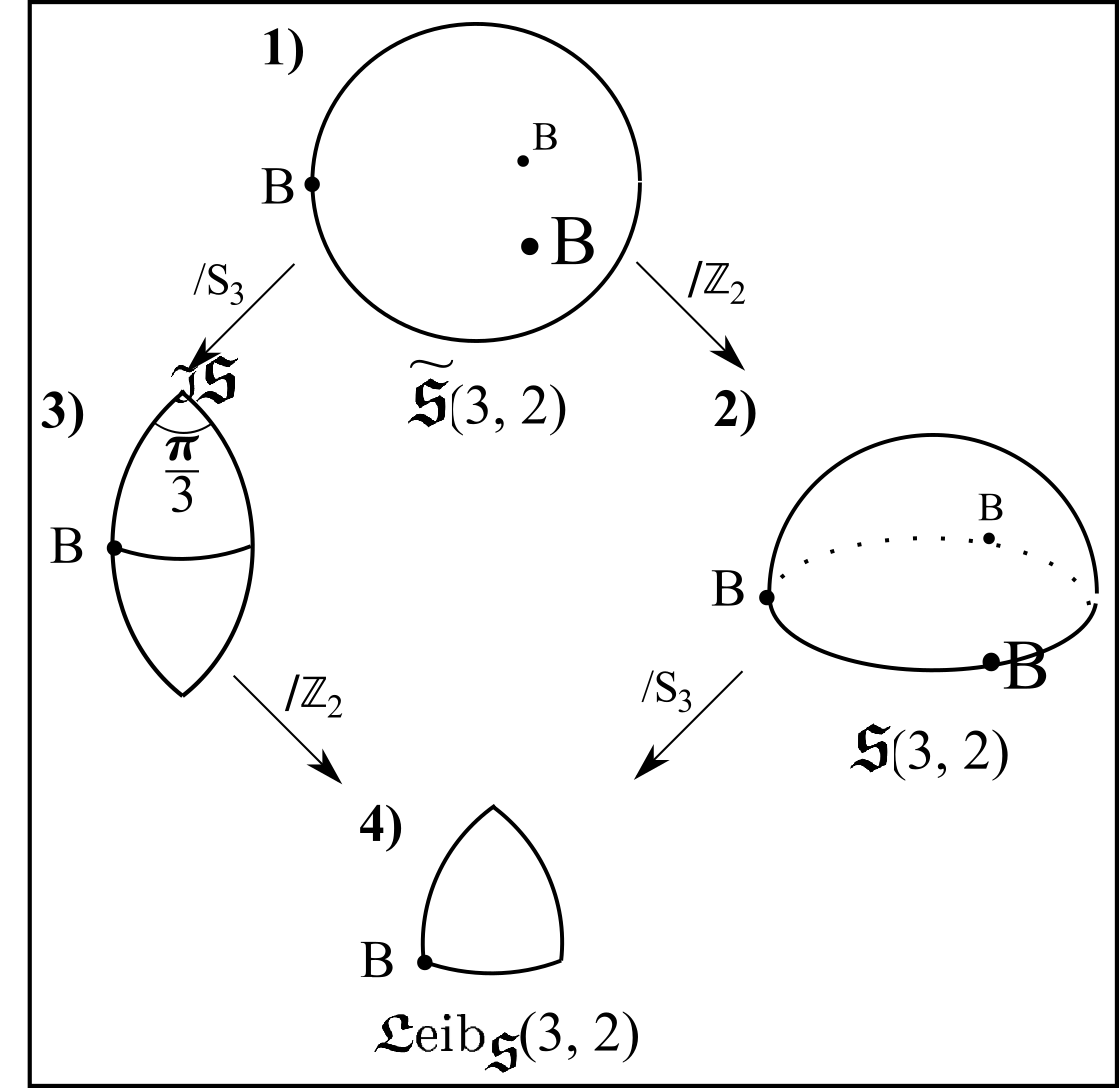}
\caption[Text der im Bilderverzeichnis auftaucht]{        \footnotesize{Topological-level configurations and configuration spaces for 3 particles in 2-$d$.  

\m

\n We include 5) and 6) for comparison and clarification, since these have on occasion been discussed as well. 
Since these correspond to a type of partial distinguishability \c{A-Monopoles}, they do not feature further in the current treatise. 
While 6) and 3) both cover 1/6 of the shape sphere, 
it suffices to observe that 6) involves 2 labellings of the equilateral triangle whereas 3) involves just one to establish that these are not equivalent.
If the B's in 1) are excised, one obtains the quite well-known `pair of pants' \c{ArchRat, Montgomery2}.
See \c{A-Monopoles} for a wider variety of modelling situations' configuration spaces.  } }
\l{(3,2)-Config} \end{figure}          }

\n Among the normalizable shapes, this is just 3 points, corresponding to the 3 different labellings of the binary coincidence-or-collision, 
or just one point in the case in which these labellings are indistinguishable (Fig \r{(3,2)-Config}).  

\vspace{10in}

\section{Triangular shapes at the Lagrangian metric level.}\l{Metric-(3,2)}

\subsection{Metric shapes of triangle that are characterized by symmetries in space}

At this point, much material familiar from elementary Flat Geometry is encountered.  
This is since triangles have been much more studied than 3 or 4 points on a line. 
Benefits from reconcile this material with how we have been developing Shape Theory so far are considerable and mutual as regards a number of further insights, variables and techniques.  

\m

\n Let us first recollect which variety of notions of triangles can be envisaged in terms of symmetries in space 

\m

\n{\bf Qualitative type 1)} The {\it equilateral triangle} E is the most symmetric triangle in space; it possesses a $D_3$ symmetry (order 6).  

\m

\n{\bf Qualitative type 2)} {\it Isosceles triangles} I are the next most symmetric in space, possessing $\mathbb{Z}_2$ reflection symmetry (order 2); 
       note that this characterization of isosceles includes B.    

\m

\n{\bf Qualitative type 3)} {\it Scalene triangles} S have no element of symmetry in space at all.

\subsection{Metric shapes of triangle that are characterized by relative separations}\l{rel-sep}
%
{            \begin{figure}[!ht]
\centering
\includegraphics[width=0.27\textwidth]{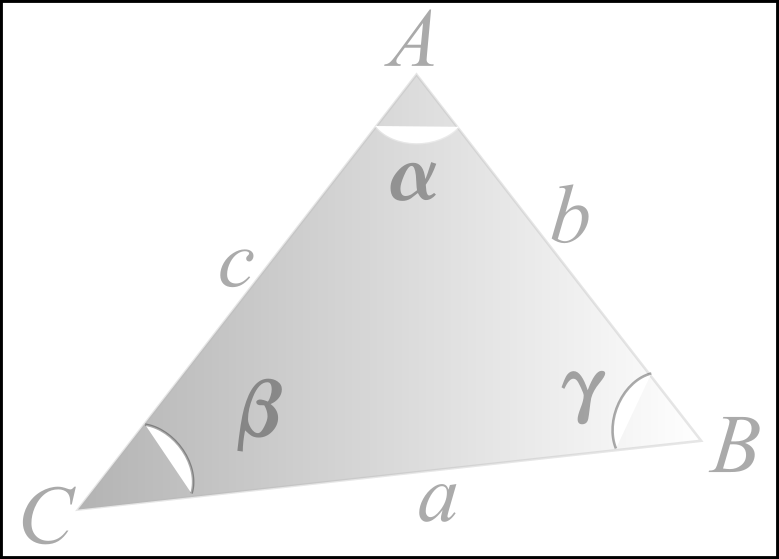}
\caption[Text der im Bilderverzeichnis auftaucht]{        \footnotesize{Cyclic notation for a triangle's vertices, sides and angles.
It is also useful for us to denote $a, b, c$ by $s_i$, $i = 1$ to $3$. } }
\l{abc0} \end{figure}          }

\n{\bf Qualitative type 0)} A zero edge, $a = 0$ say (see Fig \r{abc0} for the triangle notation in use) 
returns the binary coincidence-or-collision shape B which already had topological-level meaning.  

\m

\n At the metric level this is moreover now accompanied by basic and familiar and intuitive notions of non-coincident vertex triangles, 
1) to 3) below, of which admit a relative-separation characterization.   
For triangles, relative separations reduce entirely to edges rather than diagonals.  
All in all, we consider here the notion of edge-uniformity.

\m

\n{\bf Qualitative type 1)} Maximal edge uniformity consists of all three edges being equal in length, $a = b = c$, giving the equilateral triangle E. 

\m

\n{\bf Qualitative type 2)} The minimal element of edge uniformity is for two edges to be equal in length, $a \neq b = c$ say, giving the isosceles triangles, I. 

\m

\n{\bf Qualitative type 3)} Finally, if no element of edge uniformity is present, $a \neq b \neq c$, one has a scalene triangle, S.  

\m

\n{\bf Remark 2} Uniformity is nontrivially realized by E.  
Note that one has no notion of linearly-ordered separations in non-degenerately 2-$d$ configurations.  
However for $N = 3$, there is still a notion of adjacent separations: all separations are adjacent and without having to specify any ordering.  
We can thus use separations to qualify maximum uniformity. 

\m

\n{\bf Remark 3} We can view E's features as arising from two clusters being isosceles (this implies that the third is as well).  
\be
\mE =  \mI \caps  \mI 
    =  \mI \caps  \mI \caps  \mI   \m . 
\l{E=II}
\ee
\n{\bf Remark 4} E also plays the role of most uniform state U(3, 2) for triangleland: 
\be
\mU(3, 2) = \mE   \m .
\ee
\n{\bf Remark 5} These shapes have, respectively, the symmetry groups $S_2$ and $S_3$ acting upon 
\be 
\mbox{(separation space)} = \mbox{(side space)}  \m . 
\ee
\n{\bf Qualitative type 4)} The relative separation characterization of collinearity is that one separation is the sum of the other two, say 
\be
a = b + c   \m .
\l{a=b+c}
\ee  
\n{\bf Remark 6} For all (3, 2) shapes, moreover, 
\be
a \leqs b + c \mbox{ and cycles } :  \mbox{ the {\it triangle inequality }} .
\l{Triangle-Ineq}
\ee 
Collinear shapes C are then the only shapes that saturate this, returning (\r{a=b+c}).  
The normalized such both have the same amount of diversity as the mirror image identified (3, 1) shapes of Part I. 
Thus these include B and U(3, 1).  

\m

\n{\bf Remark 7} This U(3, 1) moreover posseses an element of relative separation uniformity, by which it is a subcase of isoscelesness I according to the current section's criterion.   
B can moreover also be viewed as another limiting case of I; see Sec 3.6 for further comments.  
%
%
Overall, 
\be
\mI \caps\, \mC = \{ \mB, \, \mU(3, 1)\} \m .  
\ee
\n{\bf Definition 1} The {\it perimeter} of a triangle is 
\be
S := a + b + c  \m ; 
\ee 
the {\it semi-perimeter} is then 
\be
s \:= \frac{S}{2}  \m .
\ee

\subsection{Metric shapes of triangle that are characterized by relative angles}\l{Rel-angle}

\n Qualitative types 1) to 3) admit yet another standard characterization in terms of uniformity of relative angles (which, for triangles, are between edges).  

\m

\n{\bf Qualitative type 1)} The equilateral triangle E has maximal angle uniformity: all three angles are equal ({\it equiangularity} condition):  
\be
\alpha = \beta = \gamma \es \frac{\pi}{3} \m .
\ee  

\m

\n{\bf Qualitative type 2)} The isosceles triangles I have the minimal angle uniformity: two angles equal (let us refer to this as `isoangularity'): 
$\alpha \neq \beta = \gamma$, say. 

\m

\n{\bf Qualitative type 3)} Scalene triangles S have no element of angle uniformity: $\alpha \neq \beta \neq \gamma$.

\m

\n{\bf Remark 1} Isoscelesness $\Leftrightarrow$ `isoangularity' and equilaterality $\Leftrightarrow$ equiangularity, 
by which each of these notions can be thought of either in pure-ratio terms or in pure-angle terms.  

\m

\n{\bf Qualitative type 4)} The relative angle manifestation of collinearity\footnote{$N = 3$ supports only the overlapping index case of Remark 7 of Sec I.6.3, 
so the entirety of its parallelism equations are in this case collinearity equations. 
It takes Part IV's consideration of quadrilaterals to realize parallel but not collinear configurations: trapeziums and parallelograms.  
The corresponding $\epsilon$-tolerant collinearity inequalities have been considered by Kendall \c{Kendall}.}
is that one of the relative angles is 0 and another is $\pi$.
This includes U(3, 1) but now not B, since no relative angles are defined within B, so one is now really characterizing C$^1$ rather than C.  

\m

\n{\bf Remark 2} Secs 3.1-3's criteria only agree if degenerate triangles -- C including B and U(3, 1) -- are excluded.

\subsection{Summary figure so far}

See Fig \r{Summary-Fig-1}. 

{\begin{figure}[ht]
\centering
\includegraphics[width=0.65\textwidth]{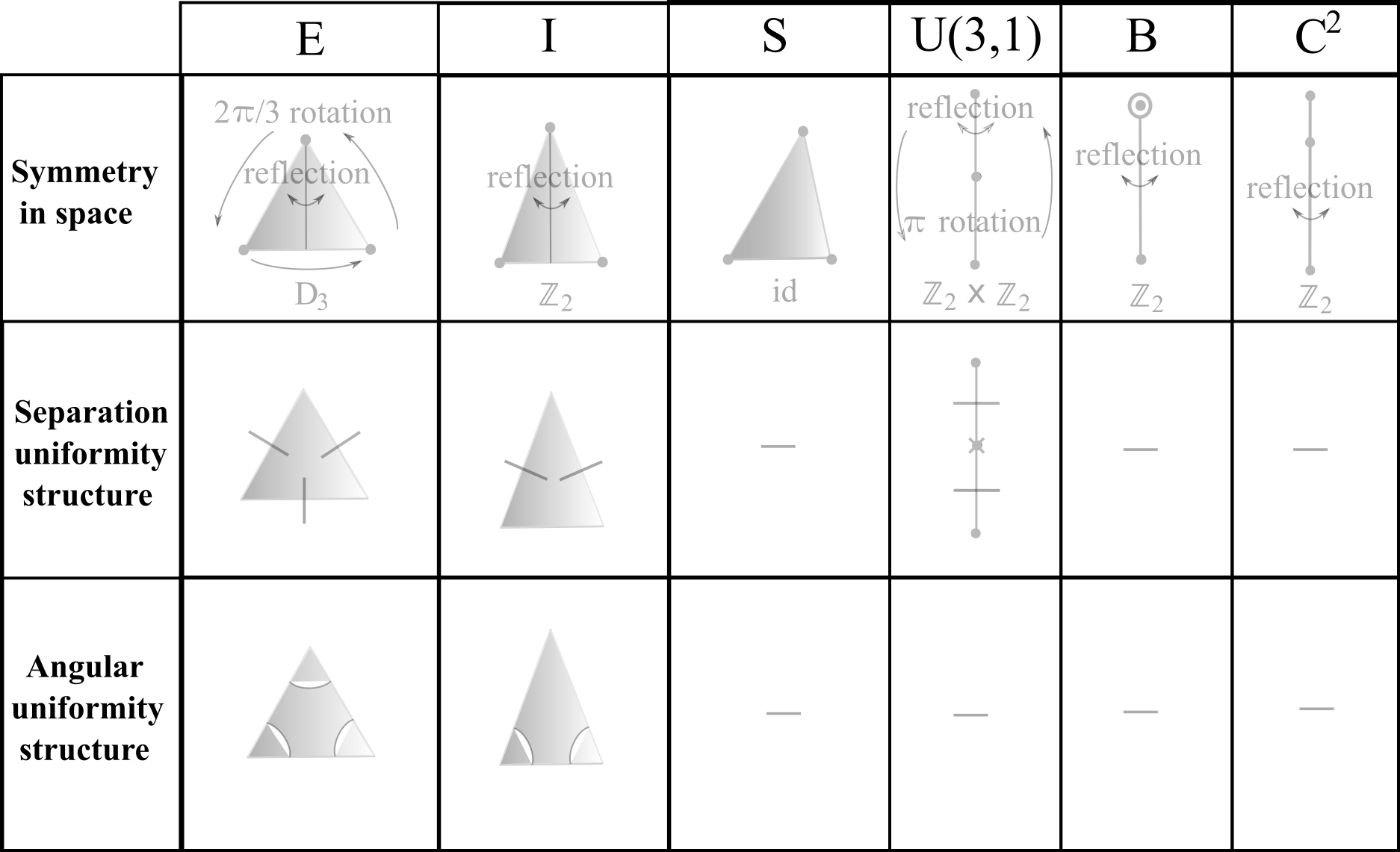}
\caption[Text der im Bilderverzeichnis auftaucht]{\footnotesize{Lagrangian properties of equilateral E, 
                                                                                           isosceles I, 
																						 and scalene S triangles, 
                                                                                   alongside binary coincidence-or-collision B, 
																				   collinear-uniform U(3, 1), 
																				   and other collinear $\mC^2$ shapes.}} 
\l{Summary-Fig-1}\end{figure} } 

\subsection{A first few standard congruence and similarity conditions for triangles}\l{C-S}

\n For later comparison and use in proofs, we remind the reader of the usual congruence and similarity conditions for triangles.

\m

\n{\bf Congruence condition 1)} s-s-s: each of a pair of triangles shares the same three side lengths.

\m                                                                  

\n{\bf Congruence condition 2)}  s-$\alpha$-s: each of a pair of triangles shares 2 side lengths and angle included between these.  
	
\m	

\n{\bf Congruence condition 3)}  $\alpha$-s-$\alpha$:  each of a pair of triangles shares two angles and the side included between these. 
	
\m
	
\n{\bf Congruence condition 4)}  $\alpha$-$\alpha$-s:  each of a pair of triangles shares two angles and a non-included side. 
														      
\m
	
\n{\bf Similarity condition 1)}  r-r:  each of a pair of triangles shares two side-ratios. 
    
\m
	
\n{\bf Similarity conditions 2) and 3)} r-$\alpha$:  each of a pair of triangles shares a side ratio and angle, whether included or not. 

\m

\n{\bf Similarity condition 4)} $\alpha-\alpha$  each of a pair of triangles shares two angles (and thus all three, by angles in a triangle adding up to $\pi$).   
		
\m
		   
\n{\bf Remark 1} Congruence conditions are Euclidean-relational, whereas similarity conditions are similarity-relational.  
The above and subsequent enumeration of congruence and similarity conditions match upon quotienting scale out of the former. 
Moreover, all of the above conditions are Lagrangian, 
whereas the current Part III shall furthermore point to further shape-theoretically useful {\sl Jacobian} characterizations of congruence and similarity.

\m

\n{\bf Remark 2} In the current treatise's context, we view the above conditions as {\sl well-posed data sets}, 
from which the entirety of the features of a given geometrical conception of a triangle can be determined.
In particular, congruency conditions are data to fully solve a Euclidean triangle, 
whereas similarity conditions are data to fully solve a triangle in the sense of similarity geometry.\f{\c{Affine-Shape-2} extends this to equiareal and affine notions of triangle.} 
These equations for which these are data sets are moreover algebraic, so well-posedness reduces to well-determinedness.
This has a counting element to it: 3 parts to a congruency condition versus 2 to a similarity condition; 
this counting corresponds to the dimension of the corresponding configuration space: $\mbox{dim}({\cal R}(3, 2)) = 3$ and $\mbox{dim}(\FrS(3, 2)) = 2$.
Such counting need not however suffice to tie down existence or uniqueness.
For instance, not all triples of lengths constitute s-s-s data since some violate the triangle inequality.
Also, not all triples of angles correspond to a Euclidean triangle; they must sum to $\pi$ in order to do so. 
This then gives a dependency equation by which $\alpha$-$\alpha$-$\alpha$ reduces to $\alpha$-$\alpha$, 
which cannot uniquely specify Euclidean triangles out of not restricting size.  
To relate to general relativist and physicist readers' experiences, 
well-posedness for triangles is indeed analogous to well-posedness for physical laws, though, these in general being modelled by PDEs, 
there are further PDE-specific aspects to well-posedness there. 
More specifically well-posedness for instantaneous laws such as Gauss' Law of electrostatics or the GR initial-value problem is furtherly analogous:   
configurational problems, for which far from all functions solve these problems due to their constraints.\footnote{Conversely, 
for readers whose background is geometry, topology, shape theory, or its applications, 
Gauss's law is here cast in a gauge in which it is a constraint, whereas data for the GR initial-value problem need satisfy the Hamiltonian and momentum contraints: 
the time--time and time-space components of the Einstein field equations.
`Constraint' is here characterized as an equation which is less than the habitual second-order in its time derivatives.}

\m

\n{\bf Remark 3} The motivation for the current section is then clear: we will be providing {\sl relationally adapted} well-determined data for 
similarity shape and Euclidean shape-and-scale triangles, and this will be in the form of new congruence and similarity conditions. 
(Their novelty is rather probably at the level of these being {\sl motivated} as the primary congruence and similarity conditions 
from the Jacobi $3$-body, Hopf bundle and Kendall Shape Theory points of view.)
Secs \r{Overview}, \r{Cong-Sim-Disc}, \r{Jac-Cong-Sim} and \r{Hopf-Cong-Sim} continue this discussion after more structure has been laid out.

\subsection{Labelling and mirror image distinctions}

\n{\bf Remark 1} scalene, isosceles and equilateral triangles admit 6, 3 and 1 labellings respectively.

\m

\n{\bf Remark 2} Non-collinear triangles admit 2 choices of orientation ({\it clockwise} and {\it anticlockwise}), 
whereas collinear ones admit just a single undifferentiated orientation.  
For instance, one can distinguish between 2 choices of orientation for the equilateral triangle (denoted by E and $\overline{\mE}$ in Paper III).  

\m

\n{\bf Remark 3} The effect of labelling on collinear shapes C is as per Part I. 
All C shapes are immune to the clockwise--anticlockwise distinction which doubles the number of all non-degenerate (non-collinear) triangles.
%

\m

\n{\bf Remark 4} Isosceles triangles' label choice can be thought of as picking a base pair of vertices.

\m

\n{\bf Remark 5} Triangles with mirror image distinction support a meaningful difference between {\it left-leaning} and {\it right-leaning} triangles, 
with isosceles triangles separating these (Fig \r{BIEMC-3}.1).  
%
{            \begin{figure}[!ht]
\centering
\includegraphics[width=0.8\textwidth]{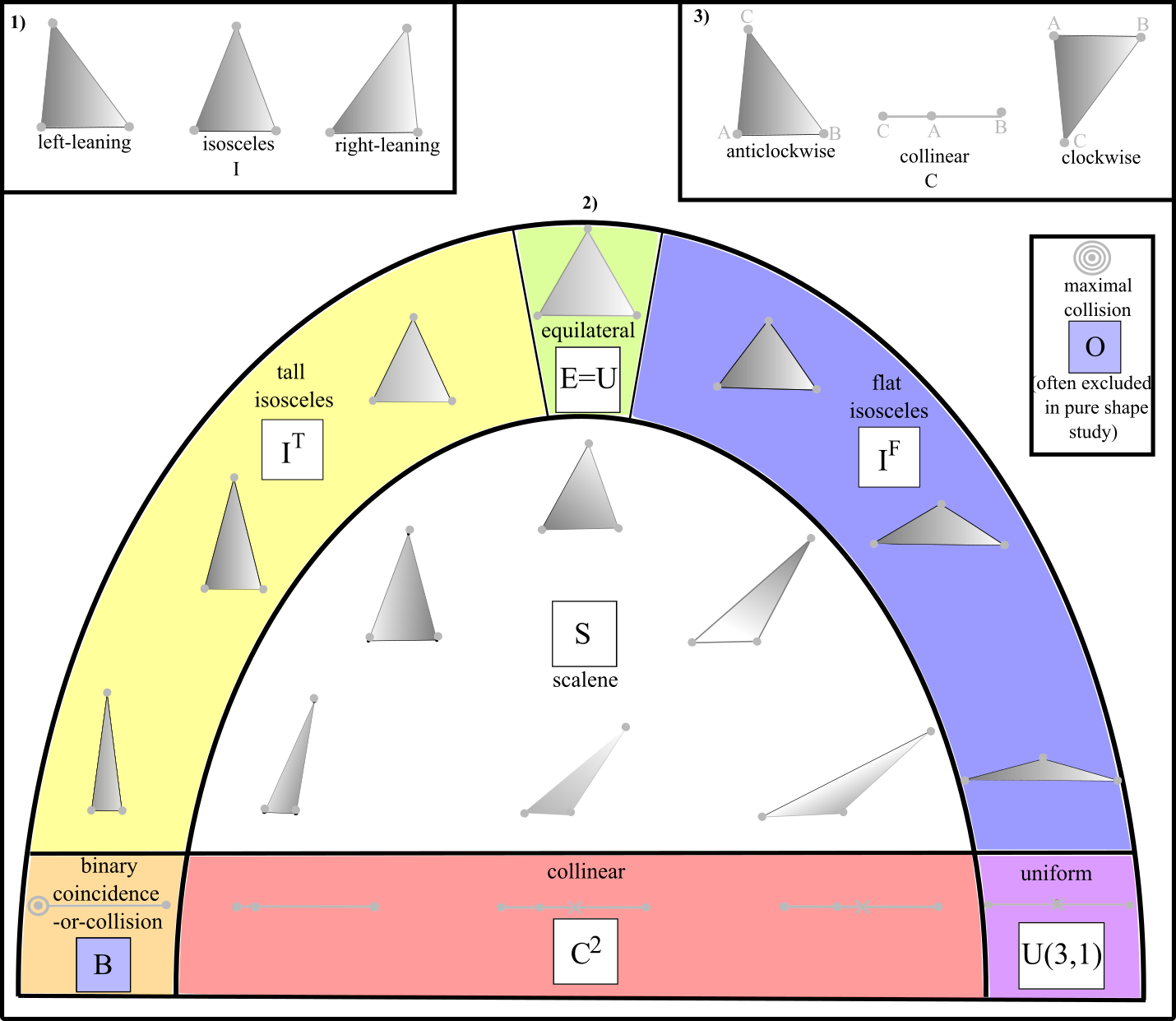}
\caption[Text der im Bilderverzeichnis auftaucht]{        \footnotesize{1) Isosceles triangles separate left-leaning and right-leaning triangles. 

\m

\n 2) As Sec \r{Met-Shape-Spaces} will characterize further, isoscelesness is split by the equilateral triangle.
The lower corners where isocelesness I meets collinearity C are the B and U(3, 1) shapes.   
The interior of this so-far heuristic wedge of notions of triangles is itself made up of the generic scalene triangles. 
For now, this is a shape-space-topological-manifold-level diagram of the metric types of shape; an accurate metric counterpart of it is set up in the next Section.  

\m

\n 3) Furthermore, for labelled triangles, collinear configurations separate clockwise and anticlockwise labelled triangles.} }
\l{BIEMC-3} \end{figure}          }

\subsection{Splits so far of G, C and I notions of shape}

In support of Fig \r{BIEMC-3}.3, 
firstly the collinear configurations have been split to the exclusion of the topologically distinct binary coincidence-or-collision B: 
\be
\mC = \mC^1 \coprods  \mB  \m, 
\ee
followed by the further split 
\be
\mC^1 = \mC^2 \coprods  \mU(3, 1)  \m . 
\ee
Also, the topological-level generic configuration G has been decomposed according to 
\be
\mG = \mI \coprods  \mS \coprods  \mC^1  \m,
\ee
for I isosceles and S scalene.
Finally I has been furthermore decomposed into 
\be
\mI = \mI^{\sT} \coprods  \mE \coprods  \mI^{\sF}  \m .
\ee
The T and F superscripts here refer to tall and flat triangles, a notion we further detail in Sec \r{Jac-Shape-in-Space}. 
For now, we note that for isosceles triangles, tallness and flatness are relative to the equilateral shape, 
which splits the isosceles part of $\Leib_{\sFrS}(3, 2)$'s perimeter into its $\mI^{\sT}$ and $\mI^{\sF}$ parts.

\subsection{Lagrangian angle uniformity}

\n{\bf Remark 1} For triangleland, the element of angle-uniformity is the same as the element of separation-uniformity, 
i.e.\ isoangularity -- two angles equal -- is equivalent to isocelesness: two sides equal. 
 
\m
 
\n{\bf Remark 2} Moreover the presence of two elements of angle uniformity imposes a third: the equiangularity characterization of equilaterality.  

\m

\n{\bf Remark 3} Isoangularity and equiangularity involve, respectively, the groups $S_2$ and $S_3$ acting on angle space. 

\m

\n{\bf Remark 4} For triangles, the groups acting on space, relative separations and relative angles are moreover isomorphic in each case.

\section{Shape spaces of triangles at the Lagrangian metric level}\l{Met-Shape-Spaces}
%
{            \begin{figure}[!ht]
\centering
\includegraphics[width=0.9\textwidth]{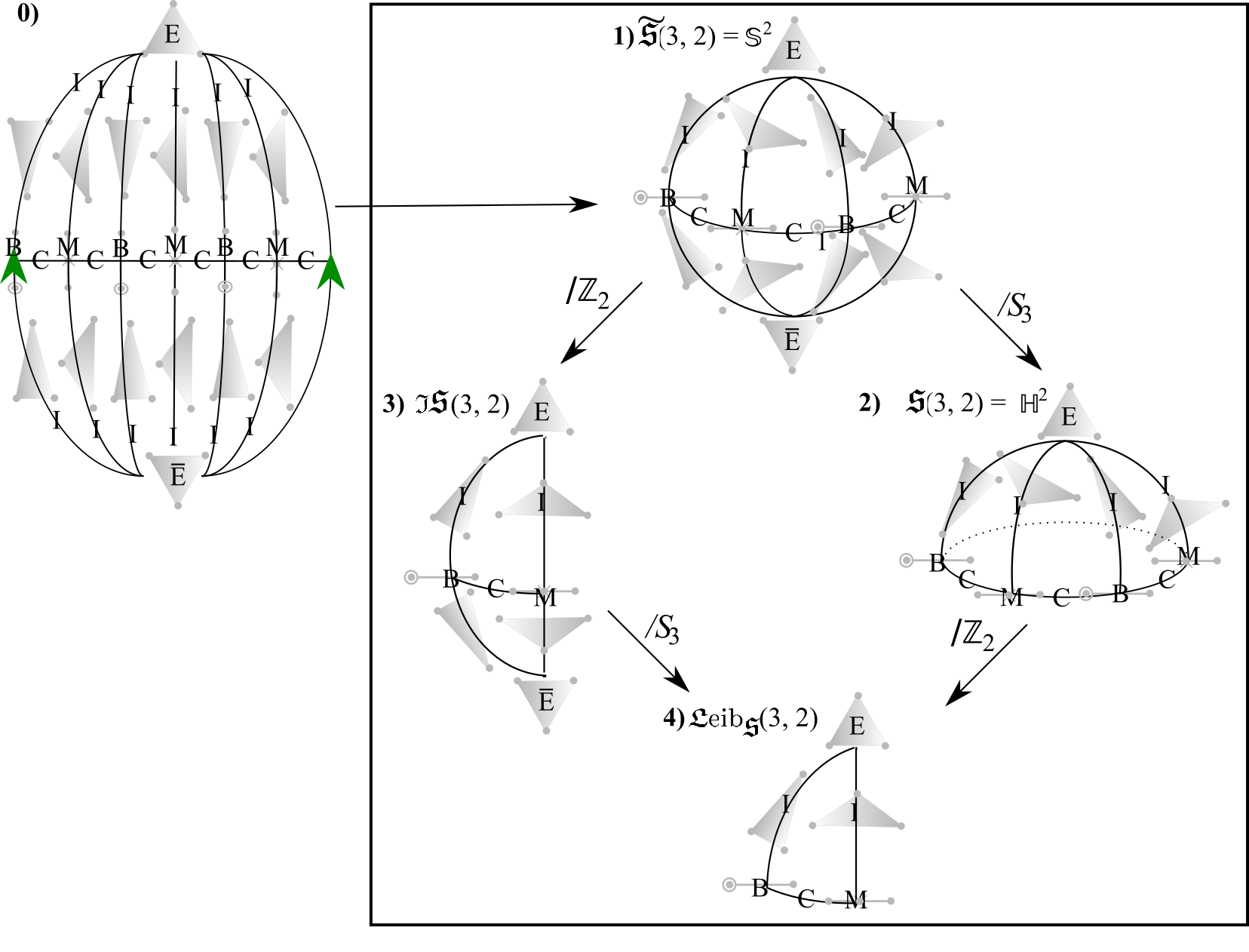}
\caption[Text der im Bilderverzeichnis auftaucht]{        \footnotesize{0) The continuity method gives the shape sphere 1).
This Figure is moreover drawn to be metrically correct with respect to the naturally induced spherical metric, 
with azimuthal and polar angles as given by eqs (I.171-172) and Fig I.18.a).   
Isoscelesness and collinearity decorate this as a 6-segmented orange cut in half perpendicular to its segments.
%

\m

\n 2) Applying mirror image identification kills off one hemisphere while preserving the equator, thus giving a hemisphere with edge. 
[This is often misinterpreted as $\mathbb{RP}^2$ in the literature, out of not detailedly investigating the shapes in question 
and which action in configuration space the mirror image $\mathbb{Z}_2$ symmetry in space corresponds to.]   

\m

\n 3) Applying label indistinguishability instead leaves one with a lune of width $\frac{\pi}{3}$.

\m

\n 4) Applying both at once leaves one with a the Leibniz space demilune of width $\frac{\pi}{3}$: {\it Kendall's spherical blackboard} \c{Kendall89}.} }
\l{S(3,2)-Level-1} \end{figure}          }

The previous section's metric-level observations are accounted for by the following topological shape space structure.

\m
 
\n{\bf Proposition 1} The shape space topology topological features are as per Fig \r{(3,2)-Config}
and these and the Lagrangian-level metric features obtained by the continuity method, as per Fig \r{S(3,2)-Level-1}.

\vspace{10in}

\subsection{Axis systems}

We point to the following two axis systems, for now postponing naming the third mutually perpendicular axis until Sec 6.1. 

\m

\n{\bf Axis System 1)} BE: with B as principal axis and E as secondary axis. 
As we shall see in Sec 6, this is natural in Jacobi coordinates and thus commonly arises as an adapted coordinate system in 3-body problems. 
The simplest spherical angle interpretations are for the corresponding angles, denoted by $\Theta, \Phi$

\m

\n{\bf Axis System 2)} EB: with E as principal axis and B as secondary axis.
This is more geometrically natural and also accords axis principality to the democratic concept. 
The price to pay is that its spherical angles, 
denoted $\w{\Theta}, \w{\Phi}$, has a more complicated in interpretation in spatial (formulae in terms of $\uq^I$). 
These are of course however related to $\Theta$ and $\Phi$ by basic spherical trigonometry formulae \c{FileR}.

\m

\n{\bf Remark 1} It turns out to often be useful to check whether one has remembered to pass from tilded to untilded variables.  
Failure to emphasize this point can cause arguments to sound wrong. 
This is an important point to make in an introductory account, as it avoids much confusion including obscuring what would elsewise be encouraging intuitions. 
This lacuna is moreover compounded by an axis switch usually being irrelevant in geometrical problems, 
but none the less making a huge interpretational difference in passing between shape space and shapes-realized-in-physical-space, 
since the axes now mean very different things: a binary collision to an equilateral triangle. 
For axis switches usually being irrelevant in practise means that one may not spot that an undeclared axis swich is why one's interpretation 
of a presentation's words do not match one's intuitions using what {\sl one thinks} the paper or seminar's notation is referring to...

\m

\n{\bf Example 1} For instance, the equator of collinearity C corresponds to         $\Phi = 0$ or $\pi$, or, 
                                                           in tilded coordinates, to $\w{\Theta} = \pi/2$.

\m

\n{\bf Remark 2} On the other hand, one of the bimeridians of isoscelesness I is at $\Phi = \frac{\pi}{2}$ or $\frac{3 \, \pi}{2}$, or, 
                                                           in tilded coordinates, at $\w{\Phi} = 0$ or $\pi$.

														   \m

\n{\bf Remark 3} We postpone physical interpretations of $\Phi$, $\Theta$ to Sec 6.

\subsection{Coincidence, collinearity, symmetry and uniformity structures over the shape sphere.}

\n{\bf Proposition 1} Triangleland's {\it metric-level coincidence-or-collision structure} -- consult Fig \r{(3,2)-Full-Structure-Web}.a) -- 
consists of 3 points at $\frac{2 \, \pi}{3}$ to each other on the equator: the B's. 

\m

\n{\bf Proposition 2} Triangleland's {\it collinearity structure} -- Fig \r{(3,2)-Full-Structure-Web}.d) -- is a purely metric-level concept, 
and consists of the equator of collinearity C.

\m

\n{\bf Remark 1} This realizes $\FrS(3, 1)$, as another example of the Shape-Theoretic Aufbau Principle.
The current model's second spatial dimension automatically identifies mirror image collinear configurations, explaining this realization (rather than that of $\FrS(3, 1)$).
This contains a total of six alternating equally spaced out B and M configurations. 
\m

\n{\bf Remark 2} C separates northern and southern hemispheres of clockwise and anticlockwise triangles.  

\m

\n{\bf Proposition 3} The (Lagrangian) symmetric configurations are distributed as follows.  

\m

\n The isosceles configurations I form bimeridian great circles at $\frac{\pi}{3}$ to each other.  
Each clustering choice of I separates the triangleland sphere into hemispheres of corresponding right-leaning and left-leaning triangles with respect to the same clustering choice.

\m

\n These intersect at the poles (in the new coordinates) which are the equilateral triangle shapes E. 

\m

\n This observation, and computation of Part II's symmetry element counting function over the shape sphere, establishes the following.

\m

\n{\bf Proposition 4} E (and $\overline{\mE}$ when distinct) maximizes 

\m

\n i) symmetry in space. 

\m

\n ii) separation uniformity.  

\m

\n iii) angle uniformity.  

\m

\n{\bf Remark 3} ii) and iii) make use of Part II's separation and angle-uniformity counting functions respectively. 

\m

\n{\bf Remark 4} See Figures \r{(3, 2)-Structures-Top} and \r{(3, 2)-Jac-Structures-Top} 
for each level of structure's full topological and metric structure over shape space respectively.  
These are arrived at by solving the coincidence equations, 
                                    collinearity equations redux of the parallelism equations, 
									symmetry in space equations, 
									separation uniformity equations, 
							    and angle uniformity equations respectively. 
In particular, 

\m

\n i) the symmetry structure is topologically the disc with five punctures and metrically the edges of the `orange of 3 segments' tessellation.

\m

\n ii) The Lagrangian separation uniformity structure is topologically and metrically the symmetry structure with the collision structure removed. 
This is because isosceles triangles have two sides of the same length -- an element of uniformity -- 
excepting the coincidence-or-collision points in shape spaces, since we defined uniformity to not include features already implied by coincidences-or-collisions.  

\m

\n iii) The Lagrangian angle uniformity structure coincides with ii). 
This is by Remark 1 of Sec \r{Rel-angle}'s basic observation that isoscelesness  and equilaterality 
                                         can be reformulated as `isoangularity' and equiangularity respectively.  

\m

\n{\bf Remark 5} For triangleland, the pure-ratio structure coincides with the uniformity structure: Fig \r{(3,2)-Full-Structure-Web}.c).  

\m

\n{\bf Remark 6} The pure-angle collinearity--perpendicularity structure is the 3 circles linking at the B's cut in half (Fig \r{(3,2)-Full-Structure-Web}.d);  
the M are not distinguished points here.                                       

\m

\n{\bf Remark 7} The full pure-angle structure is the 3 circles linking at the B's cut in half and the I meridians (Fig \r{(3,2)-Full-Structure-Web}.j), 
since I is an angular as well as side-ratio concept.

\subsection{Mirror image identified counterpart}

The figure looks the same from the bird's eye view but lower hemisphere part is of course now missing.
This leads to the topological differences of row 2 of Fig \r{(3, 2)-Structures-Top} relative to row 1 (see Appendix A for how this computes out).

\subsection{Indistinguishable shape space and Leibniz space counterparts}\l{I-Leib-Counter}

\n Indistinguishable shape space is a $\frac{\pi}{3}$ lune with corners E and $\bar{\mE}$ and edges I$^{\sT}$ and I$^{\sF}$. 

\m

\n{\bf Proposition 1} The coincidence-or-collision, 
collinearity, symmetry and uniformity structures at the level of indistinguishable space are as per Fig \r{(3,2)-I-Structure-Web}.
Topologically, these are as per row 3 of Fig \r{(3, 2)-Structures-Top}. 

\m

\n Leibniz space is an isosceles spherical triangle with corners E, B, M and edges I$^{\sT}$, I$^{\sF}$ and C.  

\m

\n{\bf Proposition 2} The collision, collinearity, symmetry and uniformity structures at the level of Leibniz space are as per Fig \r{(3,2)-Structure-Web}.
Topologically, these are as per row 4 of Fig \r{(3, 2)-Structures-Top}.

\m

\n{\bf Remark 1} As regards levels of structure on $\Leib_{\sFrS}(3, 2)$ and $\FrI\FrS(3, 2)$, 
see now Fig \r{(3,2)-Structure-Web} and \r{(3,2)-I-Structure-Web} in place of Fig \r{(3,2)-Full-Structure-Web}.

\subsection{Count of qualitative types so far}\l{T-F-Isosceles}
%
{            \begin{figure}[!ht]
\centering
\includegraphics[width=1\textwidth]{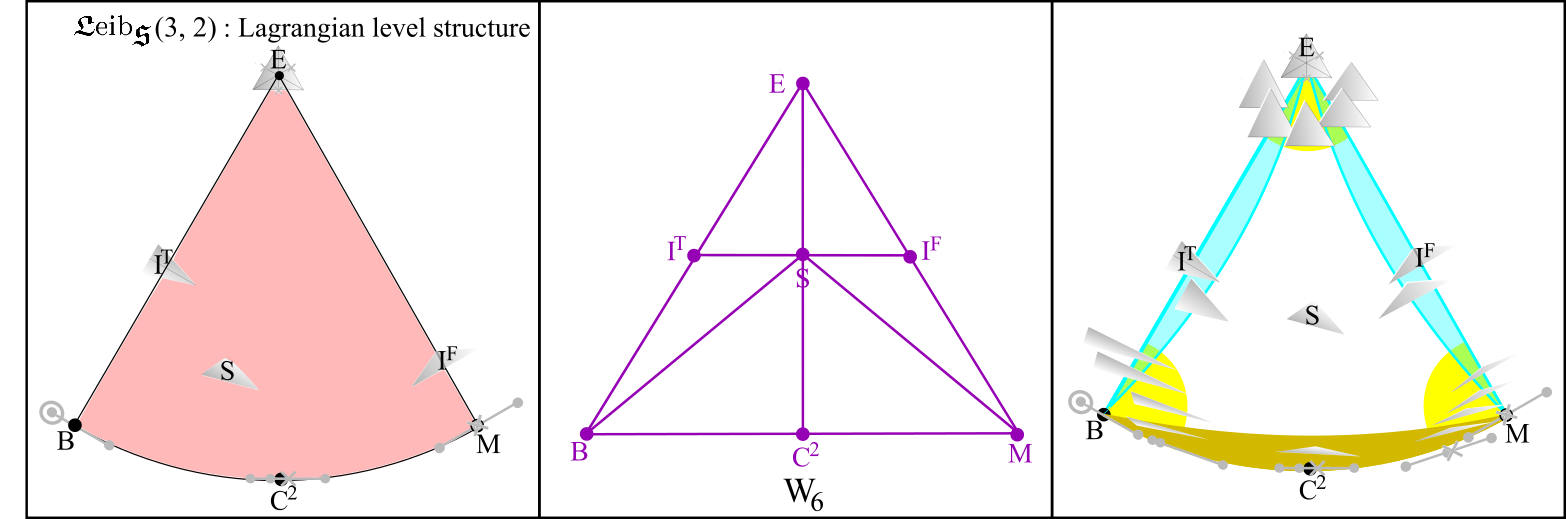}
\caption[Text der im Bilderverzeichnis auftaucht]{        \footnotesize{Shape-space-metric-level detail of $\Leib_{\tFrS}(3, 2)$ 
for shapes-in-space at the metric Lagrangian separation level.} }
\l{Leib(3, 2)-Lag} \end{figure}          }

\n{\bf Remark 1} At the level of Lagrangian separations, there are $V = 3$ types of special points: E, B and U(3, 1).

\m

\n{\bf Remark 2} These are also $E = 3$ types of edges $\mC^2$, I$^{\sT}$ and I$^{\sF}$.

\m

\n{\bf Remark 3} There is just the one type of face ($F = 1$): the generic non-collinear scalene triangles, S.  

\m

\n{\bf Remark 4} These can be located, and counted off, as per Fig \r{Leib(3, 2)-Lag}.a).

\m

\n{\bf Remark 5} Thus the number of qualitative types in Leibniz space is
\be
Q := V + E + F = 3 + 3 + 1 = 7 \m .
\ee
\n{\bf Remark 6} The adjacency graph Fig \r{Leib(3, 2)-Lag}.b), $V(\mbox{Adj}) = Q = 7$ and $E(\mbox{Adj}) = 12$, and is moreover the 6-wheel $\mW_{6}$.  

\m

\n{\bf Remark 7} See Figs  \r{(3,2)-Structure-Web}--\r{(3,2)-Full-Structure-Web} for an end-summary of special arcs and special points among the (3, 2) shapes.  
See Fig \r{(3, 2)-Q-Lag} for further counts of qualitative types in the four (3, 2) shape spaces, as exhibited in Fig \r{(3, 2)-Approx-Detail} in the case of Leibniz space.   
%
{            \begin{figure}[!ht]
\centering
\includegraphics[width=0.75\textwidth]{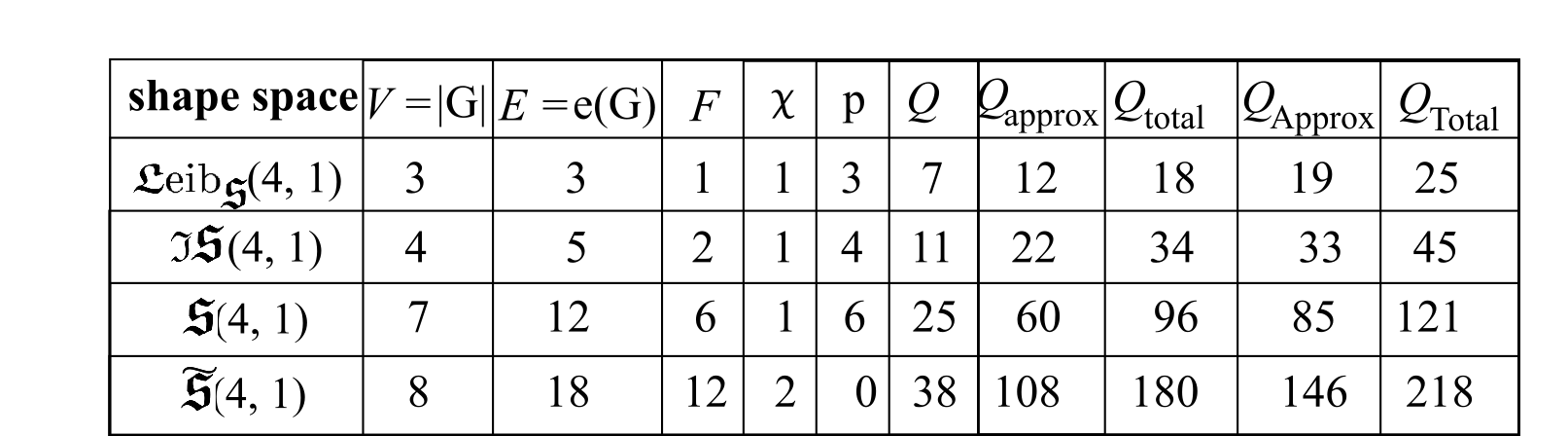}
\caption[Text der im Bilderverzeichnis auftaucht]{        \footnotesize{Number of qualitative types with Lagrangian separation decor.
Note that the first rows the current Figure and of Fig II.18 are the same due to underlying isomorphic graphs, and that the same applies again to their second rows.
}}
\l{(3, 2)-Q-Lag} \end{figure}          }

\section{Right, acute and obtuse triangles}\l{Right}

\subsection{Overview}\l{Overview}

\n{\bf Qualitative type 5)} The relative angle notion of perpendicularity supports the notion of {\it right-angled triangles}, $\perp$.\footnote{As all relative separations are edges for 
triangles, these right angles are between edges. 
On the other hand, Part IV's quadrilaterals will also have to contend with edge-diagonal and diagonal-diagonal perpendiculatities.}  

\m

\n{\bf Qualitative type 6)} Moreover, if a triangle contains an angle which is larger than the right angle, 
it is an {\it obtuse triangle}, $\obtuse$.  

\m

\n{\bf Qualitative type 7)} Finally if none of a triangle's angles are as large as a right angle, it is an {\it acute triangle}, $\angle$.  

\m

\n{\bf Remark 1} Obtuse comes in three labelling choices, but acute is a labelling-independent concept. 

\m

\n{\bf Remark 2} For a right-angled triangle, the side opposite the right angle, say a, is known as the {\it hypotenuse}, and the other two sides as the {\it legs}. 
Pythagoras' Theorem applies:
\be 
a^2 = b^2 + c^2 \m .
\ee
%
%
Given one of the non-right angles therein, say $\alpha$ the legs are designated {\it opposite} and {\it adjacent} relative to it (Fig \r{Special-Triangles}.a).  
Trigonometric definitions ensue
\be 
\sin \, \alpha \:= \frac{\mbox{opposite}}{\mbox{hypotenuse}} \m,  \m \m
\cos \, \alpha \:= \frac{\mbox{adjacent}}{\mbox{hypotenuse}} \m,  \m \m
\tan \, \alpha \:= \frac{\mbox{opposite}}{\mbox{adjacent}}   \m ;  \m \m
\ee  
by Pythagoras, 
\be 
\sin^2 \alpha + \cos^2 \alpha = 1 \m . 
\l{s+c=1}
\ee 
\n{\bf Remark 2} For the general triangle, Pythagoras generalizes to the {\it cosine rule}
\be 
a^2 = b^2 + c^2 - 2 \, b \, c \, \cos \, \alpha  \m ; 
\ee
One also has the {\it sine rule} 
\be 
\frac{a}{\sin \, \alpha}  \es
\frac{b}{\sin \, \beta}   \es
\frac{c}{\sin \, \gamma}       \m .
\ee
\n{\bf Remark 3} Given s-s-s, the perimeter $S$ follows immediately.   
Given s-$\alpha$-s, the cosine rule supplies the final side, thus giving another formula for $S$.  

\m

\n{\bf Remark 4} Right angles involve picking a hypotenuse pair of vertices, now with mirror image distinction (except for the isosceles right triangle).  

\m
 
\n{\bf Remark 5} The right isosceles triangle, detailed in Fig \r{Special-Triangles}.b), will play a significant role in this treatise. 
%
{\begin{figure}[ht]
\centering
\includegraphics[width=0.15\textwidth]{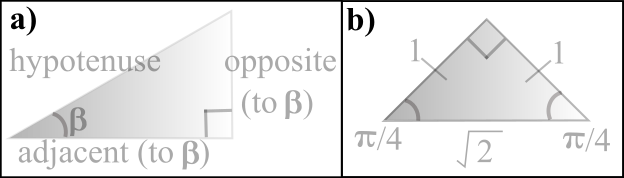}
\caption[Text der im Bilderverzeichnis auftaucht]{\footnotesize{a) Notation for right-angled triangles. 
b) The right isosceles triangle $\mI^{\perp}$.  
Its symmetry, separation uniformity and angular uniformity are as for any other isosceles triangle.}} 
\l{Special-Triangles}\end{figure} } 

\subsection{Right angled triangle specific congruence and similarity conditions}\l{RHC-S}

\n{\bf Congruence condition 5)}  $\perp$-h-s: two right angled triangles with shared-length hypotenuses and another pair of equal side-lengths.  

\m	
	
\n{\bf Similarity condition 5)}  $\perp$-r: two right angled triangles with a shared hypotenuse-to-side ratio. 
 
\m

\n{\bf Remark 2} Qualitative types 5) to 7) give splits for, firstly, the scalene triangles (see also Fig \r{BIEMC-4}),
\be
\mS =  \acute \coprods  \perp \coprods  \obtuse  \m,
\ee
and, secondly, for the isosceles triangles, 
\be
\mI = \mI^{\acute} \coprods \mI^{\perp} \coprods  \mI^{\obtuse}  \m . 
\ee
(see Fig \r{Special-Triangles} for $\mI^{\perp}$).
This split moreover affects exclusively the flat isosceles triangles -- 
\be
\mI^{\sF} = \mI^{\sF\angle} \coprods \mI^{\perp} \coprods \mI^{\obtuse} \m 
\ee
-- since the equilateral triangle and all the tall isosceles triangles are acute. 

\m

\n{\bf Remark 3} $\perp$ itself is further split into
\be
\perp \es \perp^1 \coprods  \mI^{\perp}  \m .  
\ee
The furthermore degenerate case B is arbitrarily close to all of obtuse, right and acute triangles, though there is a sense in which is the 2-right-angle limit 
of the arbitrarily tall isosceles triangle, by which it belongs to the right and isosceles arcs. 

\m

\n{\bf Remark 4} The degenerate shapes $\mC^1$ can moreover be adjoined to the obtuse triangles to make a slightly more general notion of obtuse.
%
{            \begin{figure}[!ht]
\centering
\includegraphics[width=0.8\textwidth]{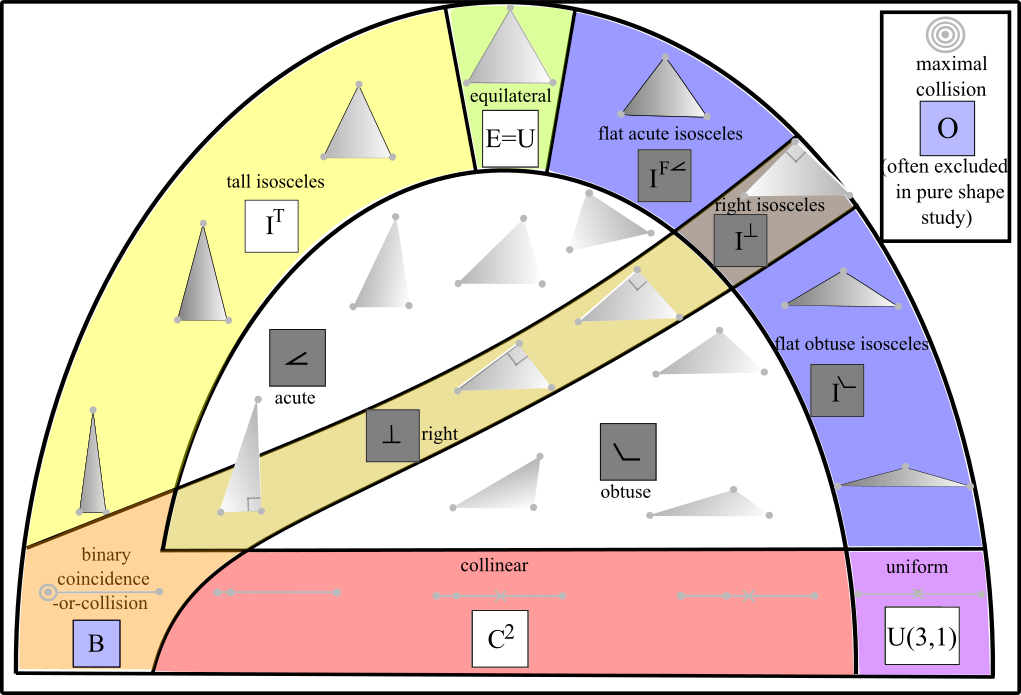}
\caption[Text der im Bilderverzeichnis auftaucht]{        \footnotesize{Shape-space-topological-manifold-level diagram of the 
shapes-in-space metric Lagrangian level decor, now updated to include rightness and the ensuing acute-to-obtuse distinction.}}
\l{BIEMC-4} \end{figure}          }

\m

\n{\bf Remark 5} Let us also comment on $\perp$-h-s and $\perp$-r only applying to some triangles: right-angled ones, which are moreover not generic.  
This is of limited value when the aim is to treat the entirety of the triangles on a common footing, as is the case in Shape Theory through setting up the (scale-and-)shape space.

\subsection{On which congruence and similarity conditions are complete as regards degenerate configurations}\l{Cong-Sim-Disc}

\n{\bf Remark 1} This is a good point at which to assess which of the remaining, generically applicable standard congruence and similarity conditions 
extend well-determinedly to include the non-generic degenerate cases. 
Namely, the collinear C and binary coincidence-or-collision B normalizable shapes, and, in the Euclidean case, the maximal collision O as well.  
This is motivated by most pure-shape considerations including the former, and most shape-and-scale considerations additionally including the latter.

\m

\n{\bf Remark 2} Let us start by considering C.   

\m

\n{\bf Congruence condition 1)} s-s-s's triangle inequality is now saturated, so s-s data is fully determining. 

\m

\n{\bf Congruence condition 2)} s-$\alpha$-s data functions as usual, for a taking a collinear value (i.e.\ 0 or $\pi$).

\m

\n{\bf Congruence conditions 3) and 4)} $\alpha$-s-$\alpha$ and $\alpha$-$\alpha$-s data have both their a's take collinear values, 
but their remaining s information does not suffice to uniquely determine the shape-and-scale.

\m

\n{\bf Similarity condition 1)} r-r inherits a dependency from the saturated triangle inequality, so r data is fully determining. 

\m

\n{\bf Similarity conditions 2) and 3)} r-$\alpha$ functions as usual, for a taking a collinear value.

\m

\n{\bf Similarity condition 4)} $\alpha$-$\alpha$ takes collinear values but, being bereft of length ratio information, does not uniquely determine the configuration.  

\m

\n So s-s-s remains operational subject to an interpretational caveat, whereas s-$\alpha$-s carries over {\sl without comment}.  
r-r and r-$\alpha$ then follow suit. 

\m

\n{\bf Remark 3} We next comment on B.   

\m

\n{\bf Congruence condition} 1) s-s-s here reduces to s-s-0 which is determined by a single s datum.  

\m

\n{\bf Congruence condition 2)} s-$\alpha$-s reduces to a's collinear value and 1 nonzero side datum (the nonzero s's are equal). 

\m

\n{\bf Congruence conditions 3)} and 4) $\alpha$-s-$\alpha$ and $\alpha$-$\alpha$-s break down as only 1 relative angle is defined.  

\m

\n{\bf Similarity condition 1)} r-r reduces here to a single finite nonzero ratio r = 1.

\m

\n{\bf Similarity condition 2)} r-$\alpha$ included reduces to r = 1 and $\alpha = 0$. 

\m

\n{\bf Similarity condition 3)} r-$\alpha$ not included reduces to r = 0 or $\infty$ and $\alpha = 0$. 

\m

\n{\bf Similarity condition 4)} $\alpha$-$\alpha$ breaks down as only 1 relative angle is defined.  

\m

\n Thus, by somewhat different arguments, s-$\alpha$-s and r-$\alpha$ (now specifically included) are the most desirable for the extension, 
whereas s-s-s and r-r also pass muster with interpretational caveats.  
This suffices to crown s-$\alpha$-s as the conventional conguence condition that is most suitable to the entirety of the shape space of triangles.

\m

\n{\bf Remark 4} For the shape-and-scale case, we also need consider the maximal coincidence-or-collision O.

\m

\n{\bf Congruence condition 1)} s-s-s here reduces to 0-0-0.  

\m

\n{\bf Congruence conditions 2) to 4)} all break down since now no relative angles are defined.  

\m

\n Thus for Shape-and-Scale Theory, it is s-s-s which extends to the final point. 
One interpretation of this is to make do with s-s-s for all shapes; 
another is to view s-s-s as a valid shape-and-scale space coordinate chart to approach O, whereas s-$\alpha$-s could be in use everywhere else.

\section{Medians and further Cevian structure}\l{Cevians}

\n{\bf Definition 1} The {\it medians} of a triangle are as per \r{Cevians-SB}.a)-b).  
It will also be useful for us to use $m_i$, $i = 1$ to $3$ to denote $m_a$, $m_b$, $m_c$.  
%
{            \begin{figure}[!ht]
\centering
\includegraphics[width=0.84\textwidth]{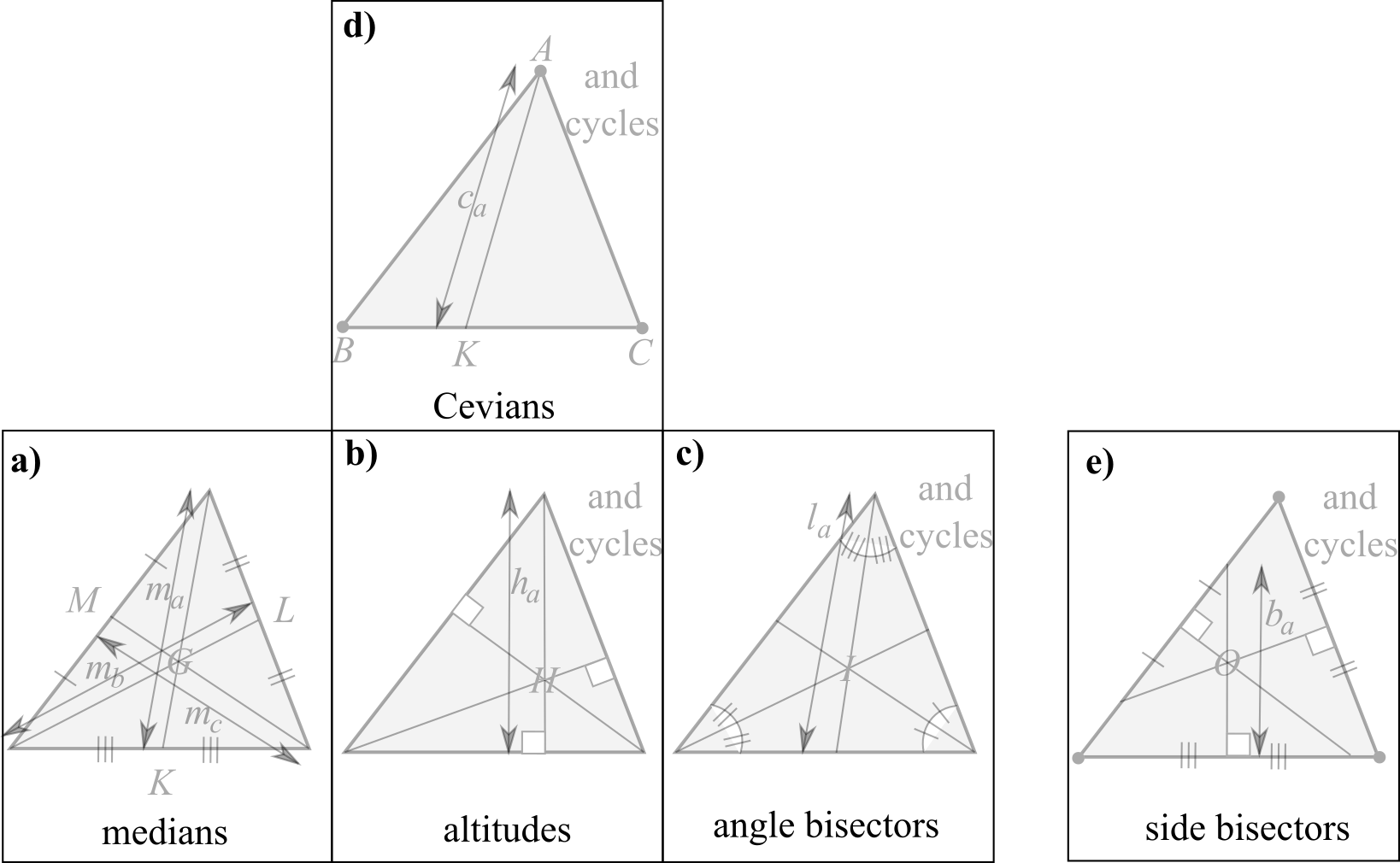}
\caption[Text der im Bilderverzeichnis auftaucht]{        \footnotesize{a) Definition of the medians; these concur at the {\it centroid}, alias {\it centre of mass}, $G$.
b) Definition of the altitudes, which intersect at the {\it orthocentre} $H$.  
c) The internal angle bisectors $h_a$ from each vertex to its opposite side; these intersect at the {\it incentre} $I$ 
d) General Cevian set-up, of which medians and internal angle bisectors are a particular subcase. 
e) The sides' perpendicular bisectors; 
note that these are not in general Cevians, via their not going through the vertex opposite the side; 
these moreover intersect at the {\it circumcentre} O. }  } 
\l{Cevians-SB} \end{figure}           }

\m

\n{\bf Remark 1} By treating the sides and the medians on an equal footing -- which we motivate in Secs (\r{MSIIDFF}) and (\r{RJSI}) -- 
we have more definitions (or at least accordances of equal significance) than in hitherto standard treatments of triangles, starting with the following.  

\m

\n{\bf Definition 2} The {\it medimeter} $M$ is 
\be
M  :=  m_a + m_b + m_c 
  \es \sum_i m_i          \m,   
\ee
whereas the {\it semi-medimeter} is   
\be
s = \frac{M}{2}  \m .  
\ee
\n{\bf Remark 2} The perimeter and medimeter can furthermore be viewed as first moments of sides and medians respectively.
The second moments counterparts of each of these also enter the current treatise, as follows. 

\m

{            \begin{figure}[!ht]
\centering
\includegraphics[width=0.85\textwidth]{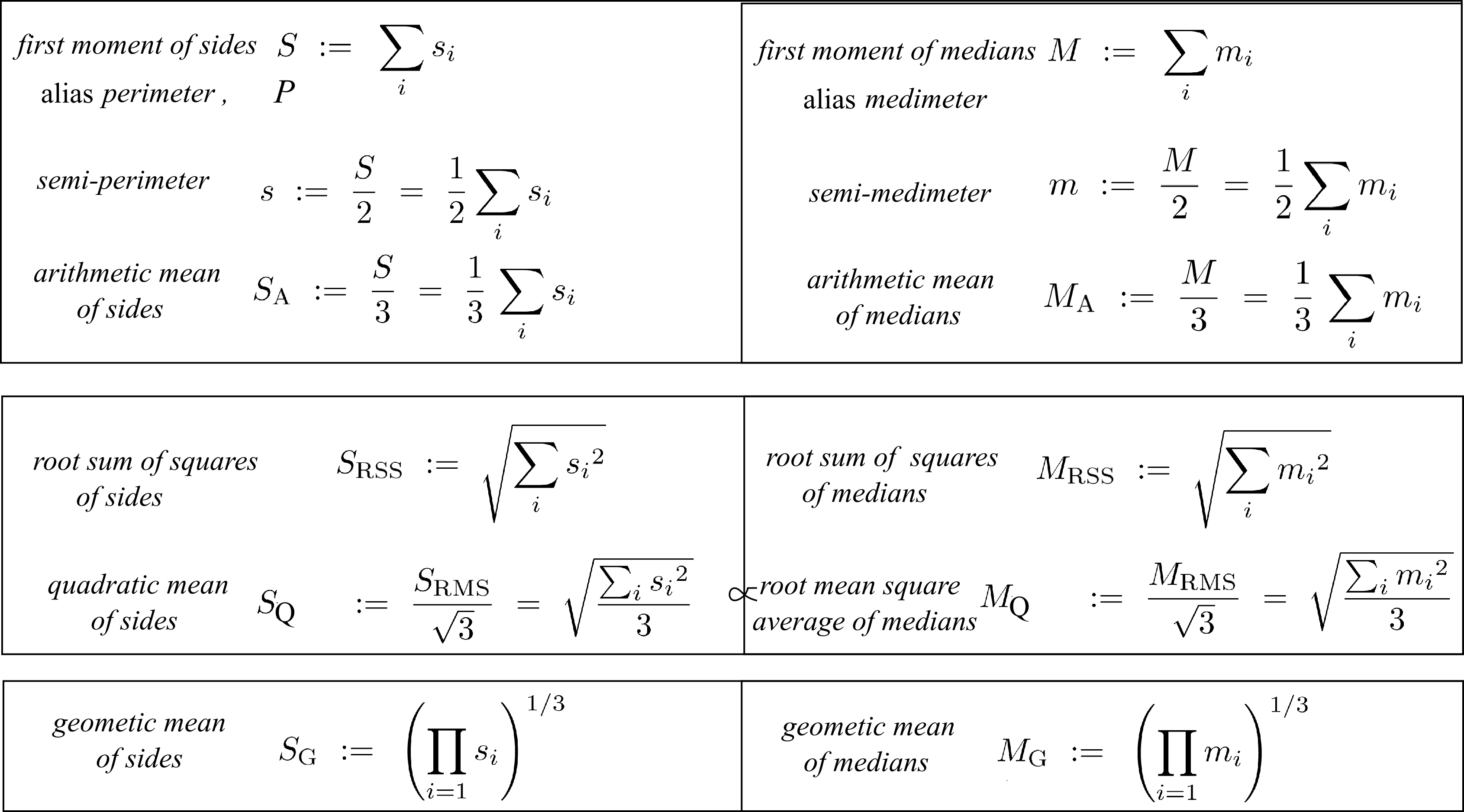}
\caption[Text der im Bilderverzeichnis auftaucht]{        \footnotesize{Three pairs of variables, with one inter-relation and various useful and conceptually meaningful rescalings. 
Note that we are using the letters 'M' and 'S' to denote quantites which are dimensionally lengths, while also keeping track of which are side-based and which are median-based.}  } 
\l{6-variables} \end{figure}           }

\m 

\n{\bf Definition 4} The segment 
\be 
c_a := AK
\ee 
from a triangle's vertex $A$ to a point $K$ on the opposite side $BC$ is known as a {\it Cevian} (pronounced `Chevian', and depicted in Fig \r{Cevians-SB}.d). 
We also denote $c_a$, $c_b$ and $c_c$ collectively by $c_i$.  

\m 

\n{\bf Remark 3} This is a more general concept than a median; further examples of Cevian are Fig \r{Cevians-SB}.b)'s {\it altitudes}                $h_i$ 
                                                                                           and Fig \r{Cevians-SB}.c)'s {\it internal angle bisectors} $l_i$.

\m
																						   																						   
\n{\bf Remark 4} The {\it Cevian-based ratios} 
\be 
\frac{KC}{BC} \m \mbox{ and } \m \frac{BK}{BC} 
\ee
moreover have affine significance, entering the following Theorem by Ceva himself. 

\m

\n{\bf Theorem 1 (Ceva's Concurrence Theorem)} $AK$, $BL$ and $CM$ are collinear iff 
\be
\frac{BK}{KC} \, \cdot \, \frac{CL}{LA} \, \cdot \, \frac{AM}{MB} \es - 1  \m .  
\ee 
\n(See e.g. \c{PS70, Silvester} for proofs.)  

\m

\n{\bf Exercise 1} Use Ceva's Concurrence Theorem to show that each of medians, altitudes, and internal angle bisectors are concurrent.  
Also give a simpler physical argument for median concurence.  

\m

\n{\bf Remark 5}  In Euclidean geometry, moreover, Cevians have lengths as well as affine properties, and it is in fact the following `Cevian length theorem' that we make use of.  

\m

\n{\bf Theorem 2 (Stewart's Cevian-length Theorem)} Let $\triangle ABC$ be a triangle with $K$ an arbitrary point on side $BC$.  
Then
\be
{AK}^2 \es \frac{KC}{BC} \, {AB}^2 \m + \m \frac{BK}{BC} \, {AC}^2 - BK \, KC  \m .  
\ee 
\n(See e.g.\ \c{PS70} for a proof.)

\m

\n{\bf Corollary 1} i) The squares of the medians' lengths are given by 
\be
m_a\mbox{}^2 \es \frac{2 \, b^2 + 2 \, c^2 - a^2}{4}  \m \mbox{ and cycles } .
\l{med-2}
\ee
\n ii) The quadratic means of sides and of medians are related by 
\be 
S_{\sQ} \es \frac{2}{\sqrt{3}} \, M_{\sQ} 
         =        \kappa       \, M_{\sQ}  \m, 
\l{0.2}
\ee
for 
\be 
\kappa \:= \frac{2}{\sqrt{3}}              \m . 
\ee 
\Proof i)  This readily follows from Stewart's Theorem, as per Problem 1 of \c{IMO}. 

\m

\n ii) then follows immediately from summing i) over all cycles, and square-rooting. $\Box$

\m

\n{\bf Remark 6} Thus the middle pair of quantities defined in Fig \r{6-variables} are in fact not independent from each other, leaving us with five independent quantities.  

\m

\n{\bf Remark 7} $\kappa$ is for now to be treated as a number, powers of which are recurrent in the geometrical theory of the triangle. 
A conceptual and physical meaning for $\kappa$ will moreover be elucidated in Sec \r{RJSI}. 

\m

\n{\bf Corollary 2} i) An incipient Linear Algebra form for this is \c{2-Herons}  
\be
m_i\mbox{}^2 \es \frac{1}{4} \, B_{ij} s_j\mbox{}^2 \m,
\l{0.4}
\ee
for  
\be
\Bmat \:=   \left( \s{  \s{  \mbox{$   -              1  \m \m \m  2 \m \m \m \m  2$}  }  
                                  {  \mbox{$\m \m \m 2  \m    -               1 \m \m \m  2$}  }  }
						    	  {  \mbox{$\m \m \m 2  \m \m \m  \m 2 \m    -               1$}  } \right)
\l{0.5}
\ee
a symmetric matrix 
\be
\Bmat = \Bmat^{\sT} \mma \mbox{i.e.\ in components } \m B_{ij} = B_{ji}  \m .
\l{Sym-B}
\ee
\n ii) Inverting, 
\be
s_i\mbox{}^2 \es \frac{4}{9} \, B_{ij} m_j\mbox{}^2  \m . 
\l{0.6}
\ee
\n{\bf Remark 8} That the same $B_{ij}$ appears in the inverted expression indicates that $B_{ij}$ is proportional to an {\it involution} $J_{ij}$, 
i.e.\ it is a matrix such that 
\be
\invol^2 = \1mat: \mbox{ the identity matrix } .  
\l{0.7}
\ee
We can thereby further tidy up \c{2-Herons} Corollary 2 Linear Algebra formulation by identifying and using $\invol$, as follows. 

\m

\n{\bf Corollary 3 (Sides--Medians involution)} \c{2-Herons} i) 
\be
m_i\mbox{}^2 = \kappa^{-2} \, J_{ij} s_j\mbox{}^2  \m, 
\l{0.8}
\ee
where 
\be
\invol \:= \frac{1}{3}   \left( \s{  \s{  \mbox{$   -              1  \m \m \m  2 \m \m \m \,  2$}  }  
                                               {  \mbox{$\m \m \, \, 2  \m    -               1 \m \m  \m \,  2$}  }  }
								       		   {  \mbox{$\m \m \, 2  \m \m \m  2 \m    -               1$}  } \right)             \m 
\ee
is the {\it sides-medians involution}. 
This of course remains symmetric, 
\be
\invol = \invol^{\sT}  \m, \m  \mbox{ i.e.\ in components } 
\m J_{ij} = J_{ji}     \m .
\l{Sym}
\ee
\n ii) Inverting, 
\be
s_i\mbox{}^2 = \kappa^2 \, J_{ij} m_j\mbox{}^2  \m .   
\l{0.9}
\ee
(See \c{Ineq} for a proof.)

\m

\n{\bf Lemma 1 (Perimeter bounds on the sum of the medians $M$} 
\be
\kappa^{-2} S \leqs M 
              \leqs S \m . 
\l{Ineq-1}
\ee 
\n(See \c{Court} for a proof.)  

\m

\n{\bf Remark 6} Including degenerate triangles -- as is required in Shape Theory -- means that indeed the saturated version of these inequalities is required.
The value 1 occurs at the binary coincidence-or-collision B, and $\kappa^{-2}$ at the uniform collinear configuration U(3, 1).

\section{Associated circles}
%
{            \begin{figure}[!ht]
\centering
\includegraphics[width=0.66\textwidth]{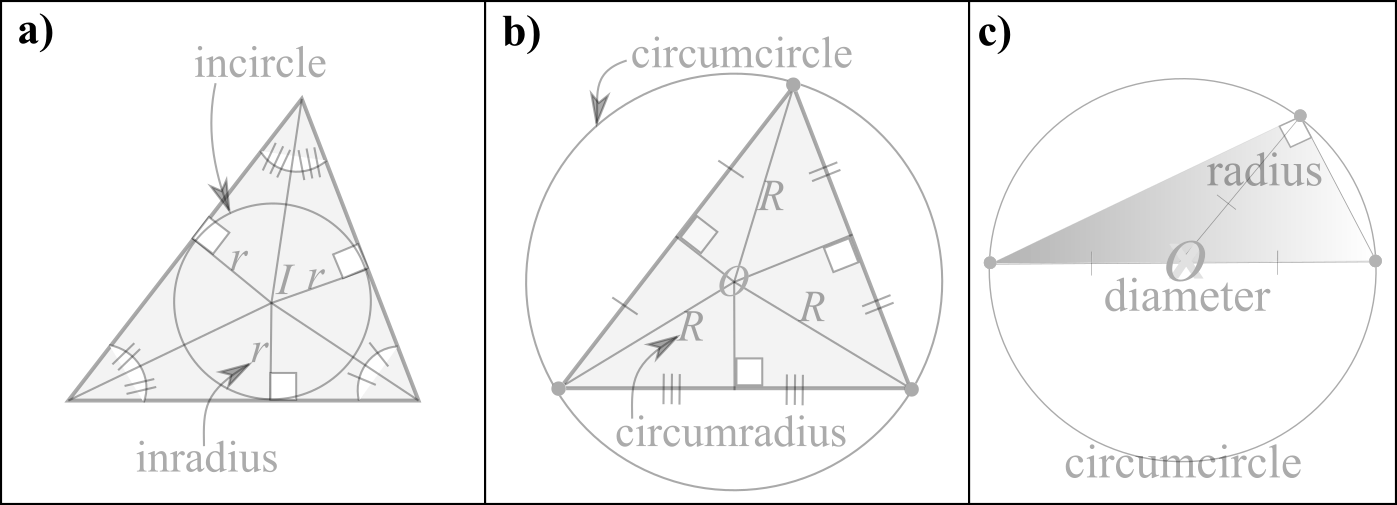}
\caption[Text der im Bilderverzeichnis auftaucht]{        \footnotesize{
a) The angle bisectors define the incircle as indicated, with incentre $I$ and inradius $r$. 

\m

\n b) The perpendicular bisectors of the sides define the circumcircle as indicated, with circumcentre $O$ and circumradius $R$. 

\m

\n c) Diameter subtends a right angle, which can be reinterpreted as the circumcircle of a right-angled triangle, for which the hypotenuse is a diameter, $2 \, R$.
The corresponding median is, moreover, the corresponding radius $R$, and thus is in 1: 2 proportion with the hypotenuse.}  } 
\l{In-Circum-DR} \end{figure}           }

\n {\bf Definition 1} The {\it inradius}     $r$ is defined as per Fig \r{In-Circum-DR}.a), 
                  and the {\it circumradius} $R$            as per Fig \r{In-Circum-DR}.b).

\m

\n {\bf Remark 1}  At least prima facie, these are secondary features extrinsic to the triangle itself.

\m

\n {\bf Remark 2} Right-angled triangles' circumcircles play a subsequent role in the current Part as per Fig \r{In-Circum-DR}.c).

\m

\n {\bf Remark 3} Inradius and circumradius moreover provide bounding inequalities on sums of (squares of) sides or medians; these are either very simple or quite well-known.  

\m

\n{\bf Lemma 1 (Preliminary crude joint bounds on sides and medians)}  
\be
2 \,  r    \ls       s_i                 \mma      m_i                      \leqs 2 \, R \m, 
\l{Crude-1}
\ee  
\be
6 \,  r    \ls        S                  \mma      M                        \leqs 6 \, R \mma \mbox{and }
\l{Crude-2}
\ee 
\be 
12 \, r^2 \leqs S_{\sR\sS\sS}\mbox{}^2   \mma      M_{\sR\sS\sS}\mbox{}^2    
          \leqs 12 \, R \m .
\l{Crude-3}
\ee 
(See \c{Ineq} for a proof.)

\m

\n{\bf Remark 4} (\r{Crude-2}) and (\r{Crude-3}) are mostly provided for comparison with subsequent sharper inequalities, which do discern between sides and medians.   

\m

\n{\bf Lemma 2} The root sum of squares is bounded by the square of the semi-perimeter length, $s$, according to 
\be
s^2 \leqs \kappa^{-2} \, S_{\sR\sM\sS}\mbox{}^2  
     \es                 M_{\sR\sM\sS}\mbox{}^2                                 \m .
\l{0.10}
\ee 
(See \c{Ineq} for a proof.)

\section{The (tetra-)area variable}\label{Tet-1}

\n{\bf Definition 1} We use $\mbox{Area}(ABC)$, or $\mbox{Area}$ for short when unambiguous, to denote the area of $\triangle\,ABC$.

\m

\n{\bf Lemma 1} $Area$ is given by, in terms of s-$\alpha$-s data, 
\be
Area \es \frac{1}{2} \, a \, b \, \sin \, \gamma \m \mbox{ or cycles } .
\l{Basic-Area}
\ee 
\n{\bf Theorem 1 (Heron's formula)} $Area$ is given by, in terms of 3-sides data
\be
Area = \sqrt{s(s - a)(s - b)(s - c)}  \m . 
\l{Heron}
\ee 
This is a classical result, known since the first century A.D. \c{Hero}; see e.g.\ \c{Coxeter} for a modern-era proof.   

\m

\n{\bf Corollary 1 (Expanded Heron's formula)}  The square of the {\it tetra-area} $T$ is given by 
\be
T^2 :=   \{4 \times Area  \}^2 
    \es  (2 \, a^2 b^2 - c^4) + \mbox{ cycles}  \m . 
\l{Exp-Heron} 
\ee
\n{\bf Corollary 2} The expanded form (\r{Exp-Heron}) of Heron's formula (\r{Heron}) can be recast in Linear Algebra terms as the quadratic form 
-- in squares $a_i\mbox{}^2$, so it is quartic in the $a_i$ themselves --
\be 
T^2 = H_{ij} s_i\mbox{}^2 s_j\mbox{}^2
\l{Heron-form}
\ee 
for `{\it Heron matrix}' 
\be
\Heron \:= \frac{1}{3} \left( \s{  \s{  \mbox{$   -             1  \m \m \m  1 \m \m \m  1$}  }  
                                             {  \mbox{$\m \m \, 1  \m    -               1 \m \m \m  1$}  }  }
										     {  \mbox{$\m \m \, 1  \m \m \m  1 \m    -               1$}  } \right)  \m .
\ee
\n{\bf Theorem 2 (Medians' Heron formula)} \cite{DCKay}, alias `area from median data' formula. 
\be
\mbox{Area} \es \kappa^2 \, \sqrt{\ms(\ms - m_a)(\ms - m_b)(\ms - m_c)}   \m . 
\l{Median-Heron}
\ee
(See \c{2-Herons} for a proof.) 

\m

\n{\bf Remark 2} While traditional geometric proofs of this are not uncommon \c{Benyi}, my Linear Algebra proof above contains an insight that these miss. 
Namely, that the side--median involution matrix $\invol$ and the `Heron matrix' $\Heron$ {\sl commute}, 
\be
\mbox{\bf [}  \invol \mbox{\bf ,} \, \m \Heron \mbox{\bf ]} \es 0 \mma \mbox{i.e.\ }  \m 
              \invol \, \Heron = \Heron \, \invol                                     \m .
\ee
\n{\bf Remark 3} It is because of this that the `median-Heron' matrix $\u{\u{M}}$ in the a priori conceptual form of the medians-data Heron's formula, 
\be 
Area^2 = M_{ij} m_i\mbox{}^2 m_j\mbox{}^2  \m, 
\l{LA-Median-Heron-Conceptual}
\ee 
is just proportional to the `Heron matrix' itself, 
\be
\Mmat \es \frac{1}{9} \, \Heron            \m . 
\ee 
\n{\bf Remark 4} In summary, the sides-Heron and medians-Heron formulae are 
\be
\sqrt{s(s - a)(s - b)(s - c)}  =  \mbox{Area} 
                              \es \kappa^2 \sqrt{\ms(\ms - m_a)(\ms - m_b)(\ms - m_c)}  \m .  
\l{Summary-1}
\ee 
\n{\bf Lemma 2} ($r$ and $R$ recast in terms of intrinsic variables) i) 
\be
Area = r \, s                          \m .  
\l{Area-r-s}
\ee
ii) 
\be
Area = r_a (s  - a) \m \mbox{ and cycles } .    
\l{Area-ra-s}
\ee
iii) 
\be
S_{\sG}\mbox{}^3 = a \, b \, c = 4 \, s \, r \, R  = 4 \, Area \, R  = R \, T  \m .  
\ee 
\Proof These are all standard; see e.g.\ \c{Coxeter} for i), ii), as well as for a neat way of obtaining Heron's formula (\r{Heron}) from ii).  
The first equality of iii) is Fig \r{6-variables}.e)'s definition, 
the third follows from the extended sine rule and formula (\r{Basic-Area}) for the area, 
the second then follows from i). 
The new fourth equality then just follows from the third by the definition of tetra-area, $T$. $\Box$

\subsection{Inequalities reformulated intrinsically}\l{IRI}

\n{\bf Lemma 1 (Inradius--circumradius bounds)} i) The medimeter is bounded according to  
\be
r \leqs \frac{M}{9} 
  \leqs \frac{R}{2}                                                        \m . 
\l{0.14}
\ee
ii) The sum squared of medians is bounded according to 
\be 
r^2 \leqs \frac{  M_{\sR\sM\sS}\mbox{}^2  }{  27  } 
    \leqs \frac{            R^2           }{   4  }                        \m .  
\l{0.15} 
\ee
\Proof This is a subset of worked Problems 1 and 2 of \c{IMO-2}. $\Box$

\m

\n{\bf Remark 1} On the one hand,    i) tightens Lemma 2.ii) from $6 \, r$ to $9 \, r$ -- a factor of 3/2 -- and from $6 \, R$ to $9 \, R/2$: a factor of 4/3. 

\m

\n               On the other hand, ii) tightens Lemma 2.iii) from $12 \, r$ to $27 \, r$ --  a factor of $9/4  = (3/2)^2$ -- 
                                                           and from $12 \, R^2$ to $27 R^2/4$: a factor of $16/9 = (4/3)^2$.

\m

\n{\bf Remark 2} Inradius $r$ and circumradius $R$ are moreover not natural or primary constructs from a shape-theoretic point of view, being `extrinsic' features of triangles. 
We can instead reformulate these quantities as per Lemma 1.i) and iii), giving the following further form for the inequalities.

\m

\n{\bf Corollary 1}  
\be
\frac{T}{S}                 \ls          s_i     \mma  
                                         m_i     \mma  
		    							 S_A     \mma  
										 M_A     \leqs   2 \, \frac{S_{\sG}\mbox{}^3}{T}      \mma \mbox{and } 
\ee  
\be 
\left(\frac{T}{S}\right)^2  \ls       S_{\sQ}\mbox{}^2 ,  M_{\sQ}\mbox{}^2 
                            \leqs  4 \, \left(\frac{S_{\sG}\mbox{}^3}{T}\right)^2                                               \m .
\ee
\n{\bf Corollary 2}
\be
\sqrt{3} \, \frac{T}{S} \leqs    \kappa \, M_{\sA}    
                        \leqs    \sqrt{3} \, \frac{S_{\sG}\mbox{}^3}{T}              \mma \mbox{and } 
\l{0.14-b}
\ee
\be 
3 \left(\frac{T}{S}\right)^2   \leqs              S_{\sQ}\mbox{}^2 
                                  =   \kappa^2 \, M_{\sQ}\mbox{}^2 
                               \leqs  3 \left(\frac{  S_{\sG}\mbox{}^2  }{  T  }\right)^2                                        \m .  
\l{0.15-b} 
\ee
\n{\bf Theorem 1} These two inequalities furthermore admit a joint repackaging, as 
\be
\sqrt{3} \, \frac{T}{S} \leqs \kappa \, M_{\sA} 
                        \leqs \kappa \, M_{\sQ} 
					      =             S_{\sQ}
                        \leqs \sqrt{3} \, \frac{S_{\sG}\mbox{}^3}{T}                                                              \m . 
\l{0.14-c}
\ee
(See \c{Ineq} for a proof.)

\subsection{Ratio quantities}

\n{\bf Structure 1} We started with seven quantities, but showed two to be equal up to proportion.
Thus five independent ratios are supported.
For reasons which later become clear, we pick two side-quantity ratios, two median-quantity ratios and the isoperimetric ratio for this purpose, as follows.  
%
{            \begin{figure}[!ht]
\centering
\includegraphics[width=0.65\textwidth]{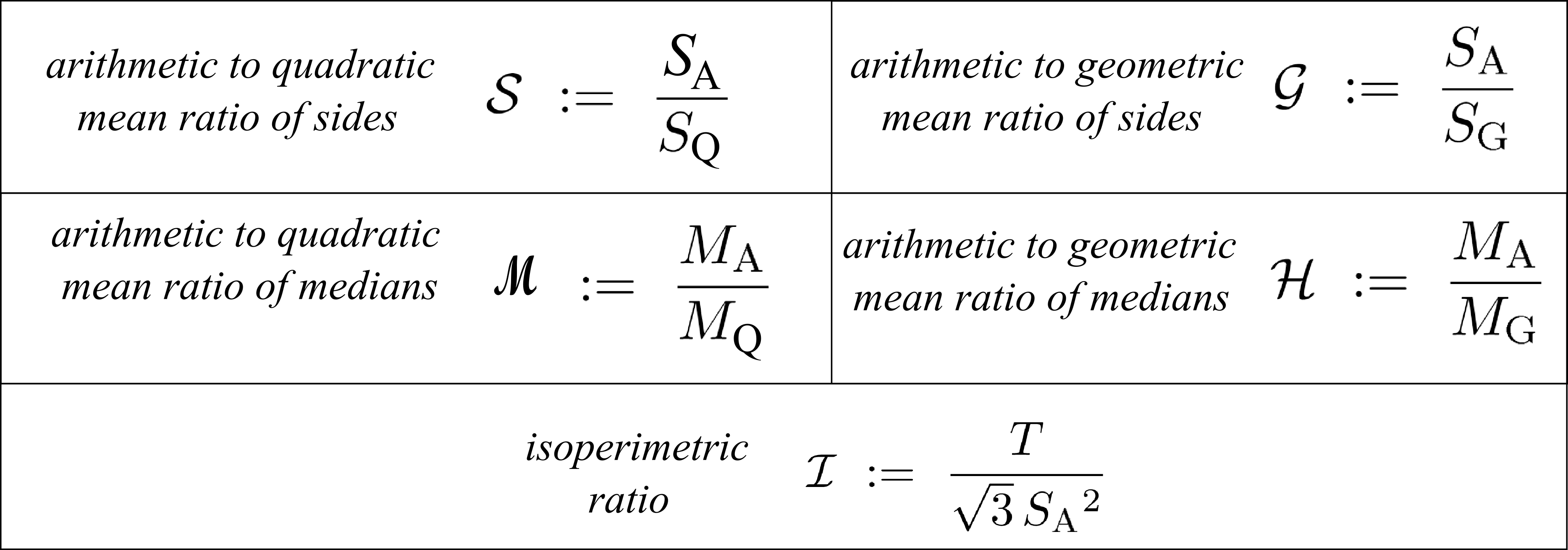}
\caption[Text der im Bilderverzeichnis auftaucht]{        \footnotesize{A choice of five independent ratios, as supported by $Area$ and Figure \r{6-variables} 
and five other independent quantities.}   } 
\l{5-ratios} \end{figure}           }

\m

\n{\bf Definition 1} The five independent ratios the current treatise conceives in terms of are given in Fig \r{5-ratios}.  
  
\m

\n{\bf Remark 1} These are all scaled to have range 0 to 1.  
Setting this up requires finding and evaluating their maxima (see \c{A-Perimeter} for this and further extremization calculations.  

\m

\n{\bf Remark 2} To carry over to Shape Theory, we first recast all the inequalities in Sec 2 in terms of ratios.  
We first define the quantities to appear in these ratios, which are composites of some of the above five `most natural' ratios.  

\m

\n{\bf Definition 2} 
\be 
{\cal F} \:=  \frac{  M  }{  S  } 
         \es  \frac{  M_{\sA}  }{  S_{\sA}  } 
\ee 
is the {\it medimeter to perimeter ratio}.  

\m

\n{\bf Remark 3} 
\be 
\frac{  M_{\sQ}  }{  S_{\sQ}  }
\ee 
is sterile by the constancy implied by (\r{7}).  
The following alternative quantity, however, does have nontrivial shape-theoretic content (and is what occurs in the below inequalities). 

\m

\n{\bf Definition 3} The second ratio featuring in our inequalities is the {\it quadratic-to-arithmetic mean ratio of medians}, 
\be
{\cal L} \:= \frac{  M_{\sQ}  }{  S_{\sA}  } .  
\ee 
\n{\bf Definition 4 and Proposition 1} The third and fourth ratios featuring in the ratio version of the traditional form of the inequalities are the {\it inradius per unit perimeter}
\be 
\scN     \:= \frac{  r  }{  S  } 
         \es \frac{  T  }{  2 \, S^2  } 
		 \es \frac{  {\cal I}  }{  6 \sqrt{3}  }  \m,  
\l{n}
\ee
and the {\it circumradius per unit perimeter}.  
\be 
{\cal N} \:= \frac{  R  }{  S  } 
         \es \frac{  S_{\sG}\mbox{}^3  }{  T \, S  } 
         \es \frac{1}{3\sqrt3 \, {\cal I} \, {\cal G}^3}  \m .
\l{N}
\ee

\subsection{Rational form of inequalities}

\n{\bf Lemma 1}) 
\be 
\kappa^{-2} \leqs {\cal F}  
            \leqs 1                                                \m .  
\l{Rat-Ineq-1}
\ee

\m

\n{\bf Remark 1} The maximum of ${\cal F}$ is thus also 1 in accord with our standardization convention, but its range is $\left[\frac{3}{4}, \, 1\right]$ rather than [0, 1].
As \c{A-Perimeter} shows, the range for ${\cal L}$ is $\left[\frac{1}{2\sqrt{2}}, 1\right]$.
${\cal F}_{\sm\sa\sx}$ is at the binary coincidence-or-collision B, whereas ${\cal L}_{\sm\sa\sx}$ is at the equilateral triangle E.
Both ${\cal F}_{\sm\si\sn}$ and ${\cal L}_{\sm\si\sn}$ are at the uniform collinear configuration U(3, 1). 

\m

\n{\bf Theorem 1}  
\be
{\cal I} \leqs \kappa   \, {\cal F}  
         \leqs \kappa   \, {\cal L}
           =               {\cal S}^{-1}		 
		 \leqs             {\cal I}^{-1} {\cal G}^{-3}  \m . 
\l{Rat-Ineq-2}
\ee 
\n{\bf Remark 3} This is an intrinsic reconceptualization of in-and-circumradius bounds on median moments of Sec \r{IRI}, 
to now be in terms of the isoperimetric ratio and arithmetic-to-geometric sides ratio shape quantities.

\section{Lagrangian-level shape space structures including rightness}\l{Lag-Str-Right}
%
{            \begin{figure}[!ht]
\centering
\includegraphics[width=1\textwidth]{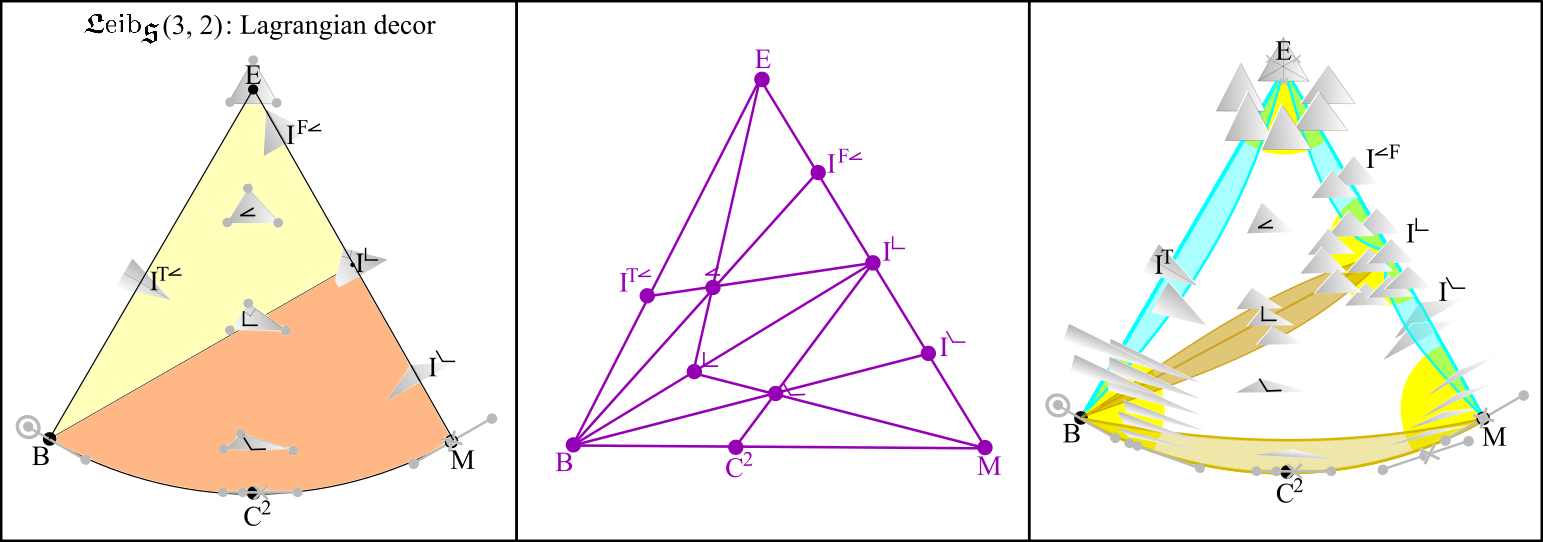}
\caption[Text der im Bilderverzeichnis auftaucht]{        \footnotesize{a) Metric-level shape space structure of $\Leib_{\tFrS}(3, 2)$ 
corresponding to the full metric Lagrangian level structure of triangular shapes-in-space at the Lagrangian level.} }
\l{Leib(3, 2)-Detail} \end{figure}          }
 
\n{\bf Remark 1} Including rightness, there are now $V = 4$ types of special vertices: E, B, M, I$^{\perp}$.

\m

\n{\bf Remark 2} There are also $E = 5$ edges: C, $\perp$, I$^{\sT}$, I$^{\sF\angle}$, and $I^{\sF\obtuse}$ which can be denoted just $I^{\obtuse}$.  

\m

\n{\bf Remark 3} Finally there are now $F = 2$ faces: obtuse and acute triangles, denoted by $\obtuse$ and $\angle$ respectively (explaining also 2 of the previous item's suffices). 

\m

\n{\bf Remark 4} These vertices, edges and faces can be located, and counted off, as per Fig \r{Leib(3, 2)-Detail}.a).

\m

\n{\bf Remark 5} The number of exact qualitative types is thus  
\be
Q := V + E + F = 4 + 5 + 2 = 11 \m .
\ee 
\n{\bf Remark 6} The adjacency graph \r{Leib(3, 2)-Detail}.b) has $V(\mbox{Adj}) = Q = 11$, and is moreover 2 $\mW_{6}$ wheel subgraphs with 2 common edges, so 
\be 
E(\mbox{Adj}) = 12 \times 2 - 2 
              = 22               \m .
\ee
\n{\bf Remark 7} See Figs  \r{(3,2)-Structure-Web}--\r{(3,2)-Full-Structure-Web} for an end-summary of special arcs and special points among the (3, 2) shapes.  
See Fig \r{(3, 2)-Q-Right} for further counts of qualitative types in the four (3, 2) shape spaces, as exhibited in Fig \r{(3, 2)-Approx-Detail} in the case of Leibniz space.   
%
%
{\begin{figure}[ht]
\centering
\includegraphics[width=1.0\textwidth]{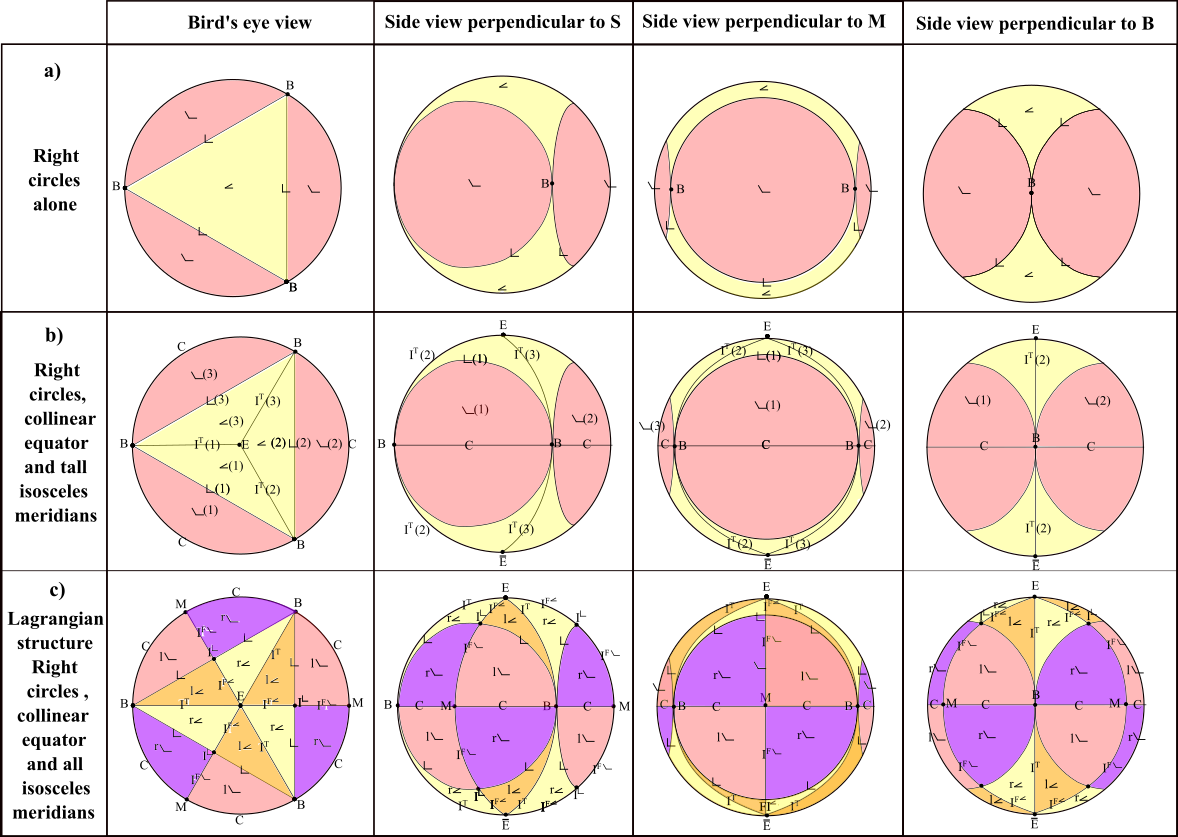}
\caption[Text der im Bilderverzeichnis auftaucht]{\footnotesize{Each shape space decoration 
is given a bird's eye view from above the pole in the left column and an equatorial view in the right column.  

\m

\n a) Cap-circles of rightness, kissing in pairs at the B points. 
The bird's eye view of these is the inscribed equilateral triangle.

\m

\n b) The meridians of tall isoscelessness are moreover related to this as the bird's eye view's equilateral triangle's medians; 
the numbers in round brackets here serve to distinguish between the various clustering choices.  

\m

\n c) Considering flat isosceles triangles as well yields a 24-face pattern, the bird's eye view of which is the smaller of the current treatise's two `kaleidoscopes'.}} 
\l{S(3, 2)-Detail}\end{figure} } 
%
{            \begin{figure}[!ht]
\centering
\includegraphics[width=0.75\textwidth]{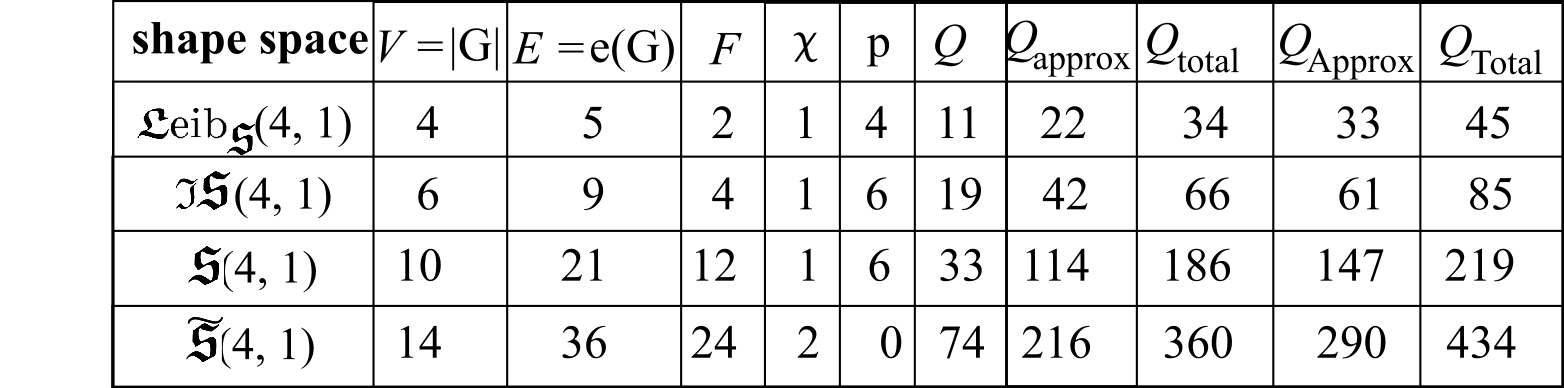}
\caption[Text der im Bilderverzeichnis auftaucht]{        \footnotesize{Number of qualitative types including the full Jacobian decor.
Note that the first row is the same as row 2 of Fig II.18 due to underlying isomorphic graphs; the second row's graph is isomorphic to the 4-fan, $\mF_4$.
}}
\l{(3, 2)-Q-Right} \end{figure}          }

\vspace{10in}

\vspace{10in}

\section{Jacobi level of structure for shapes in space}\l{Jac-Shape-in-Space}

\subsection{Mass-unweighted version}

\n{\bf Definition 1} For triangleland, we furthermore denote the Jacobi masses $\mu_1$ by $\mu_{\sss}$ -- {\it side Jacobi mass}    -- 
                                        and $\mu_2$ by $\mu_{\sm}$:   {\it median Jacobi mass}.  
This is {\sl possible} since the $\mu_i$ are cluster choice independent, 
and {\sl useful} by its replacing the 1 and 2 labels with more conceptually meaningful and memorable labels, $s$ for side and $m$ for median.  

\m

\n{\bf Definition 2} We follow suit by calling the triangle model's first and second relative Jacobi vectors the {\it side} and {\it median} vectors 
(for all that these {\sl are} cluster-dependent).  
I.e.\ 
\be 
R_{1}^{(a_i)} = s_i \mma
R_{2}^{(a_i)} = m_i \m . 
\ee 
\n{\bf Corollary 1} 
\be
R_2^{(i)\,2} \es \kappa^{-2} J_{ij} R_1^{(i)\,2} \mma \mbox{inverting to } \m
R_1^{(i)\,2} \es \kappa^2    J_{ij} R_2^{(i)\,2} \m . 
\l{2}
\ee
{\u{Proof}} Substitute $R_1 = a$ and $R_2 = m_a$ into the Linear Algebra form of the sides--medians relation (\r{0.8}). $\Box$

\subsection{Relative Jacobi separations inter-relations}\l{RJSI}

We now work in terms of the mass-weighted relative Jacobi coordinates provided in eq. (I.67).

\m

\n{\bf Corollary 1} 
\be
\rho_2^{(i)\,2} \es J_{ij} \rho_1^{(i)\,2} \mma \mbox{inverting to } \m
\rho_1^{(i)\,2} \es J_{ij} \rho_2^{(i)\,2} \m . 
\l{mw2}
\ee
\n{\bf Remark 1} Note that the mass-weighting cleans out the awkward numerical factor of $\frac{4}{3} = \kappa^2$ in eqs (\r{2}), revealing this factor to be 
\be
\kappa^2 \es \frac{4}{3} 
         \es \frac{\mu_{m}}{\mu_{s}}  
		 \es  \frac{(\mbox{median Jacobi mass})}{(\mbox{side Jacobi mass}} \m .                                                                          
\l{4/3} 
\ee 
\n{\bf Lemma 1} i) The mass-weighted version of the formula for median lengths (\r{med-2}) is 
\be
\rho_2^{(1)\,2} \es \frac{  2  \left(  \rho_1^{(2)\,2} + \rho_1^{(3)\,2}  \right) - \rho_1^{(1)\,2}  }{  3  } \mbox{ and cycles } .  
\ee
\Proof Apply definition (I.34) and mass weighting (I.47) to Corollary 1 of Sec \r{Cevians}. $\Box$

\m

\n{\bf Remark 2} In \c{2-Herons}, we furthermore interpret this as a mass-weighted counterpart to Stewart's Theorem, 
in the special case in which the Cevians under consideration are medians. 

\m

\n{\bf Definition 1} The {\it first mass-weighted moment sums} are, for $a$ = 1, 2, 
\be 
F_a \:= \sum_i\rho_a^{(i)}                                               \m .
\ee 
%
%
\n{\bf Remark 1} The mass-weighted Jacobi separations are moreover related to the more widely used partial moments of inertia $\bigiota_a$ by   
\be 
\bigiota_a = \rho_a\mbox{}^2                                                                     \m .
\l{I-rho} 
\ee 
In particular, with clustering labels explicit,  
\be 
\bigiota_1^{(a)}  =          \rho_1^2 
                  =          \mu_1 R_1\mbox{}^2 
                 \es \frac{a^2}{2} \m \mbox{ and cycles } , 
\ee 
and 
\be 
\bigiota_2^{(a)}  =           \rho_2^2 
                  =           \mu_2 R_2\mbox{}^2 
		         \es \frac{m_a\mbox{}^2}{2} \m \mbox{ and cycles } .  
\ee
\n{\bf Definition 2} More familiarly, summing over disjoint partial moments rather than over clusters, the {\it total moment of inertia} is  
\be 
\bigiota^{(a)} := \bigiota_1^{(a)} + \bigiota_2^{(a)}                                                       \m . 
\ee 
The definition here is that the total object is the sum of all disjoint partial contributions.  

\m

\n{\bf Lemma 2} i) 
\be
\bigiota = \rho_2^2 (1 + {\cal R}^2)                                                           \m .  
\l{IrhoR}
\ee 
ii) (Democratic formula for the moment of inertia)
\be
\bigiota = \frac{  a^2 + b^2 + c^2  }{  3  } 
         = \langle side^2 \rangle                         \m , 
\l{Iabc}
\ee
where $\langle \m \rangle$ denotes `democratic average'.   

\m

\n(See \c{2-Herons} for a proof.)  

\m

\n{\bf Remark 3} The right hand side of the first equality being cluster-independent, we are entitled to rewrite 
\be 
\bigiota := \bigiota_1^{(a)} + \bigiota_2^{(a)} \mbox{  or cycles } . 
\ee 
\n While trivial to prove, the Lemma's second equality none the less has {\it conceptual content} 
as regards giving a sharp interpretation for the computational right hand side of the first equality. 

\m

\n{\bf Definition 3} The {\it second mass-weighted moment sums} are, for $a$ = 1, 2, 
\be 
Z_a \:= \sum_i\rho_a^{(i)2} 
    \es \sum_i \bigiota_a^{(i)}                                              \m . 
\ee
\n{\bf Definition 4} Let $\bigsigma$, $\bigmu$  $\bigsigma_{\sR\sS\sS}$, $\bigmu_{\sR\sS\sS}$ denote the mass-weighted version of the perimeter $S$, medimeter $M$, 
                         $S_{\sR\sS\sS}$ and $M_{\sR\sS\sS}$ respectively.  

\m				  
						 
\n{\bf Corollary 1} 
\be 
\bigiota = 2 \, \bigsigma_{\sQ}\mbox{}^2 
         = 2 \, \mbox{\Large$\mu$}_{\sQ}\mbox{}^2                                                                   \mma \mbox{or, inverting } \mma
\ee
\be 
\bigsigma_{\sQ}\mbox{}^2   =           \mbox{\Large$\mu$}_{\sQ}\mbox{}^2  
                          \es \frac{\bigiota}{2}                                                                      \m .
\ee
\n{\bf Remark 4} This signifies that, in mass-weighted space, cluster summed or averaged second moments of side or of median are equivalent, 
and both are readily convertible to the moment of inertia up to a constant of proportionality.  

\m  

\n{\bf Remark 5} Thus summing up over the clusterings gives 
\be
Z_1 \:=  \sum_i     \rho_1^{(i)\,2} 
    \es  \sum_i     \bigiota_1^{(i)} 
	\es  \m  \sum_i \bigiota_2^{(i)} 
	\es  \m  \sum_i \rho_2^{(i)\,2} 
	 =:  Z_2                                                                                    \m, 
\l{7}
\ee
which is also now free of numerical factors [as compared to (\r{0.10})'s factor of $\frac{3}{4} \es \kappa^{-2}$].  
This signifies that median partial moments of inertia contribute -- over all clusterings -- the {\sl same amount} as side moments of inertia.  
This amounts to a statement of {\it Jacobi-isotropy} among the sums over all clusters of the partial moments of inertia. 
Furthermore, dividing both sides by 3, 
\be
\left\langle \, \rho_1^{2} \, \right\rangle = \left\langle \, \rho_2^{2} \, \right\rangle \m  
\l{7.b}
\ee
casts this interpretation in the form of equality between (cluster-averaged median partial moment of inertia) 
                                                      and (cluster-averaged  side  partial moment of inertia).

\m

\n{\bf Definition 5} The {\it mass-weighted inertia quadric} (I.71) is 
\be 
\bigiota(\urho\mbox{}_a) \es \sum_a \rho_a\mbox{}^2 
                         \es \rho_1\mbox{}^2 + \rho_2\mbox{}^2       \m , 
\ee 
where the last equality is an $N = 3$ specialization. 
Computationally, this amounts to returning to the previous section's tilded formulation. 
So one motivation for the mass-weighted relative Jacobi coordinates is that they are what drops out of the Linear Algebra approach. 
Another follows from the matrix in the quadric being the identity, alongside the following interpretation. 

\m

\n{\bf Remark 7} The Cartesian equivalence in this (mass-weighted notion of) relative space of $\uR_1$ and $\uR_2$ moreover provides the following. 

\m

\n{\bf `Jacobian' first motivation for median coprimality}. 
Mass-weighted medians and mass-weighted sides are on an identical geometrical footing in (mass-weighted) relative space. 
These are moreover what drops out of the linear algebra most directly, in obtaining Jacobi coordinates by diagonalization. 
Motivating relative Jacobi coordinates themselves has moreover further parts to it. 
For, aside from their usefulness in treating the $N$-body problem, they turn out to be coordinates in terms of which the shape space's natural coordinates are simple 
(spherical coordinates for $N = 3$ or inhomogeneous coordinates for $N \geq 4$).

\subsection{Jacobi congruence and similarity conditions}\l{Jac-Cong-Sim}

\n{\bf Remark 1} Mechanical naturality of medians points to various further congruence and similarity conditions.

\m

\n{\bf Definition and Proposition 1} i) s-S-m -- a side s,                   say $a = R_1$, 
                                                  the corresponding median m, say $m_a = R_2$, 
                                              and the Swiss-army-knife angle      $\mS = \Phi$ included between the corresponding vectors, 
                                              as per Fig I.18.a) and eq (I.171) -- 
is a further congruence condition, which we term the {\bf Jacobi congruence condition, 6)}.  

\m

\n ii) R-S, for R a side-to-corresponding median ratio and $S = \Phi$ the Swiss-army-knife angle included between these is a further similarity condition, 
which we term the {\bf Jacobi similarity condition, 6)}.  
									
\m
 
\n  Let us next consider how we {\sl solve} for the scaled or pure-shape triangle in question given s-m-S or R-S data respectively.  
If $a$ and $m_a$ are given, the problem turns out to be symmetric for $b$, $c$ and for $\gamma$, $\beta$ 
by which it is useful to denote these respectively by $b_{\pm}$ and $\gamma_{\pm}$

\m

\n{\bf Proposition 2} i) With reference to Fig \r{abc0}, prior to mass weighting, 
the scaled triangle is solved in terms of the Euclidean Jacobi data: side-length $R_1$, half-side length $H_1$, corresponding median-length $R_2$, and Swiss-army-knife angle $\Phi$,   
\be
a = R_1                                                                                                                         \mma
b_{\pm}   =   \sqrt{  R_2\mbox{}^2 \pm R_1 R_2 \, \cos\,\Phi + R_1\mbox{}^2/4  } = ||\u{H}_1 \pm \u{R}_2||                                         \m, 
\l{R-c}
\ee
\be                                                                                                                             
\cos \, \gamma_{\pm}   \es   \frac{    R_1 \pm 2 \, R_2 \cos \, \Phi    }{    \sqrt{  4 \, R_2\mbox{}^2 \pm 4 \, R_1 \, \cos \, \Phi + R_1\mbox{}^2  }    }  
                       \es   \frac{    H_1 \pm      R_2 \cos \, \Phi    }{    ||\u{H}_1 \pm \u{R}_2||     }                                              \m, 
\l{R-gamma}
\ee
\be
\cos \, \alpha   \es   \frac{    4 \, R_2\mbox{}^2 - R_1\mbox{}^2    }{    \sqrt{  8 \, R_2\mbox{}^2 [ 2 \, R_2\mbox{}^2 + R_1\mbox{}^2 ( 1 - 2 \, \cos^2\Phi)] + R_1\mbox{}^4 }  }  
                 \es   \frac{         R_2\mbox{}^2 - H_1\mbox{}^2    }{    \sqrt{  (  R_2\mbox{}^2 + H_2\mbox{}^2  )^2 - 2 \, (  \u{H}_1 \cdot \u{R}_2  )^2     }    }          \m .  
\l{R-alpha}
\ee
\n ii) The mass-weighted versions of this solution is  
\be
a       =   \sqrt{2} \, \rho_1                                                                                                  \mma 
b_{\pm} =    
        =   ||\urho_1 \pm \sqrt{3} \, \urho_2||/\sqrt{2}                                                                         \m, 
\l{rho-c}
\ee
\be
\cos \, \gamma_{\pm} 
\frac{  \rho_1 \pm \sqrt{3} \, \rho_2 \cos \, \Phi    }{    \sqrt{  \rho_2 ( 3  \, \rho_2 \pm 2 \sqrt{3} \rho_1 \, \cos \, \Phi) + \rho_1\mbox{}^2  }    }  
                            \es  \frac{  \rho_1 \pm \sqrt{3} \, \rho_2 \cos \, \Phi    }{    ||\urho_1 \pm \sqrt{3} \, \urho_2||  }                          \m, 
\l{rho-gamma}
\ee
\be
\cos \, \alpha   
                \es   \frac{  3 \, \rho_2\mbox{}^2 - \rho_1\mbox{}^2  }{  \sqrt{  (3 \, \rho_2\mbox{}^2 + \rho_1\mbox{}^2)^2 - 4 \sqrt{3} \, (\urho_1 \cdot \urho_2)^2  }  }           \m . 
\l{rho-alpha}
\ee
\n iii) The pure-shape triangle is solved in terms of the similarity Jacobi data: median-to-base ratio ${\cal R}$ and Swiss-army-knife angle $\Phi$, 
\be
\frac{b_{\pm}}{a} \es \frac{1}{2}  \sqrt{ 3 \, {\cal R}^2 \pm 2 \sqrt{3} \, {\cal R} \, \cos\,\Phi + 1 }                     
                  \es \frac{1}{2}  \sqrt{      {\cal V}^2 \pm 2          \, {\cal V} \, \cos\,\Phi + 1 }                                                                          \m, 
\l{calR-ca}
\ee
\be
\cos \, \gamma_{\pm} \es  \frac{    1 \pm \sqrt{3} \, {\cal R} \, \cos \, \Phi    }{    \sqrt{ 3 \, {\cal R}^2   \pm   2 \sqrt{3} \, {\cal R} \, \cos \, \Phi + 1  }    }            
                     \es  \frac{    1 \pm             {\cal V} \, \cos \, \Phi    }{    \sqrt{      {\cal V}^2   \pm   2          \, {\cal V} \, \cos \, \Phi + 1  }    } \m, 
\l{calR-gamma}
\ee 
\be
\cos \, \alpha \es  \frac{    3 \, {\cal R}^2 - 1    }{    \sqrt{ (3 \, {\cal R}^2  + 1)^2 - 12 \, {\cal R}^2 \cos^2\Phi  )   ] + 1  }    }      
               \es  \frac{         {\cal V}^2 - 1    }{    \sqrt{ ({\cal V}^2 + 1)^2 - 4 \, {\cal V}^2 \cos^2\Phi  }    }                                                        \m , 
\l{calR-alpha}
\ee
where we have tidied up by introducing the transformation 
\be
{\cal V} := \sqrt{3} \, {\cal R}                                                                                                                                                 \m. 
\l{V-R}
\ee 
\Proof Five sequential uses of the cosine rule give i). 
Then mass-weight as per (I.34) to obtain ii).
Finally form length ratios and divide tops and bottoms of right-hand-side expressions by suitable powers of $\rho_i$ to recast them in terms of the ratio ${\cal R}$.  $\Box$  

\m

\n{\bf Remark 2} The transformation (\ref{V-R}) is moreover revealed in Sec \r{Right-SS} to have further shape space significance.   

\m

\n{\bf Remark 3} The Jacobi conditions s-S-m and R-S moreover closely emulate the s-$\alpha$-s and r-$\alpha$ (included) conditions argued for in Sec \r{Cong-Sim-Disc}. 
Indeed, the arguments by which the latter extend to C (including B) largely carry over; the Swiss-army-knife angle now takes a collinear value 0 or $\pi$
[One now has to exclude U(3, 1) since the Swiss-army-knife angle is not defined there; one can probe near there using s-s-s or r-r, 
by which the use of a s-s-s patch to deal with O covers this moderate inconvenience with U(3, 1) as well.] 

\m

\n{\bf Remark 4} Having obtained the above angle formulae, (\r{s+c=1}) and cycles give that 
\be
\sin\, \gamma_{\pm}   \es   \frac{  {\cal V} \, \sin \, \Phi}{\sqrt{{\cal V}^2 \pm 2 \, {\cal V} \, \cos \, \Phi + 1}  }                     \m \m \m   \mbox{ and } \m \m \m 
\sin\, \alpha   \es   \frac{   2 \, {\cal V} \, \sin \, \Phi    }{  \sqrt{{\cal V}^4 + 2 \, {\cal V}^2 ( 1 - \cos \, 2 \, \Phi ) + 1}  }     \m . 
\l{sin-alpha}
\ee 
The simplest expressions for these angles are moreover 
\be
\cot \, \gamma_{\pm} = {\cal V}^{-1} \cosec \, \Phi \pm \cot \, \Phi                                                                         \m \m \m   \mbox{ and } \m \m \m 
\cot \, \alpha = 2^{-1}\cosec \, \Phi  \left(  {\cal V} - {\cal V}^{-1}  \right)                                                             \m .
\l{cot-alpha}
\ee

\section{Theory of mass-weighted area}

\subsection{Sides-to-medians symmetric Jacobi--Heron formula}

\n{\bf Definition 1} The {\it mass-weighted area} is, in terms of relative Jacobi vectors,  
\be
\Area \:= \frac{1}{2}|\rho_1 \cr \rho_2| \m .  
\ee
We also denote the {\it mass-weighted tetra-area} by 
\be 
\bigtau \:= 4 \times \Area \mbox{ } . 
\ee 
\n{\bf Definition 2} The {\it mass-weighted area per unit moment of inertia} is 
\be
\area \:= \frac{  \mbox{(mass-weighted area)}  }{  \mbox{(moment of inertia)}  } 
      \es \frac{    |\rho_1 \cr \rho_2|    }{ 2  (\rho_1^2 + \rho_2^2)    }               \m ;  
\ee
We also denote the {\it mass-weighted tetra-area per unit moment of inertia} by 
\be 
\tau \:= 4 \times \area \mbox{ } . 
\ee 
\n{\bf Remark 1} $\area$ and $\tau$ are shape quantities (and thus dimensionless).    

\m

\n{\bf Remark 2} Upon reaching preshape space, moreover, $\bigiota$ is constant, so $\area$ and $\tau$ continue to be proportional to the geometrical area variable.  

\m

\n{\bf Definition 3} Let us denote the {\it mass-weighted semi-perimeter} denoted by $\sigma$, and the {\it mass-weighted semi-medimeter} by $\varsigma$.  

\m

\n{\bf Remark 3} Various subsequent sections of this Part require knowing furthermore how the quantities in question scale.  

\m

\n{\bf Lemma 1} i) Mass-weighted semi-perimeter is a mass-weighting side-vector,\footnote{See \c{Ineq} for further exposition of what 
`side-vector', `median-vector' and `side-median bivector' mean.}  
\be
\sigma  :=          \sqrt{\mu_{\sss}}s 
        \es \frac{s}{\sqrt{2}}             \m .
\ee 
ii) {\it Mass-weighted semi-medimeter} is a mass-weighting median-vector, 
\be
\varsigma :=           \sqrt{\mu_{\sm}}s 
          \es \sqrt{\frac{2}{3}} \, s      \m .
\ee 
\n iii) Area is a mass-weighting side--median bivector,
\be
\Area          =    \sqrt{\mu_{\sss}\mu_{\sm}} \, \mbox{Area} 
              \es   \frac{   \mbox{Area}   }{   \sqrt{3}   }     \m .
\ee
(See \c{2-Herons} for a proof.) 

\m

\n{\bf Theorem 1} Mass-weighted area's {\it Heron--Jacobi formulae} in terms of each of mass-weighted sides and mass-weighted medians is 
\be 
\sqrt{    \sigma     \left(  \sigma - \rho_1^{(a)}  \right)   
                     \left(  \sigma - \rho_1^{(b)}  \right)   
					 \left(  \sigma - \rho_1^{(c)}  \right)   } 
\es 
\Area \es  \sqrt{    \varsigma  \left(  \varsigma - \rho_2^{(a)}  \right)   
                                \left(  \varsigma - \rho_2^{(b)}  \right)   
					            \left(  \varsigma - \rho_2^{(c)}  \right)   }  \m .
\l{HHF}
\ee
(See \c{2-Herons} for a proof.) 

\m

\n{\bf Remark 4} In allocating primary ratios, we already used a maximal amount of what would be mass-weighted scalars, for subsequent convenience. 
Thus no further mass-weighted versions of these need to be introduced at this point.  

\m

\n{\bf Remark 5} Note that mass-weighting places the sides and medians versions of Heron on an equal footing. 
I.e.\ without any constant prefactor difference like the $\kappa^2$ in the case of the mass-unweighted medians  
                                                           relative to the case of the mass-unweighted sides   of (\r{Summary-1}).
This amounts to explaining the $\kappa^2$ factor discrepancy between the medians' and sides' Heron's formulae 
as resulting from formulation in terms of area rather than mass-weighted area.
This is a further way in which the Jacobi formulation exhibits naturality.  
It is moreover a {\sl common} explanation of the numerical factors in the sides-to-medians Heron formulae and the sides-to-medians involution, as per eq. (\r{4/3}).

\subsection{Diagonalizing the Heron matrix gives the Hopf Map and Kendall's Theorem}\l{HoKe}

{\bf Remark 1} Furthermore, setting  
\be
0 = \mbox{det}\left(\Heron - \lambda \Imat \right) 
%
%
\ee
$\lambda = -2$ with multiplicity 2, with respective orthonormal eigenvectors  
\be 
\frac{1}{\sqrt{3}}    \left(    \s{           \mbox{1}  }{  \s{  \mbox{1}     }{  1  }  }  \right)  \mma 
\frac{1}{\sqrt{2}}    \left(    \s{  \m \mbox{1}  }{  \s{  \mbox{$-1$}  }{ \m 0  }  }  \right)  \mma 
\frac{1}{\sqrt{6}}    \left(    \s{  \m \mbox{1}  }{  \s{  \m \mbox{1}     }{ -2  }  }  \right)  \m .  
\l{H-evectors}
\ee 
%
%
The diagonalizing variables are thus 
\be
\overline{a}^2 \es \frac{a^2 + b^2 + c^2}{\sqrt{3}}       \mma
\overline{b}^2 \es \frac{a^2 - b^2}{\sqrt{2}}             \mma \mbox{and } \m  
\overline{c}^2 \es \frac{a^2 + b^2 - 2 \, c^2}{\sqrt{2}}                   \m .   
\ee 
%
%
The Heron quadratic form (\r{Heron-form}) is thus equal to 
\be
\sum_i \Lambda_{ii} \overline{a}_i\mbox{}^2 \overline{a}_i\mbox{}^2 
\ee 
for {\it diagonalized Heron matrix}
\be 
\Lambda_{ij} = \mbox{diag}(1, \, - 2, \, - 2)                        \m .  
\ee 
This is related to the original Heron matrix $H_{ij}$ by conjugation with the $P_{ij}$ formed by using (\r{H-evectors})'s orthonormal eigenvectors as columns.  
Thus we have derived that Heron's formula also takes the following form. 

\m

\n{\bf Theorem 1 (Diagonal Heron formula)} 
\be 
T = \sqrt{  \overline{a}\mbox{}^4 - 2(\overline{b}\mbox{}^2 + \overline{c}\mbox{}^2)  }  \m .  
\ee 
\n {\bf Corollary 1 (Ratio form of Diagonal Heron formula)} 
%
%
\be 
\frac{T}{\overline{a}^2}                         = \sqrt{  1 - \left[    \sqrt{2}\left(    \frac{  \overline{b} }{  \overline{a}  }    \right)^2    \right]^2 
                                                 - \left[    \sqrt{2}\left(    \frac{  \overline{c} }{  \overline{a}  }    \right)^2    \right]^2    }          \m .  
\ee
\n{\bf Definition 1} The rescaled {\it ratio variables} 
\be 
z \:=  \sqrt{2}  \left(  \frac{  \overline{c}  }{  \overline{a}  }  \right)^2 
  \es      \frac{  a^2 + b^2 - 2 \, c^2  }{  a^2 + b^2 + c^2 }                                                                                                  \m .  
\l{z}
\ee 
\be 
x \:=  \sqrt{2}  \left(  \frac{\overline{b}}{\overline{a}}  \right)^2 
  \es  \frac{\sqrt{3}(a^2 - b^2)  }{  a^2 + b^2 + c^2 }                                                                                                     \m \m \mbox{ and } 
\l{x}
\ee 
\be 
y \:=  \frac{\sqrt{3} \, T}{a^2 + b^2 + c^2}                                                                                                                 \m  
\l{y}
\ee 
are then natural. 

\m 

\n{\bf Remark 2} The denominator of the ratio is proportional to $\bigiota$ by Lemma 2.ii) of Sec \r{RJSI}.   
$y$ is moreover to be interpreted precisely as {\it mass-weighted tetra-area per unit moment of inertia}, $\tau$

\m

\n{\bf Corollary 1} In terms of the rescaled ratio variables, the diagonal Heron formula has been reduced to  
\be 
x^2 + y^2 + z^2 = 1  \m .  
\l{x+y+z=1}
\ee 
This is mathematically just the on-2-sphere condition.  
This moreover constitutes a derivation of the Kendall shape sphere from new first principles. 

\m

\n{\bf Remark 3} With the following extra computation, this also amounts to a derivation of the Hopf map ${\cal H}_{\sC}$ following on from (I.180).
This work amounts to identifying $x$, $y$, $z$ in terms of the four $\urho_i$
Firstly, it is straightforward from (\r{y}) and Lemma 2.ii) that 
\be 
y = 2 \, ( \u{\nu}_1 \cr \u{\nu}_2 )_3    \m . 
\ee 
Secondly, using the first form of (\r{calR-ca}) and (I.174) in (\r{x}-\r{z}) 
\be 
x = 2 \, \u{\nu}_1 \cdot \u{\nu}_2   \m, 
\ee 
\be 
z = \nu_1\mbox{}^2 - \nu_2\mbox{}^2  \m .  
\ee 
\n{\bf Remark 4} $x$ and $z$ are to be interpreted as Heron map eigenvectors. 
$x$ is furthermore to be interpreted as anisoscelesness, $\aniso$: departure from isoscelesness. 
\be
\aniso \es  \frac{  \sqrt{3}  (  a^2 - b^2  )  }{  a^2 + b^2 + c^2  }   \m ,
\ee 
and 
$z$                                  as ellipticity, $\ellip$: departure from regularity.  
\be
\ellip \es  \frac{a^2 + b^2 - 2 \, c^2}{a^2 + b^2 + c^2}                \m .
\ee 
We will have more to say about these quantifiers in Sec \r{Hopf}.

\m

\n{\bf Remark 5} See \c{Kendall, FORD, FileR} for further derivations of the shape sphere: geometry and associated mechanical action reduction.

\subsection{Median--sides interchange form invariance of diagonal Heron--Hopf formula}\l{MSIIDFF}

\n{\bf Corollary 1} In terms of the medians, i) 
\be 
\ellip \es  - \frac{m_a\mbox{}^2 + m_b\mbox{}^2 - 2 \, m_c\mbox{}^2  }{  m_a\mbox{}^2 + m_b\mbox{}^2 + m_c\mbox{}^2  }   \m , 
\ee 
\be
\aniso \es  - \frac{  m_a\mbox{}^2 - m_b\mbox{}^2  }{\sqrt{3} \{ m_a\mbox{}^2 + m_b\mbox{}^2 + m_c\mbox{}^2 \} }         \m , \mbox{ check downstairsness of this 3 }
\ee
i.e.\ equality of form up to sign choice (which is an allowed ambiguity in setting up Cartesian axis systems).

\m

\Proof Use the below Lemma and the sides--medians involution. $\Box$ 

\m

\n{\bf Lemma 1} (Democratic medians form of total moment of inertia)
\be 
\bigiota \es  \kappa^2 \langle m_i\mbox{}^2 \rangle 
         \es  \kappa^2 M_{\sQ}\mbox{}^2                                                                                  \m .  
\ee
(See \c{Ineq} for a proof.) 

\mbox{  }

\n{\bf Remark 1} The above Corollary can be interpreted as follows.  
Expressing $\ellip$ and $\aniso$ in terms of medians instead does not moreover affect the diagonality.  
The Heron--Hopf formula is thus independent of whether one is conceiving in terms of sides or of medians; 
the Hopf quantities offer a third point of view that is side--median symmetric.  
Thus we have arrived at the 

\m

\n{\bf `Hopfian' second motivation for median coprmality}.

\subsection{Heron--Hopf--Kendall, Heron--Hopf and two concomitant formulae}\l{HHK}

We conclude the previous three sections as follows.
Sec \r{HoKe}'s workings readily imply the following Theorems.     

\m

\n{\bf Theorem 1} The diagonalized form of the mass-weighted Heron formula is 
\be
\tau = \sqrt{1 - \aniso^2 - \ellip^2}  \m . 
\l{HH}
\ee
This `{\it Heron--Hopf}' formula is moreover sides--to-medians symmetric.

\m

\n{\bf Theorem 2} The most symmetrical presentation of the diagonalized mass-weighted Heron formula is, from \r{x+y+z=1},  
\be 
\tau^2 + \aniso^2 + \ellip^2 = 1       \m .  
\l{EATI}
\ee 
This `{\it Heron--Hopf--Kendall}' formula is mathematically just the on-sphere condition; 
moreover observing this amounts to a {\sl recovery} of {\it Kendall's Theorem} that the shape space of all triangles in 2-$d$ is a sphere. 

\m

\n{\bf Remark 1} As considered in current treatise, Kendall's Theorem is in the context of distinctly labelled point-or-particle vertices without mirror image identification. 
A conceptual name for this theorem is `{\it triangleland sphere theorem}', whereas one for the more general Kendall theorem is `{\it N-a-gonland complex projective space theorem}'. 

\m

\n{\bf Remark 2} The Heron--Hopf formula, as the area-subject Hopf formula in terms of ellipticity and anisoscelesness data, now has two concomitant formulae in many senses.
These are, firstly, the ellipticity-subject Hopf formula in terms of anisoscelesness and area data, 
\be
\ellip = \sqrt{1 - \aniso^2 - \tau^2}  \m . 
\ee 
Secondly, the anisoscelesness-subject Hopf formula in terms of elliptiity and area data,  
\be
\aniso = \sqrt{1 - \ellip^2 - \tau^2}  \m .   
\ee
The sense in which Heron--Hopf has a further quality than these two other formulae is that it is the only one among these which has a democratic subject 
(in the clustering alias Jacobi sense).

\subsection{Maximum and minimum $\area$}

\n{\bf Remark 1} $\area$ is a natural 2-$d$ analogue of $(N, 1)$'s $\delta iam$ of eq. (I.260).

\m

\n{\bf Remark 2} On the one hand, for mirror image identified versions,   
\be
\area_{\sm\sa\sx} \mbox{ is at the equilateral triangle E } 
\l{A-Max} 
\ee
and 
\be
\area_{\sm\si\sn} \mbox{ is shared by all collinear configurations C } .   
\ee
{\bf Remark 3} On the other hand, for the mirror image distinct versions, using the unsigned version of $\area$, 
\be
\area_{\sm\sa\sx} \mbox{ is at the equilateral points E and $\bar{\mE}$ } ,    
\ee
and $\area_{\sm\si\sn}$ is as above. 
For the signed version, however, $\area_{\sm\sa\sx}$ is at E and $\area_{\sm\si\sn}$ is at $\bar{\mE}$.

\m

\n{\bf Remark 4} Collinear configurations $\area = 0$ moreover bound clockwise triangles obeying the signed-area inequality,  
\be
\area \gs 0  \m,
\l{A>0}
\ee
from anticlockwise triangles obeying the opposite signed-area inequality,  
\be
\area \ls 0  \m . 
\l{A<0}  
\ee
{\bf Remark 5}  
\be
{\cal P} \:=  \frac{\bigsigma}{\sqrt{\bigiota}} 
         \es  \frac{\bigsigma}{\rho}
\ee 
-- {\it mass-weighted perimeter per square root of moment of inertia} or {\it per (mass-weighted) configuration space radius} -- is a comparably significant quantity.
This however takes more work to extremize, so we leave this task to a further paper \c{A-Perimeter}.

\m

\n{\bf Remark 7} 
\be
{\cal J}  \:=  \frac{Area}{S^2} 
\ee
is the {\it isoperimetric ratio} that features in the classical {\it isoperimetric problem}; for 3 points-or-particles, this is also maximized for the equilateral triangle 
(see e.g. \cite{Kazarinoff} for 3 elementary proofs).
Moreover, the three ratios above clearly only have two degrees of freedom between them. 
(There is proportionality to the mass-weighted quantity 
\be 
\left. \frac{\Tau}{\bigsigma^2} \m \right) .  
\ee 
What has happened is that naturality of moment of inertia in Shape Theory has rendered the classical isoperimetric ratio less primary, 
by which it is now to be regarded as the composite ration of the previous two more shape-theoretic ratios.  
On these grounds, extremizing the first two of these ratios has significance as a shape-theoretic replacement of the isoperimetric problem 
by two more primary problems whose ratio is the insoperimetric quantitiy itself.  

\m

\n{\bf Remark 8} See \c{A-Perimeter} for further study of shape-theoretically significant ratios for triangles.

\subsection{Jacobi separation inequalities}\l{JSI}

\n Passing to mass weighted Jacobi versions of Sec 3's inequalities requires knowing how the quantities involved scale under mass-weighting.  

\m

\n{\bf Lemma 1} i)  In Fig \r{6-variables}, everything in column 1 is a side-vector and median-scalar, 
                     whereas                   everything in column 2 is a median-vector and side-scalar.

\m 					
					
\n              ii) ${\cal S}$, ${\cal G}$, ${\cal M}$ and ${\cal H}$ are mass-weighting scalars.

\m

\n               iii) ${\cal F}$, ${\cal E}$ and ${\cal I}$ are all median-vectors and side-covectors. 

\m

\n(See \c{Ineq} for a proof.) 

\m

\n{\bf Remark 2} As set up, while working at the level of ratios, everything is a biscalar or a vector-covector, whose weight is moreover $\kappa$. 

\m

\n{\bf Definition 1} Denote the {\it mass-weighted medimeter per unit perimeter} by   
\be
\Psi     := \kappa \,  {\cal F}  \m,  
\ee 
the {\it mass-weighted root sum of medians per unit perimeter} by  
\be
\Lambda  := \kappa \,  {\cal L}  \m,  
\ee 
We do not use the {\it mass-weighted isoperimetric ratio} 
\be
\Upsilon := \kappa \,  {\cal I}  \m, 
\ee
so as to maintain the [0, 1] range. 
To be clear, in each case here, `mass weighted' means that both the numerator and the denominator are mass-weighted.  

\m

\n{\bf Theorem 1}   
\be
\kappa^{-1}  \leqs  \Psi 
             \leqs  \kappa   \m .
\l{mw-Rat-Ineq-1}
\ee 
\Proof Mass-weight (\r{Rat-Ineq-1}). $\Box$.  

\m

\n{\bf Theorem 4}  
\be
{\cal I} \leqs \Psi 
         \leqs \Lambda 
           =   {\cal S}^{-1} 
		 \leqs {\cal I}^{-1}{\cal G}^{-3} \m .
\l{mw-rat-ineq}
\ee
\Proof Mass-weight (\r{Rat-Ineq-2}).                  $\Box$

\m

\n{\bf Remark 4} On the one hand, Theorem 1 places shape-independent bounds on the mass-weighted medimeter-to-perimeter ratio. 
Mass-weighting has multiplicatively centred this inequalities' bounds, meaning that one bound is now the reciprocal of the other. 
On the other hand, Theorem 2 places shape-dependent bounds on this and the root sum median to perimeter ratio, 
the shape dependence of which is manifested as powers of the isoperimetric ratio and arithmetic-to-geometric mean side raito shape variables.
As regards lower bounds, for 
\be 
{\cal I} \ls \kappa^{-1}  \m, 
\ee 
Theorem 2's shape-dependent one is more stringent, whereas for 
\be 
{\cal I} \gs \kappa^{-1}  \m, 
\ee
Theorem 1's shape-independent one is. 
As regards upper bounds, for  
\be 
{\cal I} \, {\cal G}^3 \gs \kappa^{-1}  \m,
\ee 
Theorem 2's shape-dependent one is more stringent, whereas for 
\be 
{\cal I} \, {\cal G}^3 \ls \kappa^{-1}  \m,
\ee
Theorem 1's shape-independent one is. 
The corresponding cross-over's critical values, and consideration of the set of shapes entailed by each case, is part of \c{A-Perimeter}.

\section{Jacobian recharacterization of some familiar types of triangle}

\subsection{Jacobi-perpendicularity recharacterization of isosceles and equilateral triangles} 

\n{\bf Remark 1} Isoscelesness can furthermore be characterized by an edge being perpendicular to its median (Ex. 3.13 of \c{Schaum}).  
[This is by the shared side-equal side-right os r-h-s congruence $\Rightarrow$ other sides equal $\Rightarrow$ isosceles.]    
In the current treatise's relational language, this result can be rephrased as the corresponding Swiss-army-knife angle being right: 
a Jacobi perpendicularity criterion, 
\be
\mbox{(isosceles)} \Leftrightarrows \left( \mbox{at least on Swiss-army-knife angle is right}: \m \Phi = \frac{\pi}{2} \right) \m . 
\ee
\n{\bf Remark 2} The sole degree of freedom isosceles triangles are left with in this Jacobi characterization is the median-to-base ratio R.    

\m

\n{\bf Remark 3} At the Jacobian level of structure, 
equilaterality's characterization as isosceles with respect to two clustering choices      (implying the third as well) 
                                                 translates to two                    (also implying the third as well) 
												                   edges being perpendicular to their respective median.  
This converts (\r{E=II}) to being a Jacobi perpendicularity criterion for equilaterality.

\m

\n{\bf Remark 4} Note that the above is base--median perpendicularity within a given clustering. 
Triangleland moreover also supports Jacobi perpendicularity between one clustering's base and another's median, 
and median--median mutual perpendicularity between clusters; see Fig \r{Jac-perp-types} (in extension of Fig I.16).    
%
{\begin{figure}[ht]
\centering
\includegraphics[width=0.8\textwidth]{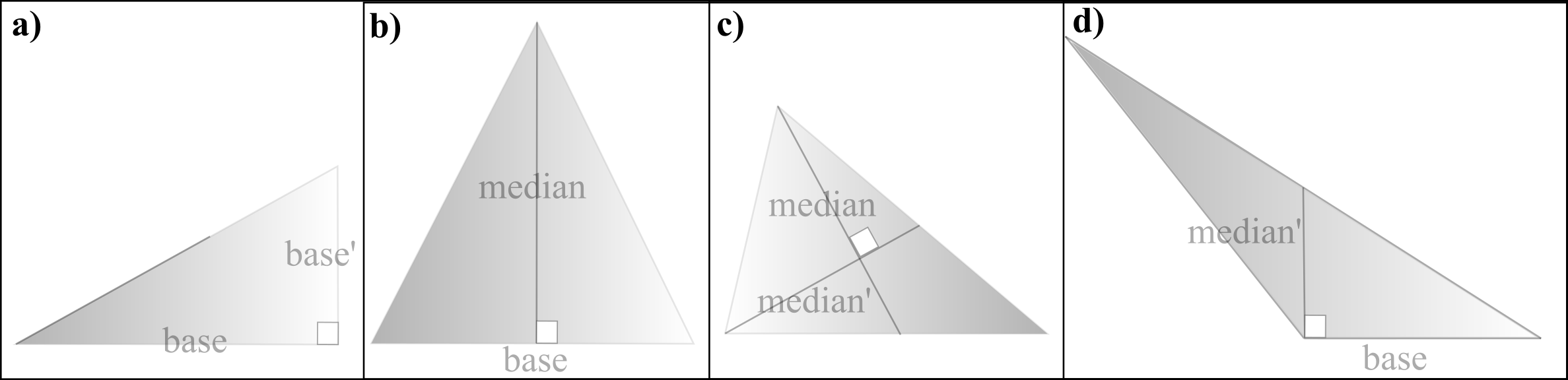}
\caption[Text der im Bilderverzeichnis auftaucht]{\footnotesize{a) An obvious alias Lagrange right triangle.

\m

\n b) Jacobi rightness is a further isoscelesness condition.

\m

\n c) Median-median perpendiculatity.  

\m

\n d) Perpendicularity between one clustering's base and another's median.}} 
\l{Jac-perp-types}\end{figure} } 
	
\subsection{Mergers, cluster hierarchy and hierarchical coincidence structure}
%
{\begin{figure}[ht]
\centering
\includegraphics[width=0.85\textwidth]{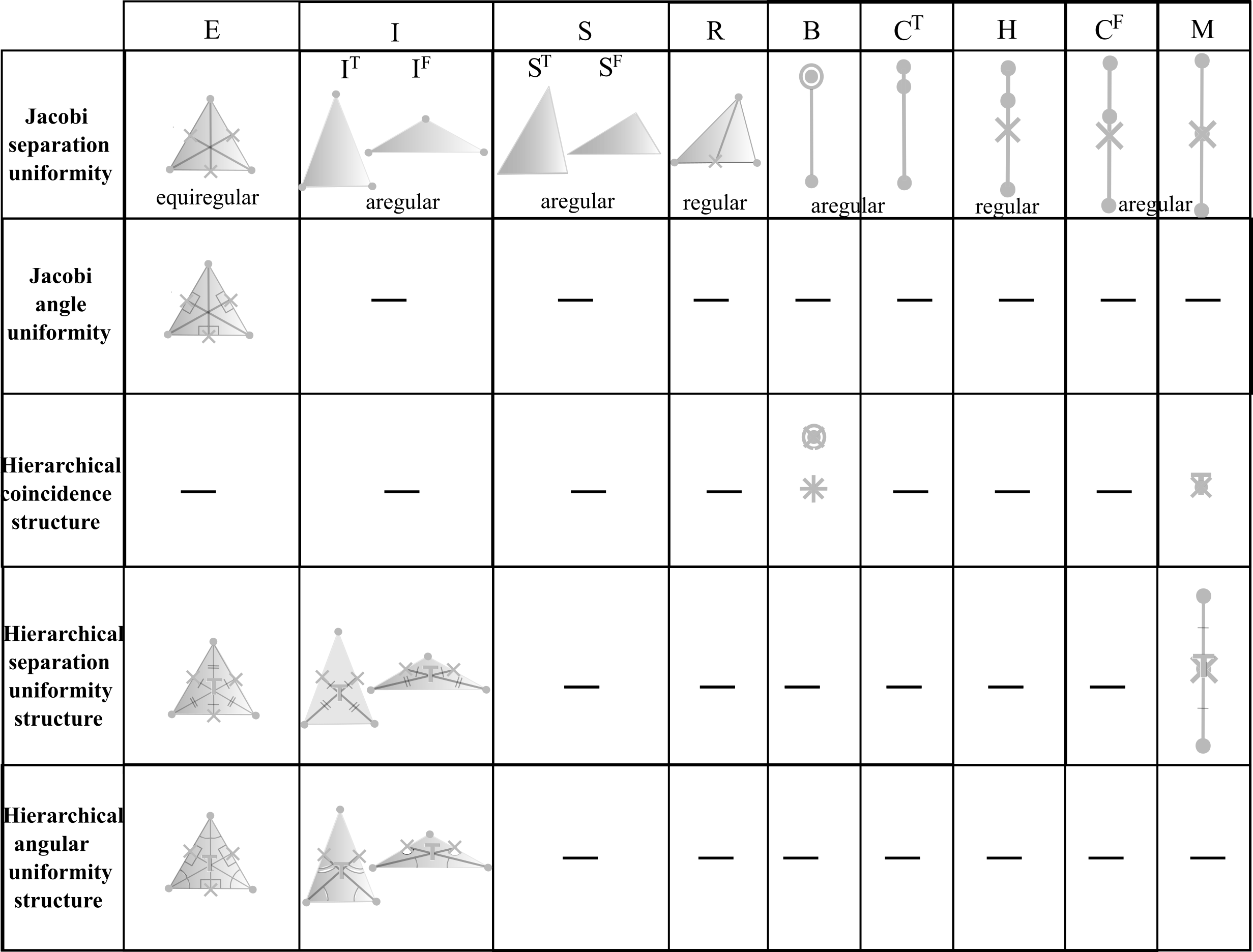}
\caption[Text der im Bilderverzeichnis auftaucht]{\footnotesize{a) Regularity, hierarchical coincidences and hierarchical uniformity for (3, 2) shapes.}} 
\l{(3, 2)-Jac-Hier}\end{figure} } 

\n This is covered in Fig \r{(3, 2)-Jac-Hier}.

\subsection{Jacobi separation uniformity characterization: tall, flat and regular triangles}\l{Hopf-S}

\n{\bf Remark 1} Due to its pure length-ratio content, Sec \r{T-F-Isosceles}'s tall-versus-flat distinction extends to scalene triangles as well. 
This requires bringing in the following bounding notion.  

\m

\n {\bf Definition 1} A triangle is {\it regular} if its base and median partial moments of inertia are equal:  
\be 
\bigiota_{s}          = \bigiota_{m}  \Leftrightarrows  
\frac{\rho_1}{\rho_2} = 1                                \m .  
\ee 
The latter form is clearly an example of Part I's Jacobi separation uniformity criterion.  
Other useful descriptions or aliases for regular are {\it Jacobi-isotropic} -- meaning side-to-median isotropy in partial moments of inertia -- 
and {\it circular} (explained in Sec \r{Ellip-Sec}).

\m

\n{\bf Remark 2} There are moreover 3 notions of regularity, corresponding to the 3 clustering choices, 
\be 
\bigiota_{s}^{(i)} = \bigiota_{m}^{(i)}  \m . 
\ee
\n{\bf Remark 3} In space itself, due to intervention of the Jacobi masses, the regularity condition translates to 
\be
\frac{    R_1    }{    R_2    } = 
%
%
\kappa                                                                                           \m , 
\l{Regular}
\ee  
which is immediately recognizable as the equilateral triangle's base-to-height ratio.  
This gives the following further Jacobi-separation-uniformity characterization.

\m

\n{\bf Proposition 1} A triangle is equilateral iff base--median equality holds in two of its Jacobi clusterings.

\m

\n{\bf Remark 4} E clearly has base--median condition hold in {\sl all three} of its clusterings, but by the Proposition, 
this holding in two {\sl implies} it holds in the third as well. 

\m

\n{\bf Remark 5} Regularity can thus be reformulated in entirely flat space geometrical terms 
-- without any reference to mass weighting or moments of inertia -- as `the base and median are in the proportion found in the equilateral triangle'.  

\m

\n{\bf Corollary 1} The equilateral triangle E is not only twofold $\Rightarrow$ threefold isosceles I, 
                                                   but also twofold $\Rightarrow$ threefold regular R,  
%
%
\be 
\mE =  \mR \caps\,  \mR = \mR \caps\,  \mR \caps\,  \mR   \m . 
\ee
\n{\bf Definition 2} Let us term this twofold $\Rightarrow$ threefold regularity {\it equiregularity}. 

\m

\n{\bf Remark 6} This is clearly a pure-ratio characterization of equilaterality E; it involves multiple clusterings. 

\m

\n{\bf Remark 7} Moroever, 
\be
\mE = \mI \caps\, \mI \caps\, \mI \caps\mR \caps\, \mR \caps\, \mR   \m, 
\ee
which is a full characterization that is unique on the shape hemisphere and gives a pair of antipodal conjugate points E and $\overline{\mE}$ on the shape sphere.

\m

\n{\bf Remark 8} On the other hand, the following is the unique Leibniz space level characterization of equilaterality, E, 
as a mixed length-ratio and relative-angle concept within a single cluster.   

\m

\n {\bf Proposition 2 (angle-ratio symmetric, alias Jacobi--Leibniz, criterion for equilaterality)} Just being one $\mI$ and one $\mR$ implies being $\mE$: 
\be
\mE = \mI \caps\,  \mR \m : \m \mbox{ (equilateral)} = \mbox{(isosceles)} \caps\,  \mbox{(regular)} \m .  
\ee
\Proof Case 1) The median involved in the regularity condition is not relative to either of the ab initio equal sides (Fig  \r{Equiregular}.a). 
Here, isoscelesness forces this median to be perpendicular to its base.
Also regularity fixes the height in space of the isosceles triangle to be $\kappa^{-1}$ times the base. 
So applying Pythagoras' Theorem to the triangle $ABK$, equality of the third side is enforced. 

\m

\n Case 2) The median involved in the regularity condition is relative to one of the ab initio equal sides (Fig \r{Equiregular}.b).  
This requires the mass-weighted triangle $ABL$ to have sides in 1 : 1 : $\frac{1}{2}$ ratio. 
Thus removing the mass-weighting, $ABL$'s sides are in ratio $1 : \frac{\sqrt{3}}{2} :  \frac{1}{2}$, 
by which we recognize it to be the `set square triangle' with $\frac{\pi}{6}$, $\frac{\pi}{3}$, $\frac{\pi}{2}$ angles.
We then note this means that a second median is perpendicular to its corresponding side, so it is twofold isosceles and thus threefold isosceles: equilateral. 
Alternatively, we know it is isosceles with $\frac{\pi}{3}$ angle included between its equal sides, so its remaining angles are $(\pi - \pi/3)/2 = \pi/3$, 
so it is equiangular and thus equilateral.                                                                                                          $\Box{ }$   
%
{\begin{figure}[ht]
\centering
\includegraphics[width=1.0\textwidth]{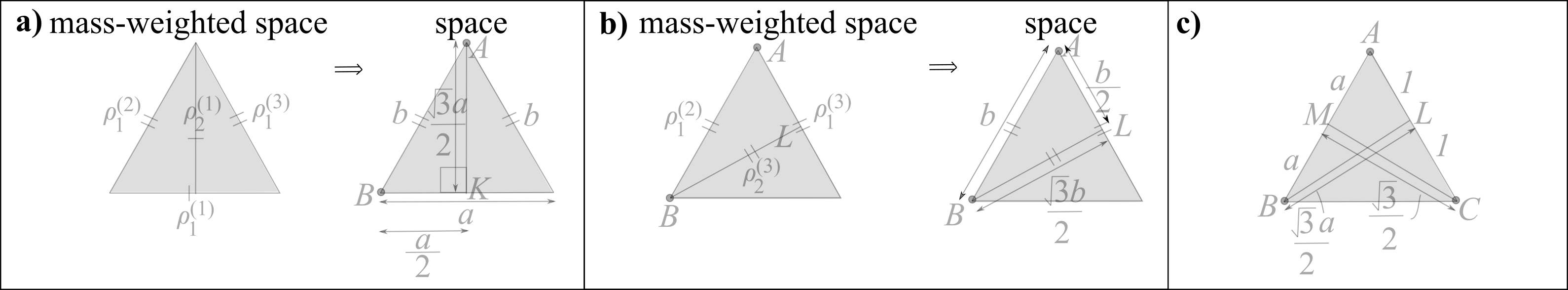}
\caption[Text der im Bilderverzeichnis auftaucht]{\footnotesize{a) and b) are the two cases in the derivation of the equiregularity criterion for equilaterality in Proposition 1.
c) gives Proposition 1's set-up.}} 
\l{Equiregular}\end{figure} } 

\m

\n{\u{Proof of Proposition 1}} In space, we have the data in Fig \r{Equiregular}.c)  
Then $\triangle ALB$ is similar to $\triangle AMC$ by r-$\alpha$ (non-included), since the angle at $A$ is shared and 
\be 
\frac{LB}{AL} \es \frac{\sqrt{3}}{2} \es \frac{MC}{AM}
\ee 
This means that
\be 
\frac{1}{a} \es \frac{AM}{AL} \es \frac{MC}{LB} \es \frac{2 \, a}{2} = a \m . 
\ee 
So, with $a \geq 0$ since it is a distance, $a = 1$.  
Thus $\triangle ABC$ is isosceles. 
But $\triangle ABC$ is certainly also regular (in either cluster it is not currently proven to be isosceles in).
Moreover, we proved in Proposition 2.ii) that this condition implies equilaterality. $\Box$ 

\m

\n{\bf Definition 3} A triangle is {\it aregular} if it is not regular with respect to any clustering. 

\m

\n{\bf Remark 9} Note the parallel between equiregularity, regularity   and aregularity, on the one   hand, 
                                        and equilaterality, isosceleness and scaleneness  on the other hand (see also Fig \r{(3, 2)-Jac-Hier} in this regard). 

\m

\n {\bf Definition 4} A triangle is {\it tall} -- denoted $\mT$ -- with respect to a given cluster if the inequality 
\be 
\bigiota_{1} \ls \bigiota_{2}    \Leftrightarrows 
      \rho_1 \ls \rho_2  
\ee 
holds. 
It is {\it flat} -- denoted $\mF$ -- if the inequality 
\be
\bigiota_{1} \gs \bigiota_{2}   \Leftrightarrows   
      \rho_1 \gs \rho_2  
\ee
holds instead. 

\m

\n{\bf Remark 10} Barbour calls very tall configurations `{\it needles}', 
while Kendall refers collectively to near-collinear configurations uniformly (bereft of distinct two-tail quantification of tall or flat content) 
as `{\it splinters}' \c{Kendall89}. 
This leaves us needing a specifically very flat (including highly bilaterally symmetric) conceptual name: {\it pickaxe heads} 
(paralleling \c{QuadI, QuadII}, which also use {\it spear head} in place of `needle'.)
See Fig \r{S(3, 2)-Splinters}.a) for the diversity of types of splinter, and Fig \r{S(3, 2)-Splinters}.b) for where these are located in the shape sphere. 
Kendall furthermore devised \c{Kendall89, Kendall} statistical tests for splinters as configurations of approximately-collinear triples of points, 
by which his treatment of splinters is a valuable quantitative prototype of Shape Statistics.    
%
{            \begin{figure}[!ht]
\centering
\includegraphics[width=0.75\textwidth]{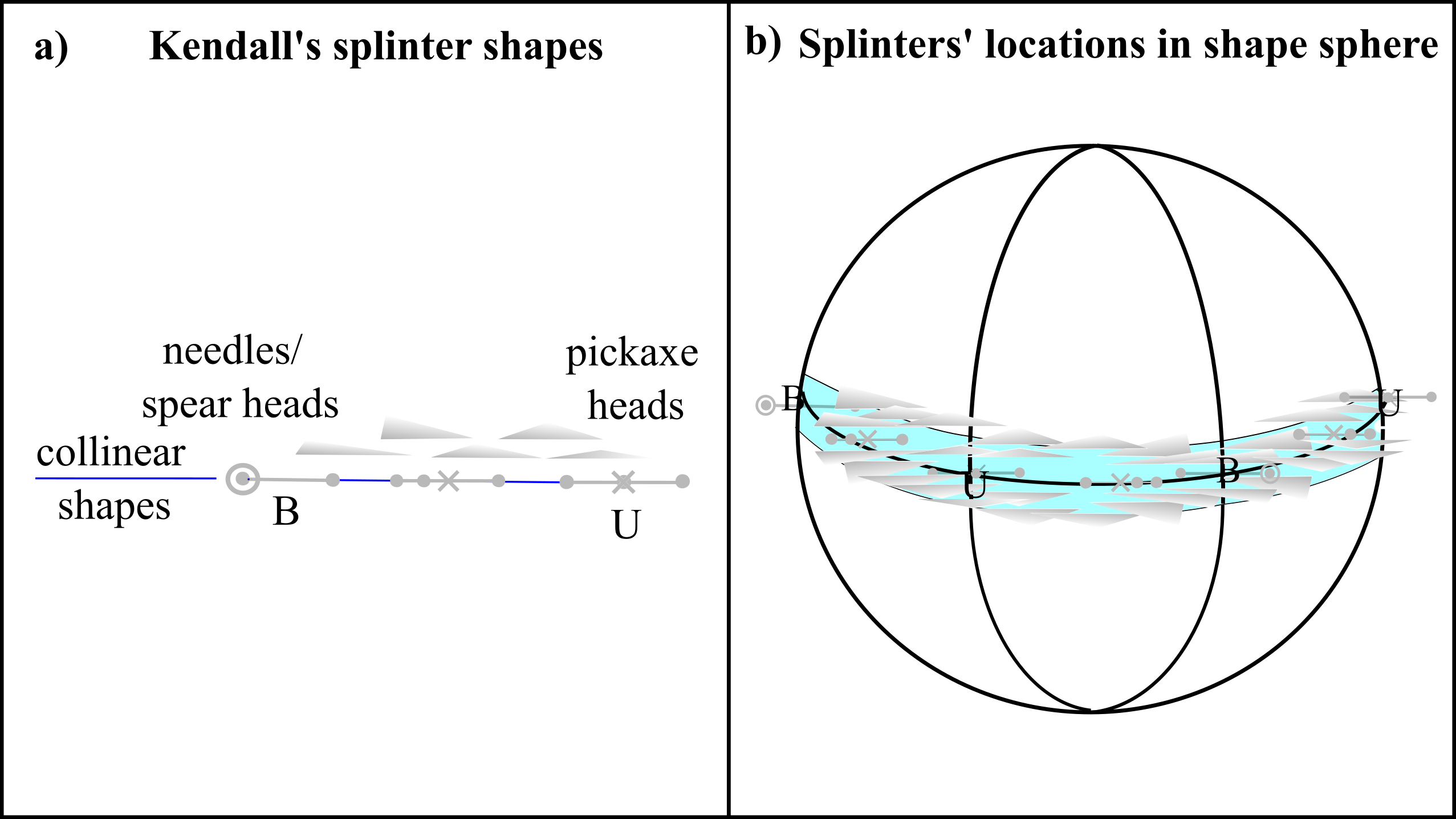}
\caption[Text der im Bilderverzeichnis auftaucht]{        \footnotesize{a)  Types of splinter shapes, and b) where they are to be found in the triangleland shape sphere.
%
} }
\l{S(3, 2)-Splinters} \end{figure}          }

\m

\n{\bf Remark 11} The halfway shape H for (3, 2) corresponds to
\be
\mH = \mC \caps  \mR  \m .
\ee
This notion moreover indeed corresponds to Part I's (3, 1) notion of H shape.  

\m 

\n{\bf Remark 12} See Fig \r{BIEMC-6} for a topological-level diagram of how this subsection's notions fit together.
%
{            \begin{figure}[!ht]
\centering
\includegraphics[width=0.8\textwidth]{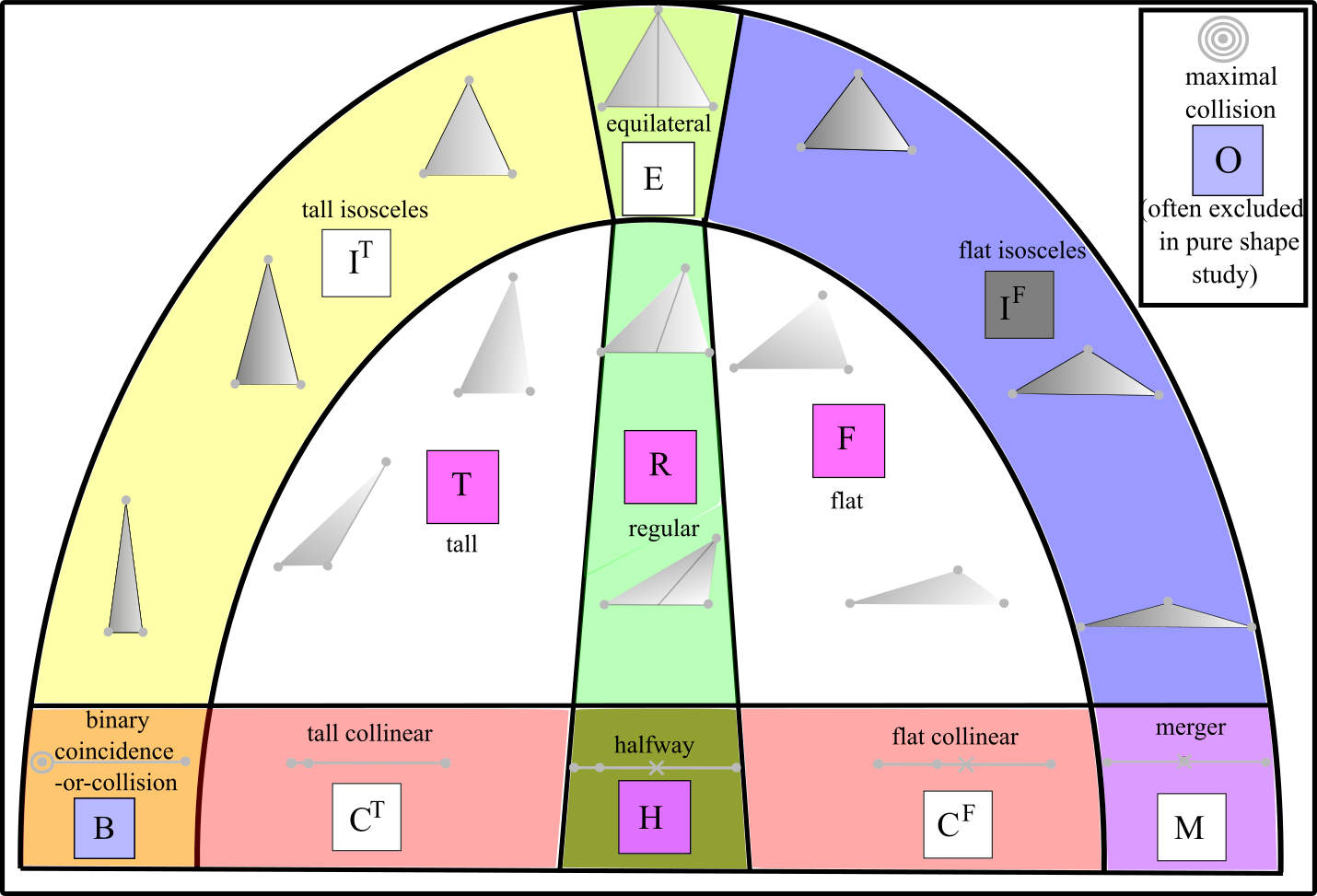}
\caption[Text der im Bilderverzeichnis auftaucht]{        \footnotesize{Shape-space-topological-manifold-level diagram for 
combined metric Lagrangian-and-Jacobian relative separation shapes-in-space. 
In Sec \r{Hopf}, we justify that this combination further merits the further name of Hopf structure.}}
\l{BIEMC-6} \end{figure}          }

\subsection{Jacobian recharacterization of right, acute and obtuse}\l{Jac-Right}

\n While we have already characterized rightness in Lagrangian terms, it turns out to admit a shape-theoretically useful Jacobian reformulation as follows. 

\m

\n{\bf Proposition 1} Right triangles are those for which 
\be
\frac{    R_2    }{    R_1    }                                  = \frac{    1    }{    2    }               \Leftrightarrows
\frac{    \rho_2    }{    \rho_1    }                            = \frac{    1    }{    \sqrt{3}    }        \Leftrightarrows
\frac{    \bigiota_{2}   }{    \bigiota_{1}    }  = \frac{    1    }{    3    }                                    \m .  
\l{rho-rho}
\ee
\Proof The angle subtended by a diameter is right.
Reformulate as the median protruding from the right angle having the same length as half of the opposite base.
This is in plain rather than mass-unweighted space, giving the first equation.
The second then follows from mass weighting as per (I.34), and the third by squaring.  $\Box$

\m

\n{\bf Remark 1} The second sentence is principle 5 of Chapter 5 of \c{Schaum}; Proposition 1 boils down to pinning a Jacobian interpretation on this result. 

\m

\n{\bf Corollary 1} Acuteness and obtuseness are then characterized in Jacobian terms by the following inequalities:     
\be
\frac{    R_2    }{    R_1    }      \m \gs \m \frac{    1    }{    2    }        \Leftrightarrows
\frac{    \rho_2    }{    \rho_1   } \m \gs \m \frac{    1    }{    \sqrt{3}    } \Leftrightarrows
\frac{\bigiota_{2}}{\bigiota_{1}}  = \frac{    1    }{    3    }                               \m,
\ee
and 
\be
\frac{    R_2    }{    R_1    }      \m \ls \m \frac{    1    }{    2    }        \Leftrightarrows
\frac{   \rho_2    }{    \rho_1    } \m \ls \m \frac{    1    }{    \sqrt{3}    } \Leftrightarrows
\frac{\bigiota_{2}}{\bigiota_{1}}  = \frac{    1    }{    3    }                               \m,   
\ee
respectively. 

\m

\n{\bf Remark 2} B is a limiting case of rightness, as one of the other angles becomes arbitrarily acute.  

\m

\n{\bf Remark 3} There is also a special right-and-regular triangle, which we denote by  
\be
\mR^{\perp} = \perp \caps\,  \mR \m ;
\ee
details of this shape can be found in Fig \r{Special-Triangles-2}. 
%
{\begin{figure}[ht]
\centering
\includegraphics[width=0.15\textwidth]{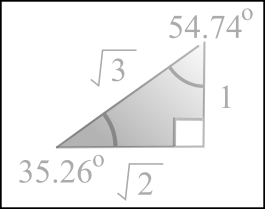}
\caption[Text der im Bilderverzeichnis auftaucht]{\footnotesize{The right-and-regular triangle $\mR^{\perp}$.
It has no uniformities of any kind.}} 
\l{Special-Triangles-2}\end{figure} } 

\m 

\n{\bf Remark 8} Combining the right and regular decor, we have the final topological picture of shape space structure in Fig \r{BIEMC-5}.
%
{            \begin{figure}[!ht]
\centering
\includegraphics[width=0.8\textwidth]{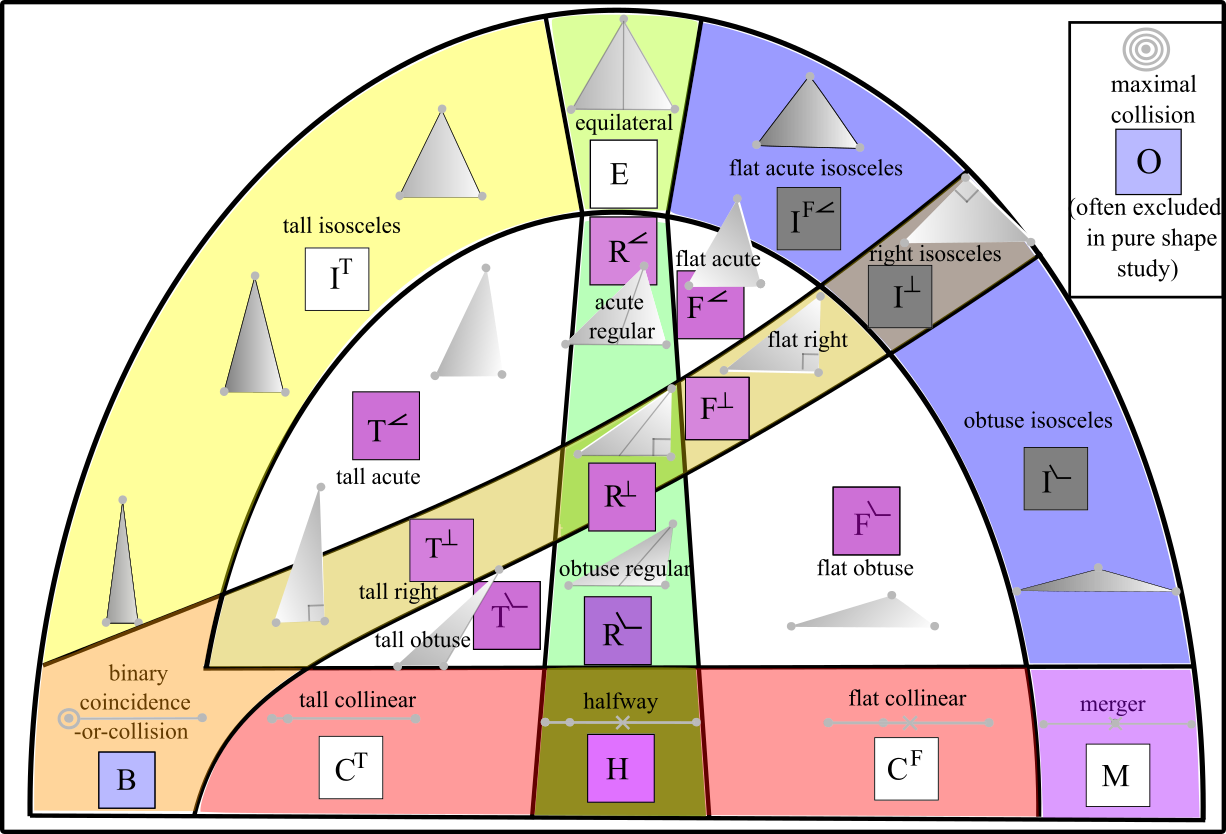}
\caption[Text der im Bilderverzeichnis auftaucht]{        \footnotesize{Shape-space-topological-manifold-level diagram 
of the full metric Jacobian-level shapes-in-space.}}
\l{BIEMC-5} \end{figure}          }

\section{Jacobi-level analysis of shape space}\l{Jac-Str}
%
{            \begin{figure}[!ht]
\centering
\includegraphics[width=0.6\textwidth]{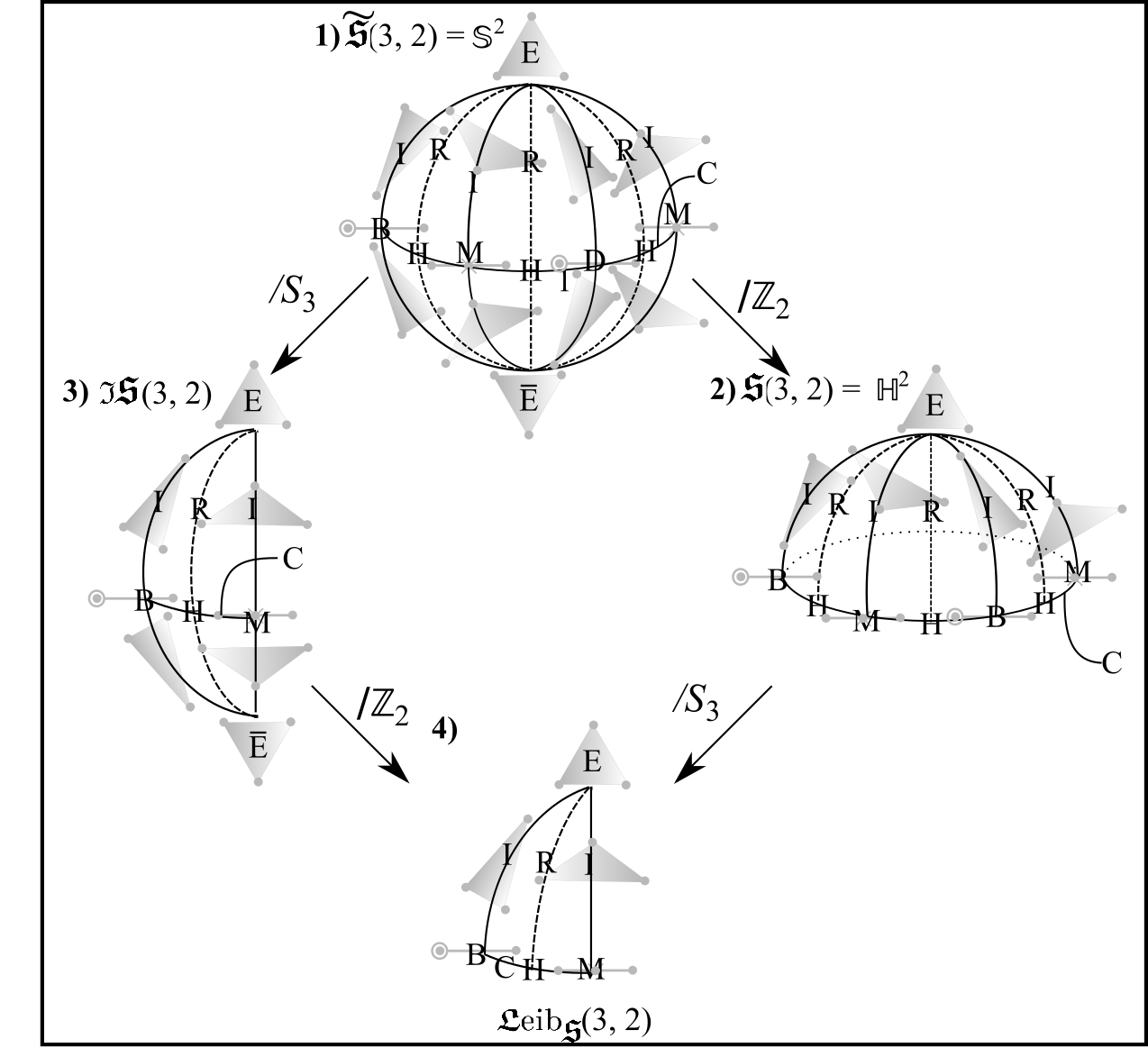}
\caption[Text der im Bilderverzeichnis auftaucht]{        \footnotesize{Triangleland configuration spaces at the metric level.

\m

\n 1) The metric-level shape sphere, now including the Hopf equal status collinear C, isosceles I and regular R shapes. 
The overall pattern is now that of a 12-segmented orange cut in half perpendicular to its segments. 
The whole and cut-in-half $k$-segment oranges are the `unexceptional' tessellations of the sphere \c{Magnus}. 
Among these, the orange of 12 segments admits the more vivid description as the Zodiac subdivision of the night sky, 
and this cut in half is the Zodiac split into northern and southern skies.   

\m

\n 2), 3) and 4) then give the mirror image indentified shape hemisphere, indistinguishable labels lune of width $\frac{\pi}{3}$, 
and Kendall's spherical blackboard alias Leibniz space $\Leib_{\sFrS}(3, 2)$. 
The C, I, R decor for these form the northern Zodiac, two star signs' worth of Zodiac, and two star sign's worth of northern Zodiac respectively.} }
\l{S(3, 2)-Metric} \end{figure}          }

\n{\bf Remark 1} The R's form 3 bimeridians which intersect E (and $\overline{\mE}$.
Thus they intersect the three bimeridians of isoscelesness I there as well (Fig \r{S(3, 2)-Metric}).  
By symmetry, the 3 R bimeridians are at $\frac{\pi}{3}$ to each other.
They are also midway between the I bimeridians.  
Thus we have a total of six significant bimeridians through E at $\frac{\pi}{6}$ to each other. 

\m 

\n{\bf Remark 2} Each clustering choice's bimeridian of regularity R separates the shape sphere into hemispheres of tall and flat triangles (relative to the same clustering).      

\m

\n{\bf Proposition 2} The Jacobi-uniform (and Jacobi-symmetric) structure consists of the R bimeridians alongside the I bimeridians (Fig \r{(3, 2)-Structures-Top}).  

\m

\n {\bf Remark 3} This forms the same structure as the Lagrangian uniformity structre, except that it is at $\frac{\pi}{6}$ as Lagrange uniformity.  

\m

\n{\bf Remark 4} Values at which B and M are that ${\cal V}$ sends the non-axial M's to 1 (and the nonaxial B's to 3), used in \c{A-Perimeter}.  

\m

\n{\bf Proposition 3} i) The full Lagrange--Jacobi uniformity structure (Fig \r{Leib(3, 2)-Hopf}.1) is the orange of 12 segments, alias Zodiac split of the sky.

\m

\n ii) The full Lagrange--Jacobi structure is the orange of twelve segments cut perpendicularly in half, 
with three additional cap-circles of rightness which kiss pairwise at the 3 B points. 

\m

\n{\bf Remark 5} Together, these intersecting arcs form the topology of the disc with eleven punctures (Appendix B).   
These intersecting arcs also provide the `orange of 12 segments' tessellation (not cut in half).

\m

\n{\bf Remark 6} The above discussion has moreover established why there are two notions of isosceleness split by E: 
these are {\it tall isosceles} I$^{\sT}$ versus {flat isosceles} triangles I$^{\sF}$.   
In this way, regularness R is needed even to finish explaining the features of the shape space distribution of the equilateral E, isosceles I, and scalene S triangles.  

\m

\n{\bf Proposition 1} The equilateral triangle E is uniquely characterized in Leibniz space as being 
\be
\mE =  \mI \caps \mR  \m .
\ee
\n{\bf Remark 7} Splinters lie near the arc of equator of collinearity, whereas needles can be found at the B end of this and pickaxe heads at the opposite M end.  

\m

\n{\bf Remark 8} Together, the Lagrangian and Jacobian separation structure form the topology of the disc with 11 points removed 
(consult Figs \r{(3,2)-Structure-Web} and Fig \r{(3, 2)-Structures-Top}) and are the basis of the `orange of 6 segments' tessellation.
The joint uniformity structure is the same minus the coincidence-or-collision points in shape space, which is topologically the same as the symmetry structure. 
The Lagrangian and Jacobian separation structure alongside collinearity --has the of the disc with 22 points removed (c.f.\ the same figures).
and corresponds to the `6-segmented orange cut in half' tessellation.
Note that motivation for considering Lagrangian and Jacobian separations and collinearity together (or pairs thereof) will be greatly enhanced in Sec \r{Hopf} 
where these are jointly identified as the Hopf structure.  

\m

\n{\bf Remark 9} This is a good point at which to recap that Sec \r{HHK} contained a shape space level result, namely that diagonalizing the Heron map yields the shape sphere.

\subsection{Maximum area per unit moment of inertia plotted over shape space}\l{A/I-Plot}
%
{            \begin{figure}[!ht]
\centering
\includegraphics[width=0.78\textwidth]{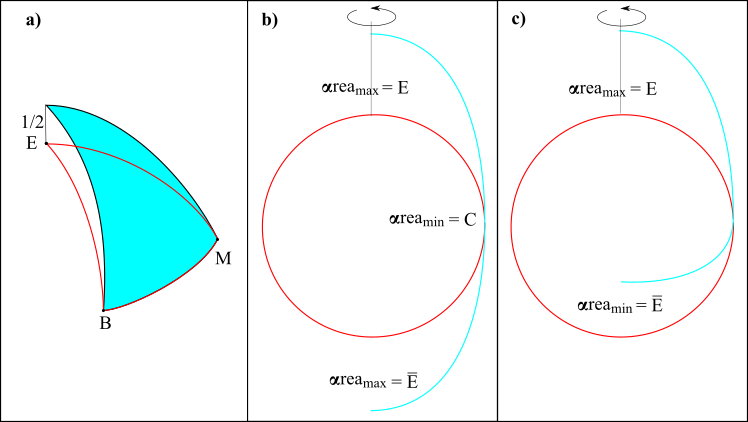}
\caption[Text der im Bilderverzeichnis auftaucht]{        \footnotesize{$\area_{\sm\sa\sx}$ is the corresponding surface of revolution about the E axis 
a) on Leibniz space, and b), c) on the shape sphere in the unsigned and signed cases respectively.}}
\l{Amax} \end{figure}          }

\n{\bf Proposition 1} i) At the level of shape space, for mirror image identified versions,  
\be
\area_{\sm\sa\sx} \mbox{ is at the equilateral point E } ,  
\l{A-Max-2} 
\ee
%
%
and 
\be
\area_{\sm\si\sn} \mbox{ is shared by the whole collinearity arc C } .   
\ee
ii) For the mirror image distinct versions, for the unsigned version of $\area$, 
\be
\area_{\sm\sa\sx} \mbox{ is at the equilateral points E and $\bar{\mE}$ } ,    
\ee
and its minimum is as above. 
For the signed version, (\r{A-Max-2}) holds and $\area_{\sm\si\sn}$ is at $\bar{\mE}$.

\m

\n{\bf Remark 1} This is axisymmetric, as is most easily seen in the EM-axis system: 
\be
\mbox{(normalized area)} = \cos \, \w{\Theta}                                                       \m,
\ee
and Fig \r{Amax}.

\subsection{Completing the main axis systems}
%
{\begin{figure}[ht]
\centering
\includegraphics[width=0.17\textwidth]{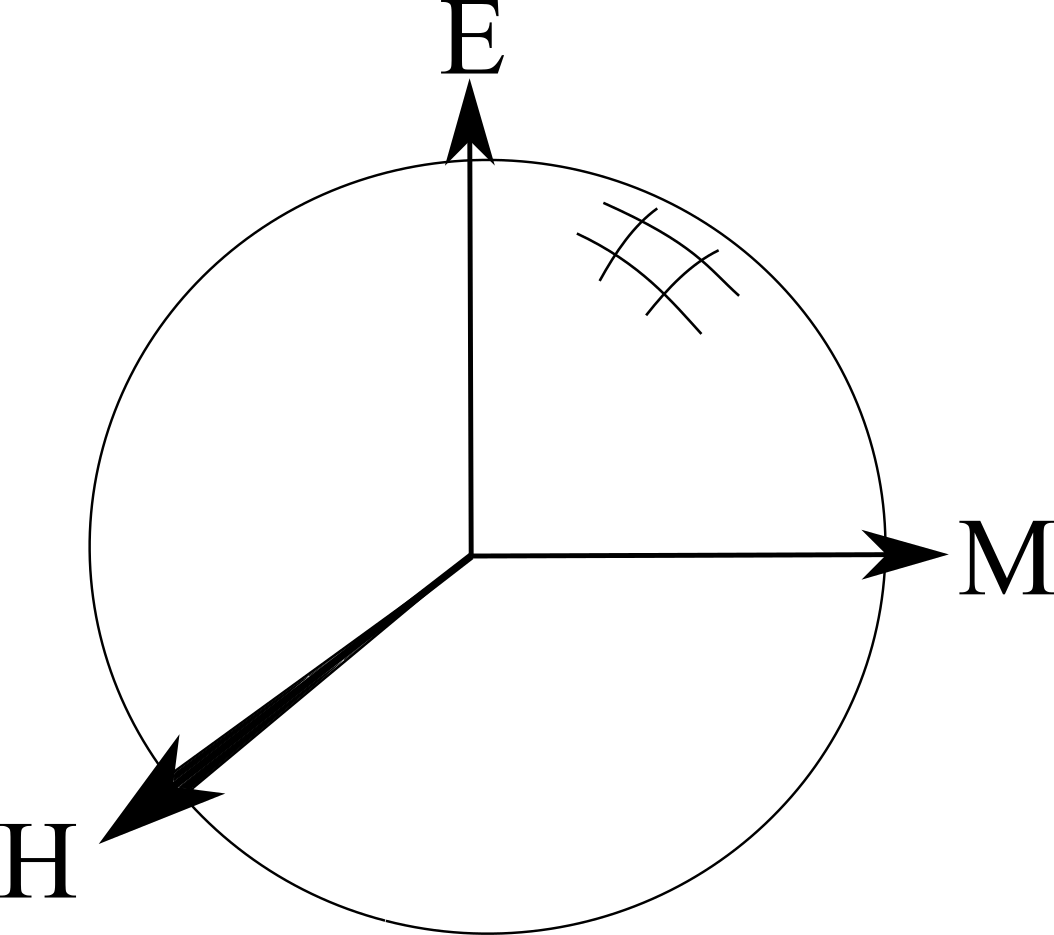}
\caption[Text der im Bilderverzeichnis auftaucht]{\footnotesize{EUH axis system.}} 
\l{E-U-H}\end{figure} } 

\n{\bf Remark 1} E$\overline{\mE}$ and MB are natural candidates for Cartesian axes for an ambient $\mathbb{R}^3$ for the triangleland shape sphere.
While three different clustering's MB axes are available, only one can be used in any given axis system since they are not perpendicular 
(being rather at $\frac{2 \, \pi}{3}$ to each other).
On the other hand, E$\overline{\mE}$ is perpendicular to all the UB axes.  
The final perpendicular direction picks out a HH axis (Fig \r{E-U-H}).  
The incipient Jacobi coordinates pick out the underlying cluster's BM axis as its principal axis, 
but this axis system can be relabelled so that the unique and more natural EE axis is the principal axis.  

\m

\n{\bf Remark 2} Following up Sec \r{Jac-Cong-Sim}, the triangle can moreover be explicitly solved in a straightforward manner in terms of shape space spherical polar data 
$\Theta$, $\Phi$ with UB as principal axis, by using the stereographic--azimuthal relation (I.174) in (I.172).  
The further passage to shape space spherical data with E$\bar{\mE}$ principal axis remains analytic, albeit with somewhat more elaborate formulae.

\subsection{Spherical, Legendre and parabolic-type variables}

\n{\bf Remark 1} Let us first recollect that the azimuthal shperical angle $\Theta$ is related to the mass-weighted relative Jacobi separations ratio 
\be
{\cal R} \es \frac{\rho_2}{\rho_1} 
\l{R-rho-rho}
\ee
by  
\be
{\cal R} \es \tan\frac{\Theta}{2}  \m .
\l{R-Theta} 
\ee 
In shape space geometry terms, ${\cal R}$ is moreover the stereographic radius.  

\m

\n{\bf Definition 1} Another useful variable in the study of spherical geometry is the {\it Legendre variable} \cite{CH}; 
this features for instance in simplifying the azimuthal part of the (now triangleland shape) spherical harmonics equation \cite{+Tri, FileR}.
The shape sphere realization of this relative to the UEH axis system is 
\be
{\cal X} := \cos\,\Theta = \ellip \m . 
\label{Legendre}
\ee 
\n{\bf Lemma 1} Thus its relation to ${\cal R}$ and to the $\rho_i$ is 
\be
\\ellip  =  {\cal X} 
        \es \frac{1 - {\cal R}^2}{1 + {\cal R}^2} 
        \es	\frac{\rho_1\mbox{}^2 - \rho_2\mbox{}^2}{\rho_1\mbox{}^2 + \rho_2\mbox{}^2}	 \m . 
\l{X-R}
\ee 
The first of these forms moreover inverts to 
\be 
{\cal R} \es \sqrt{\frac{1 - {\cal X}}{1 + {\cal X}}} 
         \es \sqrt{\frac{1 - \ellip}{1 + \ellip}}      \m .
\ee 
\n{\bf Definition 2} The Legendre variable corresponding to the EUH axis system is 
\be
\w{\cal X} := \cos\,\w{\Theta} = \tau               \m . 
\label{w-Legendre}
\ee 
\n{\bf Remark 2} While this has the added virtues of being democratic and closely related to the intuitive area variable, its link to the relative variables is less direct.  

\m

\n{\bf Remark 3} At the level of relational space, each of $(\bigiota, \, \Theta, \, \Phi)$ and $(\bigiota, \, \w{\Theta}, \, \w{\Phi})$ constitute spherical polar coordinates. 

\m

\n{\bf Remark 4} The incipient relational coordinates $(\rho_1, \, \rho_2, \Phi)$ of Fig I.19.a), on the other hand, 
turn out to be closely related to {\it parabolic coordinates} \cite{MF}: 
\be
(\xi_1, \xi_2, \Phi) := (\sqrt{2}\rho_1, \sqrt{2}\rho_2, \Phi)   \m. 
\ee
This is the way that the base-and-apex alias side-and-corresponding-median split is encoded into the mathematics of triangleland.

\subsection{Solving the triangle in terms of shape space spherical polar data}

From (\r{calR-ca}), 
\be
\frac{b_{\pm}}{a} \es \sqrt{    \frac{  2 - \cos \, \Theta    \pm    \sqrt{3} \, \sin \, \Theta \, \cos \, \Phi    }{    1 + \cos \, \Theta   }     }     \m ,    
\ee 
whereas from (\r{cot-alpha}), 
\be
\cot\,\gamma_{\pm} \es \frac{1}{\sqrt{3}} \, \cosec\,\Phi \, \cot \, \frac{\Theta}{2} \m \pm \m \cot\,\Phi \m \mbox{ and }
\ee
\be
\cot\,\alpha = \frac{   1   }{   2   } \, \cosec\,\Phi \left( \sqrt{3} \, \tan\frac{\Theta}{2} - \frac{1}{\sqrt{3}} \, \cot\frac{\Theta}{2}  \right) 
                   = \frac{   1 - 2 \, \cos \, \Theta   }{   \sqrt{3} \, \sin \,\Theta \, \sin\,\Phi  }                                          \m .
\ee

\subsection{Hopf-level count of qualitative types}\l{Hopf-SS}
%
{            \begin{figure}[!ht]
\centering
\includegraphics[width=1\textwidth]{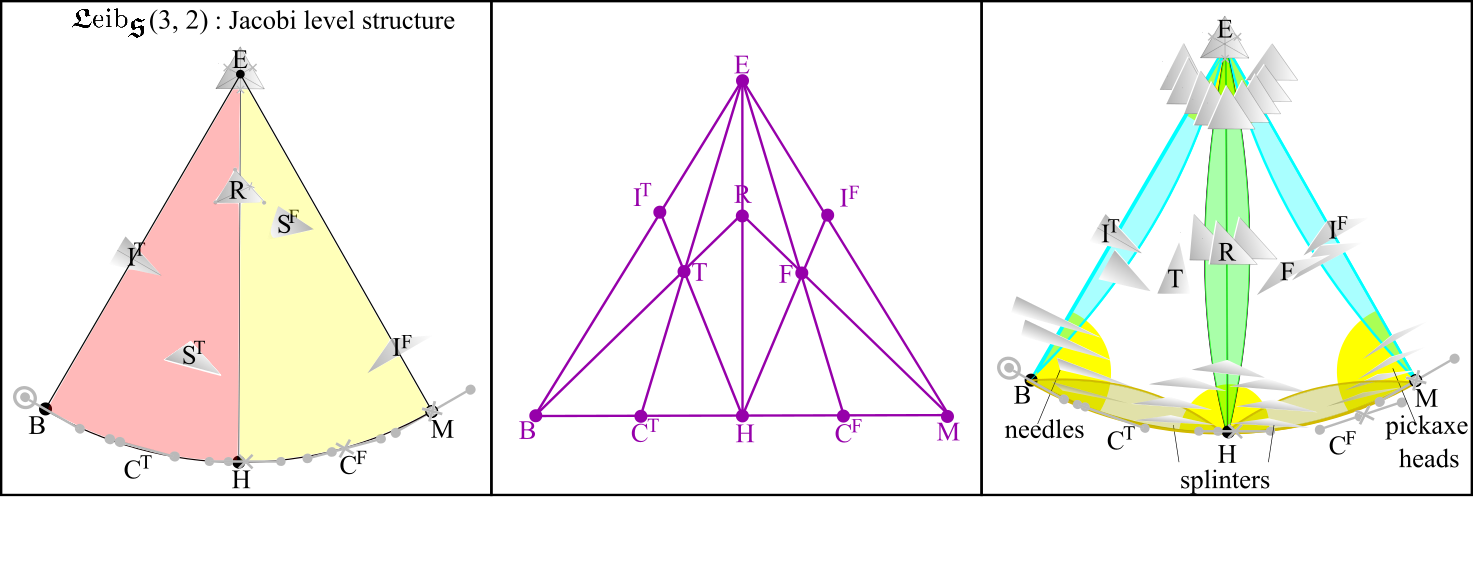}
\caption[Text der im Bilderverzeichnis auftaucht]{        \footnotesize{a) Metric-level detail of the shape space $\Leib_{\tFrS}(3, 2)$ 
marking what will be identified as the Hopf-level metric shapes-in-space.   

\m

\n b) Corresponding adjacency graph.  

\m

\n c) Corresponding exact and approximate qualitative types.} }
\l{Leib(3, 2)-Hopf} \end{figure}          }

\n We firstly consider this without rightness, since this case will later become geometrically significant as the Hopf structure (Sec \r{Hopf}).

\m 

\n{\bf Remark 1} This has $V = 4$ types of special vertices: E, B, M, and H.

\m

\n{\bf Remark 2} This also has $E = 5$ edges: $\mI^{\sT}$, $\mI^{\sF}$ as before, a new $\mR$ and the $\mC^{\sT}$, $\mC^{\sF}$ subdivision of the $\mM$ arc.  
%
{            \begin{figure}[!ht]
\centering
\includegraphics[width=0.75\textwidth]{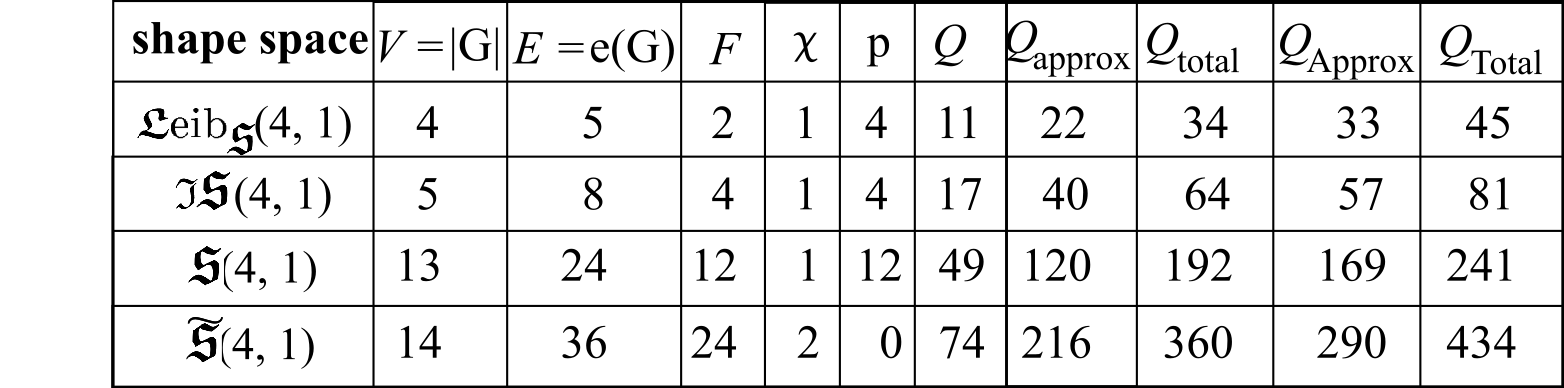}
\caption[Text der im Bilderverzeichnis auftaucht]{        \footnotesize{Number of qualitative types including the full Jacobian decor.
Note that the first and fourth of these are the same as for the full Lagrangian decor, due to underlying isomorphic graphs.
The second row's graph is isomorphic to the 4-wheel, $\mW_4$.  }}
\l{(3, 2)-Q-Hopf} \end{figure}          }

\m

\n{\bf Remark 3} Finally, there are now $F = 2$ faces: the tall $\mT$ and flat $\mF$ subdivision of the scalene triangles $\mS$.

\m

\n{\bf Remark 4} These vertices, edges and faces can be located, and counted off, as per Fig \r{Leib(3, 2)-Hopf}.b).

\m

\n{\bf Remark 5} Thus the number of exact qualitative types is
\be
Q := V + E + F = 4 + 5 + 2 = 11  \m .
\ee
\n{\bf Remark 6} The adjacency graph \r{Leib(3, 2)-Hopf}.b) has $V(\mbox{Adj}) = Q = 11$ and $E(\mbox{Adj}) = 12$ 
by being isomorphic to Sec \r{Lag-Str-Right}'s graph.  

\m

\n{\bf Remark 7} See Figs \r{(3,2)-Structure-Web}--\r{(3,2)-Full-Structure-Web} for an end-summary of special arcs and special points among the (3, 2) shapes.  
See Fig \r{(3, 2)-Q-Hopf} for further counts of qualitative types in the four (3, 2) shape spaces, as exhibited in Fig \r{(3, 2)-Approx-Detail} in the case of Leibniz space.   
%
{\begin{figure}[ht]
\centering
\includegraphics[width=0.52\textwidth]{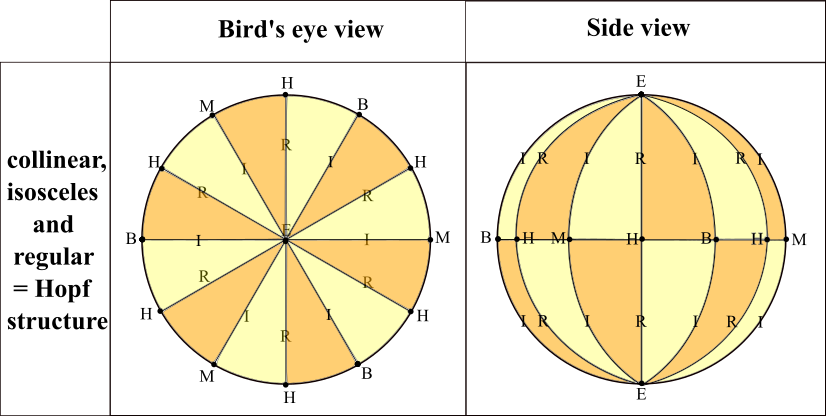}
\caption[Text der im Bilderverzeichnis auftaucht]{\footnotesize{The Hopf structure's 12-segment orange cut in half, 
alias Zodiac split into northern and southern skies.
The bird's eye view of this presents a clock face.}} 
\l{S(3, 2)-Hopf-Detail}\end{figure} } 

\subsection{Centre of Mass rigidity in triangles}

\n Hierarchical excess is interestingly trivial for triangleland; see Part IV for its further nontriviality for quadrilateralland.  

\m

\n {\bf Theorem 0 (Median Concurrence Theorem.)}  A triangle's three medians meet at a point.  

\m

\n{\bf Remark 1} This is elementary geometry, and also clear in the context of hanging triangles from each vertex.
The concurrence point is the total CoM. 
While perhaps more often stated for triangular laminae, this result also holds in the constellational context of equal-mass points at each vertex. 

\m

\n{\bf Remark 2} This takes the form of the below constellation clumping hierarchy reinterpretation (see also Fig \r{Euclid-in-CoMs}) 
of the following geometrical result already known to Euclid as a basic consequence of the Parallel Postulate. 

\m

\n{\bf Theorem 1}  For any triangle figure in general position (rather than collinear or with coincidences), 
the midpoints of the sides form a similar triangle to that of the points, reflected and of half the size.   

\m

\n{\bf Theorem 2} The total centre of mass of three point masses is at the intersection of the three medians of the triangle formed by the points.

\m

\n{\bf Theorem 3} The 2-particle CoMs form a similar triangle, reflected, of half the size.
The total CoM is then at the common centroid of the these two similar triangles. 

\m

\n{\bf Remark 3} This is a rigidity result: it says that there is no capacity for extra coincidences in CoM hierarchy for 3 particles in nondegenerate positions in 2-$d$.  
For degenerate -- collinear -- shapes, however, U(3, 1) offers the only way out.

\m

\n{\bf Remark 4} Like (3, 1), triangleland still only has one M notion corresponding to one M-point. 
But this is no longer aligned with the maximal U.   
So we need to discern between these two simplifying features. 

\m

{\begin{figure}[ht]
\centering
\includegraphics[width=0.85\textwidth]{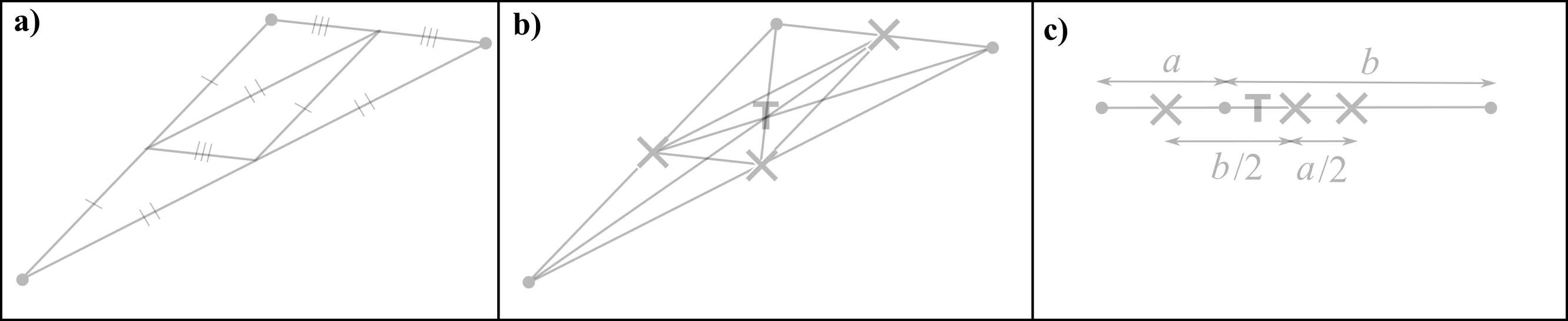}
\caption[Text der im Bilderverzeichnis auftaucht]{\footnotesize{a) A theorem of Euclid. 

\m

\n b) Its slight extension and reformulation in CoM terms.

\m

\n c) The collinear limit of this then recovers the (3, 1) CoM hierarchy rigidity result, by which there is just one M shape for (3, 1).
This recovers Part I's result (previously featuring in Fig I.34).
}} 
\l{Euclid-in-CoMs}\end{figure} } 

\n{\bf Remark 5} We now recognize the (3, 1) result of Fig I.34.    
While this theorem is usually stated for a general position triangle, it also holds for collineations and for coincidences-or-collisions. 

\m

\n{\bf Remark 6} We already know (from Part I) that for triangleland, parallelism always reduces to collinearity.  
Moreover, the only a priori distinct pure-Jacobi possibilities -- self- or mutual base--median collinearity and median--median mutual collinearity -- 
just reduce to the usual notion of collinearity.  
Part IV's quadrilaterals however suffice for some of these other cases to become nontrivial.  

\m

\n{\bf Remark 6} Secondly, once collinear, $\fT$ can coincide with a $\fP$.  
If this happens, it must additionally coincide with an $\fX$, according to `if particle 3 is at CoM(12), then CoM(123) coincides with these'.
So constellations of 3 possess precisely one merger configuration, which is collinear and a fortiori M.  

\m

\n{\bf Remark 7} A further useful 2-$d$ insight is to look for not just coincidences of particles and CoMs, 
but also for collineations of such (and eventually for right angles between such collineations). 

\m

\n{\bf Remark 8} Geometry's side midpoints are rephrased as the equal-masses 3-body problem's binary centres of mass.  
\n By this, another form of Theorem 1 is as follows. 

\m

\n{\bf Theorem 1$^{\prime}$} the binary centres of mass of any 3-point constellation form a similar constellation. 

\m

\n This is to be augmented as follows by the simple observation of where the total CoM is, and consequent collineations. 

\m

\n{\bf Theorem 1$^{\prime\prime}$} The binary centres of mass of any 3 point constellation form a similar constellation, 
with the centres of the two coincident and giving the position of the total centre of mass.  
Thus any constellation in G has 
a) three P$\fX$P collineations, 
b) three P$\fT$P collineations and 
c) three $\fX\fT\fX$ collineations.
All of b) and c) moreover intersect at $\fT$, and the angles between the three PP's are the same as those between the three $\fX\fX$'s by Theorem 1. 

\m

\n This means that none of these features are remarkable when examining more special triangles such as isosceles, equilateral, right or regular, 
since {\sl any} triangle has these features.  

\m

\n{\bf Remark 9} these coincidences determine where all the $\fX$'s and the $\fT$ are in relation to the three P's.  
This leaves no room for any excess clustering.

\subsection{Rightness, obtuseness and acuteness in shape space}\l{Right-SS}

\n We now address the question of where in the shape sphere the right triangles are located.

\m

\n{\bf Theorem 1} \c{Small, MIT, A-Pillow} The right triangles form three equal cap-spheres, each with azimuthal coordinate $\frac{\pi}{3}$ about the corresponding U point. 
These kiss at each B point.  

\m

\Proof 
Begin with Fig \r{In-Circum-DR}.c)'s 1 : 2 ratio.  
Reinterpret the hypotenuse as $R_1$ and the corresponding median as $R_2$, to obtain (\r{rho-rho})'s 
Jacobi-magnitude and mass-weighted Jacobi magnitude ratios. 
Combining with the definition (\r{R-rho-rho}) of ${\cal R}$ and its relation to $\Theta$ (\r{R-Theta}), this gives 
\be
\tan\frac{\Theta}{2} \es {\cal R} \es \frac{1}{\sqrt{3}}  \m .  
\l{R=1/sqrt3}
\ee
Thus, inverting,  
\be
\Theta  =  2 \, \arctan \, \frac{1}{\sqrt{3}} 
       \es 2 \times \frac{\pi}{6} 
       \es \frac{\pi}{3}                                  \m . 
\l{Theta=pi/3}
\ee  
This is sketched in Fig \r{1-Cap-Kissing-Square}.a) for one cluster choice's cap-circle of rightness. 
Repeat for each of the three clusters to obtain three cap-circles. 
Each is centred about the corresponding U-point pole, and kisses the other two at the two B points which are not antipodal to that U-point: 
{\it kissing cap-circles of rightness} as viewed in Fig \r{1-Cap-Kissing-Square}.b-e) from a variety of directions.                                             $\Box$
%
{            \begin{figure}[!ht]
\centering
\includegraphics[width=0.8\textwidth]{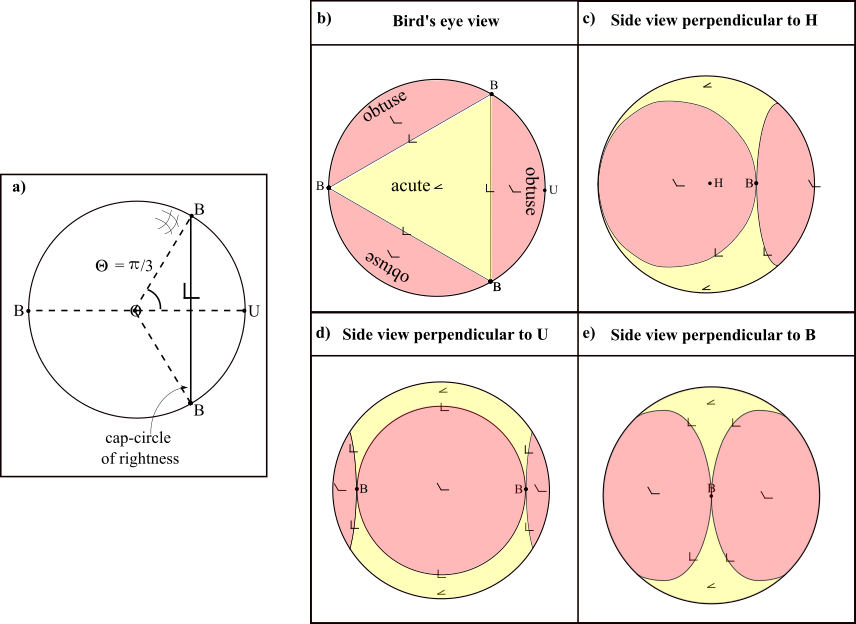}
\caption[Text der im Bilderverzeichnis auftaucht]{        \footnotesize{a) A given labelling of right triangles corresponds as indicated 
to a cap-circle of constant azimuthal value $\frac{\pi}{3}$.  

\mbox{ }

\n b)-e) Kissing cap-circles of rightness, bounding kissing caps of obtuseness (shaded red), as viewed from various directions.} }
\l{1-Cap-Kissing-Square} \end{figure}          }

\m

\n{\bf Remark 1} In terms of the useful variable ${\cal V}$, the rightness condition comes out in the normalized form ${\cal V} = 1$; 
this geometrically underlies why ${\cal V}$ is a useful variable.

\mbox{ }

\n{\bf Remark 2} The regions outside of these caps are acute, 
as seen e.g.\ from the observation that the equilateral triangles E residing at the poles are outside of these caps and are acute, and then applying a continuity argument.  
Conversely, the interiors of the caps are the obtuse triangles, so we term the caps themselves the {\it kissing caps of obtuseness}. 
There being three caps is commeasurate with there being three ways a labelled triangle can be obtuse.

\m

\n{\bf Remark 3} This cycle of pairwise-kissing circles can be deformed at the topological level by lengthening one attachment point to make the 3-finger knuckleduster.
In turn, this can be deformed by dragging one of the lengthening's endpoints across to the same circle as the other endpoint.
This forms the 4-bouquet: a topologically standard structure within the scope of Fig \r{Top-App-Fig}.  
This figure moreover shows this to be the deformation retract of the genus-2 surface. 
Alternatively, one can pass to the disc with 4 punctures and furthermore to the sphere with 5 punctures (both of these are  also in Fig \r{Top-App-Fig}).
On these grounds, the set of rightness has a well-understood general and algebraic topology.

\m

\n{\bf Remark 4} Overall, the shape sphere is split into three lunes of width $\frac{2 \, \pi}{3}$, 1 per cluster with equator-centred caps cutting each into three. 
Here clockwise and anticlockwise obtuse triangles are back-to-back about the equator with clockwise and anticlockwise acute disconnected from each other.  

\m

\n{\bf Remark 5} Noting the tall isosceles lines (Fig \r{Special-Triangles}), the acute regions of the shape sphere are also tripartitioned. 

\m

\n{\bf Remark 6} See row 1 of Fig \r{S(3, 2)-Detail} for the shape space decorated with the circles of rightness only, 
and rows 2 and 3 for the inter-relation between these and the meridians of isoscelesness and the equator of collinearity.

\subsection{Full Jacobi analysis of qualitative types for triangles}  
%
{\begin{figure}[ht]
\centering
\includegraphics[width=0.7\textwidth]{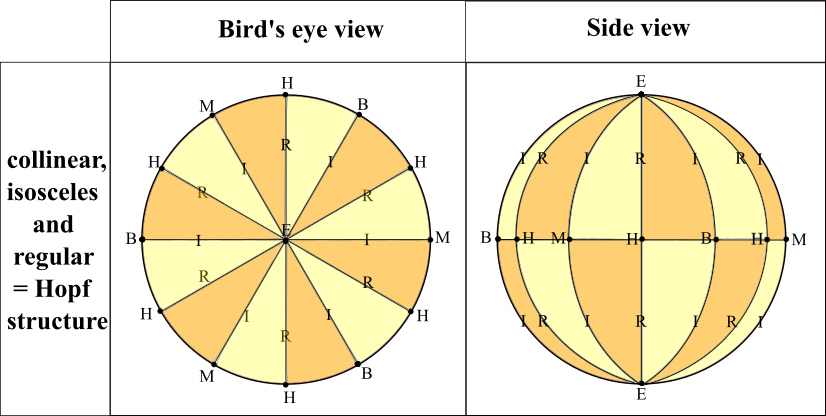}
\caption[Text der im Bilderverzeichnis auftaucht]{\footnotesize{Considering regular triangles in addition to Fig \r{S(3, 2)-Detail}'s structure gives a 48-face pattern, 
the bird's eye view of which is the current treatise's `large kaleidoscope'. 
It would be too crowded to label the lines and faces here as well, 
though we do display this for the $\Leib_{\sFrS}(3, 2)$ 1/12-sphere in the further Fig \r{Leib(3, 2)-Full-Jac}. } } 
\l{S(3, 2)-Detail-2}\end{figure} } 

\n We now include rightness so as to analyze the qualitative types for the full Jacobian structure.
This combines the right and regular structure (Fig \r{S(3, 2)-Detail-2}).

\m

\n{\bf Remark 1} This case has $V = 6$ vertices: E, B, M, H, I$^{\perp}$ and R$^{\perp}$.

\m

\n{\bf Remark 2} It also has $E = 9$ edges: $\mI^{\sT}$, $\mC^{\sT}$, $\mC^{\sF}$ as before, 
                      a further subdivision $\mI^{\sF \acute}$, $\mI^{\sF\obtuse}$ of $\mI^{\sF}$, 
                      a further subdivision $\mR^{\acute}$, $\mR^{\obtuse}$ of $\mR$, 
                              and new edges $\mT^{\perp}$ and $\mF^{\perp}$. 

\m

\n{\bf Remark 3} It finally has $F = 4$ faces: the (tall and flat) $\times$ (acute and obtuse) quartering of the scalene triangles, 
                                                giving $\mT^{\acute}$, $\mF^{\acute}$, $\mT^{\obtuse}$ and $\mF^{\obtuse}$.

\m

\n{\bf Remark 4} These vertices, edges and faces can be located, and counted off, as per Fig \r{Leib(3, 2)-Full-Jac}.a).

\m

\n{\bf Remark 5} Thus there is a total of 
\be
Q := V + E + F = 6 + 9 + 4 = 19  \m .
\ee
exact qualitative types in Leibniz space with full Jacobi decor. 

\m

\n{\bf Remark 6} In the adjacency graph of Fig  \r{Leib(3, 2)-Full-Jac}.b), $V(\mbox{Adj}) = Q = 19$ and $E(\mbox{Adj}) = 44$. 
This graph is moreover three 6-wheels $\mW_6$ and one 8-wheel $\mW_8$ joined cyclically, with one shared central vertex and four pairs of shared vertices. 
Thus 
\be
E(\mbox{Adj}) = 3 \times 12 + 16 - 4 \times 2 
              = 44                             \m . 
\ee 

\m

\n{\bf Remark 7} See Figs  \r{(3,2)-Structure-Web}--\r{(3,2)-Full-Structure-Web} for an end-summary of special arcs and special points among the (3, 2) shapes.  
See Fig \r{(3, 2)-Q-Jac} for further counts of qualitative types in the four (3, 2) shape spaces, 
as exhibited in Fig \r{(3, 2)-Approx-Detail} in the case of Leibniz space.   
%
{            \begin{figure}[!ht]
\centering
\includegraphics[width=0.75\textwidth]{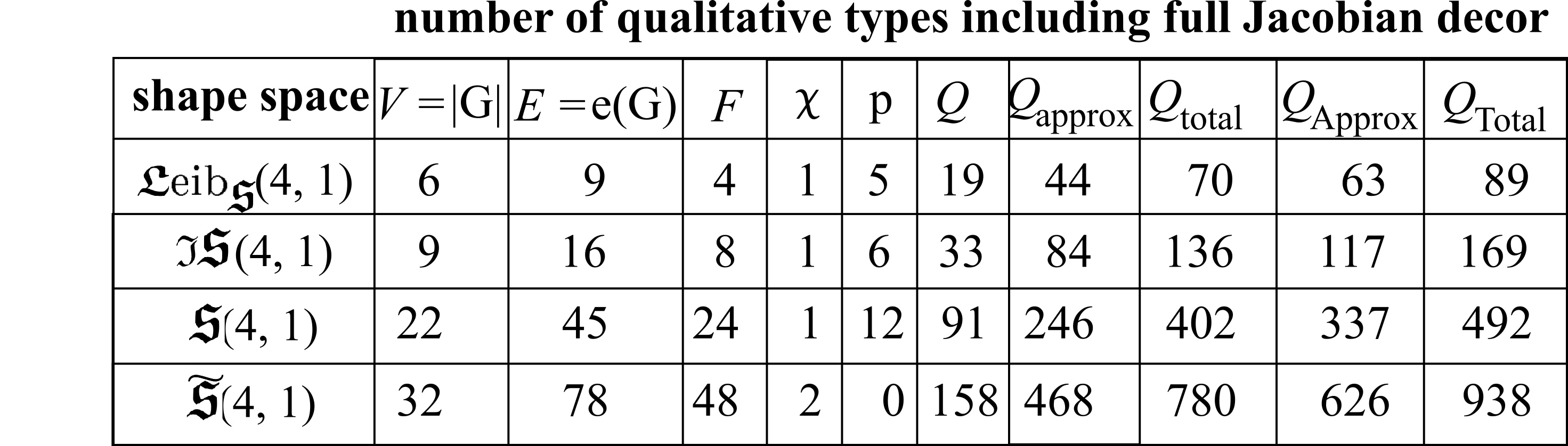}
\caption[Text der im Bilderverzeichnis auftaucht]{        \footnotesize{Number of qualitative types including the full Jacobian decor.}}
\l{(3, 2)-Q-Jac} \end{figure}          }
%
{            \begin{figure}[!ht]
\centering
\includegraphics[width=1.0\textwidth]{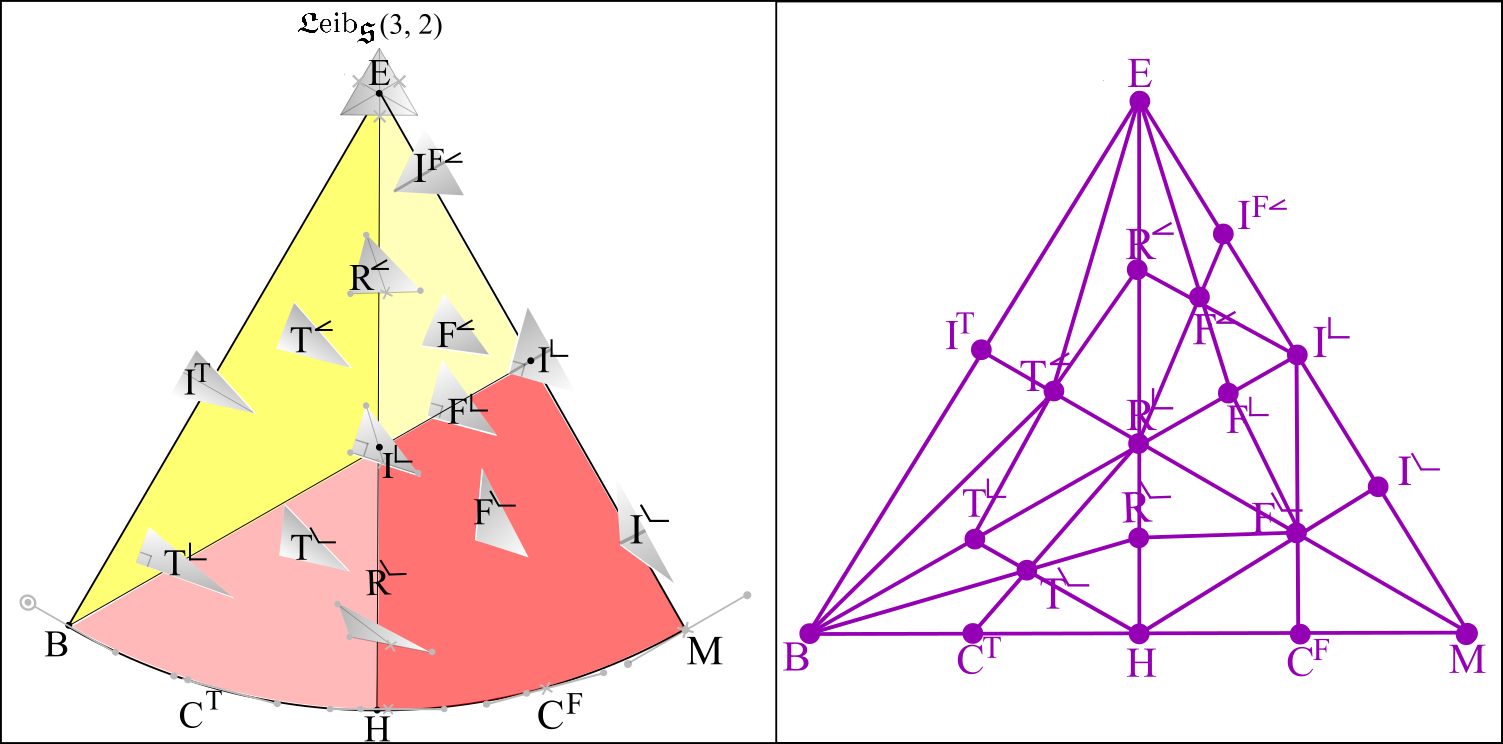}
\caption[Text der im Bilderverzeichnis auftaucht]{        \footnotesize{a) Full Jacobian-level detail of $\Leib_{\tFrS}(3, 2)$. 
b) The corresponding adjacency graph.}}
\l{Leib(3, 2)-Full-Jac} \end{figure}          }
%
{            \begin{figure}[!ht]
\centering
\includegraphics[width=0.5\textwidth]{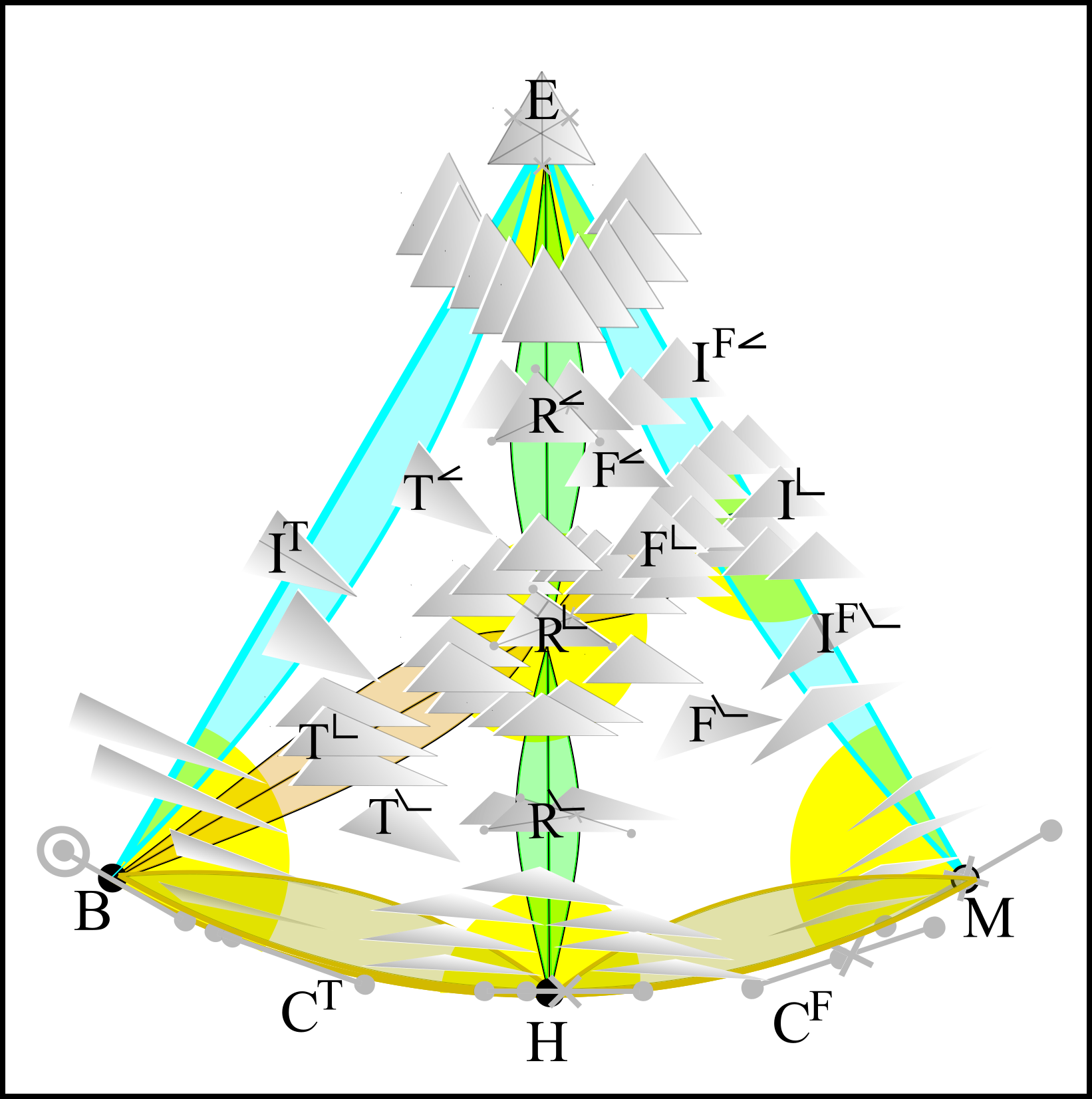}
\caption[Text der im Bilderverzeichnis auftaucht]{        \footnotesize{Exact and approximate qualitative types of triangles 
at the full Jacobian level of structure.}}
\l{(3, 2)-Approx-Detail} \end{figure}          }

\subsection{Shape space centres}
%
{            \begin{figure}[!ht]
\centering
\includegraphics[width=0.5\textwidth]{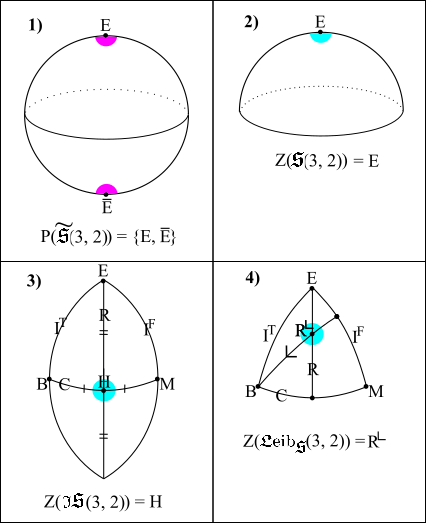}
\caption[Text der im Bilderverzeichnis auftaucht]{        \footnotesize{Shape space centres $Z$ for the various trianglelands; $P$ denotes poles.} }
\l{Z(GS(3,2))} \end{figure}          }

\vspace{10in}

\n{\bf Observation 1} $\FrS(3, 2)$ has a single centre, $Z(\FrS(3, 2)) = \mE$ (Fig \r{Z(GS(3,2))}.2).  

\m

\n{\bf Observation 2} On the other hand, $\w{\FrS}(3, 2)$ has two centres: the equilateral triangles at the poles (Fig \r{Z(GS(3,2))}.1). 
Having two centres is an indication of being a double cover.

\m

\n{\bf Observation 3} $\mH = Z(\FrI\FrS(3, 2))$ is the indisputible centre of $\FrI\FrS(3, 2)$, 
due to this possessing two perpendicular lines of symmetry, R and C, which intersect at H (Fig \r{Z(GS(3,2))}.3).  

\m

\n{\bf Observation 4} In $\Leib_{\sFrS}(3, 2)$, however, E is a vertex.
$\Leib_{\sFrS}(3, 2)$ is moreover an isosceles spherical triangle, so its notions of centre must lie on its line of symmetry, the semimeridian R.  
The point H -- also on the R meridian -- however features just as an edge midpoint for $\Leib_{\sFrS}(3, 2)$, 
so it is not available as a shape-theoretic centre for $\Leib_{\sFrS}(3, 2)$.  
Thus E and H are not available as a notion of centre. 
We then suggest the intersection of $\perp$ and R to be a shape theoretically significant notion of centre (Fig \r{Z(GS(3,2))}.4).

\subsection{The maximal angle flow}\l{MAF}

The current treatise's first new result is that the maximal angle curves on the shape sphere are underlied by the following constant-angle curves.  

\m

\n{\bf Theorem 2} The curves of constant angle 
\be
\alpha \neq \frac{\pi}{2}
\ee 
-- the angle defined in Fig \r{abc0} -- are given by  
\be
\Phi   \es   \arcsin \left(  \frac{    1 - 2 \, \cos \, \Theta    }{    \sqrt{3} \, k \, \sin \, \Theta }  \right)   
       \es   \arcsin  \left(  \frac{    1 - 2 \, {\cal X}    }{    \sqrt{3} \, k \, \sqrt{   1 - {\cal X}^2  }   }  \right)    \m . 
\l{Phi-eq-2}
\ee 
where 
\be 
k := \cot \, \alpha   \m . 
\ee 
\n(See \c{Max-Angle-Flow} for a proof.) 

\m 

\n{\bf Remark 1} We sketch and interpret the maximal-$\alpha$ curves that follow from these constant-$\alpha$ curves in Fig \r{Maximal-Angle-Flow}. 
Note that the kissing cap-circles of rightness recur within this flow as a separatrix.  
See \c{Max-Angle-Flow} for further analysis and discussion of this flow.  

{            \begin{figure}[!ht]
\centering
\includegraphics[width=0.8\textwidth]{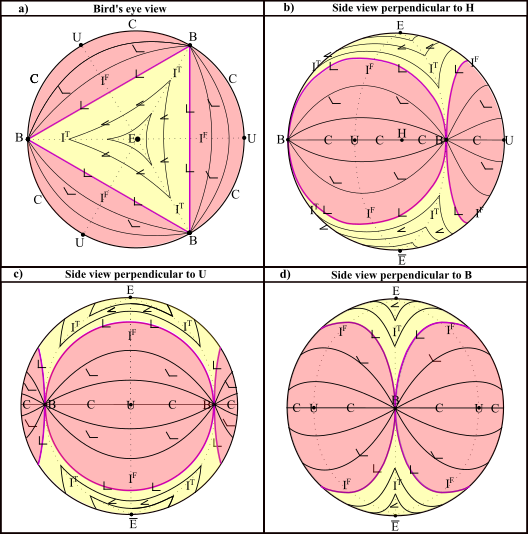}
\caption[Text der im Bilderverzeichnis auftaucht]{        \footnotesize{The maximal angle flow over the shape sphere of triangles as seen from various directions.
The kissing cap-circles of rightness separatrix is highlighted in indigo, with the two qualitatively different types of region separated by this shaded in red and yellow. 
The dotted lines indicate alignments of cusps and of stationary points.} }
\l{Maximal-Angle-Flow} \end{figure}          }

\section{Some new results concerning random triangles}

\subsection{Probability that a triangle is obtuse}

{\bf Corollary 1} 
\be
\mbox{Prob}(\mbox{obtuse}) \es \frac{3}{4}   \m, \mbox{ so, complementarily } \m \m
\mbox{Prob}(\mbox{acute})  \es \frac{1}{4}   \m . 
\ee
\Proof 
\n Each cap of obtuseness contributes an area (or more basically, surface of revolution) integral\footnote{Archimedes 
gave an even more basic derivation for the area of a spherical cap, by which Calculus could be avoided altogether in this derivation.  } 
\be
\int_{\Phi = 0}^{2\,\pi} \int_{\Theta = 0}^{\frac{\pi}{3}} \sin \, \Theta \, \d \Theta \, \d \Phi  
\es  2 \, \pi \int_{\Theta = 0}^{\frac{\pi}{3}} \sin \, \Theta \, \d \Theta 
         =           2 \, \pi \left[ - \cos \, \Theta \right]_{\Theta = 0}^{\frac{\pi}{3}}  
\es  2 \, \pi \left( - \frac{1}{2} + 1 \right)                                                           
\es       \pi                                                                                      \m . 
\ee
Thus, between them, the three caps of obtuseness contribute an area of $3 \, \pi$.
On the other hand, the area of the whole 2-sphere is $4 \, \pi$, so 
\be
\mbox{Prob}(\mbox{obtuse})  \es  \frac{3 \, \pi}{4 \, \pi} 
                            \es  \frac{3}{4}                        \m . 
\ee
\n Finally, complementarily 
\be
\mbox{Prob}(\mbox{acute})          =           1 - \mbox{Prob}(\mbox{obtuse})         
                          \es  1 - \frac{3}{4}               
						  \es      \frac{1}{4}                           \m . \m \Box 
\ee
\n{\bf Remark 1} This answer is in the context of interpreting the question's probability distribution to be the {\sl uniform} one {\sl on the corresponding shape space}, 
as equipped with the standard spherical metric that Kendall \c{Kendall84, Kendall89, Kendall} showed to be natural thereupon. 
The uniqueness of this metric is furtherly intuitively obvious from its being induced as a quotient of a simpler structure (position space or relative space: flat spaces), 
or from its arising correspondingly by reduction of the corresponding mechanical actions. 

\m
 
\n Also note that $\frac{3}{4}$ is in fact one \c{Guy, Portnoy} of the more common answers to Lewis Carroll's pillow problem, now obtained moreover on shape-theoretic premises. 
Small \cite{Small} and Edelman and Strang \c{MIT} obtained this result using Kendall's Shape Theory, 
while \c{A-Pillow} gave a simplified proof along the lines provided above. 
In conjunction with various elementary derivations of Kendall's shape sphere \c{A-Pillow, Forthcoming}, this reduces the shape-theoretic answer to Lewis Carroll's pillow 
problem to just elementary algebra and (pre)calculus.  
This renders it open,     both to anybody studying STEM subjects at University (or even High School in many countries) 
                      and also to considerable generalization to a whole class of geometrical problems, 
					  of which \c{A-Pillow}, the current treatise and \c{IV} provide the first few. 

\m

\n{\bf Remark 3} See also \c{Guy, Portnoy} for various other answers' values for $\mbox{Prob(Obtuse)}$, and a general principle 
behind the methods which share the answer 3/4 being called for.   
Shape Theory as per above surely provides a principle,  
though whether {\sl all} known methods which yield 3/4 can be shown to follow from this shape-theoretic principle is left as a good topic for a further paper.

\subsection{Probability that an isosceles triangle is obtuse}\l{Isosceles}

\n{\bf Example 1)} Let us next pose and solve a further pillow problem variant that requires no further definitions, 
\be
\mbox{Prob(Obtuse $\, | \,$ Isosceles)} \es \frac{\mbox{length}(\mI^{\acute})}{\mbox{length}(\mI)}
                                        \es  \frac{    \frac{\pi}{3}    }{    2 \times \frac{\pi}{2}    }									
									    \es   \frac{1}{3}                                                    \m,  
\ee			
using that $\mI^{\perp}$ is $\frac{\pi}{3}$ along the $\mI^{\sF}$ semi-meridian. 
Then complementarily 
\be
\mbox{Prob(Acute $\, | \,$ Isosceles)}  =  1 - \mbox{Prob(Obtuse $\, | \,$ Isosceles)} 
                                       \es \frac{2}{3}  \m . 
\ee

\subsection{Probability that a tall triangle is obtuse, and related examples}\l{TF-OA}

\n{\bf Remark 1} It is clear in shape space from the bimeridians of regularlity running along the line of reflection symmetry of the minimum cell in shape space 
(the 1/6th hemisphere bounded by arcs of isoscelesness and collinearity) that 
\be
\mbox{Prob(Flat)} \es  \frac{1}{2} 
                  \es  \mbox{Prob(Tall)} \m .
\ee
So introducing tallness and flatness {\sl by themselves} do not prompt further interesting variants of the pillow problem. 
Statements combining tallness or flatness with isoscelesness or with acuteness and obtuseness, however, do make for interesting variants, as follows.  

\m

\n{\bf Remark 2} Given that we know Prob(Obtuse) and Prob(Flat), the following four probabilities have only one degree of freedom between them, 
so they form a connected quadruple of pillow problems, as follows.  

\m

\n What are Prob(Acute and Tall), Prob(Acute and Flat), Prob(Obtuse and Tall), and Prob(Obtuse and Flat)?  

\m

\n{\bf Remark 3} Upon evaluating these, we can furthermore answer conditioned questions along the following lines. 

\m

\n What are Prob(Acute | Flat), Prob(Acute | Tall), Prob(Obtuse | Flat) and Prob(Obtuse | Tall)? 

\m

\n Conditioning conversely, what are Prob(Flat | Acute), Prob(Flat | Obtuse), Prob(Tall | Acute) and Prob(Tall | Obtuse )? 

\m

\n{\bf Remark 4} Finally, we also consider the complementary pairs Prob(Isosceles is Flat) and Prob(Isosceles is Tall),  
                                                 Prob(Right is Flat)     and Prob(Right is Tall),
				                             and Prob(Regular is Obtuse) and Prob(Regular is Acute).

\m

\n These are all questions concerning generic triangles, meaning they correspond to regions whose configuration space dimension is maximal, i.e.\ not lying on C, I, R or $\perp$.  

\m

\n{\bf Remark 5} Our treatment makes use of  
the {\it departure from independence}
\be
\delta(A, B) := \mbox{Prob}(A) \, \mbox{Prob}(B) - \mbox{Prob}(A \caps\,   B) 
\ee
of random variables $A$ and $B$.
For our $2 \times 2$-value discrete bivariate joint distributions, this takes a single value, so we can simplify notation from $\delta(A, B)$ to just $\delta$.  
Formulated in terms of this value, the four probabilities in question are cast in a symmetric form.  

\m 

\n {\bf Example 1}  \c{A-Pillow} showed that      
\be
\delta     \es          3  \left(  \frac{1}{8} - \frac{1}{\pi}    \left(  \arcsin \sqrt{\frac{2}{3}} - \frac{1}{2} \, \arcsin \, \frac{1}{3}  \right)  
\right) \approx 0.0505 \m . 
\ee
In terms of this, moreover, 
\be
\mbox{Prob}\left(\mF^{\acute}\right) \es  \frac{1}{8} - \delta  \m \approx \m  0.1755          \mma 
\mbox{Prob}\left(\mT^{\acute}\right) \es  \frac{1}{8} + \delta  \m \approx \m  0.0745          \m,
\ee 
\be
\mbox{Prob}\left(\mF^{\obtuse}\right) \es  \frac{3}{8} + \delta  \m \approx \m 0.4255          \mma 
\mbox{Prob}\left(\mT^{\obtuse}\right) \es  \frac{3}{8} - \delta  \m \approx \m 0.3245          \m .  
\ee 
So $\mbox{Prob}\left(\mT^{\obtuse}\right)$ is the most probable and $\mbox{Prob}\left(\mT^{\acute}\right)$ is the least probable, in approximately a 6 : 1 ratio. 

\m

\n{\bf Example 1)} \c{A-Pillow}
\be 
%
%
\mbox{Prob}(\acute \mbox{$\, | \,$} \mF)                 
					                          \es    \frac{1}{4} - 2 \, \delta
                                           \approxs  0.149 					                                         \mma 
\mbox{Prob}(\obtuse \mbox{$\, | \,$}          \es    \frac{3}{4} + 2 \, \delta
                                           \approxs  0.851 						                                     \m,
\ee 
\be
\mbox{Prob}(\acute \mbox{$\, | \,$} \mT) 
						                      \es     \frac{1}{4} + 2 \, \delta 
						                    \approxs  0.351                                                          \mma 
\mbox{Prob}(\obtuse \mbox{$\, | \,$} \mT) 
 						                      \es    \frac{3}{4} - 2 \, \delta 
						                   \approxs  0.649 \m, 
\ee 
and 
\be            
\mbox{Prob}(\mF \mbox{$\, | \,$} \acute) 
						                      \es    \frac{1}{2} - 4 \, \delta 
						                   \approxs  0.298                                                                 \mma  
\mbox{Prob}(\mT \mbox{$\, | \,$} \acute) 
						                      \es       \frac{1}{2} + 4 \, \delta 
						                    \approxs  0.702                                                              \m, 
\ee
\be 
\mbox{Prob}(\mF \mbox{$\, | \,$} \obtuse) 
							                  \es      \frac{1}{2} + \frac{4}{3}\delta 
							                \approxs   0.567                                                                 \mma               
\mbox{Prob}\left(\mT \mbox{$\,|\,$} \obtuse\right)  
			        				          \es      \frac{1}{2}      - \frac{4}{3}\delta 
					     		            \approxs   0.433                                                                                        \m . 
\ee

\subsection{Solution of further restricted pillow problem variants}

\n By `restricted', we mean non-generic.  
The cases considered in particular concern assuming a property that is restricted to lie on an arc (or collection of arcs) on the shape sphere.

\m

\n{\bf Example 1)} \c{A-Pillow}  
%
%
%
\be
\mbox{Prob}(\mF \, | \, \perp)          \es     \frac{    \mbox{length}  \left(   \mF^{\perp}  \right)    }{    \mbox{length}(\perp)    }   
                                        \es     \frac{2}{\pi} \arcsin \, \frac{1}{3} 
									  \approxs  0.2163                                                                            \mma \mbox{and }
\ee
\be
\mbox{Prob}(\mT \, | \, \perp)           =      1 - \mbox{Prob(Flat $\, | \, \perp$)}    
                                        \es     \frac{    \pi - 2 \, \arcsin \, \mbox{$\frac{1}{3}$}    }{    \pi    } 
								      \approxs  0.7837                                                                                                     \m .
\ee
\be
\mbox{Prob}(\mF \, | \, \mC)   \es  \frac{1}{2} 
                               \es  \mbox{Prob(\mT \, | \, \mC)} 
\ee			
is obvious by reflection symmetry.  

\m

\n{\bf Example 2} \c{A-Pillow},  
\be
\mbox{Prob}(\acute \, | \, \mR)     \es     \frac{    \mbox{length}(\mR^{\acute})    }{    \mbox{length}(\mR)    } 
                                    \es     \frac{2}{\pi} \,    \arcsin \frac{1}{\sqrt{3}} 
								  \approxs  0.3918                                                                  \m .   
\ee
Complementarily,
\be
\mbox{Prob}(\obtuse \, | \, \mR)     =      1 - \mbox{Prob}(\acute \, | \, \mR) 
                                    \es     \frac{   \pi - 2 \, \arcsin \frac{1}{\sqrt{3}}   }{    \pi    } 
                                  \approxs  0.6082                                                                  \m .
\ee

\subsection{Probability that a triangle is Fermat-acute}\l{Fermat}

\be 
\alpha_{\sm\sa\sx} \es \frac{2 \, \pi}{3} 
\ee 
is a meaningful critical value as regards Fermat's problem (Fig \r{Fermat-Points}.a) 
of which point $F$ minimizes the total distance from itself to a triangle's three vertices (Fig \r{Fermat-Points}).  
This is a critical value for Fermat's problem because for 
\be
\alpha_{\sm\sa\sx} \geqs  \frac{2 \, \pi}{3}  \m, 
\ee 
the Fermat (alias Torricelli) point $F$ is just at the obtuse angle's vertex (Fig \r{Fermat-Points}.b), whereas for 
\be
\alpha_{\sm\sa\sx}  \ls   \frac{2 \, \pi}{3}  \m, 
\ee 
the Fermat point $F$ is a nontrivial point in the interior of the triangle (Fig \r{Fermat-Points}.c).  
These are clearly qualitatively different regimes.
On these grounds, we make the following definitions.
%
{            \begin{figure}[!ht]
\centering
\includegraphics[width=0.85\textwidth]{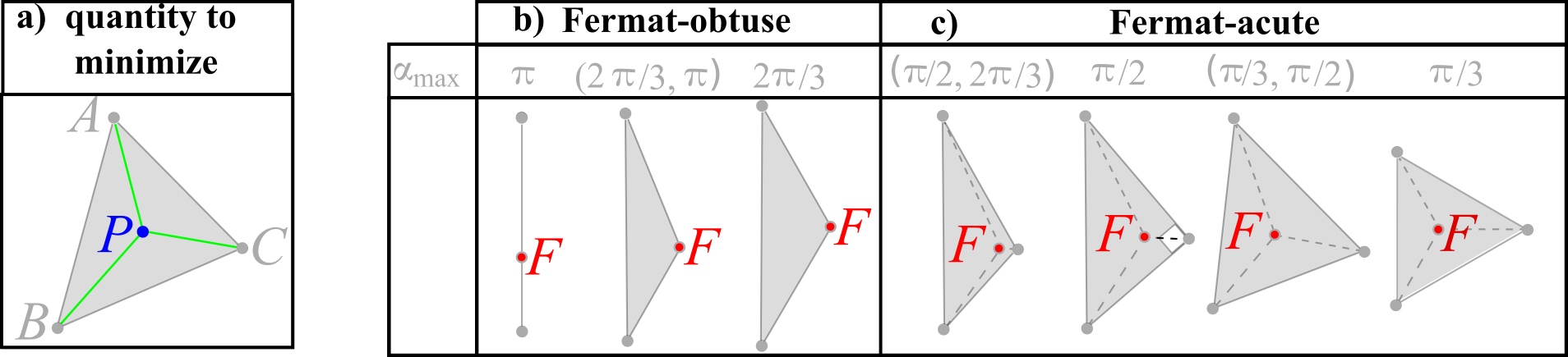}
\caption[Text der im Bilderverzeichnis auftaucht]{        \footnotesize{a) Fermat's problem is, given a triangle ABC, to find the position of the point P (in blue) 
such that the sum of the lengths (in green) from it to the three vertices is minimized. 
b) This is solved by the Fermat point $F$ (in red), which is moreover just at the obtuse vertex for triangles with $\alpha_{\sm\sa\sx} \geq \frac{2 \, \pi}{3}$, which we term Fermat-obtuse. 
c) On the other hand, the Fermat point $F$ has a nontrivial position for triangles with $\alpha_{\sm\sa\sx} <  \frac{2 \, \pi}{3}$.  
In this case, $F$ is such that the vertices are spread out at $2 \, \frac{\pi}{3}$ from each other relative to $F$, as indicated by the dashed lines. 
Moreover, for $\alpha_{\sm\sa\sx} =  \frac{\pi}{3}$, the position of $F$ reduces to just that of the centre of symmetry of the equilateral triangle.} }
\l{Fermat-Points}\end{figure}            }

\m

\n{\bf Corollary 1} \c{Max-Angle-Flow} For each cluster i) 
\be
Area \es 2 \, \left| \int_{{\cal X} = 0}^{1/2} \d {\cal X} \, \arcsin \left(  \frac{    1 - 2 \, {\cal X}    }{    \sqrt{3} \, k \, \sqrt{   1 - {\cal X}^2  }   }  \right) \right| 
     \m \s{\mbox{\scriptsize Fermat case}}{=} \m
         2 \, \left| \int_{{\cal X} = 0}^{1/2} \d {\cal X} \, \arcsin \left(  \frac{    2 \, {\cal X} - 1    }{   \, \sqrt{   1 - {\cal X}^2  }   }  \right) \right|            \m . 
\ee
This can moreover be evaluated using Maple \c{Maple}. 

\m

\n{\bf Remark 1} Furthermore, once again normalizing by the total area of the sphere, 
\be
\mbox{Prob($\alpha$-obtuse)} \es  \frac{3 \, Area}{4 \, \pi}         \m . 
\ee
Thus 
\be 
\mbox{Prob(Fermat-obtuse)} = 0.1394                                  \m . 
\ee
Complementarily, 
\be
\mbox{Prob($\alpha$-acute)} =  1 - \mbox{Prob($\alpha$-obtuse)} 
                            = \frac{4 \, \pi - 3 \, Area}{4 \, \pi}  \m, 
\ee 
So 
\be 
\mbox{Prob(Fermat-acute)} = 0.8606                                   \m, 
\ee
which is the quantity of particular interest due to its correspondence to the triangles whose Fermat points are nontrivially positioned.  

\vspace{10in}

\section{The Hopf map in the shape-theoretic context}\l{Hopf}

\n{\bf Structure 1} There is a Hopf map is from the 3-sphere to the 2-sphere (I.180), which is moreover realized by the reduction from the preshape 3-sphere to the shape sphere.
Indeed, this provides a derivation that the triangleland shape space is a sphere.

\m

\n{\bf Remark 1} On the one hand, it is immediately clear in 1-$d$ 
how to represent $\mathbb{S}^{n - 1} = \FrS(N, 1) = \FrP(N, 1)$ within ${\cal R}(N, 1) = \Frr(N, 1) = \mathbb{R}^n$.
On the other hand, for 3 points-or-particles in $d \geq 2$, such representation is more complicated.   
While there is a flat relative space $\Frr(3, 2)$, this is $\mathbb{R}^4$ rather than $\mathbb{R}^3$.  
This does not yet account for how to embed a 2-sphere in a flat 4-space in an {\sl equable manner}, i.e.\ making equal use of each component of $\urho_1$ and $\urho_2$.  
One geometrical way of arriving at this answer to this is to concatenate the Hopf map of spheres ${\cal H}_{\sS}$ 
with the obvious codimension-1 embeddings of $\mathbb{S}^2$ and $\mathbb{S}^3$: 
\beq
\mathbb{R}^4  \m \s{    \mbox{\scriptsize unit sphere map}    }{    \longrightarrow    }                 \m
\mathbb{S}^3  \m \s{    \mbox{\scriptsize Hopf fibration (I.159)}    }{    \longrightarrow    }          \m 
\mathbb{S}^2  \m \s{    \mbox{\scriptsize embedding map}    }{    \longrightarrow    }                   \m
\mathbb{R}^3                                                                                             \m . 
\eeq 
This forms the `extended Hopf map of Cartesian directions' ${\cal H}_{\sC}$. 

\m 

\n{\bf Remark 2} A somewhat more elegant reformulation: the `Hopf Cartesian coordinates map', ${\cal H}_{\sC}$ 
is related to the Hopf spheres map ${\cal H}_{\sS}$ by the composition of maps  
\be
{\cal H}_{\sC} = U_3 \, \circ \,  {\cal H}_{\sS} \, \circ \, C_2  \m,   
\ee
as per Fig \r{Hopf-Map}. 
%
{            \begin{figure}[!ht]
\centering
\includegraphics[width=0.25\textwidth]{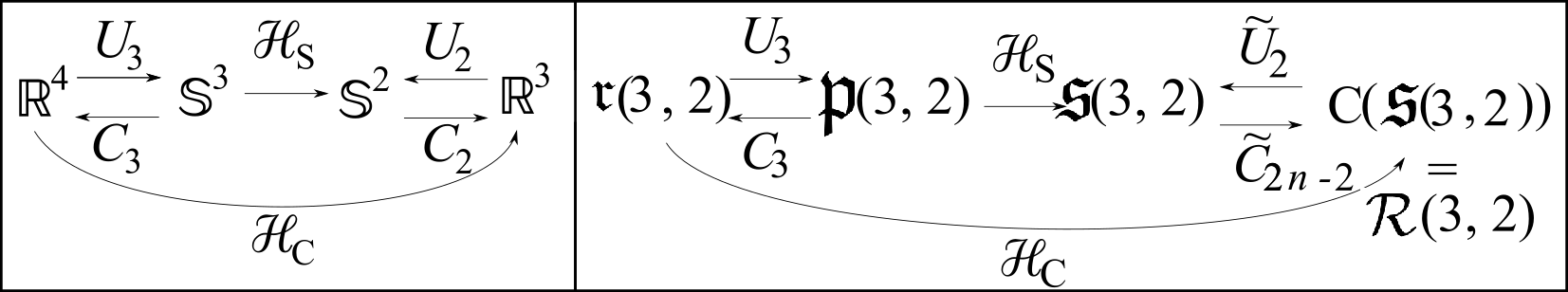}
\caption[Text der im Bilderverzeichnis auftaucht]{        \footnotesize{a) Hopf map ${\cal H}_{\sS}$ from $\mathbb{S}^3$ to $\mathbb{S}^2$, 
including also mapping the natural ambient $\mathbb{R}^4$ for the $\mathbb{S}^3$ to a less obviously realized natural ambient $\mathbb{R}^3$ for the $\mathbb{S}^2$ 
by the Hopf Cartesian map ${\cal H}_{\sC}$. 
$U_k$ are here unit vector maps to the $k$-sphere, and $C_k$ are cone maps from the $k$-sphere.
b) Its shape-theoretic implementation.}  } 
\l{Hopf-Map} \end{figure}           }

\m 

\n{\bf Remark 3} As already outlined in Part I, ${\cal H}_{\sS}$ is realized by the reduction from the preshape 3-sphere to the shape sphere.  
This and the accompanying metric-level calculation indeed constitutes an elegant derivation that the (3, 2) shape space is a sphere, 
i.e. of the special case of Kendall's Theorem that the current Part III of this treatise is built around.  

\m

\n{\bf Remark 4} As also already outlined in Part I, this unit sphere map is realized by the reduction of relative space to preshape space, 
and this embedding map by the coning of the shape space to produce the relational space. 
This indeed constitutes an elegant derivation that the (3, 2) relational space is topologically $\mathbb{R}^3$, 
though now the metric part of the calculation reveals that this does not carry the flat metric. 
It is conformally flat (if the maximal collision is excised, since the conformal transformation in question is invalid there).  

\m

\n{\bf Structure 2} In this shape-theoretic setting, and with respect to a choice of clustering to define the relative Jacobi coordinates underpinning the 
$\u{\rho}_i$, $\Theta$ and $\Phi$, the extended Hopf map takes the form given in Fig \ref{Hopf-Fig}.  

\m

{            \begin{figure}[!ht]
\centering
\includegraphics[width=0.66\textwidth]{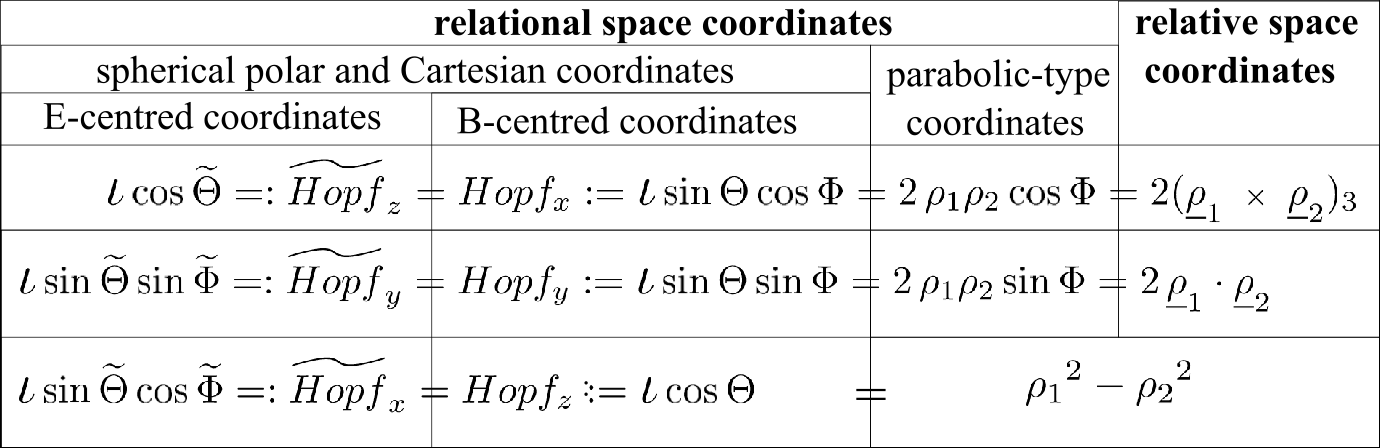}
\caption[Text der im Bilderverzeichnis auftaucht]{        \footnotesize{Here each equation's first forward equality just recasts the Hopf Cartesian coordinates 
in standard spherical polar coordinates, with the moment of inertia $\bigiota = \rho^2$ playing the role of radius.     
Each equation's second forward equality amounts to the shape-theoretic relations between the $\rho_i$ and $\bigiota$, $\Theta$ and $\Phi$.  
The third forward equality, when present, just incorporates standard basic formulae for dot and cross products.
To get parabolic coordinates, just use $\xi_i = \sqrt{2} \, \rho_i$. 
Thus Hopf and parabolic coordinates are very closely related. 
The remaining conceptual leap is to regard the $\xi_i$ as magnitudes of 2-$d$ vectors, 
thus passing from relating three squares to the square of two squares to the square of four squares in (\r{HHHI}).  
The first backward equality is a switch of which axis is primary, with the second backward equality then giving the corresponding expression in tilded spherical polar coordinates. 
}  } 
\l{Hopf-Fig} \end{figure}           }

\m

\n{\bf Exercise 1} These can be readily checked to obey 
\be
Hopf_x\mbox{}^2 + Hopf_y\mbox{}^2 + Hopf_z\mbox{}^2 = \bigiota^2 
                                                    = (\rho_1\mbox{}^2 + \rho_2\mbox{}^2)^2 
													= (\rho_{1x}\mbox{}^2 + \rho_{1y}\mbox{}^2 + \rho_{2x}\mbox{}^2 + \rho_{2y}\mbox{}^2)^2      \m .
\l{HHHI}
\ee
\n{\bf Structure 3} There is moreover a unit Cartesian direction version of this obtained by dividing each term by $\bigiota$.
Using unit relative Jacobi coordinates, this gives Fig \ref{Hopf-Fig-2}.
%
{            \begin{figure}[!ht]
\centering
\includegraphics[width=0.66\textwidth]{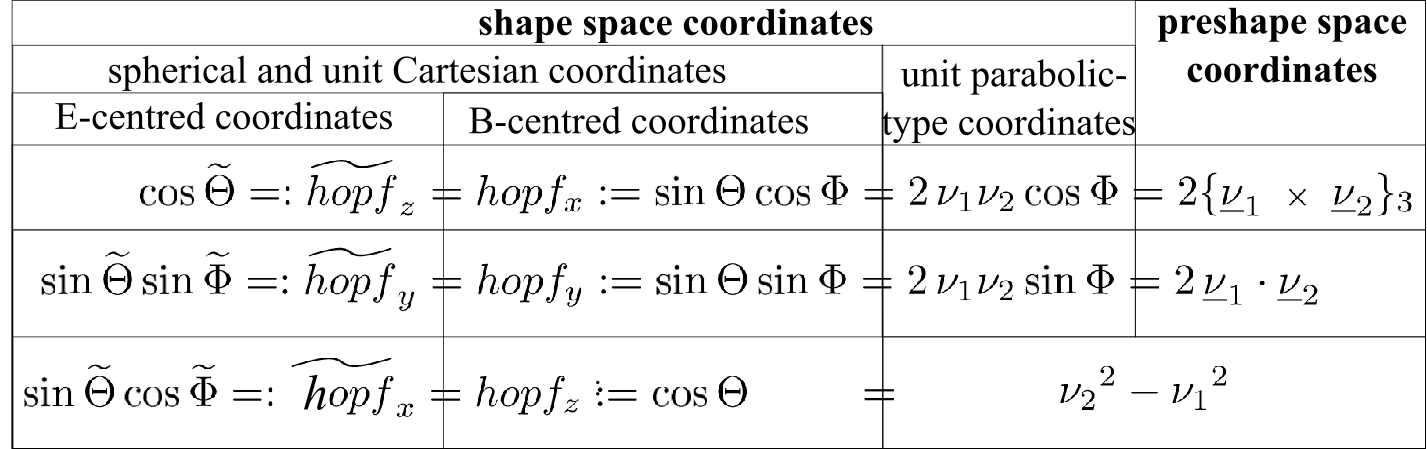}
\caption[Text der im Bilderverzeichnis auftaucht]{        \footnotesize{Moment of inertia normalized version of the previous, giving shape variables.}  } 
\l{Hopf-Fig-2} \end{figure}           }

\m

\n{\bf Exercise 2} The counterpart of (\r{HHHI}) now readily gives the on-2-sphere condition
\be
hopf_x\mbox{}^2 + hopf_y\mbox{}^2 + hopf_z\mbox{}^2 = 1  \m . 
\l{hhh1}   
\ee
\n{\bf Lemma 1} In terms of the shape-theoretic stereographic radius coordinate 
\be
hopf_z \es  \frac{  1 - {\cal R}^2     }{ 1  +   {\cal R}^2    }                             \m, 
\label{R-ellip}
\ee 
\be
hopf_x \es  \frac{  2 \, {\cal R} \, \cos \, \Phi }{1 + {\cal R}^2}                          \m, 
\ee 
\be
hopf_y \es  \frac{  2 \, {\cal R} \, \sin \, \Phi }{1 +  {\cal R}^2}                         \m .   
\ee
{\u{Proof} Use (I.174) and (\r{IrhoR}) in rows 1 and 2 of Fig \ref{Hopf-Fig-2}. The first of these is also a repackaging of (\r{X-R}). $\Box$

\m

\n{\bf Exercise 3} The reader can furthermore check that these three expressions indeed square to 1.  

\mbox{ }

\n{\bf End-Comment 1} This subsection is well-known and applied in Molecular Physics \c{LR97}, but has received little attention in Shape Statistics.

\subsection{Mirror image identified and indistinguishable versions}

\n{\bf Structure 1} The ${\cal R}(3, 2)$ case was considered by physicist Alex Dragt in the 1960's \c{Dragt} in the Molecular Physics literature 
(see also \c{Smith62}, and \c{PP87, Iwai87, LR97, ML99} for some applications). 
In this case, $\Phi = 0$ to $\pi$ or $\w{\Theta} = 0$ to $\frac{\pi}{2}$. 
This case has the advantage of being obligatory in 3-$d$ to the Hopf case itself being disallowed.

\m

\n{\bf Remark 1} This subsection's Hopf quantities are well-known and applied in Molecular Physics (often under the name of `Dragt coordinates', 
but has received little attention in Shape Statistics.
The remaining subsections gives further shape-theoretic interpretation and a start on useful applications.

\m

\n{\bf Structure 2} The $\FrI{\cal R}$ case of Hopf-type map is new to the current Part. 
Here $\w{\Phi} = 0$ to $\frac{\pi}{3}$  or $\Phi = 0$ to $\pi$ and $\Theta = 0$ to $\frac{\pi}{3}$.  

\m

\n{\bf Structure 3} The $\Leib_{\sFrR}(3, 2)$ case of Hopf-type map 
-- corresponding to Kendall's spherical blackboard from the Shape Statistics literature -- is also new to the current treatise.  
Here, $\w{\Phi}$ 0 to $\frac{\pi}{3}$  or $\Phi = 0$ to $\pi$ and $\Theta = 0$ to $\frac{\pi}{3}$.  

\m

\n This section's Hopf quantities furthermore admit the following lucid shape-theoretic interpretation \c{+Tri, FileR, ABook}.

\subsection{Shape theoretic interpretation of the Hopf-type quantities. I. Tetra-area}\l{4-times-Area}

\n{\bf Remark 1} $hopf_x$'s shape-theoretic true name is mass weighted `tetra-area' (per unit moment of inertia), $\tau$.  
This is moreover a quantifier of noncollinearity, i.e.\ departure from collinearity.  

\m

\n{\bf Remark 2} In shape space, the plane through the origin that is perpendicular to the $\tau$ = E$\overline{\mE}$ axis is indeed the plane of collinearity, C.  

\m

\n{\bf Remark 3} Interpretation using the Heron form in the square root, zero area results from zero semi-perimeter: O, 
or from semi-perimeter being equal to a side, without loss of generality $a$: 
\be
a  =   s 
  \es  \frac{a + b + c}{2}                  \Rightarrows   
a  = b + c 
                                            \Rightarrows   
\mbox{(triangle inequality is  saturated)}  \Rightarrows   \mbox{degenerate}   \m .  
\ee 
Sec \ref{Tet-1} showed how Jacobi coordinates cast the similar medians counterpart of Heron's formula in identical functional form. 

\m

\n{\bf Remark 4} $\tau$ is moreover clustering-independent, i.e.\ democracy-invariant in the Jacobian sense.  

\m

\n{\bf Remark 5} $\tau$ is a signed quantifier, with clockwise and anticlockwise configurations contributing positive and negative signs, as per  
inequalities (\r{A>0}) and (\r{A<0}). 

\m 

\n{\bf Remark 6} One can moreover define an unsigned version, $|\tau|$, which is insensitive to the clockwise--anticlockwise distinction.  
The identification $|\tau|$ gives another way of ascertaining that, firstly, the equilateral triangles E and $\overline{\mE}$ maximize and minimize $\tau$ respectively.
Secondly, that $|\tau|$ is co-maximized by E and $\overline{\mE}$ and jointly minimized by the collinear shapes C.

\subsection{Shape theoretic interpretation of the Hopf-type quantities. II. Ellipticity}\l{Ellip-Sec}

\n{\bf Interpretation 1} The ellipticity $\ellip$ is the difference of the two `normalized' partial moments of inertia involved in the clustering in question: 
that of the base and that of the median.

\m

\n{\bf Remark 1} In terms of the sides of the triangle, 
\be
\Ellip \es  \frac{a^2 + b^2 - 2 \, c^2}{3}                \m , 
\l{Ellip}
\ee 
whereas, in terms of the side-ratios of the triangle, 
\be
\ellip \es  \frac{a^2 + b^2 - 2 \, c^2}{a^2 + b^2 + c^2}  \m .   
\ee
\n These formulae single out the base side of the cluster, so $\ellip$ is a cluster-dependent quantity.  

\m

\n{\bf Interpretation 2} $\ellip$ is a Heron map eigenvector and Cartesian axis direction in relational space. 

\m

\n{\bf Interpretation 3} The further intutive reformulations  
\be
\Ellip \es \frac{    a^2 + b^2 + c^2    }{    3    } - c^2 
       \es \bigiota - c^2  
        =   \langle \, \mbox{side}^2 \, \rangle  - (c\mbox{--side})^2 \mma \mbox{and }
\ee 
\be
\ellip \es     1 - \frac{    (c\mbox{--side})^2   }{    \langle \, \mbox{side}^2 \, \rangle   } 
       \es     \frac{         \langle \, \mbox{side}^2 \, \rangle - (c\mbox{--side})^2     }{    \langle \, \mbox{side}^2 \, \rangle    }   \m, 
\ee 
are in terms of what is conceptually a {\it contrast} object, from the point of view that the democratic average `is the norm' and thus to be used to normalize.
In this way, the Hopf bundle structure provides a natural and significant contrast variable.   

\m

\n{\bf Remark 3} Setting $a = b = c$, the numerator is zero for the equilateral triangle, E, but it is not unique as a such.  

\m

\n{\bf Interpretation 4} The plane through the origin that is perpendicular to the $\ellip$ = MB axis is indeed the plane of regularity, R.  
Thus $\ellip$ is a measure of departure from regularity, R (hence regularity's `circular' alias in Sec \r{Hopf-S}).
$\ellip$ is thus conceptually  an anisotropy, more concretely a partial moments of inertia or Jacobi anisotropy.  

\m

\n{\bf Remark 5} $\ellip$ is also a signed quantifier, in that it distinguishes between tall and flat departures from regularity (Fig \r{BIEMC-3}.1).  
\be
\ellip \gs 0 \m \mbox{ is flat} \mma \mbox{ and }
\ee
\be
\ellip \ls 0 \m \mbox{ is tall } .  
\ee
\n{\bf Remark 6} One can moreover define an unsigned version, $|\ellip|$, which is insensitive to the tallness-flatness distinction, 
by using $|a^2 + b^2 - 2 \, c^2|$ in place of $a^2 + b^2 - 2 \, c^2$.  

\m

\n{\bf Remark 7} $\ellip$ is minimal for B and maximal for M, whereas $|\ellip|$ is co-maximal for B and M shapes and jointly minimal for all regular triangles, R.

\subsection{Shape theoretic interpretation of the Hopf-type quantities. III. Anisoscelesness}\l{Aniso-Sec}

{\bf Interpretation 1} $\aniso$ is a a quantifier of `anisoscelesness'  departure from being isosceles.   

\m

\n{\bf Remark 1} In terms of the sides of the triangle, 
\be
\Aniso =  \frac{  a^2 - b^2  }{  \sqrt{3}  }  \m , 
\l{Aniso}
\ee
whereas, in terms of the side-ratios of the triangle, 
\be
\aniso \es  \frac{  \sqrt{3}  (  a^2 - b^2  )  }{  a^2 + b^2 + c^2  } 
       \es  \frac{  a^2 - b^2  }{  \sqrt{3}\langle \, side^2 \, \rangle  }  \m . 
\ee
\n This is clearly a clustering-dependent notion.  

\m 

\n{\bf Remark 2} Anisoscelesness compares the remaining 2 sides of the cluster, in which way it is a measure of `contents inhomogeneity' \cite{AF, II} within the given cluster.  

\m

\n{\bf Remark 3} $a = b$ -- the basic isoscelesness condition -- indeed returns zero anisoscelesness.
\be
\aniso = 0 \m \mbox{ for isosceles triangles } .
\ee
%
%
\n{\bf Remark 4} The plane through the origin that is perpendicular to the $\aniso$ = HH axis is indeed the plane of isoscelesness, I.  

\m 

\n{\bf Interpretation 2} $\aniso$ is also a Heron map eigenvector and Cartesian axis direction in relational space. 

\m

\n{\bf Remark 5} $\aniso$ is a signed quantifier, in that it distinguishes between left-leaning and right-leaning deviations from isoscelesness 
(Fig \r{BIEMC-3}.1).  
\be
\aniso \ls 0 \m \mbox{ is right-leaning } , \m 
\ee
and 
\be
\aniso \gs 0 \m \mbox{ is left-leaning } . 
\ee
\n{\bf Remark 6} One can moreover define an unsigned version, $|\aniso|$, by using $|a^2 - b^2|$ instead of $a^2 - b^2$; 
this is insensitive to the left-or-right leaning distinction. 

\m

\n{\bf Remark 7} $\aniso$ is maximal and minimal for H shapes, whereas $|\aniso|$ is maximal for H shapes and jointly minimal for all isosceles triangles I.

\subsection{Comment on the previous three subsections' structural similarnesses}

\n{\bf Structural Parallel 1} Each concerns an axis perpendicular to a plane through the relational space origin 0, 
with moreover the three of them being mutually perpendicular, i.e.\ forming a Cartesian axis system.  

\m

\n{\bf Structural Parallel 2} Each plane divides the shape sphere into two meaningful hemispheres (Fig \r{Hopf-Axes}) characterized by geometrical inequalities.
%
{            \begin{figure}[!ht]
\centering
\includegraphics[width=0.85\textwidth]{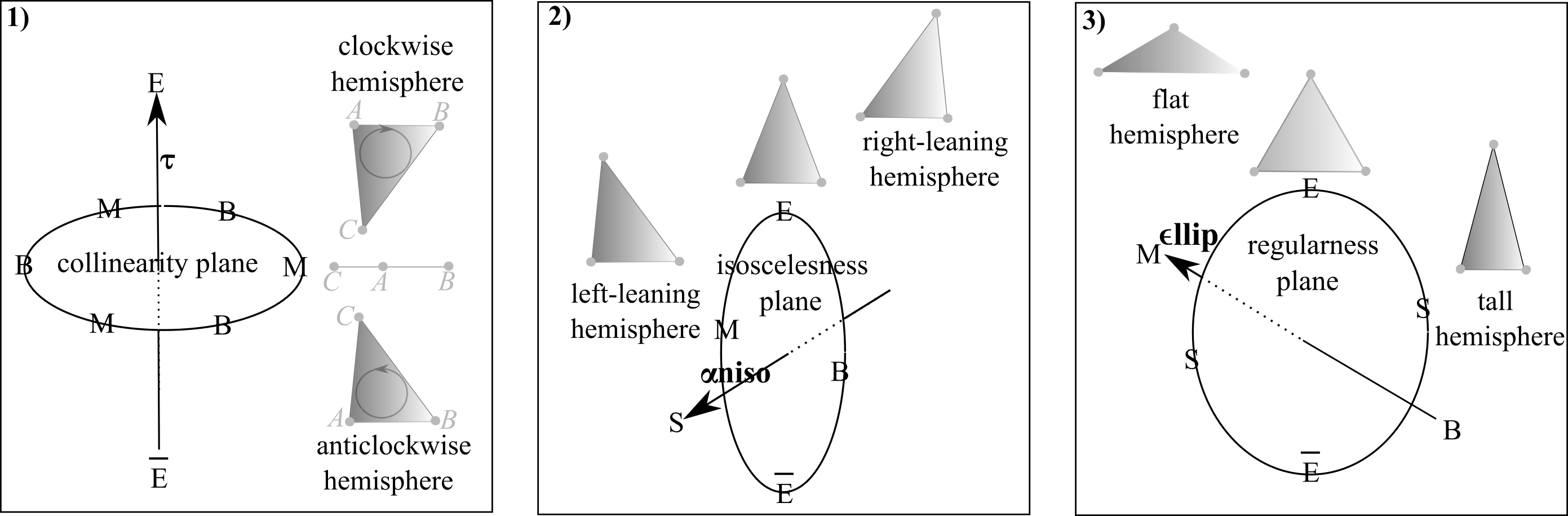}
\caption[Text der im Bilderverzeichnis auftaucht]{        \footnotesize{1) Collinear shapes separate hemispheres of clockwise and anticlockwise triangles. 

\m

\n 2) Isosceles triangles separate hemispheres of left-leaning and right-leaning triangles.

\m

\n 3) Regular triangles separate hemispheres of tall and flat triangles.} }
\l{Hopf-Axes} \end{figure}          }

\m

\n{\bf Structural Parallel 3} Each axis passes through two antipodal points on the shape sphere.
These maximize the maximize-and-minimize the signed version of the quantifier and co-maximize the normed version, 
which is in turn jointly minimized on the perpendicular plane throught the origin.  

\m

\n{\bf Difference/Primality 1} There is a difference in the labelled case (but not the unlabelled case) 
as regards one pair of axes coming with three possible clustering choices, while the other -- $\tau$ -- is always the same: clustering-democratic. 

\m

\n{\bf Structural Parallel 4} The three Hopf quantities in question moreover enter much further bundle mathematics as per below.
Due to this, $\ellip$, $\aniso$ and $\tau$ are a very geometrically structured and meaningful triple to consider.   
This justifies the combination of features exposited in Secs \r{Hopf-S} and \r{Hopf-SS}.  

\m

\n{\bf Application 1} One can also arrive at the BM, E$\overline{\mE}$ and HH axis system -- in some order -- 
by considering the three Hopf quantities $\ellip$, $\tau$ along E$\overline{\mE}$ and $\aniso$.  
Hopf's fibre bundle mathematics moreover ascribes {\sl equal status} to collinearity C, 
                                                                        isoscelesness I, 
																    and regularity R, 
so it does not source a reason for picking principal, second and third axes. 
Selection principles for these are, rather, the above democratic principle for E$\overline{\mE}$ and the next subsection's arguments for BM.

\subsection{$\tau$ versus $\ellip$'s claims to primality}

\n{\bf Difference/Primality 1} $\tau$, unlike $\ellip$ and $\aniso$, is a `democracy invariant'.   
This is attained by all three clustering choices sharing one of the Cartesian axes picked out by the corresponding Hopf map 
(their other axes all lying at $\frac{\pi}{6}$ to each other in the common plane of collinearity).  

\m

\n{\bf Remark 1} As formulated above, $\aniso$ is a pure ratio of relative separations. 
Sec \r{Rel-angle}'s isosceles--isoangular duality moreover indicates that it can be rephrased in relative angle terms: it is a mixed angles and length ratios notion of anisotropy.  
$\ellip$ is also a function of pure ratio of relative separations. 
However, on this occasion, it cannot be rephrased in angular terms; it is a pure-ratio anisotropy.  

\m

\n Another point of view is to use the spherical angles that are most directly interpreted in shape-in-space terms: the untilded $\Theta$ and $\Phi$. 
From their definitions I.(171-172) it is clear that 
\be
\Theta \m \mbox{ is pure ratio of relative separations } 
\ee
and
\be
\Phi \m \mbox{ is pure relative angle } .
\ee
Then, concatenating $\ellip$'s definition and the third row of Fig \r{Hopf-Fig-2}, 
\be
\ellip = \cos \, \Theta 
       = \mbox{pure ratio of relative separations}  \m .  
\l{ellip-cos}
\ee
The first equality here moreover means that $\ellip$ plays the mathematical role of Legendre variable, as per eq. (\ref{Legendre}). 
In the tilded variables, however, it is $\tau$ that plays this role, as per eq. (\ref{w-Legendre})

\m

\n In contrast, both $\aniso$ and $\tau$ provide mixed (ratio of relative separations) and (relative angle) information. 
On the one hand, the `ratio of relative separations' information in both of these of these is a factor of 
\beq
\sin \, \Theta = \sqrt{1 - \cos^2 \Theta} = \sqrt{1 - \ellip^2} 
               = \sqrt{\aniso^2 + \tau^2}                        \m .
\eeq
Here the second equality follows from (\r{ellip-cos}) and the third from (\r{EATI}).
On the other hand, the relative angle information is in the $\cos \, \Phi$ and $\sin \, \Phi$ factors respectively. 
This can moreover be envisaged as a parametrization of the corresponding perpendicular circle of regularity, R.

\m

\n{\bf Difference/Primality 2} the $\ellip$ = MB axis corresponds to {\sl irreducibly} (pure ratio of relative separation) information.  
BM's claim to be principal axis is rooted on this, having the simplest shape-in-space to shape space map, 
                                                   its involvement in the relative Jacobi coordinates, 
												   and its irreducibly signed nature explained next below.

\m
												   
\n{\bf Difference/Primality 3} While mirror image identification and particle indistinguishability can do away with clockwise versus anticlockwise and 
left- versus right-leaning triangles, $\ellip$'s two signs correspond to inherently different shapes: tall versus flat triangles.  
												   
\m

\n{\bf End-Point 1} So $\ellip$ = MB axis primality is also strongly supported, 
especially for label-indistinguishable modelling for which the objection of there being three clustering choices for $\ellip$ = MB axis evaporates. 
Given that spheres are 2-$d$ and so require two coordinates, 
$\tau$ and $\ellip$ havin the above privileges suggests using (\r{hhh1}) to eliminate $\aniso$ from intrinsic workings, so that these be in terms of $\tau$ and $\ellip$ alone.

\subsection{Partially angular reformulations of $\aniso$ and $\tau$}

\n{\bf Proposition 1} A partially angular characterization of anisoscelesness is  
\be
\aniso \es \frac{    4 - 3 \, \ellip    }{    2 \sqrt{3}    }  \, \frac{    \cos^2\gamma - \cos^2\beta    }{    \sin^2\gamma + \sin^2\beta    }   \m .  
\l{ellip-angle}
\ee
{\u{Proof}} subtract the squared cosine form of the first and second equations in (\r{calR-gamma}). $\Box$

\m

\n{\bf Remark 1} So $\beta = \gamma$ -- the isoangularity characterization of isoscelesness -- returns $\aniso = 0$; 
the other zeros lie outside the range of validity of the angles.

\m

\n{\bf Remark 2} That (\r{ellip-angle})'s pure-ratio prefactor is not capable of generating any zeros is clear from 
\be 
4 - \, 3 \, \ellip \es  \frac{  1 + 3 \, {\cal R}^2 + 1  }{  1 + {\cal R}^2  }  \m, 
\ee 
which follows from (\r{R-ellip}).  

\m

\n{\bf Proposition 2} A partially angular characterization of tetra-area is 
\be
T \es 2 \, a \, b \, \sin \, \gamma   = \mbox{cycles}   \m, 
\ee
or their democratic average, 
\be 
T \es \frac{2}{3} \left( a \, b \, \sin \, \gamma + \mbox{cycles} \right)  \m .  
\ee 
The normalized mass-weighted form of this is 
\be
\tau \es  2 \sqrt{3} \, \frac{a \, b \, \sin \, \gamma}{a^2 + b^2 + c^2} 
     \es  \mbox{cycles}  
     \es  \kappa  \, \frac{a \, b \, \sin \, \gamma + \mbox{cycles} } {a^2 + b^2 + c^2}  \m .  
\ee 
\n{\bf Remark 3} The non-democratic form permits C and B case zeros to just be read off (zero angle, a zero side respectively).  

\m

\n{\bf Remark 4} Finally, there is no partially angular characterization of elliptity, since this is a pure-ratio notion.

\subsection{Further structural parallels} 

\n{\bf Remark 1} As regards tilded coordinates, the azimuthal angle $\w{\Theta}$ encodes area content via 
\be
\tau = \cos \, \,\w{\Theta} \m, 
\ee
whereas its polar angle partner $\w{\Phi}$ is the mirror image identified (3, 1) shape space coordinate (Shape-Theoretic Aufbau Principle!).  

\m

\n{\bf Remark 2} While there is less use for HH-centric spherical coordinates, 
let us comment here that their azimuth is a pure function of anisoscelesness and the polar angle parametrizes the isosceles-space circle.

\m

\n{\bf Structural Parallel 5} The previous subsection's Remark 5 completes the general pattern of the principal axis having a pure quantity to its perpendicular 
plane having two mixed quantities and a coordinate that is conceptually complementary to the principal axis's azimuthal significance.
I.e.\             (area content) plus (collinearity   parametrization), 
    versus (ellipticity content) plus (regularity     parametrization), 
and versus (anisosceles content) plus (isoscelessness parametrization).  

\m

\n{\bf Structural Parallel 6 and Application 2} The Hopf map structure picks out three uniformity notions.
I.e.\ a uniform angle notion: collinearity, 
a pure ratio uniformity notion: regularity, 
and a notion which is dual angle-separation uniformity: isoscelesness.

\subsection{Further Linear Algebra, now of Hopf Quantities}

\n{\bf Remark 1} Anisoscelesness and ellipticity have the further interpretation of being the eigenvectors of the Heron map $\Heron$.
[This follows from Sec \r{HoKe} alongside the identification of $x$ with anisoscelesness and $z$ with ellipticity.]

\m

\n{\bf Remark 2} Due to $\Heron$ and $\invol$ commuting with each other, these maps moreover share their eigenvectors.\f{Sharing of 
eigenvectors under these circumstances is well-known in Quantum Mechanics, under the name of `complete set of commuting observables' (CSCO), and in Methods of Mathematical Physics.} 
Anisoscelesness and ellipticity are thus also eigenvectors of the sides--medians involution $\invol$.

\m

\n{\bf Structure 1} Let us further formulate anisoscelesness and ellipticity in Linear Algebra terms as follows.  
\be
\Aniso =   {\cal A}_i s_i\mbox{}^2 
       =   {\cal A}_i J_{ij} m_i\mbox{}^2 
	   = - {\cal A}_i m_i\mbox{}^2          \m   
\ee 
and 
\be
\Ellip = {\cal E}_i s_i\mbox{}^2 = {\cal E}_i J_{ij} m_i\mbox{}^2 = - {\cal E}_i m_i\mbox{}^2   \m .  
\ee 
for vectors 
\be
\u{\cal A} := \left( \s{\m \m \mbox{1}}{\s{\mbox{$-1$}}{\m \m 0}} \right) \mbox{ and }          \m 
\u{\cal E} := \left( \s{\m \m \mbox{1}}{\s{\m \m \mbox{$1$}}{-2}} \right)                       \m .
\ee 
\n{\bf Structure 2} Introduce furthermore $\Ellmat$ and $\Animat$ matrices for ellipticity squared and anisoscelesness squared,
\be
\Ellip^2 = E_{ij} s_i\mbox{}^2 s_j\mbox{}^2   \m 
\ee 
and 
\be 
\Aniso^2 = A_{ij} s_i\mbox{}^2 s_j\mbox{}^2   \m, 
\ee 
for 
\be
\Ellmat \:= \left( \s{  \s{  \mbox{$4  \m                            -2 \m                  -2$}  }  
                                                                 {  \mbox{$          2  \m \m \m \m  1 \m \m \m \m 1$}  }  }
										                         {  \mbox{$          2  \m \m \m \m  1 \m \m \m \m 1$}  }    \right)    \m, \mbox{ and } \m 
\Animat \:= \left( \s{  \s{  \mbox{$0  \m \m \m \m   0 \m  \m  \m \m           0$}  }  
                                                                 {  \mbox{$          0  \m \m \m \m  1 \m          -1$}  }  }
										                         {  \mbox{$          0  \m          -1 \m \m  \m  1$}  }     \right)    \m . 																 
\ee
Motivation for doing this for the squares is that it is these which are on a Hopf bundle-theoretic par with the Heron map $\Heron$ of Sec 2. 
We then observe the following remarkable commutativity theorem. 

\m

\n{\bf Theorem 1} All three of the $Hopf^2$ quantities' matrices commute 

\m

\n i) with each other, 
\be
\mbox{\bf [} \Heron \mbox{\bf ,}  \, \Ellmat \mbox{\bf ]} \es   
\mbox{\bf [} \Heron \mbox{\bf ,}  \, \Animat \mbox{\bf ]} \es
\mbox{\bf [} \Ellmat \mbox{\bf ,} \, \Animat \mbox{\bf ]} \es 0   \mma  \mbox{ and }
\ee 
\n ii) with the sides-median involution $\invol$, 
\be
\mbox{\bf [} \Heron \mbox{\bf ,}  \, \invol \mbox{\bf ]} \es   
\mbox{\bf [} \Ellmat \mbox{\bf ,} \, \invol \mbox{\bf ]} \es 
\mbox{\bf [} \Animat \mbox{\bf ,} \, \invol \mbox{\bf ]} \es 0   \m .    
\ee 
\n(See \c{2-Herons} for a proof.) 

\m 

\n{\bf Remark 3)} A matrix form of the Hopf on-sphere condition is given by 
\be 
\Heron + \Animat + \Ellmat \es \frac{1}{9}  \, \Imat     \m .   
\ee 
Here, \Imat \m is the degenerate `all unit entries matrix' 
\be 
\Imat := \m \left( \s{  \s{  \mbox{$1  \m \m 1 \m \m 1$}  }  
                          {  \mbox{$1  \m \m 1 \m \m 1$}  }  }
						  {  \mbox{$1  \m \m 1 \m \m 1$}  }     \right)   \m . 
\ee 
%

\subsection{Hopf congruence and similarity conditions}\l{Hopf-Cong-Sim} 

Following on from Sec \r{C-S} and \r{RHC-S}'s standard congruence and similarity conditions, and then Sec \r{Jac-Cong-Sim}'s involving medians or Jacobi coordinates, 
we now provide Hopf congruence and similarity conditions for triangles.  

\m

\n{\bf Definition--Proposition 1: Hopf congruence condition 7)} $\Tau$-$\Ellip$-$\Aniso$ (for a given clustering choice's $\Ellip$ and $\Aniso$).

\m

\n{\bf Definition--Proposition 2: Hopf similarity condition 7)} $\tau$-$\ellip$ (for a given clustering choice's $\ellip$). 

\m

\n{\bf Definition--Proposition 3: Hopf similarity condition 8)} $\tau$-$\aniso$ (for a given clustering choice's $\aniso$). 

\m

\n{\bf Definition--Proposition 4: Hopf similarity condition 9)} $\ellip$-$\aniso$ (for a given clustering choice's $\ellip$ and $\aniso$). 

\m

\n {\bf Remark 1} Among these Hopf similarity conditions, 
$\tau$-$\ellip$ has further democracy and pure-ratio raisons d'\^{e}tre than the $\tau$-$\aniso$ and $\ellip$-$\aniso$ versions muster. 

\m

\n{\bf Remark 2} On the other hand, it is $\ellip$-$\aniso$ data which most readily interconverts with the standard side-angle and side-ratio variables, as per the next subsection.

\subsection{Solving a triangle given Hopf data}\l{Hopf-Data} 

\n{\bf Proposition 1} For Hopf relational data, 
\be
a        =  \sqrt{\bigiota + \Ellip}                                                                                                 \mma 
b_{\pm} \es \frac{1}{\sqrt{2}}  \sqrt{    2 \, \bigiota  - \Ellip \pm \sqrt{3} \, \Aniso    }                                        \mma  
\ee
\be 
\cot \, \gamma_{\pm}  \es  \frac{1}{\sqrt{3}}  \, \sqrt{\frac{1 - \ellip}{1 + \ellip}} \, \cosec \, \Phi \pm \cot \Phi       \m \mbox{ and } \m 
\cot \, \alpha        \es  \frac{1 - 2 \, \ellip}{\sqrt{3} \, \sqrt{1 - \ellip^2} \, \sin \, \Phi }                                          \m .   
\ee 

\m

\Proof Let us first note that 
\be
\lambda \, \bigiota_1 + \mu \, \bigiota_2 \es \frac{\lambda + \mu}{2} \, \bigiota - \frac{\mu - \lambda}{2} \, \Ellip               \m .
\l{Useful}
\ee 
As a first instance of this, 
\be
\frac{\bigiota + \Ellip}{2} \m = \bigiota_1 = \rho_1^2 = \mu_1 R_1^2 = \m \frac{1}{2} a^2                                           \m, 
\ee 
from which the first equation follows.

\m

\n The second equation follows from  eq. 1 of (\r{rho-c})'s mass-weighted consequence of the cosine rule, definition (\r{Aniso}) and subcase 
\be
3 \, \bigiota_2 + \bigiota_1 = 2 \, \bigiota - \Ellip                                                                               \m,
\ee 
of (\r{Useful}).
The third equation follows likewise from (\r{rho-c}).

\m

\n The fourth equation follows from the first equation in (\r{calR-gamma}) by use of definitions (\r{Aniso}) and (\r{Ellip}).
The fifth follows likewise from the second equation in (\r{calR-gamma}), and the sixth from (\r{calR-alpha}) $\Box{ }$ . 

\m

\n{\bf Proposition 2} For $\ellip$-$\aniso$ Hopf similarity data, 
\be 
\frac{b_{\pm}}{a} = \sqrt{\frac{2 - \ellip \pm \sqrt{3} \, \aniso}{2( 1 + \ellip )}}   \mma   
\ee
\be 
\cot \, \gamma_{\pm} =  \frac{1}{\sqrt{3} \, \area} (1 - \ellip) \pm \aniso   \m , 
\ee
\be 
\cot \, \alpha =  \frac{1 - 2 \, \ellip}{\sqrt{3} \, \area}                   \m . 
\ee

\subsection{Relational space application of the Hopf map to Shape Theory} 

\n{\bf Application 3} An ambient $\mathbb{R}^3$ can moreover be interpreted as relational space.
\beq
{\cal R}(3, 2) \m \s{\mbox{topologically}}{=} \m \mathbb{R}^3   \m . 
\eeq 
\n{\bf Remark 1} Relational space does not however admit the flat metric. 
For instance, the Ricci scalar is 
\be
Ric \es \frac{6}{\bigiota}   \m .
\ee
As well as indicating curvature, this moreover blows up at $\bigiota = 0$, i.e.\ at the maximal coincidence-or-collision O, showing this to be a singular metric geometry.    

\m

\n{\bf Remark 2} This metric is however conformally flat [c.f.\ eq. (I.186)] if one disregards the mathematically pathological O.
\n Moreover, we would expect some features would then be lost, starting with the metric-level consequences of the curvature singularity at O. 

\m

\n Thus we disrecommend solely proceeding via `passing to' a conformally flat representation for which the underlying conformal transformation, 
however, goes singular at precisely the point of main geometrical, topological and dynamical interest: O.  

\m

\n{\bf Remark 3} Mirror image identified configurations are singular for 2- and 3-$d$ RPMs.

\subsection{Hopf map for degenerate shapes}

For collinear configurations, $Area$ vanishes, whereas 
\be
\Aniso = 2 \, \rho_1 \rho_2
\ee 
and $\Ellip$ is unaffected. 
One can readily check these obey the on-circle condition, 
\be
\aniso^2 + \ellip^2 = 1   \m .  
\ee 
This amounts to 
\be
\aniso  =        \frac{2 \, {\cal R}}{1 + {\cal R}^2} 
       \es 2 \,  \frac{  \tan\frac{\Theta}{2}  }{  1 + \tan^2\frac{\Theta}{2}  } 
	   \es       \sin \, \Theta                                                    \m , 
\ee 
and consequently 
\be
\ellip = cos \, \Theta   \m . 
\ee
This inter-relates $\ellip$ and $\aniso$, for collinear configurations, with Part I's (3, 1) angular ratio variable.

\subsection{Democratic ellipticity quantifier and inequalities satisfied by it}\label{linell}

\n{\bf Definition 1} The democratic version of ellipticity is 
\be 
\left\langle \Ellip \right\rangle \:=  \bigmu_{\sQ}\mbox{}^2 - \bigsigma_{\sQ}\mbox{}^2                     \m .   
\ee 
The corresponding pure-ratio quantity is, normalizing by 
\be
Z_{\sT} :=   \bigmu_{\sQ}\mbox{}^2 + \bigsigma_{\sQ}   
        \es  \frac{1}{3} \, \sum_i \left(    \rho_1^{(i) \, 2} + \rho_2^{(i) \, 2}    \right)               \m,  
\ee 
\be
\left\langle \ellip \right\rangle  \:=   \frac{\left\langle \Ellip \right\rangle}{Z_{\sT} }                 \m .  
\ee
\n{\bf Remark 1} However, since $Z_1 = Z_2$ by (\r{7}), $\left\langle \Ellip \right\rangle$ and thus $\left\langle \ellip \right\rangle$ are zero.
Thus democratic ellipticity in the immediate senses of the preceding Sec is a sterile concept. 

\m

\n{\bf Remark 2} This suggests seeking an alternative ellipiticity quantifier whose democratic form remains shape-theoretically nontrivial. 
The obvious candidate for a such follows from considering the linear counterpart of the above construct, as follows. 

\m

\n{\bf Definition 2} The {\it linear ellipticity} is given by    
\be
\LinEllip                 := \rho_2 - \rho_1   \m .  
\ee 
\n{\bf Remark 3} $\LinEllip$'s zero correponds to
\be 
2 \, a^2 = b^2 + c^2   \m .
\ee
Thus equilaterality $a = b = c$ solves, but not uniquely \c{Ineq}. 

\m

\n{\bf Remark 4} The normalized version, $\linellip$, turns out to have no extrema, 
its {\it extremal values} occurring at the ends of its allowed range: -- 1 for the binary coincidence-or-collision shape B and + 1 for the uniform collinear configuration, U.  

\m

\n{\bf Structure 1} The democratic version of these are, firstly,  
\be 
\left\langle \LinEllip \right\rangle \:=  \bigmu_{\sA} - \bigsigma_{\sA}   \m . 
\ee 
Secondly, normalizing by 
\be
F_{\sT}  :=  \bigmu_{\sA} + \bigsigma_{\sA} 
        \es  \frac{1}{3} \sum_i \left(    \rho_1^{(i)} + \rho_1^{(i)}    \right)                                      \m,  
\ee 
where `T' stands for total average: {\sl all} Jacobi coordinates, corresponding to all mass-weighted sides {\sl and} all mass-weighted medians,
\be
\left\langle \linellip \right\rangle   \:=   \frac{\left\langle \LinEllip \right\rangle}{F_{\sT} } 
                                       \es   \frac{\bigmu - \bigsigma}{\bigmu + \bigsigma} 
								       \es   \frac{\Psi - 1}{\Psi + 1}                                                \m . 
\ee
\n{\bf Remark 5} See \c{Ineq} for a determination of the zeros of this and \c{A-Perimeter} for its extremization. 

\m

\n{\bf Corollary 1}   
\be
- \frac{2 - \sqrt{3}}{2 + \sqrt{3}} \leqs  \left\langle \linellip \right\rangle  
                                    \leqs         \frac{2 - \sqrt{3}}{2 + \sqrt{3}}                                                     \m .
\ee
\n{\bf Remark 7} This is symmetrically placed, as befits a two-tailed quantity. 
This result also prompts a renormalized re-issuing of the definition of linear ellipticity as follows, so that the range now be the standard 2-tailed one from -- 1 to +1.
\be
\left\langle \w{\linellip} \right\rangle \:= \frac{2 + \sqrt{3}}{2 - \sqrt{3}} \, \frac{\bigmu - \bigsigma}{\bigmu + \bigsigma} 
								                         \m  = \m \frac{2 + \sqrt{3}}{2 - \sqrt{3}} \, \frac{\Psi - 1}{\Psi + 1}            \m . 
\ee
\n{\bf Remark 8} For the equilateral triangle $E$, $\left\langle \w{\linellip} \right\rangle = 0$. 
$\left\langle \w{\linellip} \right\rangle_{\sm\sa\sx}$ is at the binary coincidence-or-collision B, whereas  
$\left\langle \w{\linellip} \right\rangle_{\sm\si\sn}$ is at the uniform collinear shape U.  
Among the splinters, these are (Fig \r{S(3, 2)-Splinters}.b) respectively, the extreme spear-head and the extreme pickaxe-heads. 
We finally have the following results \cite{Ineq}.  

\m

\n{\bf Theorem 1}   
\be
- 1 \leqs \left\langle \w{\linellip} \right\rangle 
    \leqs 1                                                                                                                            \m .
\ee
\n{\bf Theorem 2} 
\be
\frac{{\cal I} - 1}{{\cal I} + 1}                                             \leqs  
\frac{2 - \sqrt{3}}{2 + \sqrt{3}} \left\langle \w{\linellip} \right\rangle    \leqs 
\frac{1 - {\cal I} \, {\cal G}^3}{1 + {\cal I} \, {\cal G}^3}                                                                          \m .
\ee

\section{Configuration space automorphism groups}\l{Killing-(3,2)}

\subsection{Graph automorphisms}

\n{\bf Proposition 1} i) 
\be
Aut(\Top\mbox{--}\w{\FrS}(3, 2)) = 
Aut(\Top\mbox{--}\FrS(3, 2))     = Aut(\mbox{claw graph with equal talons as per Fig 2.1}) 
                                 = D_3 
					             = S_3                                                          \m . 
\ee
\n ii) 
\be
Aut(\Top\mbox{--}\Leib_{\sFrS}(3, 2)) = 
Aut(\Top\mbox{--}\FrI\FrS(3, 2))     = Aut(P_{2} \mbox{ as labelled in Fig 2.2)}) 
                                     = \mathbb{Z}_2                                         \m .
\ee
iii) By the argument of Sec I.20.1, the above automorphism groups remain unchanged under $\FrS \longrightarrow {\cal R}$.

\subsection{Shape momenta}

The remainder of this section is {\bf Application 4} of the Hopf map, 
due to the Cartesian axes it provides being useful in discussing shape momenta and conserved quantities.  

\m

\n{\bf Remark 1} For triangleland in spherical coordinates, the momenta are
\beq
\sJ =: p_{\Phi} = hopf_1\pi^{hopf}_2 -  hopf_1\pi^{hopf}_2  \m . 
\eeq
Since $\Phi$ is a relative angle in space, this is a relative angular momentum in space (hence the notation $\sJ$). 

\m

\n On the other hand,  
\beq
\sD\si\sll_{\triangle} := p_{\Theta}
\eeq
is indeed a relative dilational quantity since $\Theta$ is a function of a relative distance (length ratio).

\subsection{Killing vectors and isometry group}

\n{\bf Remark 1} Pure-shape triangleland has the maximal three Killing vectors, among which the `axial' $\pa/\pa\Phi$ corresponds to invariance under change of relative angle.  
Scaled triangleland has three conformal Killing vectors.  
These do not include the three 
\beq
\frac{\pa}{\pa \, hopf^i}
\eeq 
because of the origin's topological and geometrical distinctiveness.
It is rather the three 
\beq 
hopf^j \frac{\pa}{\pa \, hopf^i} - hopf^i \frac{\pa}{\pa \, hopf^j}  
\eeq  
-- $SO(3)$ quantities -- which survive.  

\m

\n\n{\bf Remark 2} In the present context these are to be interpreted as follows \c{08I}. 
One is the relative angular momentum  
\beq
\sJ      \es \frac{\bigiota_1\bigiota_2}{\bigiota}\dot{\Phi} 
         \es \frac{\bigiota_1\bigiota_2}{\bigiota}  (  \dot{\theta}_2 - \dot{\theta}_1  ) 
		 \es \frac{\bigiota_1\Lambda_2 - \bigiota_2\Lambda_1}{\bigiota} 
		 \es   \Lambda_2 
		 \es - \Lambda_1 
		 \es \frac{1}{2}  (  \Lambda_2 - \Lambda_1  )                           \m .
\l{MaxZ}
\eeq
The $\theta_i$ here are polar relative Jacobi angles and the $\Lambda_i$ are the corresponding partial angular momenta.   
The last expression can thus be interpreted as 
half of the difference between the two subsystems' angular momenta.  

\m

\n\n{\bf Remark 3} The other two Killing vectors are mixed relative dilational and angular momenta.
Such quantities have previously been termed {\it generalized angular momenta} by physicist Felix Smith \c{Smith60}, 
more descriptively {\it rational momenta} (conjugate to general ratio variables) by Anne Franzen and the Author \c{AF}, 
and finally {\it shape momenta} \c{QuadI, FileR} following discussions with Eduardo Serna.  
Hence we denote the three of them by 
\beq
\sS_i = \epsilon_{ijk} hopf_j \pi^{hopf}_k   \m,  
\eeq
and recover the pure relative angular momentum 
\beq
\sJ  :=  \sS_3 
    \es  \frac{\pa}{\pa\Phi}                  \m .
\eeq
\n{\bf Proposition 1} All in all, 
\beq
Isom(\FrS(3, 2)) = SO(3)                          \m .  
\eeq
Consult Fig \r{12-3} for the mirror image identified, indistinguishable and Leibniz versions.  
%
{\begin{figure}[ht]
\centering
\includegraphics[width=0.7\textwidth]{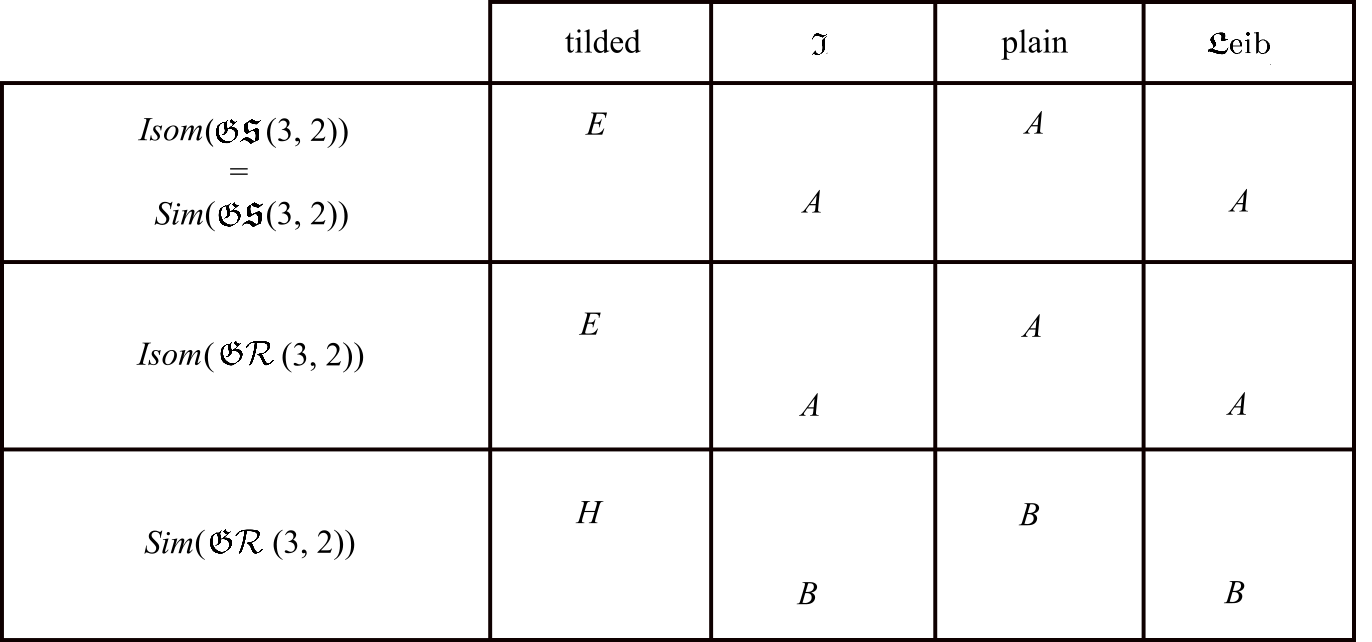}
\caption[Text der im Bilderverzeichnis auftaucht]{\footnotesize{Table of pattern of isometry and similarity groups 
                                                                for (scale and) shape spaces for 3 points in 2-$d$.}} 
\l{12-3}\end{figure} } 

\m

\n{\bf Remark 3} Finally note that the constant-$\Phi$ curves are furthermore privileged as Killing vector preserving curves.
Such `adapted' submanifolds privileged by Killing vector preservation will play a substantial role in the theory of planar quadrilateral shapes \c{IV, Affine-Shape-2}.

\section{Shape representatives}\l{Sections}

\n{\bf Structure 1} The Hopf map is moreover closely related to the Hopf fibre bundle 
a principal fibre bundle with $\mathbb{S}^2$ as base space, 
                              $\mathbb{S}^3$ as total space, 
                          and $\mathbb{S}^1 = U(1)$ as fibres and structure group. 						

\m
						  
\n{\bf Remark 1} This and the next section exploit this further feature.

\m

\n{\bf Structure 2} Let us now label $\w{\FrS}(3, 2)$, 
                     $\FrS(3, 2)$, $\FrI\FrS(3, 2)$ 
				 and $\Leib_{\sFrS}(3, 2)$ not with a representative as in Figs \r{Leib(3, 2)-Global-Section}-\r{Leib(3, 2)-Global-Section-2} 
but with what the entire space of representatives at that point is.  

\m

\n{\bf Proposition 1} In the case of the shape sphere, this is 
\be 
SO(2)       =         U(1)  
      \m \s{m}{=} \m  \mathbb{S}^1
\ee 
as its representative everywhere.  
This homogeneity of fibre objects permits a simple reformulation for the space of spaces of representatives, $\FrS\FrS\FrR(\FrS(3, 2))$, as follows. 
\be
\FrS\FrS\FrR(\FrS(3, 2)) = P(\mathbb{S}^2, \mathbb{S}^1)   \m : 
\ee
the principal bundle of the base space $\mathbb{S}^2$ and fibres = structure group $\mathbb{S}^1 = U(1)$.  

\m

\n{\bf Remark 2} This is of course the Hopf fibre bundle itself, since 
\be
\FrS\FrS\FrR(\FrS(3, 2)) \m \mbox{ is just another conception of preshape space } \m \FrP(3, 2) 
\ee
with  
\be
\FrP(3, 2) = \mathbb{S}^3 \mbox{ viewed as the total space of the } \m P(\mathbb{S}^2, \mathbb{S}^1) \mbox{ Hopf bundle }   \m .  
\ee
\n{\bf Remark 3} A major issue with this result is that the particular mathematical niceness of preshape space being a fibre bundle does not generalize particularly far. 
This causes a number of `intuitions' about the nature of Shape Theory obtained by studying these few particularly nice cases to unfortunately turn out to be atypical.  
Note firstly that the generalization which still does work out nicely is to the similarity $N$-a-gon, as exposited in Part IV.  

\m

\n{\bf Remark 4} What we show below is that mirror image identification and/or point indistinguishability partly spoil this result. 
This counts as a mathematical simplicity argument against studying these cases, but also, due to the inherent interest of these cases, 
as an argument to upgrade the mathematical methods used in Shape Theory to overcome this impasse. 

\m

\n{\bf Remark 5} As a final example, \c{Affine-Shape-2} shows a yet more substantial breakdown of this fibre bundle result in the case of affine shape spaces 
(even without mirror image identification or indistinguishable points).

\m

\n{\bf Remark 6} Choice of representative is moreover a realization of gauge freedom, whereas space of representatives is the gauge orbit corresponding to each base point space.

\subsection{Shape Theory requires general bundles}
%
{            \begin{figure}[!ht]
\centering
\includegraphics[width=0.7\textwidth]{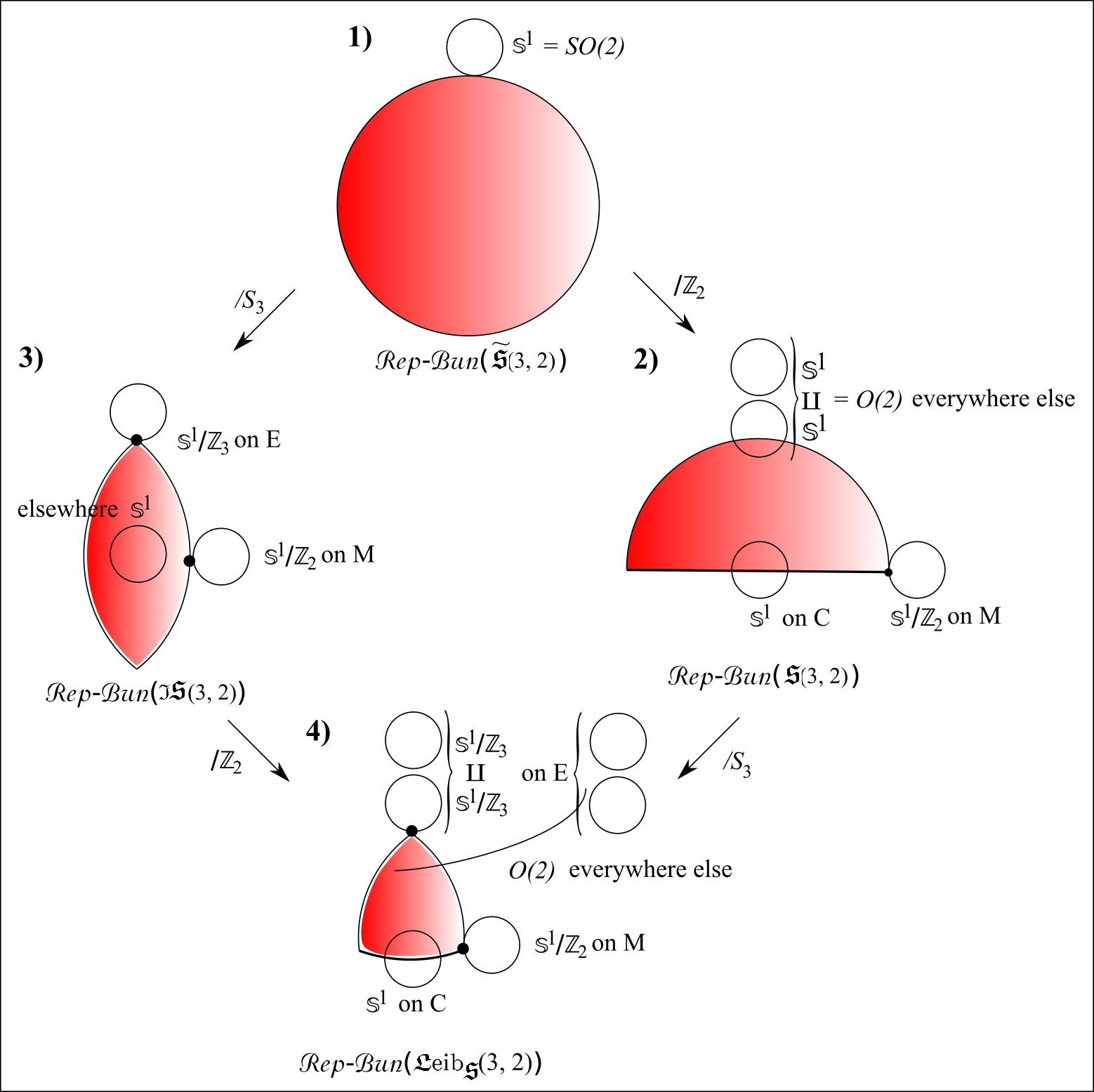}
\caption[Text der im Bilderverzeichnis auftaucht]{        \footnotesize{(3, 2) triangleland bundles of shape representatives.} }
\l{Rep-Bun} \end{figure}          }

\n{\bf Proposition 1}  For 2-$d$ similarity shape theory, the space of spaces of representatives admits decomposition into a {\it bundle of representatives}:
\be
\FrG\FrP(N, 2) = \FrS\FrS\FrR(\FrG\FrS(N, 2)) 
               =      \RepBun(\FrG\FrS(N, 2))    \m . 
\ee
\n{\bf Remark 2} This is a {\sl general bundle} and not usually more specifically a {\sl fibre bundle}, since the attached objects can differ from point to point.\footnote{E.g.\ 
differ homeomorphically  at the topological  manifold level, or a fortiori 
       diffeomorphically at the differential manifold level, or furthermore  
       isometrically     at the level of Metric Geometry.}  
%
For $\w{\FrP}(N, 2)$ itself, this is, exceptionally, a fibre bundle as well. 
The point of this generalization is, rather, covering how $\FrP(N, 2)$, $\FrI\FrP(N, 2)$ and $\Leib_{\sFrP}(N, 2)$ decompose 
as regards the corresponding base shape space and the corresponding spaces of representatives.

\m

\n{\bf Proposition 2} 
\be
\w{\FrP}(3, 2) = \RepBun(\w{\FrS}(3, 2)) 
               = P(\mathbb{S}^2, \mathbb{S}^1) 
	    	   = \mbox{Hopf fibration of } \mathbb{S}^3                                      \m .
\ee
\be
\FrP(3, 2)     =    \RepBun(\FrS(3, 2))                                                      \m,
\ee
\be 
\FrI\FrP(3, 2)    =    \RepBun(\FrI\FrS(3, 2))                                              \m \mbox{ and }
\ee
\be
\Leib_{\FrP}(3, 2) =  \RepBun(\Leib_{\sFrS}(3, 2))
\ee
are as per Fig \r{Rep-Bun}.2)-4) respectively.   

\m

\n{\bf Remark 3} Now not only C is distinguished by stratification, but also E and M = U(3, 1) are as well.  
This further justifies interest in M = U(3, 1).  
%

\m

\n{\bf Remark 4} These strata arise as follows.   
The relevant master group is now $O(2)$, which $SO(2)$ as a nontrivial subgroup, while three further discrete subgroups of this turn out to also be relevant. 
So even in 2-$d$ similarity shape theory, strata make an appearance, 
provided that one considers one or both of mirror image identification or point-or-particle indistinguishability.  

\m

\n{\bf Remark 5} General bundles can be furthermore equipped as presheaves and then as sheaves, by which stronger gluing and localization methods become available.  
Note once again that the above three general bundles' base spaces are Hausdorff, second-countable and compact, 
so that whichever of the simpler and more worked out cases in \c{Project-2}.   
(The attached entities are, moreover, also all Hausdorff, second countable and compact, with all bar one -- $O(2)$ -- additionally enjoying connectedness).

\subsection{Picking representatives is choosing a section}
%
{            \begin{figure}[!ht]
\centering
\includegraphics[width=0.5\textwidth]{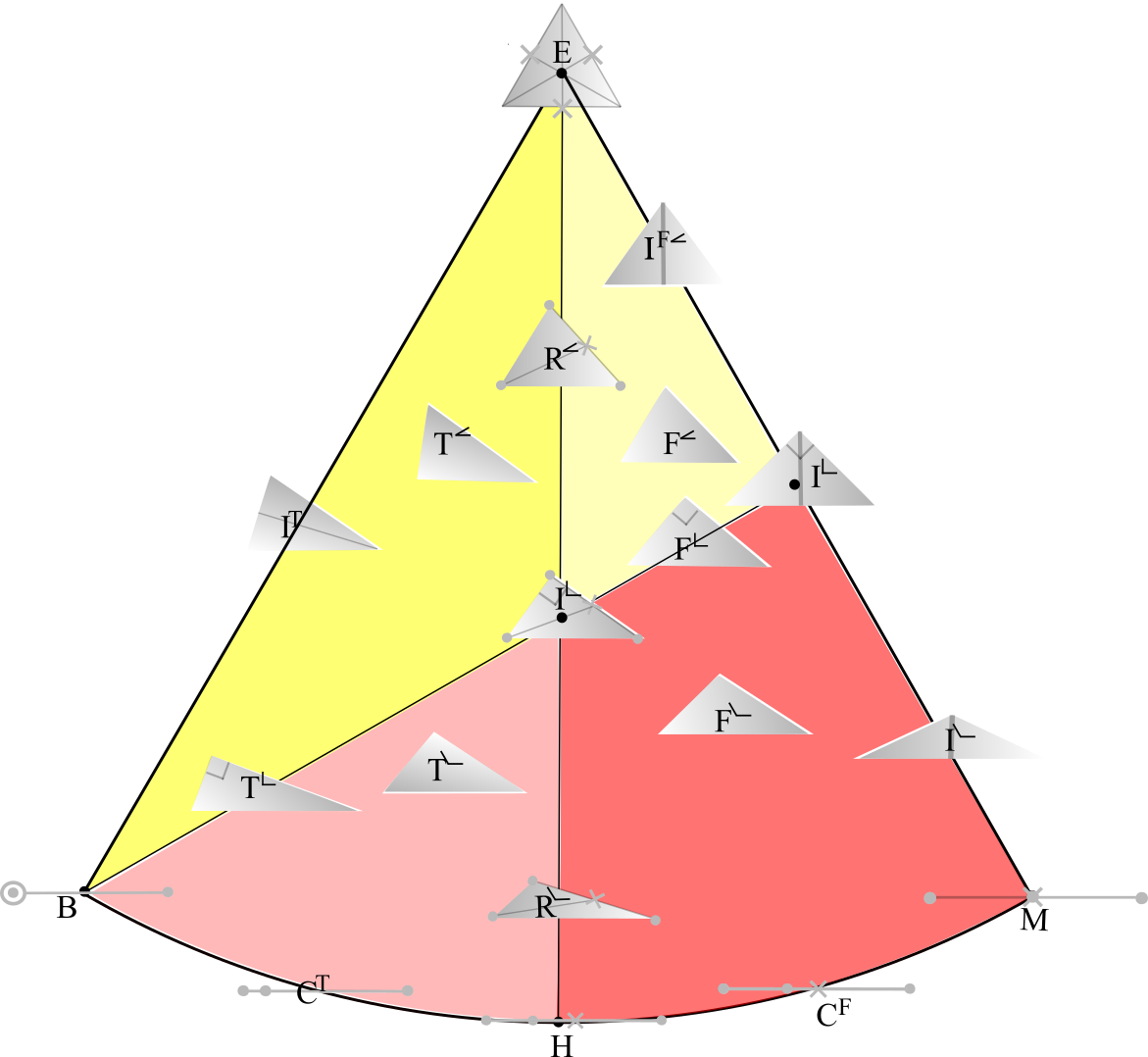}
\caption[Text der im Bilderverzeichnis auftaucht]{        \footnotesize{$\Leib_{\tFrS}(3, 2)$ `standing up' `flat isosceles favouring' global section.} }
\l{Leib(3, 2)-Global-Section} \end{figure}          }
%
{            \begin{figure}[!ht]
\centering
\includegraphics[width=0.5\textwidth]{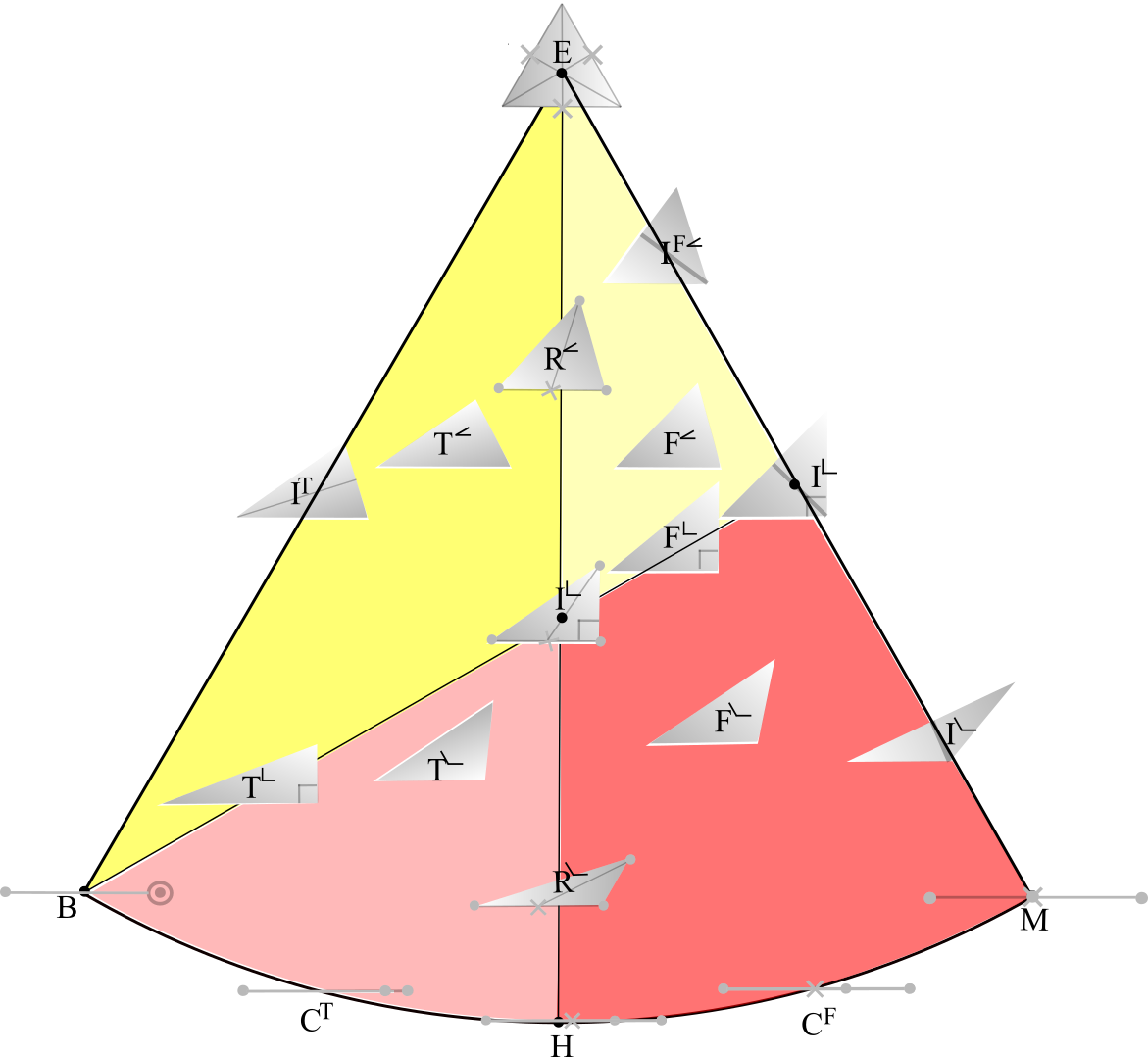}
\caption[Text der im Bilderverzeichnis auftaucht]{        \footnotesize{$\Leib_{\tFrS}(3, 2)$ `standing up' `right-and-regular favouring' global section. } }
\l{Leib(3, 2)-Global-Section-2} \end{figure}          }
%
{            \begin{figure}[!ht]
\centering
\includegraphics[width=0.5\textwidth]{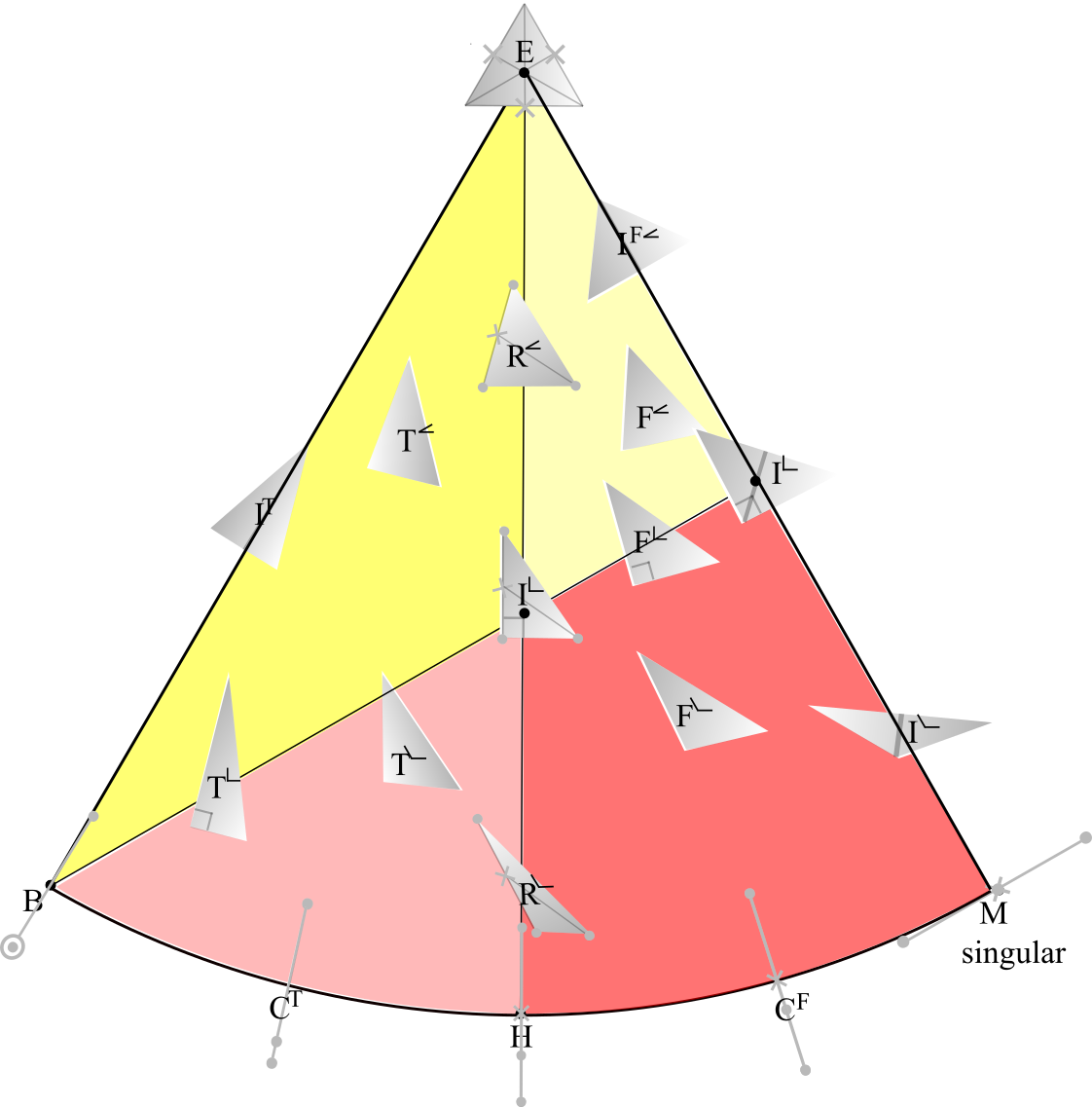}
\caption[Text der im Bilderverzeichnis auftaucht]{        \footnotesize{$\Leib_{\tFrS}(3, 2)$ non-global section which is not defined at the equilateral triangle 
and is discontinuous at M but is elsewise well-defined.} }
\l{Leib(3, 2)-Nonglobal-Section-2} \end{figure}          }

\n{\bf Remark 1} Let us now return to Kendall's labelling of points by triangles and identify it as a {\sl choice of section}.  
In $\Leib_{\sFrS}(4, 1)$ [and $\FrI\FrS(4, 1)$], while this is not a fibre bundle section, the notion of section straighforwardly extends to general bundles (and to sheaves). 
This case has the added advantage of supporting global sections. 
Figs (\r{Leib(3, 2)-Global-Section}) and (\r{Leib(3, 2)-Global-Section-2}) are such, while (\r{Leib(3, 2)-Nonglobal-Section-2}) is an example which is not globally well-defined.
$\RepBun(\FrS(3, 2))$ itself admits no global sections since it is the Hopf fibre bundle which is indeed a nontrivial fibre bundle.  

\m

\n{\bf Structure 1} All in all, the `shape-in-space to position-in-shape-space' correspondence is to be built up into a bundle and sheaf construct in \c{Project-2}.  

\vspace{10in}

\vspace{10in}

\section{Triangleland's suite of monopoles}\l{Monopoles}
%
{            \begin{figure}[!ht]
\centering
\includegraphics[width=0.45\textwidth]{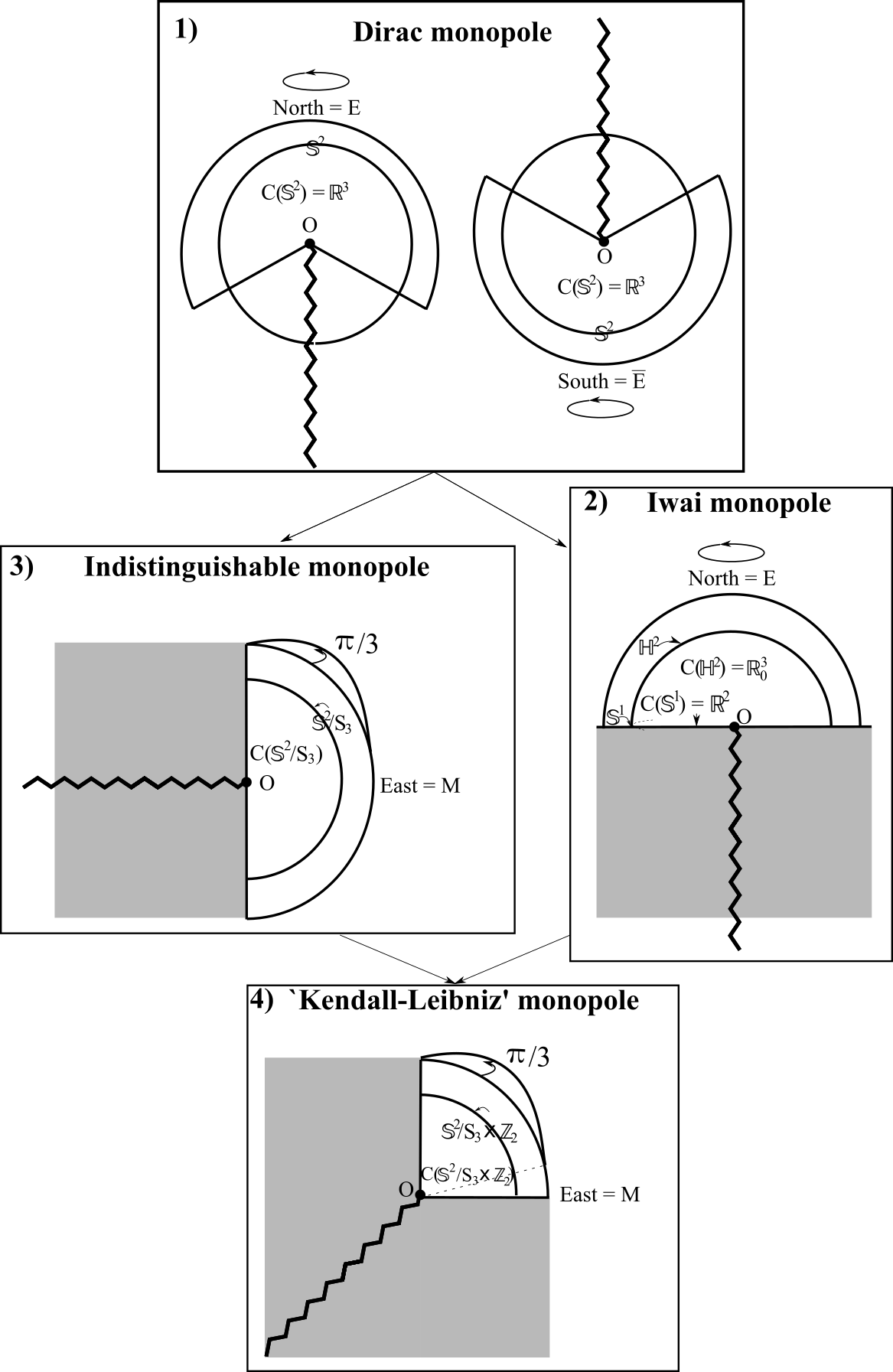}
\caption[Text der im Bilderverzeichnis auftaucht]{        \footnotesize{Relational triangleland suite of monopoles. 

\m

\n In 1), N and S are `North' and `South' stereographic coordinate chart patches which are large enough to overlap with each other.
More shape-theoretic names for these use anticlockwise and clockwise equilateral triangle pole names E and $\overline{\mE}$ rather than North and South.  

\m

\n In 2), N is now a hemisphere's worth of `North' (now orintetation-less equilateral triangle E) stereographic coordinate.  

\m

\n In 3), the chart is a third-hemisphere's worth of `East' (here the unlabelled M configuration) stereographic coordinate. 

\m

\n Finally, in 4) the chart is a sixth-hemisphere's worth of `East' (i.e.\ M) stereographic coordinate. } }
\l{Triangleland-Monopoles} \end{figure}          }

\m

\n Iwai's realization that the half-space from the 3-body problem carries a distinct monopole from Dirac's whole-space version recurs for each of 
$\FrI{\cal R}(3, 2)$ and $\Leib_{\cal R}(3, 2)$.  
Let us introduce the following terminology. 

\m

\n{\bf Definition 1} The $\w{\cal R}(3, 2)$-monopole is a configuration space realization of the {\it Dirac monopole}, which is usually realized in space instead.

\m

\n{\bf Definition 2} The ${\cal R}(3, 2)$-monopole is the {\it Iwai half-space monopole}.  

\m

\n{\bf Definition 3} The ${\cal IR}(3, 2)$-monopole is the {\it 1/6th-segment monopole}.  

\m

\n{\bf Definition 4} The $\Leib_{\cal R}(3, 2)$-monopole is the {\it half-of-a-1/6th-segment} alias {\it Leibniz--Kendall monopole}.  

\m

\n{\bf Remark 1} All four of these are depicted in Fig \r{Triangleland-Monopoles} and are further studied in \c{A-Monopoles}.   
\n As per Iwai's monopole, the ${\cal IR}(3, 2)$ and Leibniz--Kendall monopoles' Dirac string emanating from the maximal coincidence-or-collision O 
can be placed entirely in the corresponding non physically realized sector, by which a single chart or section description is afforded.   
On the other hand, for Dirac's monopole, two charts or sections are required.  
The Iwai monopole is realized -- in configuration space -- in an important $N$-body problem setting: the mirror image identified 3-body problem in $d \geq 2$,  
this identification being a fortiori non-optional in $d = 3$ (and higher; see Sec 12.4 as regards 2 versus $\geq 3$-$d$ distinction in structure).   
As far as the Author is aware, monopoles 3) and 4) are new.

\m

\n{\bf Remark 2} Following on from the previous section, another reason to not conformally transforming to the flat metric on relational space 
is that this would dispense with the monopole structure emanating from O (but this is part of the physical properties of the system).   

\m

\n{\bf Remark 3} \c{A-Monopoles} studies 7 further cases of 2-$d$ 3-body problems.

\section{Conclusion}

\subsection{Results}

\n The current Part III's triangle model's topological configurations reduce to partititions, a wider $d \geq 2$ feature in contrast with Part II's $d = 1$ feature. 
Part III's corresponding topological shape spaces are also considerably simpler than Part II's.  

\m

\n B) As special great circle arcs on triangleland shape space, we identified the equator of collinearity C and meridians of isoscelesness I.
Collinearity splits clockwise and anticlockwise labelled triangles if regarded as distinct. 
Isoscelesness splits left-leaning and right-leaning triangles. 
As special points on triangleland shape space, we firstly identified E:           equilateral triangle, 
                                                                     B:           binary coincidence-or-collision, 
															     and M = U(3, 1): the merger and most uniform collinear shape inherited from the (3, 1) model. 

\m

\n C) Bringing in the Jacobian notion of regularity R -- equal base and median partial moments of inertia -- which is also realized by meridians, 
we identify a final two types of significant point,  H:             the halfway shape also inherited from the (3, 1) model, 
                                                 and $\mR^{\perp}$: the right-and-regular triangle.

\m

\n D) The special B point maximizes clumpiness among the normalizable triangular shapes.
On the other hand, uniformity, mass-weighted area per unit moment of inertia and perimeter per unit area are maximized by the special point E.  
Finally, M is the sole notion of merger in triangleland. 
All in all, more virtues are conferred on the equilateral triangle than are conferred on any one (4, 1) shape.  

\m

\n E) In Part III, we considered placing medians of triangles on the same footing as their sides due to Jacobi and Hopf motivation, 
both of these being structures entering the Shape Theory of the space af triangles.  
We first reformulated the medians--sides inter-relation as an involution $\invol$.
We observed that $\invol$(s) factor of $\frac{4}{3}$ has the same origin as the $\frac{4}{3}$ factor 
discrepancy between sides and medians in the standard Heron's formulae in space. 
Moreover, in both the involution $\invol$ and Heron's formula, 
this factor of $\frac{4}{3}$ is removed by the passage to the mass-weighted Jacobi coordinate version of the inter-relation and of the Heron's formulae.  
Thus the mass-weighted sides and mass-weighted medians versions of Heron's formula for the now also mass-weighted area are placed on an indentical footing. 
We term these the {\it Heron--Jacobi formulae}. 
In the process, the orignal factor of $\frac{4}{3}$ is identified to be the ratio of the medians' Jacobi mass to the sides' Jacobi mass.  

\m

\n F) Triangleland is the smallest relational model with nontrivial relative angles.  
It realizes perpendicularity -- in particular but not limited to rightness -- and collinearity (but non-collinear parallelism is not possible for $N = 3$).
It also realizes angle uniformity, of two angles in isosceles triangles, or all three in the equilateral triangle.

\m

\n G) Considering right-angled triangles $\perp$ as well, we additionally identified I$^{\perp}$ -- the right isosceles triangle -- as a special point, 
and also characterized regions of acuteness and obtuseness. 
Realizing $\perp$ in shape space requires spherical cap -- rather than great -- circles.
Via this, Shape Theory moreover provides an answer to Lewis Carroll's longstanding and conceptually interesting pillow problem pillow \c{Pillow} 
of what is $\mbox{Prob(Obtuse)}$ for a triangle, namely 3/4 \c{Small, MIT,  A-Pillow}. 
We furthemore argued for this to be a prototype for mapping problems in flat space geometry\footnote{Or for constellations on other directly realized geometry, 
such as constellations on spheres \c{ASphe} or tori \c{ATorus} or real projective spaces \c{PE16, KKH16}.}
to the corresponding shape spaces, where the problems take a prescribed differential-geometric form.
Part III, we have indeed presented a substantial extension of this problem and its shape-theoretic solution.

\m

\n One extension \c{A-Pillow} is to finding the probability that an isosceles triangle is obtuse, as well as how obtuseness-or-acuteness condition with tallness-or-flatness 
and vice versa, with and without the extra assumption of isoscelesness. 

\m

\n Another extension follows from viewing right-angled triangles as triangles whose maximal angle is right, 
and then passing to consider all the other possible values of maximal angle, $\alpha_{\sm\sa\sx} = [\frac{\pi}{3}, \, \pi]$.
This gives the {\it maximal angle flow} over Kendall's shape sphere.  
We find that the three kissing cap-circles of rightness -- that so readily provide the shape-theoretic resolution of Lewis Carroll's original pillow problem -- 
furthermore play the role of separatrix in the maximal angle flow; this flow is further discussed in \cite{Max-Angle-Flow}.  
We apply our new knowledge of the maximal angle flow over the shape sphere to find the probability that a triangle has nontrivially-located Fermat point. 
This corresponds to the critical angle $\alpha_{\sm\sa\sx} = \frac{2 \, \pi}{3}$. 
This is critical in the sense that if the angle is smaller than this (`Fermat acute'), 
the Fermat point lies in the interior of the triangle, but if it is larger (`Fermat obtuse'), the Fermat point just lies on the obtuse vertex.

\m

\n H) E is moreover the centre of the $\FrS(3, 2)$ hemisphere, whereas M is the $\FrI\FrS(3, 2)$ lune centre, 
whereas $\perp^{\sR}$ provides a notion of $\Leib_{\sFrS}(3, 2)$ centre. 
As in Parts I and II of this treatise, passing to Leibniz space involves paying the price of having a weaker notion of centre.    
Neither $\FrI\FrS(3, 1)$ nor $\Leib_{\sFrS}(3, 1)$ admits a topologically characterized boundary -- only the point B therein is. 
E, isosceles I, M and collinear C are all metric-level notions.

\subsection{The Hopf map's role in Shape Theory}

\n I) The Hopf map is realized by the map from the preshape 3-sphere and the shape 2-sphere.  
In this realization, collinearity, isoscelesness and regularity occur as a package, firstly as mutually orthogonal Cartesian axes from the ambient $\mathbb{R}^3$, 
which is also the triangleland relationalspace.  
Thereby, Jacobi notions are to be considered alongside `geometrically more obvious' Lagrangian notions.  
%
%
Secondly, the Hopf map is rooted in bundle mathematics, consituting a unified geometrical treatment of some traditionally geometric notions and 
other notions arising in the $3$-body problem.
We have also given parallels of the Hopf map for the mirror image identified half-sphere $\FrS(3, 2$ (due to Dragt \c{Dragt}), 
                                                                                         $\FrI\FrS(3, 2)$, 
																				     and $\Leib_{\sFrS}(3, 2)$, 
																					 and again for their (3, 3) counterparts (see Sec 12.4).
This set of stuctures moreover transcends from fibre bundles to general bundles. 
Part IV extends this unified perspective to arbitrary $N$-a-gons.  
We moreover envisage this trend to grow with (c.f.\ Fig I.2) the affine quadrilateral, 
                                                             similarity tetrahaedron, 
												             similarity pentagon, 
												         and subsequent increases in dimension $d$, 
														              point-or-particle number $N$, 
														 and redundant group $\lFrg$'s number of generators.  
Application of the full extent of Hopf structure map has moreover only come to the attention of a subset of the literatures concerning or making use of shapes.  
By this, its prominent inclusion in the current Part is a useful complementation of some of the existing literatures.  

\m

\n J) We show that the elsewise well-known Hopf coordinates diagonalizing the Heron map $\Heron$, a fact that appears to have hitherto escaped attention.  
Indeed, in this manner, diagonalizing Heron's map $\Heron$ provides us with a new derivation both of the Hopf map, 
and of the shape space formed by the triangles being a sphere equipped with the standard round metric: Kendall's Theorem.  
What occurs at the level of the Hopf formulation of the triangle is that Heron's formula has become a `{\it Heron--Hopf}' formula that is one and the same as 
the on-sphere condition determining that the shape space of triangles is a sphere.  
The factor of 4 in the Hopf quantity that, in its 3-body problem incarnation, is the `tetra-area' is moreover accounted for in the current Part's working 
as being none other than the prefactor of 1/4 in the expanded version of Heron's formula.  

\m

\n K) The other two Hopf quantities are, in their 3-body problem incarnation, to be interpreted as ellipticity and anisoscelesness.
In the current Part, these now receive the further enlightening interpretation of being Heron map $\Heron$ eigenvectors, 
                               which are moreover also sides--medians involution $\invol$ eigenvectors by 
the commutation relation between $\Heron$ and $\invol$ also established in this Part.  

\m

\n L) Our {\it Heron--Hopf formula} is obtained by diagonalizing the expanded Heron form, keeping mass-weighted area as the subject.
This is moreover mathematically in the form of an on-sphere condition, which, if represented symmetrically, we term the {\it Heron--Hopf--Kendall formula} 
in honour of Kendall's iconic shape sphere of triangles.  
We finally argued that one can (almost) just as well interpret ellipticity or anisoscelesness as the subject, giving two further concomitants of the Heron--Hopf formula.  

\m

\n M) We are now in possession of a considerable body of shape-theoretic results \c{Kendall, GT09, PE16, KKH16, FileR, ASoS, AMech, I, II, IV, Affine-Shape-1, Affine-Shape-2}.  
Therein, topological and differential geometric methods open many doors as regards the study of figures in a wide range of geometries. 
The current treatise illustrates this for triangles in similarity geometry, 
whereas the last six references above entertain this for a wider range of problems in similarity and affine geometry. 

\m

\n N) The current treatise's consideration of globally-valid Killing vectors over $\FrS(3, 2)$, $\FrI\FrS(3, 2)$, $\Leib_{\sFrS}(3, 2)$ and their (3, 3) counterparts 
is a crucial step for all of establishing Classical Dynamics conserved quantities on these.
And also as regards finding and optimally formulating probability distribution functions over these, and quantizing over these \c{QLS, Quantum-Triangles}.   
Knowledge of the Killing vectors feeds into conserved quantities, 
which play a bigger role for $N$-a-gons, and then have the idea of not only Killing vector-respecting potentials but also of Killing vector-respecting 
formulating probability distribution functions.   
Detailed knowledge of the Killing vectors of each shape space are moreover crucial as regards giving a robust and topologically well-founded 
quantization of each of these very interesting models of triangular shapes.   
This refers both to getting the kinematical quantization right and to features of the dynamical quantization such as Killing vector respecting operator ordering, 
Killing vector invariant wavefunctions and Killing vector-invariant quantum operators. 
 
\m

\n O) We moreover reinterpret Kendall's pictures of the shape sphere and spherical blackboard as sections of the Hopf fibre bundle and of a general bundle. 
We comment on which such pictures are global sections and which need to be covered by multiple local sections. 
We also discuss the spaces of all possible representatives at each point of shape space, i.e. the in-space interpretation of the fibres 
(or the general bundle's inhomogeneous generalization thereof).  
This approach has considerable further applications for larger $N$, $d$ and $\lFrg$.  

\m

\n P) The Hopf map moreover provides the fibre bundle theoretic description of the Dirac monopole  \c{DirMon, Dirac48}, which while usually realized in space, 
is realized {\it in relational space} by the mirror images distinct triangleland.
Iwai \c{Iwai87} moreover also pointed to the mirror images identified counterpart having a distinct monopole structure.
The current Part completes this description by giving two novel monopole structures: 
the 1/6th-segment mirror-images-distinct indistinguishable-particle monopole and the `1/6th-hemisegment `Leibniz--Kendall' monopole. 

\m

\n Q) Indistinguishable $\FrI\FrS$ and Leibniz $\Leib_{\sFrS}$ shape spaces incurr stratification and bundles even for (3, 2).  
Understanding representatives as sections and the space of representatives as bundles or sheaves is an already classically useful observation, 
foundational for the 3-body problem, the geometry of triangles and Shape Statistics.

\subsection{Comparison: (3, 2) versus (4, 1) models}\l{(3, 2)-(4, 1)}
%
{            \begin{figure}[!ht]
\centering
\includegraphics[width=0.88\textwidth]{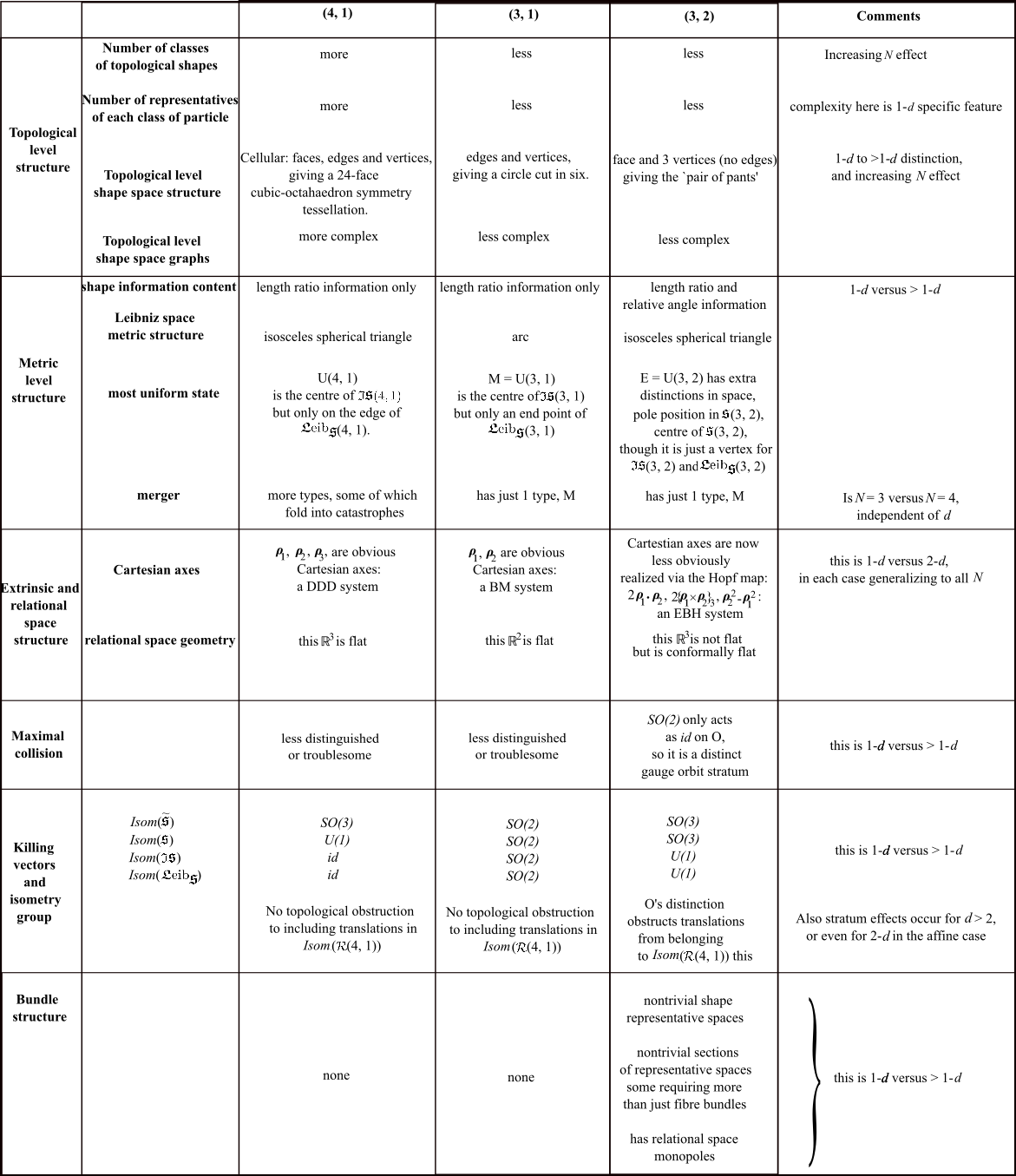}
\caption[Text der im Bilderverzeichnis auftaucht]{        \footnotesize{(3, 2) and (4,1) compared, inluding as extensions of (3, 1).} }
\l{(4, 1)-(3, 2)-Compared} \end{figure}          }

\n See the summary Figure \r{(4, 1)-(3, 2)-Compared}, including how each model extends the (3, 1) model's scope. 

\vspace{10in}

\subsection{Extensions of Part III's work}

\n Let us end by pointing to various extensions of this work.

\m

\n Extension 1) Our flow and area-ratio probability method can be applied to find further maximal critical angle dependent Flat Geometry triangles' properties' probabilities 
of occurring in the shape-theoretic sense of `random triangles'.

\m

\n Extension 2) Sec \r{MAF}'s maximal angle flow over the shape sphere's above-described features likely render the current treatise of interest as an example 
in Differential Geometry and Dynamical Systems. 

\m

\n Extension 3)} A subsequent step is to plot and optimize the shape quantities ${\cal S}$, ${\cal M}$, ${\cal G}$, ${\cal H}$, $\Psi$, $\Lambda$ defined in Sec \r{JSI} 
and $\left\langle \w{\linellip} \right\rangle$ of Sec \r{linell}, over the shape sphere, much as Sec \r{A/I-Plot}  did for mass-weighted area per unit moment of inertia.  
This is considered in the subsequent paper \c{A-Perimeter}, alongside plotting further geometrically significant functions for triangles over the shape sphere of triangles. 

\m

\n Extension 4) \c{MIT, A-Pillow, Max-Angle-Flow} and the current treatise's use of Shape Theory to answer questions about triangle inequalities and random triangles 
moreover extends to its use in answering questions about random quadrilaterals and polygons using the Differential Geometry of the $N$-a-gonland $\mathbb{CP}^{N - 2}$ 
\c{Kendall, QuadI, IV, Pentagon, Hexagon}. 
This program uses Differential Geometry on Kendall's shape spaces to pose and/or solve probabilistic questions 
involving polygons by considering these to live on Kendall's natural shape space with uniform measure thereupon.
Fermat's own problem has moreover a large number of generalizations as well -- geometric means -- which overlap well with the theory of polygons. 

\m

\n Extension 5) The current treatise's notion of similarity Shape Theory moreover itself extends to a wide variety of further notions of shape and shape space, 
such as affine, projective, conformal, M\"{o}bius... \c{GT09, AMech, ABook, PE16, KKH16, Affine-Shape-1, Affine-Shape-2}.   
Many of Extension 4)'s considerations recur in this wider arena.

\subsection{Comparison: (3, 2) versus (3, 3) models}\l{(3, 2)-(3, 3)}

\n{\bf Parallel 1} At the topological level, these have the same shapes and shape spaces.  

\m

\n{\bf Difference 1} At the metric level, however, only mirror image identified shapes are possible.
So the number of configuration spaces is halved: only the right hand side of figures \r{(3,2)-Config} and \r{S(3,2)-Level-1} survives.    

\m

\n{\bf Difference 2} Furthermore (3, 3)'s $SO(3)$ isometry group has a nontrivial subgroup, $SO(2)$, whereas (3, 2)'s isometry group is $SO(2)$.   

\m

\n This has the next four items as consequences. 

\m

\n{\bf Difference 3--Proposition 1} At the level of quotient configuration spaces, 
\be
\FrS(3, 2) = \frac{\mathbb{R}^6}{Sim(2)}
\ee 
is distinct from 
\be
\FrS(3, 3) = \frac{\mathbb{R}^9}{Sim(3)}   \m . 
\l{OS(3,3)}
\ee
Also,  
\be
{\cal R}(3, 2) = \frac{\mathbb{R}^6}{Eucl(2)}
\ee 
is distinct from 
\be
{\cal R}(3, 3) = \frac{\mathbb{R}^9}{Eucl(3)}   \m . 
\l{OR(3,3)}
\ee 
A fortiori, $\FrS(3, 3)$ and ${\cal R}(3, 3)$ are stratified, whereas $\FrS(3, 2)$ and ${\cal R}(3, 2)$ are not. 

\m

\Proof General G and collinear C configurations are on the same orbit  in the first case, so $SO(2)$ acts fully on both. 
On the other hand, they     are on distinct orbits in the second case, with $SO(3)$ acting fully on G but only an $SO(2)$ subgroup acting on C. $\Box$

\m

\n{\bf  Difference 4--Proposition 2} The inertia tensor for 3 particles in 3-$d$ is not invertible at C \c{Gergely}, 
whereas it is for 3 particles in 2-$d$ \c{FORD}. 

\m

\n{\bf  Difference 5--Corollary 1} The dynamical problem can consequently be reduced for all configurations in the 2-$d$ case, 
but only for general (non-collinear) configurations in the 3-$d$. 

\m 

\n This is a global difference rooted on the difference in stratification structure, and can be further rephrased as follows.

\m

\n{\bf Corollary 2} Wheeler's Thin Sandwich Problem's \c{BSW, WheelerGRT} Best Matching Problem \c{BB82} generalization 
is globally solvable for 3 particles in 2-$d$ but only locally solvable away from C for 3 particles in 3-$d$ \c{ABook}.  

\m

\n{\bf Remark 1)} Proposition 1's stratification is moreover particularly benign in the following two ways. 

\m

\n{\bf Proposition 3} Eq (\r{OS(3,3)}) fits together particularly nicely, in the sense that it is not just a stratified manifold but a fortiori a manifold with boundary. 
More specifically, the two strata jointly constitute the {\it hemisphere with boundary}.

\m

\n{\bf Proposition 4} The Ricci curvature scalar is regular at C \c{Iwai87}.

\m

\n{\bf Remark 2)} This is moreover in contradistinction with $\w{\cal R}(N, 3)$ for $N \geq 4$, for which the Ricci curvature {\sl is} singular at C.  

\m

\n{\bf Remark 3)} Thus it is a stratum which is straightforward to attach. 
It is obvious where it fits (Proposition 3), which condition can be formulated entirely geometrically, 
and there is furthermore no curvature singularity obstruction to placing it there. 

\m

\n {\bf Palallel 2--Proposition 5} Both $\FrS(3, 2)$ and $\FrS(3, 3)$ have the same Killing vectors and thus isometry groups: $SO(2)$.

\m

\Proof These are just ${\cal J}$ everywhere for the 2-$d$ case.
For the 3-$d$ case, ${\cal J}$ succeeds in being a Killing vector on both strata, by fitting both $\mathbb{H}^{2}$ and $\mathbb{S}^1$'s global specifications. 
The further Leibniz space boundary conditions are also met. \m $\Box$

\m

\n{\bf Remark 4} Propositions 3, 4 and 5 are key in kinematically quantizing 3 particles in 3-$d$ \c{Quantum-Triangles}.  

\m

\n{\bf Difference 6} The shape representative spaces, and bundles of representatives, are now as in Fig \r{Rep-Bun-(3, 3)} rather than as in Fig \r{Rep-Bun}. 
%
{            \begin{figure}[!ht]
\centering
\includegraphics[width=0.4\textwidth]{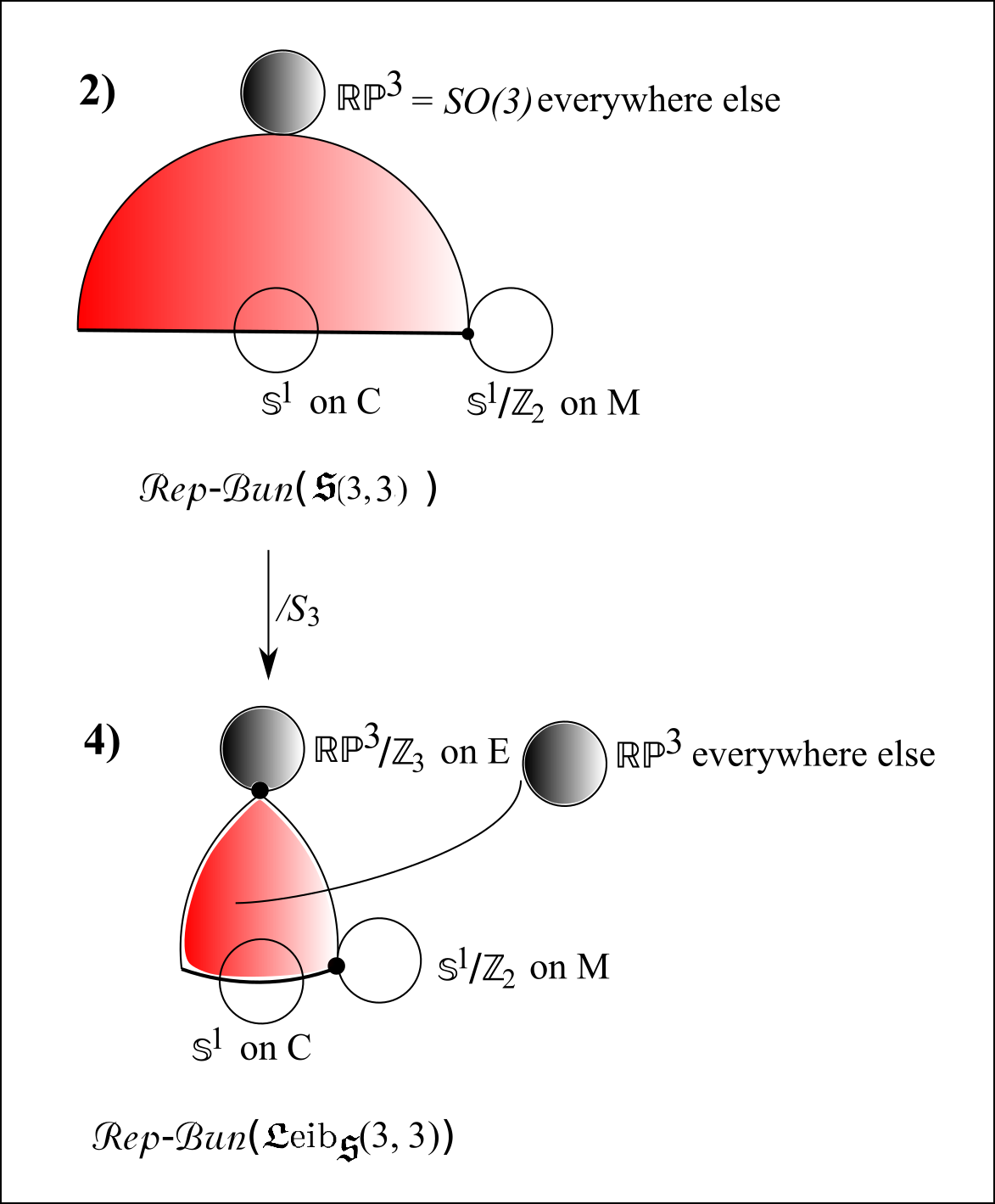}
\caption[Text der im Bilderverzeichnis auftaucht]{        \footnotesize{(3, 3) bundles of shape representatives.} }
\l{Rep-Bun-(3, 3)} \end{figure}          }

\m

\n{\bf Parallel 3} Sections of representatives, however, coincide for (3, 3) models and their two immediate (3, 2) counterparts.  

\m

\n{\bf Difference 7} Due to the above-mentioned stratification, 
(3, 3) has further need of general bundles than (3, 2) in the analysis of its spaces of shape representatives, and of shape representative sections over shape spaces.  
This is because the spaces of shape representatives of collinear configurations exceed those of generic shapes in dimension in the (3, 3) case.  

\m

\n{\bf Difference 8} For the (3, 3) model, variants of the Iwai and Kendall--Leibniz monopoles alone exist. 
These are not geometrically the same as in 2-$d$ since now their collision plane and 1/6th-plane respectively, in each case excluding the origin, 
is now a separate stratum (Fig \r{Triangleland-Monopoles-3-d}).
So the current treatise points out that, at the level of stratified manifolds, there are in fact 2 distinct Iwai monopoles and 2 distinct Kendall--Leibniz monopoles. 
We note that {\sl physical} investigation of topological defects in the presence of stratification remains in its infancy.  
%
{            \begin{figure}[!ht]
\centering
\includegraphics[width=0.3\textwidth]{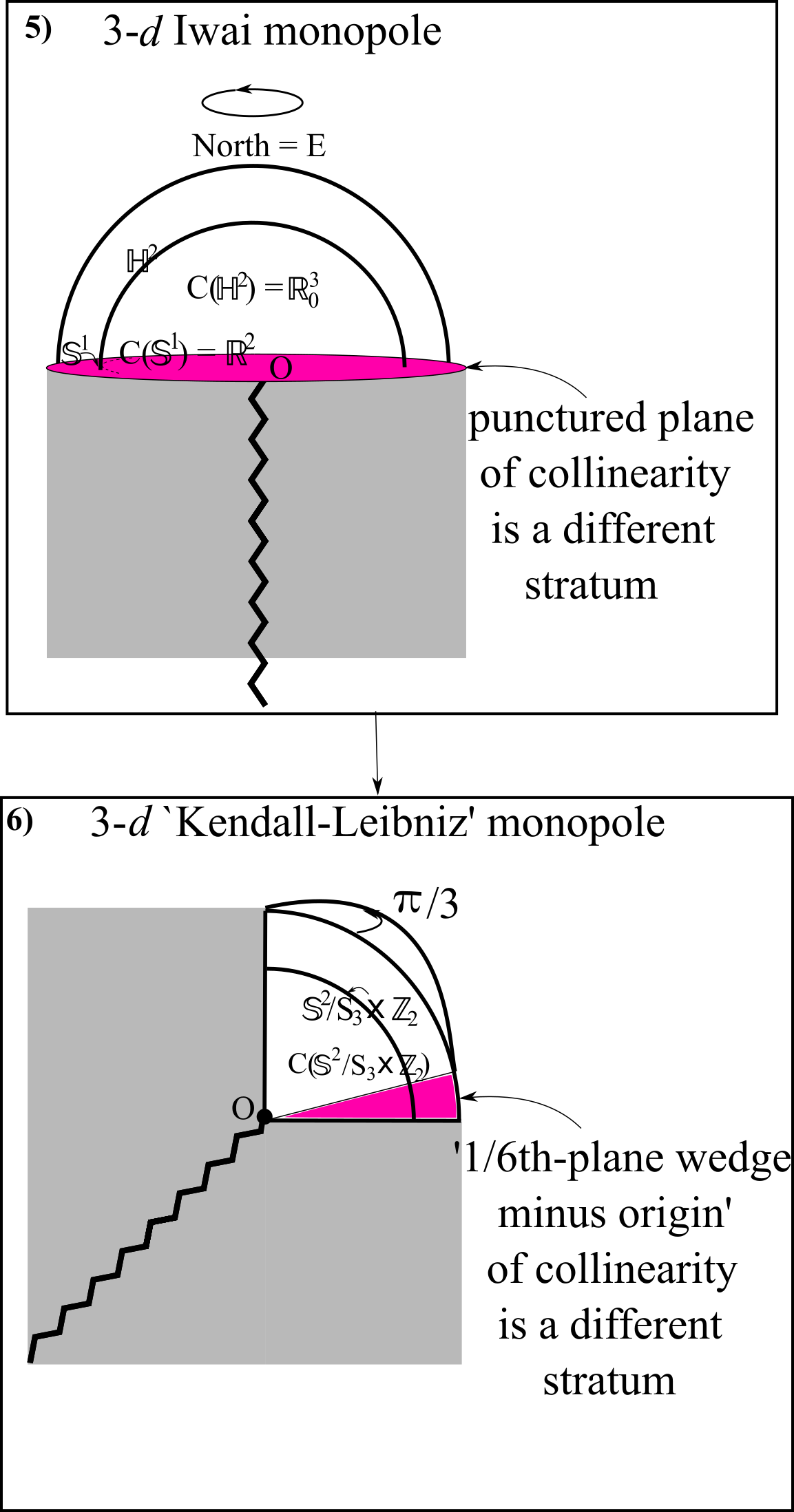}
\caption[Text der im Bilderverzeichnis auftaucht]{        \footnotesize{Relational triangleland monopoles 
in their most physically natural 3-$d$ space setting.

\m

\n 5) 3-$d$ Iwai monopole.

\m

\n 6) 3-$d$ Leibniz--Kendall monopole. 

\m

\n \c{A-Monopoles} studies 2 further cases of 2-$d$ 3-body problems.} }
\l{Triangleland-Monopoles-3-d} \end{figure}          }

\m

\n{\bf Difference 8} Stratified manifolds can furthermore be usefully paired \c{Pflaum, Kreck, ABook, Project-2} with sheaves \c{Sheaves, Sheaves1}.  
Better gluing and localization methods are then available to model the (3, 3) case's now inhomogeneous counterpart of shape representative space fibres.  
That $\FrS(3,3)$ and $\Leib_{\sFrS}(3, 3)$ are Hausdorff, second-countable and compact, 
fortunately enables whichever of the simpler and more worked out cases outlined in \c{Project-2} to apply.
This includes via 
\be
\mbox{ (Hausdorff and second countable) } \m \Rightarrow \m \mbox{ (paracompact) } 
\ee
and the very basic result that  
\be
\mbox{ (compact) } \m \Rightarrow \m \mbox{ (locally compact) } \m,    
\ee
by the definition of this last concept \c{Lee1, Munkres}.

\subsection{Advent of strata, general bundles and sheaves}

These having been mentioned in the past four subsections, and also subsequently becoming the generic core of the subject of Shape Theory, 
let us end Part III with a more unified and referenced account of them.

\m

\n{\bf Remark 1} As regards stratification, 1-$d$ has no capacity for isotropy groups of different dimension.  
On the other hand, metric shape spaces for 2-$d$ shapes avoid stratification issues.  
This is due to only involving $SO(2) = U(1)$, which acts in the same manner same on $\mC$ and non-$\mC$ configurations.  
However, in 3-$d$ the $\mC$ have only an $SO(2)$ subgroup of the $SO(3)$ acting upon them, so stratification ensues.

\m

\n{\bf Remark 2} Plain $N$-stop metroland and $N$-a-gonland avoid stratification issues due to $SO(2) = U(1)$'s particular straightforwardness, 
whereas 3-$d$ RPMs are not so fortunate.  

\m

\n See Appendices F, M and W of \c{ABook} for supporting outlines of fibre and general bundles, stratified manifolds, and sheaves respectively. 
The upcoming work \c{A-Monopoles, Project-2} provides further details.  

\m

\n{\bf Acknowledgments} I thank Chris Isham for previous discussions, Jimmy York Jr for great motivation when I was young, and Don Page for discussions.  
Reza Tavakol, Malcolm MacCallum, Enrique Alvarez and Jeremy Butterfield for support with my career.  
I also thank Angela Lahee and the typesetting editors for my book manuscript this past summer for prompt and excellent work without which the current treatise could not 
have been completed by this date. 
I finally dedicate this paper to several friends who are going through considerably difficult times.

\vspace{10in}

\begin{appendices}

\section{Figure-webs of levels of structure for triangleland}

This Appendix is a useful end-summary in support of levels of structure discussed in Secs 4 to 8.  
It is presented all in one piece so as to exhibit the inter-relations between these structures, 
which are useful in building up these various structures and in understanding and remembering their content.  

\m

\n{\bf Structure 1} Figures \r{(3,2)-Structure-Web}, 
                             \r{(3,2)-I-Structure-Web} and 
							 \r{(3,2)-Full-Structure-Web} 
				     exhibit 24 levels of structure for triangleland, in which setting they are realized as 18 different structures. 
These figures are, respectively, for Leibniz space $\Leib_{\sFrS}(3, 2)$, 
indistinguishable point-or-particle space $\FrI\FrS(3, 2)$, 
and a bird's eye view which coincides for plain and mirror-image-identified shape spaces $\w{\FrS}(3, 2)$ and $\FrS(3, 2)$.  

{            \begin{figure}[!ht]
\centering
\includegraphics[width=0.8\textwidth]{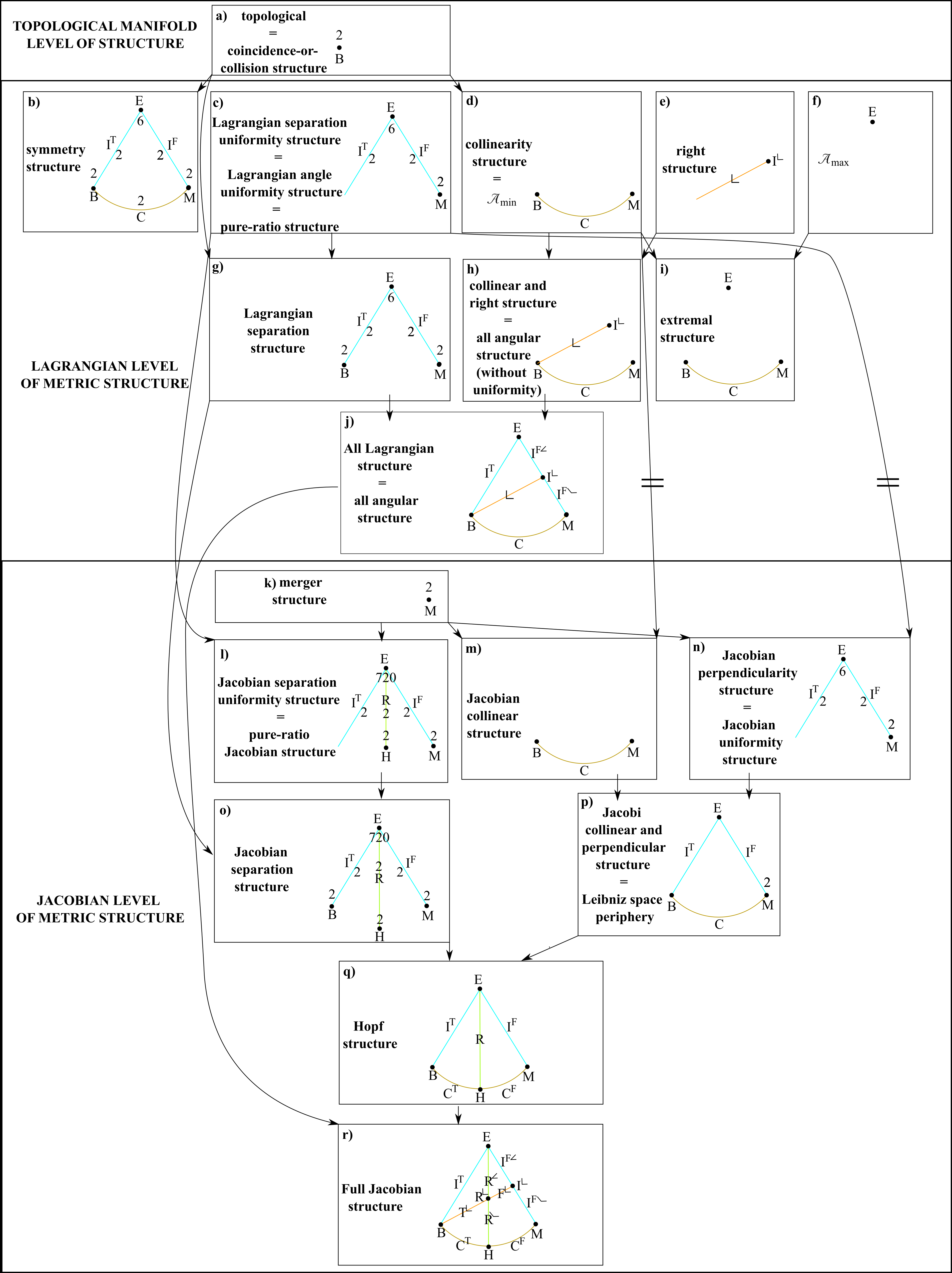}
\caption[Text der im Bilderverzeichnis auftaucht]{        \footnotesize{24 levels of structure, which $\Leib_{\sFrS}(3, 2)$ realizes as 18 
distinct networks of points and arcs. 
These are viewed equatorially in the figure. 
The numbers indicated are, upon taking their logarithms, some of Part II's Appendices' quantifiers.
This figure expands on eqs (I.255-256) for the (3, 1) model.} }
\l{(3,2)-Structure-Web} \end{figure}          }
%
{            \begin{figure}[!ht]
\centering
\includegraphics[width=0.73\textwidth]{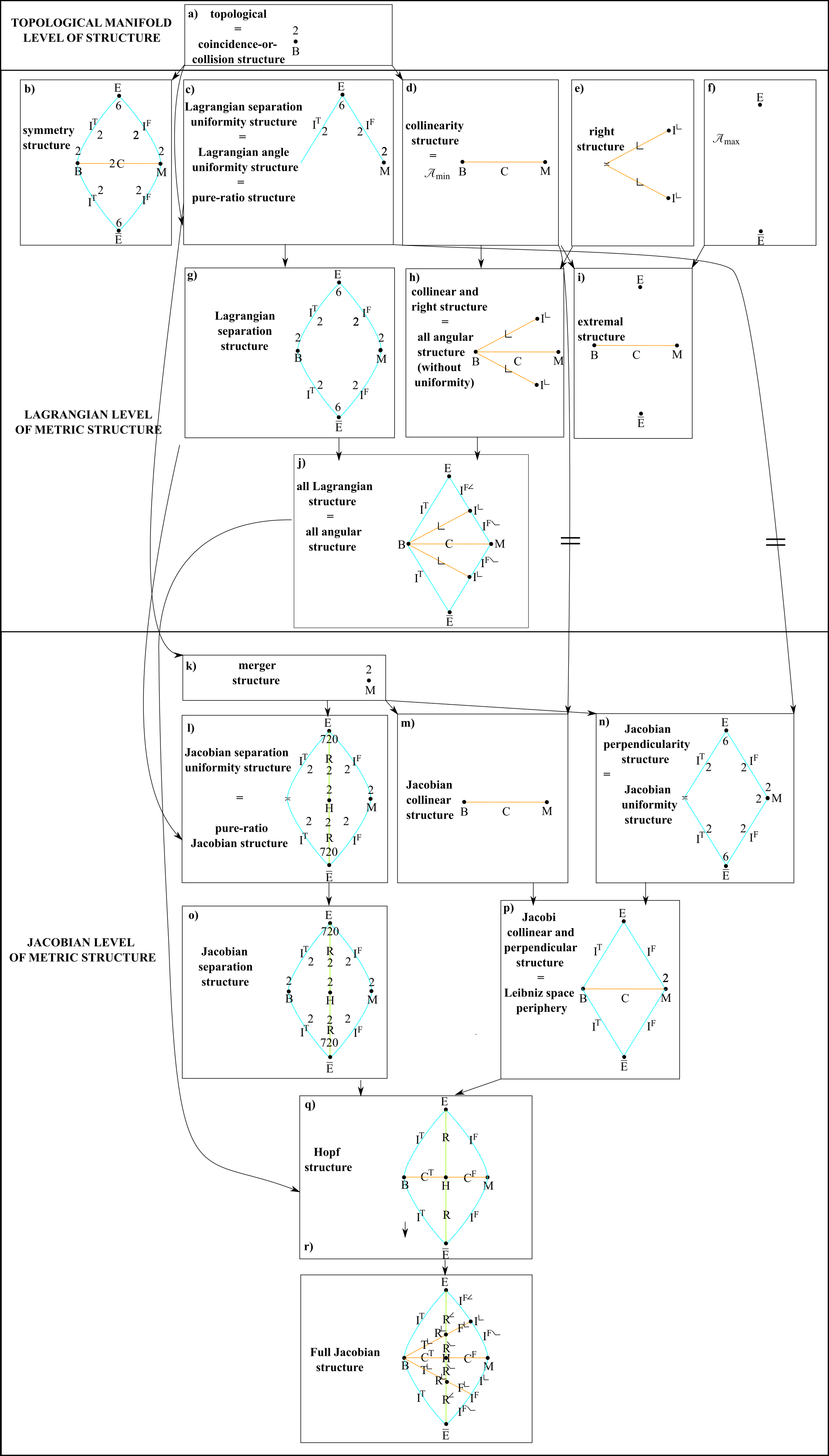}
\caption[Text der im Bilderverzeichnis auftaucht]{        \footnotesize{24 levels of structure, which $\FrI\sFrS(3, 2)$'s realizes as 18 
distinct networks of points and arcs, again presented from an equatorial point of view .} }
\l{(3,2)-I-Structure-Web} \end{figure}          }
%
{            \begin{figure}[!ht]
\centering
\includegraphics[width=0.8\textwidth]{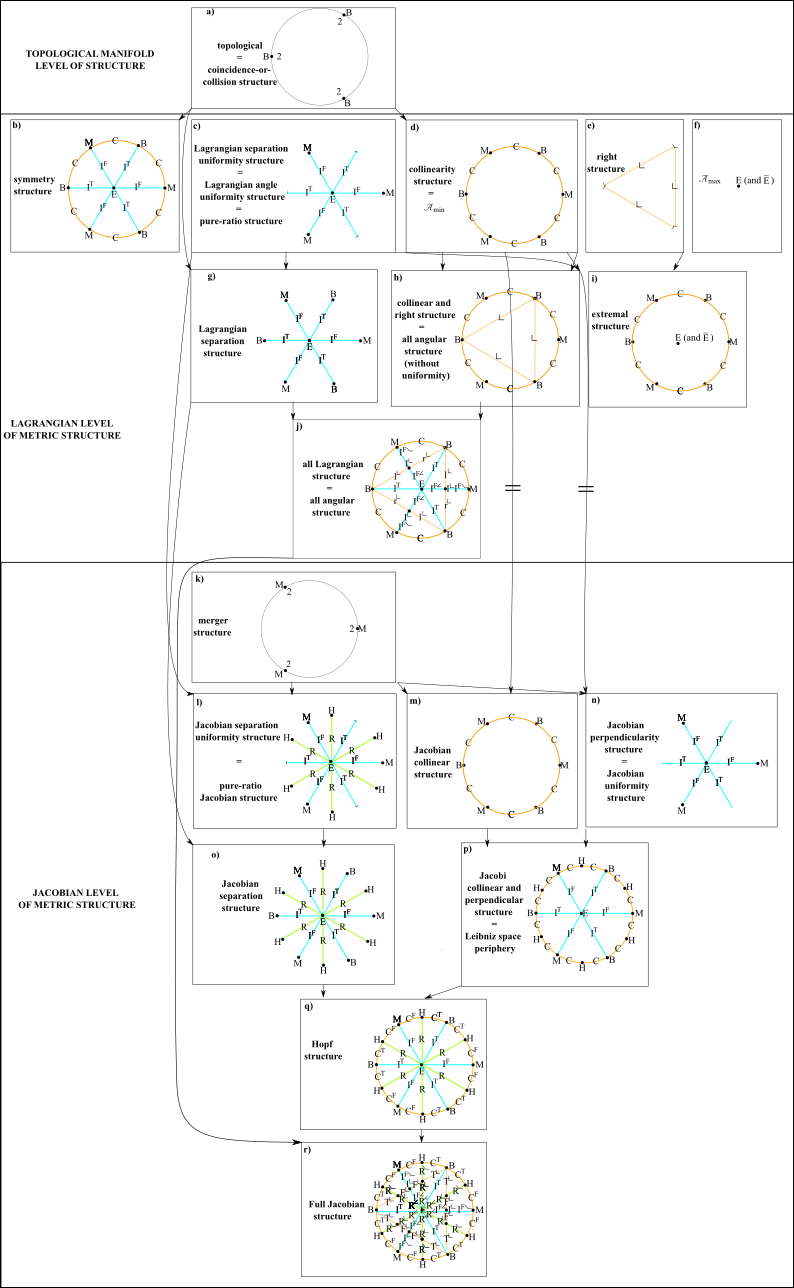}
\caption[Text der im Bilderverzeichnis auftaucht]{        \footnotesize{Bird's eye view 24 levels of structure, which are realized 
as 18 distinct networks of points and arcs in $\w{\FrS}(3, 2)$ and 18 further such in $\sFrS(3, 2)$.}  }
\l{(3,2)-Full-Structure-Web} \end{figure}           }

\vspace{10in}

\vspace{10in}

\vspace{10in}

\m

\m

\m

\section{Topological content of Part III's structures}

\n{\bf Lemma 1}
The sphere minus $K$ points is topologically equivalent to the hemisphere minus $k = K - 1$ points, the disc minus $k$ points and the $k$-bouquet.
For $K$ even ($ = 2 \, p$) these are topologically equivalent to the genus-$p$ surface, i.e.\ the $p$-torus (where the 0-torus is the sphere), in each case with 1 point deleted.
For $K$ odd ($ = 2 \, p + 1$), these are topologically equivalent to the genus-$p$ surface with a single puncture.  

\m

\Proof See Fig \r{Top-App-Fig}. $\Box$
%
{\begin{figure}[ht]
\centering
\includegraphics[width=1\textwidth]{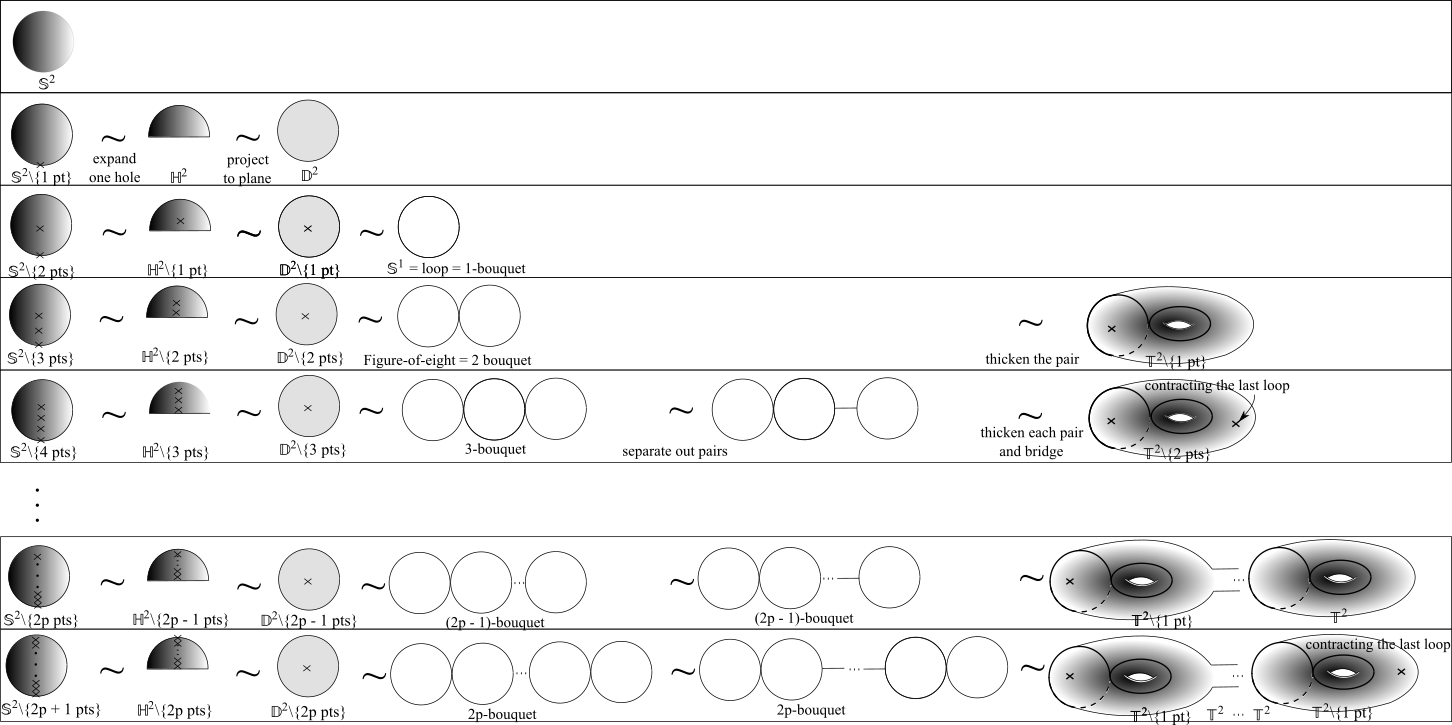}
\caption[Text der im Bilderverzeichnis auftaucht]{\footnotesize{From circle to $k$-bouquet, with relation to punctured discs and (arbitrary hole number) tori.
All of the previous section's structures are such objects or their disjoint union with further discs or points, as the next two figures tabulate.}} 
\l{Top-App-Fig}\end{figure} } 

\m

\n{\bf Corollary 1} The topological content of the previous Appendix's structures 
 is as tabulated in Fig \r{(3, 2)-Structures-Top}     for the Lagrangian-level structures, 
and as in           Fig \r{(3, 2)-Jac-Structures-Top} for the Lagrangian-level structures.  
%
{            \begin{figure}[!ht]
\centering
\includegraphics[width=0.85\textwidth]{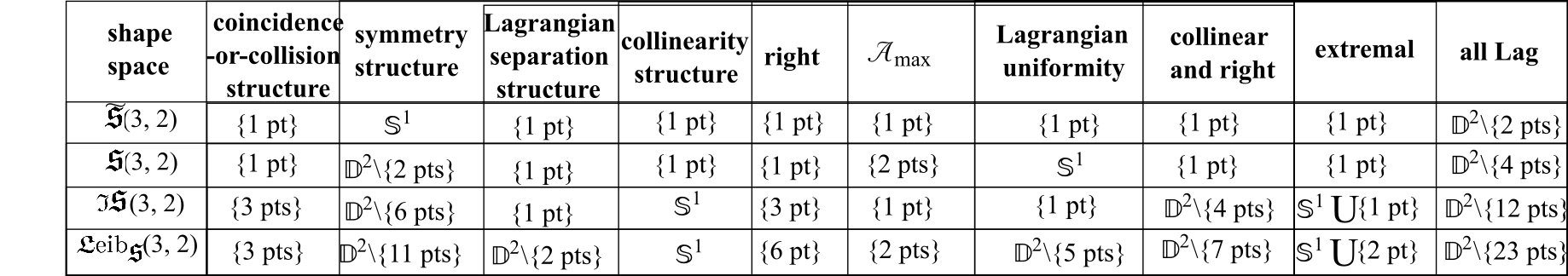}
\caption[Text der im Bilderverzeichnis auftaucht]{        \footnotesize{Table of topologies of various significant structures at the Lagrangian level}  } 
\l{(3, 2)-Structures-Top} \end{figure}           }
%
{            \begin{figure}[!ht]
\centering
\includegraphics[width=0.85\textwidth]{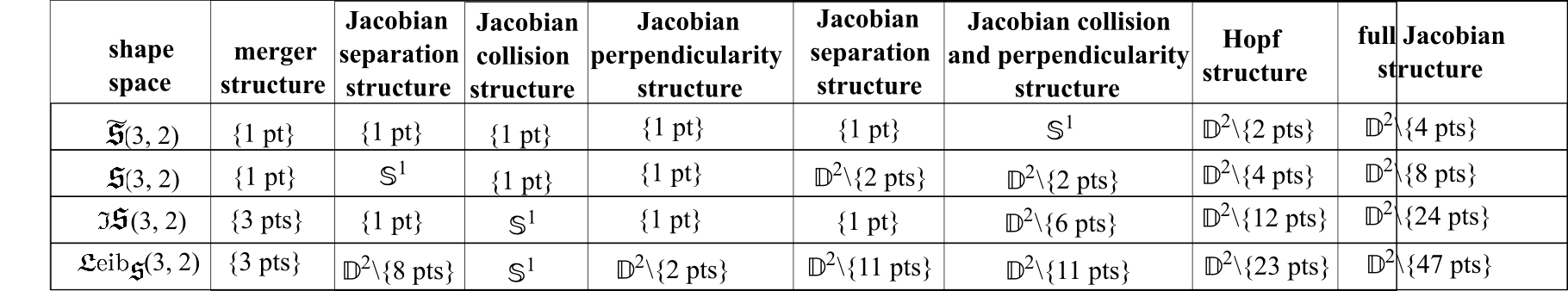}
\caption[Text der im Bilderverzeichnis auftaucht]{        \footnotesize{Table of topologies of various significant structures at the Jacobian level.}  } 
\l{(3, 2)-Jac-Structures-Top} \end{figure}           }

\vspace{10in}

\vspace{10in}

\vspace{10in}

\vspace{10in}

\end{appendices}


\end{document}